\documentclass[11pt]{article}
 \usepackage{titlesec}
\usepackage{hyperref}

\titleclass{\subsubsubsection}{straight}[\subsection]

\usepackage[utf8]{inputenc}
\usepackage{amsmath,amsfonts,amssymb,stackengine,graphicx}

\usepackage[utf8]{inputenc}
\newif\iffigs\figstrue

\usepackage[title]{appendix}
\usepackage{epsfig,latexsym}
\usepackage{hyperref}
\usepackage{amsmath}
\usepackage{verbatim}
\usepackage{color}
\usepackage{mathrsfs}
\usepackage{slashed}
\usepackage{amssymb}

\DeclareMathAlphabet{\mathpzc}{OT1}{pzc}{m}{it}

 \csname
@addtoreset\endcsname{equation}{section}

\def\gz0{\gamma^{0}}

\def\e{\epsilon}

\def\n{\nu}

\def\beq{\begin{equation}}
\def\eeq{\end{equation}}
\def\bea{\begin{eqnarray}}
\def\eea{\end{eqnarray}}
\def\ba{\begin{array}}
\def\ea{\end{array}}
\def\bec{\begin{center}}
\def\ec{\end{center}}
\def\ba{\begin{align}}
\def\ena{\end{align}}

\def\12{\frac{1}{2}}

\newcounter{subsubsubsection}[subsubsection]
\renewcommand\thesubsubsubsection{\thesubsubsection.\arabic{subsubsubsection}}

\titleformat{\subsubsubsection}
  {\normalfont\normalsize\bfseries}{\thesubsubsubsection}{1em}{}
\titlespacing*{\subsubsubsection}
{0pt}{3.25ex plus 1ex minus .2ex}{1.5ex plus .2ex}

\makeatletter
\renewcommand\paragraph{\@startsection{paragraph}{5}{\z@}%
  {3.25ex \@plus1ex \@minus.2ex}%
  {-1em}%
  {\normalfont\normalsize\bfseries}}
\renewcommand\subparagraph{\@startsection{subparagraph}{6}{\parindent}%
  {3.25ex \@plus1ex \@minus .2ex}%
  {-1em}%
  {\normalfont\normalsize\bfseries}}
\def\toclevel@subsubsubsection{4}
\def\toclevel@paragraph{5}
\def\toclevel@paragraph{6}
\def\l@subsubsubsection{\@dottedtocline{4}{7em}{4em}}
\def\l@paragraph{\@dottedtocline{5}{10em}{5em}}
\def\l@subparagraph{\@dottedtocline{6}{14em}{6em}}
\makeatother

\begin{document}

\begin{flushright}
{\today}
\end{flushright}

\vspace{10pt}

\begin{center}


{\Large\sc A 4D IIB Flux Vacuum and Supersymmetry Breaking}\vskip 12pt
{\large \sc II.~Bosonic Spectrum and Stability}


\vspace{25pt}
{\sc J.~Mourad${}^{\; a}$  \ and \ A.~Sagnotti${}^{\; b}$\\[15pt]

${}^a$\sl\small APC, UMR 7164-CNRS, Universit\'e   Paris Cit\'e  \\
10 rue Alice Domon et L\'eonie Duquet \\75205 Paris Cedex 13 \ FRANCE
\\ e-mail: {\small \it
mourad@apc.univ-paris7.fr}\vspace{10
pt}

{${}^b$\sl\small
Scuola Normale Superiore and INFN\\
Piazza dei Cavalieri, 7\\ 56126 Pisa \ ITALY \\
e-mail: {\small \it sagnotti@sns.it}}\vspace{10pt}
}

\vspace{40pt} {\sc\large Abstract}\end{center}
\noindent
We recently constructed type--IIB compactifications to four dimensions depending on a single additional coordinate, where a five--form flux $\Phi$ on an internal torus leads to a constant string coupling. Supersymmetry is fully broken when the internal manifold includes a finite interval of length $\ell$, which is spanned by a conformal coordinate in a finite range $0 < z < z_m$. Here we examine the low--lying bosonic spectra and their classical stability, paying special attention to self--adjoint boundary conditions. Special boundary conditions result in the emergence of zero modes, which are determined exactly by first--order equations. The different sectors of the spectrum can be related to Schr\"odinger operators on a finite interval, characterized by pairs of real constants $\mu$ and $\tilde{\mu}$, with $\mu$ equal to ${1}/{3}$ or ${2}/{3}$ in all cases and different values of $\tilde{\mu}$. The potentials behave as $\frac{\mu^2-1/4}{z^2}$ and $\frac{\tilde{\mu}^2-1/4}{\left(z_m-z\right)^2}$ near the ends and can be closely approximated by exactly solvable trigonometric ones. With vanishing internal momenta, one can thus identify a wide range of boundary conditions granting perturbative stability, despite the intricacies that emerge in some sectors. For the Kaluza--Klein excitations of non--singlet vectors and scalars the Schr\"odinger systems couple pairs of fields, and the stability regions, which depend on the background, widen as the ratio ${\Phi}/{\ell^4}$ decreases.

\vskip 12pt

\setcounter{page}{1}

\pagebreak

\newpage
\tableofcontents
\newpage
\baselineskip=20pt
\section{\sc  Introduction and Summary}\label{sec:intro}

Different scenarios for supersymmetry breaking in String Theory~\cite{strings} have been explored over the years. The resulting pictures are captivating, but they all entail, in one way or another, strong back reactions on the vacuum. To wit, quantum corrections to Scherk--Schwarz compactifications~\cite{ss_closed,ss_open} lead to runaway potentials (and possibly to tachyonic modes), which also emerge, already at the (projective) disk level, in the three non--tachyonic ten--dimensional orientifolds~\cite{orientifolds} of~\cite{so1616,susy95,bsb}. While supersymmetry is absent in the first two, Sugimoto's model in~\cite{bsb} rests on a non--linear realization of supersymmetry~\cite{bsb_nonlinear}, and in fact it is the simplest setting for brane supersymmetry breaking~\cite{bsb}. Fluxed $AdS$ vacua for the three non--tachyonic ten-dimensional strings of~\cite{so1616,susy95,bsb}, where curvatures and string couplings are everywhere weak, do exist~\cite{gubsermitra,ms_16,raucci_22,sethi_22}, but they generally host unstable modes~\cite{bms,selfadjoint}. In contrast with the standard Kaluza--Klein constructions that play a ubiquitous role for supersymmetric strings~\cite{strings}, the Dudas--Mourad vacua~\cite{dm_vacuum} rest on an internal interval, and are perturbatively stable~\cite{bms,selfadjoint}. However, they include regions where the expected corrections sized by the string coupling, and/or by the space--time curvature, become unbounded.

This paper concerns a class of type IIB~\cite{IIB} compactifications to four--dimensional Minkowski space with internal fluxes~\cite{ms21_1} that avoid the emergence of regions where the string coupling becomes large~\footnote{As in the original vacua of~\cite{dm_vacuum}, the strong curvatures present in these types of vacua can be also confined to small portions of the internal space with suitable choices of their free parameters~\cite{ms21_1,ms21_2}.}. It is a sequel of~\cite{ms22_1}, where a number of properties of these vacua were elucidated, including the presence of an effective BPS orientifold at one end of the internal interval. Furthermore, supersymmetry is fully broken when this orientifold lies at a finite distance from another singularity, while it is partly recovered in the limit where the two ends are separated by an infinite distance. The analysis also showed that, for a natural choice of boundary conditions, the massless Fermi modes present in these vacua are those of $N=4$ supergravity~\cite{sugra} coupled to five vector multiplets. Our task here is to investigate the nature of the corresponding bosonic modes, taking a close look at the available choices of self--adjoint boundary conditions~\cite{selfadjoint}, while also analyzing their implications for vacuum stability.

These backgrounds are characterized by a constant dilaton profile $\phi_0$, which we set to zero for simplicity, and by metric and five-form profiles that depend on a single coordinate $r$, and read
    \bea{}{}{}{}{}{}{}{}{}{}{}{}{}
ds^2 &=& \frac{\eta_{\mu\nu}\,dx^\mu\,dx^\nu}{\left[2\left|H\right|\rho\,\sinh\left(\frac{r}{\rho}\right)\right]^\frac{1}{2}} \,+\, \left[2 \left|H\right|\,\rho\,\sinh\left(\frac{r}{\rho}\right)\right]^\frac{1}{2} \left[e^{ \,- \, \frac{\sqrt{10}}{2\rho}\, r} \, dr^2 \ + \ e^{\,- \, \frac{\sqrt{10}}{10\rho}\, r} \, \left(d\,{y}^i\right)^2\right] \ , \nonumber \\
{{\cal H}_5^{(0)}} &=&
H\left\{ \frac{dx^0 \wedge ...\wedge dx^3\wedge dr}{\left[2\left|H\right|\,\rho\,\sinh\left(\frac{r}{\rho}\right)\right]^2} \ + \ dy^1 \wedge ... \wedge dy^5\right\} \ . \label{back_epos_fin2}
\eea
The $x^\mu$ are coordinates of a four--dimensional Minkowski space, and positive values of the coordinate $r$ parametri\-ze the interior of an internal interval. The five $y^i$ coordinates have a finite range,
\beq{}{}{}{}{}{}{}{}{}{}{}{}{}{}{}{}{}{}{}{}{}{}{}{}{}{}{}{}{}
0 \ \leq \ y^i \ \leq \ 2\,\pi\,R \ ,
\eeq
and parametrize an internal torus, which for simplicity we take to be the direct product of five circles of radius $R$.
These vacua thus depend on the two constants $H$ and $\rho$, and supersymmetry is fully broken for finite values of $\rho$. We shall often use the conformal coordinate
\beq
z(r) \ = \ z_0 \int_0^r d\xi \ \sinh\,\xi^\frac{1}{2} \ e^{-\,\frac{\sqrt{10}\, \xi}{4}} \ , \label{zr_i}
\eeq
where
\beq{}{}{}{}{}{}{}{}{}{}{}{}{}{}{}{}{}{}{}{}
z_0  \ = \ \left(2 H \rho^3\right)^\frac{1}{2} \ . \label{z0_i}
\eeq
The upper bound for $z$, which will play a role in the ensuing discussion, is
\beq
z_m \ \simeq \ 2.24 \ z_0 \ .
\eeq

Half of the supersymmetries originally present in ten dimensions are recovered in the $\rho \to \infty$ limit~\cite{ms22_1}, albeit within a curved spacetime that still includes the singularity at $r=0$. After the coordinate transformation
\beq{}{}{}{}{}{}{}{}{}{}{}{}{}{}{}{}{}{}{}{}{}{}{}{}{}{}{}{}{}
\xi \,H \ = \ \frac{2}{5}\,\left(2\,H\,r\right)^\frac{5}{4} \ ,
\eeq
the limiting behavior of the solution reads
\bea{}{}{}{}{}{}{}{}{}{}{}{}{}
ds^2 &=& \frac{\eta_{\mu\nu}\,dx^\mu\,dx^\nu}{\left(\frac{5}{2}\,H\,\xi\right)^\frac{2}{5}} \,+\, d\,\xi^2 \,+\, \left(\frac{5}{2}\,H\,\xi\right)^\frac{2}{5} \left(d\,y^i\right)^2   \ , \nonumber \\
{{\cal H}_5^{(0)}} &=&
H\left\{ \frac{dx^0 \wedge ...\wedge dx^3\wedge d\xi}{\left(\frac{5}{2}\,H\,\xi\right)^\frac{9}{5}} \ + \ dy^1 \wedge ... \wedge dy^5\right\} \ , \label{back_epos_susy}
\eea
and describes indeed a non--homogeneous curved background including an internal torus with an effective $\xi$--dependent size. As $\xi\to 0$, the volume of the internal torus shrinks to zero and the scale factor of the spacetime coordinates blows up, while conversely as $\xi\to \infty$ the volume of the internal torus blows up while the scale factor of the spacetime coordinates shrinks to zero. Both limits are thus delicate within supergravity~\cite{sugra} although, as we saw in detail in~\cite{ms22_1},  this limiting form of the vacuum does preserve half of the supersymmetries of ten--dimensional Minkowski space. One can also absorb the constant $H$, while rescaling the radius $R$ of the internal torus by a factor $H^\frac{1}{5}$.

Alternatively, the backgrounds of eqs.~\eqref{back_epos_fin2} can be presented in the form
\bea
ds^2 &=& \frac{\eta_{\mu\nu}\,dx^\mu\,dx^\nu}{\left[h\,\sinh\left(\tilde{r}\right)\right]^\frac{1}{2}} \,+\, \left[\sinh\left(\tilde{r}\right)\right]^\frac{1}{2}\left[\ell^2\, e^{ \,-\,\frac{\sqrt{10}}{2}\, \tilde{r}} \, d\tilde{r}^2 \,+\, \left({2}\,\Phi\,\ell\right)^\frac{2}{5} \  e^{\,-\,\frac{\sqrt{10}}{10}\, \tilde{r}}   \,\left(d\,{\tilde{y}}^{i}\right)^2 \right]\ , \nonumber \\
{{\cal H}_5^{(0)}} &=&
\frac{1}{{2\,h}}\, \frac{dx^0 \wedge ...\wedge dx^3\wedge d\tilde{r}}{\left[\sinh \left(\tilde{r}\right)\right]^2} \ + \ \Phi\, d\tilde{y}^1 \wedge ... \wedge d\tilde{y}^5  \ , \label{4d_PhiEpos2}
\eea
where $\tilde{r}>0$ is a dimensionless variable, $\ell$ is the length scale of the interval, $h$ is a dimensionless parameter and $\Phi$ is the five-form flux in the internal torus. Here
\beq{}{}{}{}{}{}
\tilde{r} \,=\, \frac{r}{\rho} \ , \qquad
\tilde{y}^i \,=\, \frac{y^i}{2\pi R}   \ , \label{xry}
\eeq
so that the $\tilde{y}^i$ are dimensionless coordinates of unit range, and
the new parameters are related to those in eqs.~\eqref{back_epos_fin2} according to
\beq{}{}{}{}{}{}
h \ = \ 2\,H\,\rho \ ,  \qquad  \ell \,=\, \left(2\,H\right)^\frac{1}{4} \ \rho^\frac{5}{4} \,=\,h^\frac{1}{4}\,\rho  \ , \qquad \Phi \,=\, H \left(2 \,\pi\,R\right)^5 \ , \qquad z_0  \ = \ \ell\, h^\frac{1}{4} \ . \label{hH5}
\eeq
The volume of the six--dimensional internal space, which comprises the torus and the $r$-interval, scales as
\beq
V_6 \ \sim \ \Phi\,\ell^2 \ \sim \ H^\frac{3}{2}\, \rho^\frac{5}{2} \,  R^5 \ ,
\eeq
and similarly the volume of the internal torus, which can be defined as $\frac{V_6}{\ell}$, scales as
\beq{}{}{}{}{}{}{}{}{}{}{}{}{}{}{}{}{}{}{}{}{}{}{}{}{}{}{}{}{}{}{}{}{}{}{}{}{}{}{}{}{}{}{}{}{}{}{}{}{}{}{}{}{}{}{}{}{}{}{}{}{}{}{}{}{}{}{}{}{}{}
V_5 \ \sim \ \Phi\,\ell \ \sim \ H^\frac{5}{4}\, \rho^\frac{5}{4} \,  R^5 \ .
\eeq
Both quantities diverge in the supersymmetric limit $\rho \to \infty$, which thus occurs in a ten--dimensional curved space with one singularity at the origin, as we had anticipated.

The fermionic zero modes present in these backgrounds were determined in~\cite{ms22_1}: for finite values of $\rho$ or $\ell$ they are four Majorana gravitini and 20 Majorana spinors, the massless fermions of $N=4$ supergravity coupled to \emph{five} $N=4$ vector multiplets, despite the breaking of supersymmetry~\footnote{More precisely, as explained in~\cite{ms22_1}, this is the case if Fermi fields are subject to identical ``$\Lambda$ projections'' at the two ends of the $r$ interval. Here we focus on this interesting option, but opposite ``$\Lambda$ projections'' would eliminate the massless Fermi modes.}. In the following sections we shall analyze in detail the corresponding bosonic zero modes and their excitations, with the aim of addressing the perturbative stability of these vacua at the classical level. As in~\cite{selfadjoint}, other parameters beyond those appearing in the background are generally encoded in the boundary conditions. Moreover, as we shall see, additional modes can be confined to boundaries. Only a subset of the possible self--adjoint boundary conditions grant stability, but in all cases we find an ample range of these choices. 

The four--dimensional bosonic modes that we shall exhibit originate from the ten--dimensional fields of type--IIB supergravity~\cite{IIB,sugra}, and have $r$--profiles and discrete momenta $\mathbf{k}$ in the internal torus. Linearized perturbations of the backgrounds can be characterized by their four--dimensional spin and their internal quantum numbers, and different representations of the residual symmetry groups can be studied independently. For example, starting from the ten--dimensional metric and the four--form gauge field, which can be decomposed as
\beq
g_{MN} = g^{(0)}_{MN} + h_{MN} \ , \qquad B_{M_1 \ldots M_4}=B^{(0)}_{M_1\dots M_2}\ +\ \delta B_{M_1\ldots M_4}\ , \label{field_pert}
\eeq
after a suitable gauge fixing the perturbations $h_{MN}$ and $\delta B_{M_1\ldots M_4}$ give rise to tensor, vector and scalar modes of different types.
The metric and the selfdual four--form are the only two fields with non--trivial vacuum profiles, and for this reason they generally mix, are more difficult to analyze and can affect other modes. In addition, the selfdual five-form is by itself somewhat unfamiliar. All modes afford Fourier decompositions in the internal space, so that for example the perturbations of the ten--dimensional metric can be decomposed according to
\bea
h_{MN}(x,r,y) &=& h_{MN}(x,r) \ + \ \sum_{\mathbf{k} \neq 0} h_{MN}^{(\mathbf{k})}(x,r)\,e^{i \mathbf{k} \cdot \mathbf{y}} \ . \label{pert_modes}
\eea
The modes with vanishing internal momentum $\mathbf{k}$, and in particular $h_{MN}(x,r)$ in this example, fill effectively multiplets of a continuous $SO(5)$ symmetry. The $SO(5)$ tangent--space symmetry endows them with the internal quantum numbers that we alluded to above, although only a discrete subgroup of $SO(5)$ is actually an isometry of the torus. On the other hand, the remaining modes carry lattice momenta $\mathbf{k}$, and fill effectively $SO(4)$ multiplets corresponding to their stability groups. In practice, these will emerge as $SO(5)$ multiplets subject to orthogonality constraints. These properties will play an important role in our analysis, and the internal quantum numbers of the modes that we shall identify will correspond indeed to $SO(5)$ representations for $\mathbf{k} = 0$, and to $SO(4)$ representations for $\mathbf{k} \neq 0$.
For brevity, however, we shall usually drop the superscript $(\mathbf{k})$ present in eq.~\eqref{pert_modes} when discussing modes with non--vanishing internal momenta. It would be interesting to address generalizations of this type of setup where the internal torus is replaced by a more general Ricci--flat internal manifold.

Since the resulting spacetime is flat for the backgrounds of interest, the vacua of eq.~\eqref{back_epos_fin2} are perturbatively unstable if $m^2 < 0$ for some four--dimensional modes, and we shall perform a detailed scrutiny to this end. As in~\cite{bms,ms22_1,selfadjoint}, the four--dimensional masses can be related to the eigenvalues of one--dimensional Schr\"odinger--like operators acting on one or two wavefunctions, whose potentials are determined by the background. 

The field equations lead, in general, to operators that are not manifestly Hermitian, so that the replacement of the independent variable $r$ with the ``conformal'' variable of eq.~\eqref{zr_i}, whose range $0 \leq z \leq z_m$ is finite and proportional to $z_0$, together with redefinitions of the different fields, will be instrumental to cast them into standard forms. Remarkably, as in the nine--dimensional vacua discussed in~\cite{selfadjoint}, in all cases the resulting potentials develop double poles at the two ends, where they behave as
\beq{}{}{}{}{}
V \ \sim \ \frac{\mu^2\,-\,\frac{1}{4}}{z^2} \ ,
\qquad\quad
V \ \sim \ \frac{\tilde{\mu}^2\,-\,\frac{1}{4}}{\left(z-z_m\right)^2} \ . \label{lim_dil_axion_intro}
\eeq
The constants $\mu$ and $\tilde{\mu}$ depend on the mode sector, while the scale dependence is only encoded in $z_m$. Moreover, $\tilde{\mu}$ is zero for $h_{\mu\nu}$, $h_{ij}$, dilaton and axion perturbations, which share the same Schr\"odinger operator, while it is a real number between zero and about 2.3 in all other sectors. On the other hand, the parameter $\mu$ associated to the $z=0$ end is curiously either $\frac{1}{3}$ of $\frac{2}{3}$ in all cases. The squared masses are eigenvalues of Hermitian operators and, as we saw in~\cite{ms22_1} for Fermi fields, these steps also determine the normalization conditions, a necessary ingredient to identify the actual physical modes. In most cases, these normalizations can be simply recovered from the four--dimensional kinetic terms determined by ten--dimensional action, but the self--dual tensor field does introduce some complications. However, despite its reduced manifest symmetry, the non--standard Henneaux--Teitelboim action of~\cite{henneaux}, when properly adapted, suffices to grant covariant descriptions in the backgrounds of eq.~\eqref{back_epos_fin2}. 

The completeness of the modes thus identified is essential to make statements on perturbative stability. It is granted if the Schr\"odinger--like operators are not only Hermitian but also self--adjoint, and this property demands judicious choices of boundary conditions. These are determined by the asymptotics of the wavefunctions at the ends of the interval~\cite{selfadjoint}, which reflects, in its turn, the singular behavior~\eqref{lim_dil_axion_intro} of the potentials. Additional sets of parameters thus emerge, which impinge on the positivity of the Hermitian Schr\"odinger--like operators, and the stability of the resulting mass spectra generally places some constraints on them~\cite{selfadjoint}. 

The massless modes are exactly calculable in most cases, with suitable boundary conditions, while in a few instances the allowed squared masses emerge as eigenvalues of operators that are manifestly positive, again with suitable boundary conditions. Two sectors with $\mathbf{k} \neq 0$, the non--singlet vector modes of Section~\ref{sec:intvectexc} and the non--singlet scalar modes of Section~\ref{sec:nonsingletscalrknot0}, did not allow exact statements, and approximation methods were necessary to address their stability. We thus relied on the variational principle of non--relativistic Quantum Mechanics, which can be adapted to the present setting and allows reliable numerical estimates of the lowest eigenvalues and of their dependence on $\mathbf{k}$  and on the boundary conditions. In this fashion, we could identify background--dependent constraints on the boundary conditions that grant stability in both cases.

In general, different boundary conditions lead to different spectra, and the residual global symmetries required for the backgrounds of interest are instrumental to discriminate among them.
These symmetries include infinitesimal translations and Lorentz rotations in spacetime, together with infinitesimal translations along the internal torus. The corresponding currents are already present in the background, and do not flow across the ends of the interval provided~\cite{ms_20,ms22_1}
\beq
\sqrt{-g}\,\left. T^{r}{}_{\mu}\right|_{\partial\,{\cal M}} \ = \ 0 \ , \qquad \sqrt{-g}\,\left. S^{r}{}_{\mu\nu}\right|_{\partial\,{\cal M}} \ = \ 0 \ , \qquad \sqrt{-g}\,\left. T^{r}{}_{i}\right|_{\partial\,{\cal M}} \ = \ 0\label{bc_bose_interval_mui} \ ,
\eeq
where $T^{r}{}_{\mu}$ and $T^{r}{}_{i}$ are components of the energy--momentum tensor and $S^{r}{}_{\mu\nu}$ are components of the spin portion of the angular momentum current. Enforcing eq.~\eqref{bc_bose_interval_mui} selects special boundary conditions for the fields. The solutions of the linearized equations that we shall examine in detail, with special emphasis on the low--lying modes, determine the leading contributions to these no--flow conditions. One can thus use them independently for the different sectors of the spectrum to identify the boundary conditions compatible with them. While in a quantum theory of gravity global symmetries are expected not to be conserved~\cite{swampland}, and therefore these requirements do not appear mandatory, they will prove nonetheless useful to characterize the choices of boundary conditions.

The contents of this paper are as follows. In Section~\ref{sec:exactschrod} we briefly review the self--adjoint boundary conditions for Bose fields at the ends of the interval in the presence of singular potentials as in eq.~\eqref{lim_dil_axion_intro}. We also describe a class of exactly solvable trigonometric potentials related to hypergeometric functions that can closely approximate the Schr\"odinger potentials of all sectors in our background.
In Section~\ref{sec:dilatonaxion} we begin our detailed analysis of bosonic modes, starting from the dilaton--axion system. We examine the possible self-adjoint boundary conditions and the stability regions for the corresponding Schr\"odinger equation, and present an analytic solution for its zero mode. The same analysis applies to the axion, and to the spin-2 $h_{\mu\nu}$ and spin-0 $h_{ij}$ modes of gravity. In section~\ref{sec:IIB3forms} we discuss the modes of the type--IIB three--forms. All these fields lack vacuum profiles, but their equations are affected by the five--form background via Chern--Simons couplings. As a result, their massless modes require a special treatment, which finally leads to third--order equations and to some unfamiliar features for their spectra. We also determine the corresponding stability regions for the boundary conditions.
In section~\ref{sec:perturbed_tensor} we first discuss a convenient parametrization for the components of the self--dual tensor field, and then present the perturbed self--duality equations, after making a convenient gauge choice.  To the best of our knowledge, this is the first time that this type of detailed analysis is carried out, to this extent, for a self--dual tensor field. We then discuss the perturbed Einstein equations, including tensor contributions. To this end, we rely extensively on Appendices~\ref{app:background}, \ref{app:fiveform} and \ref{app:Einstein}, which contain a number of useful technical details on the background and on the perturbed equations for the various fields.
In Sections~\ref{sec:tensor_no_grav}, \ref{sec:4Dgrav} and \ref{sec:hij} we examine the modes arising from the self--dual tensor or from the ten--dimensional metric, with no mixings among them.  The tensor modes of Section~\ref{sec:tensor_no_grav} arising from the five-form alone have some unfamiliar features, including the possible presence of zero modes within all possible Kaluza--Klein excitations. This would clearly seem unphysical, but actually the different occurrences correspond to different choices of boundary conditions, and thus to different vacua. These choices would also lead to tachyons for $\mathbf{k}=0$, and are thus excluded a priori by the requirement of vacuum stability.
In Sections~\ref{sec:4Dgrav} and \ref{sec:hij} we discuss the spin-2 $h_{\mu\nu}$ and spin-0 $h_{ij}$ modes, and show that the corresponding Schr\"odinger equations are identical to that of the dilaton--axion sector.
In Section~\ref{sec:singlet_vector} we analyze the vector modes that are singlets under the internal symmetries and show that there are only massive excitations among them, since the zero mode is not normalizable. In Section~\ref{sec:nonsingletvectors} we consider non--singlet vector modes. The modes of this type with $\mathbf{k}=0$, which are valued in the fundamental representation of $SO(5)$, lead to a one--dimensional Schr\"odinger system. We show that they are stable with suitable self--adjoint boundary conditions, and determine the corresponding zero mode. The sector with $\mathbf{k}\neq 0$ is more difficult to analyze, since it leads to a two--component Schr\"odinger system. After reducing it to a manifestly Hermitian form, we estimate the lowest allowed values for $m^2$ and their dependence on the lattice momentum $\mathbf{k}$ and on the choice of self--adjoint boundary conditions, resorting to the variational principle. We show the existence of self–adjoint boundary conditions that grant stability also in this case. Their parameter space is background dependent, and widens as the ratio $\frac{\Phi}{\ell^4}$, or equivalently as the ratio $\frac{R}{\rho}$, decreases. In Section~\ref{sec:nonsingletscalars} we discuss non--singlet scalar perturbations. For $\mathbf{k}=0$ these are valued in the fundamental representation of $SO(5)$, and with appropriate self--adjoint boundary conditions they are again stable, as in the preceding sectors. We also determine the corresponding massless mode. For $\mathbf{k} \neq 0$, the modes belonging to this sector are valued in the fundamental representation of SO(4) and lead again to a two--component Schr\"odinger system. After some redefinitions their squared mass can be related to the eigenvalues of an operator that, with proper self--adjoint boundary conditions, does not contain any unstable modes, but their parameter space is again background dependent, and widens as the ratio $\frac{\Phi}{\ell^4}$ decreases. In Section~\ref{sec:singletscalarmodes} we discuss singlet scalar modes. For $\mathbf{k}=0$, after suitable redefinitions, their squared masses emerge from an operator that is strictly positive with proper self--adjoint boundary conditions. We show that stability persists within  finite range of boundary conditions. We defer to a different publication the analysis of scalar singlets with $\mathbf{k} \neq 0$, which is more involved and requires different techniques, for reasons that we explain in Section~\ref{sec:conclusions}, where we collect our conclusions and some perspectives for future work.
Appendix~\ref{app:Mtildek0} contains some details on small--$\mathbf{k}$ limit for the modes discussed in Section~\ref{sec:intvectexc}, and finally
Appendix~\ref{app:variationa2} contains some details on our variational tests for Sections~\ref{sec:intvectexc} and \ref{sec:nonsingletscalrknot0}.

\section{\sc Self--Adjoint Extensions and Solvable $\left(\mu,\tilde{\mu}\right)$ Potentials} \label{sec:exactschrod}

In the Introduction we stated that, as in the nine--dimensional cases analyzed in~\cite{selfadjoint}, the potentials for the various mode sectors in this class of vacua can be characterized by a pair of parameters $(\mu,\tilde{\mu})$ that determine their leading singularities at the ends of the internal interval. 
Let us first consider the sectors where the analysis of the fluctuations leads to a single second--order equation. This case was widely studied, over the years, in the Mathematical literature~\cite{math_literature_1,math_literature_2,math_literature_3}, and in~\cite{selfadjoint} we formulated the whole setup in a way that seems particularly transparent to us, while also adapting it to the stability issue that is central to this work.

\subsection{\sc One--Dimensional Sectors}

We shall determine the values of  $\mu$ and $\tilde{\mu}$ in the following sections, but for convenience we collect the results in Table~\ref{tab:tab_munu}. Note that the values of $\mu$ are either $\frac{1}{3}$ or $\frac{2}{3}$, while the values of $\tilde{\mu}$ range from 0 to about 2.3.
\begin{table}[h!]
\centering
\begin{tabular}{||c | c c c||}
 \hline
Case & Sector & $\mu$ & $\tilde{\mu}$  \\ [0.5ex]
 \hline\hline
1 & $\varphi$, $a$, $h_{\mu\nu}$, $h_{ij}$ & $\frac{1}{3}$ &  $0$ \\ \hline
2 & $B_{\mu \nu}$ & $\frac{2}{3}$ &  1.72 \\
2 & $b_{\mu\nu}{}^{ij}(g_1)$ & $\frac{1}{3}$ & 1.09 \\
2 & $V_\mu$ & $\frac{1}{3}$ & 2.18 \\
2 & $h_{\mu i}$ & $\frac{2}{3}$ & $1.18$ \\
2 & $\phi_i$ & $\frac{2}{3}$ & 2.27 \\
2 & $\phi$ & $\frac{2}{3}$ & 1  \\ \hline
3 & $B_{\mu i}$ & $\frac{1}{3}$ & $0.54$    \\
3 & $B_{ij}$ & $\frac{2}{3}$ &  $0.63$    \\
3 & $b^{(2)}{}_\mu{}^{ij}(g_2)$ & $\frac{2}{3}$ & 0.09  \\
 [1ex]
 \hline
\end{tabular}
\caption{Values of $\mu$ and $\tilde{\mu}$ for the different sectors of Bose modes with $\mathbf{k}=0$. The notation $b_{\mu\nu}{}^{ij}(g_1)$ and $b^{(2)}{}_\mu{}^{ij}(g_2)$ is meant to stress that the values of $\mu$ and $\tilde{\mu}$ refer to the Schr\"odinger--like equations for $g_1$ or $g_2$ in Section~~\ref{sec:tensor_no_grav}.}
\label{tab:tab_munu}
\end{table}

As explained in~\cite{selfadjoint}, the possible self--adjoint extensions of Schr\"odinger--like operators in an interval terminating at a pair of singular points can be characterized via the limiting behavior of the wavefunctions at the two ends. This is determined by $\mu$ and $\tilde{\mu}$, and the condition that $H\,\psi$ be in $L^2$ constrains in general the choice of the wavefunctions~\cite{math_literature_1,math_literature_2,math_literature_3}. Referring to Table~\ref{tab:tab_munu}, two independent choices of limiting behaviors are thus allowed, in all cases of interest, at $z=0$, while two choices are allowed at $z_m$ when $0 \leq \tilde{\mu} < 1$ and a single one is allowed when $\tilde{\mu} \geq 1$. In general, the asymptotic behavior at the left can be characterized by a pair of coefficients $C_1$ and $C_2$, according to
\bea
\psi &\sim \  \frac{C_1}{\sqrt{2\,\mu}} \ \left(\frac{z}{z_m}\right)^{\frac{1}{2} \ + \ \mu} \ + \  \frac{C_2}{\sqrt{2\,\mu}} \ \left(\frac{z}{z_m}\right)^{\frac{1}{2} \ - \ \mu} \qquad\qquad &\mathrm{if} \qquad 0 \ < \ \mu \ < \ 1 \ , \nonumber \\
\psi &\sim \  C_1 \ \left(\frac{z}{z_m}\right)^{\frac{1}{2}}\,\log\left(\frac{z}{z_m}\right) \ + \ C_2 \ \left(\frac{z}{z_m}\right)^{\frac{1}{2} } \qquad\qquad &\mathrm{if} \qquad \mu \ = \ 0
\label{b.10}
\eea
while at the other there are different options, depending of the value of $\tilde{\mu}$:
\bea
\psi &\sim \ \frac{C_3}{\sqrt{2\,\tilde{\mu}}} \ \left(1\,-\,\frac{z}{z_m}\right)^{\frac{1}{2} \ + \ \tilde{\mu}} \ + \  \frac{C_4}{\sqrt{2\,\tilde{\mu}}}\ \left(1\,-\,\frac{z}{z_m}\right)^{\frac{1}{2} \ - \ \tilde{\mu}}  \qquad\qquad  & \mathrm{if} \qquad 0 \ < \ \tilde{\mu} \ < \ 1 \ ; \nonumber \\
\psi &\sim \ C_3\ \left(1\,-\,\frac{z}{z_m}\right)^{\frac{1}{2}}\,\log \left(1\,-\,\frac{z}{z_m}\right)  \ + \ C_4\ \left(1\,-\,\frac{z}{z_m}\right)^{\frac{1}{2}}  \quad &\mathrm{if} \qquad \tilde{\mu} \ = \ 0 \ ; \nonumber \\
\psi &\sim \ \frac{C_3}{\sqrt{2\,\tilde{\mu}}} \ \left(1\,-\,\frac{z}{z_m}\right)^{\frac{1}{2}+\tilde{\mu}}    \quad\qquad \qquad \qquad \qquad \qquad \qquad\qquad &\mathrm{if} \qquad \tilde{\mu} \ \geq \ 1 \ . 
\label{b.11}
\eea

 When $\tilde{\mu} \geq 1$, the possible self--adjoint boundary conditions depend on a single parameter,
\beq
\frac{C_2}{C_1} \ = \ \tan\left(\frac{\alpha}{2}\right) \ . \label{C21}
\eeq

When $0 \leq \mu < 1$ and $0 \leq \tilde{\mu}<1$, defining the two vectors
\beq
\underline{C}(0) \ = \ \left(\begin{array}{c} C_1 \\ C_2  \end{array} \right) \ , \qquad
\underline{C}\left(z_m\right) \ = \  \left(\begin{array}{c} C_4 \\ C_3  \end{array} \right)  \ , \label{mutildezero}
\eeq
the self--adjointness condition becomes
\beq
\underline{C}(z_m) \ = \ {\cal U} \ \underline{C}\left(0\right) \ , \label{P_constraint}
\eeq
where ${\cal U}$ is a generic $U(1,1)$ matrix, so that
\beq
{\cal U}^\dagger\,\sigma_2\ {\cal U} \ = \ \sigma_2 \ .
\eeq
${\cal U}$ is parametrized according to
\beq
{\cal U} \ = \ e^{i\,\beta} \ U \ ,
\eeq
where $\beta$ is a phase and $U$ is a generic $SL(2,R)$ matrix, and in the global $SL(2,R)$ parametrization
\beq
U\left(\rho,\theta_1,\theta_2\right) \ = \ \cosh\rho\left(\cos\theta_1 \, \underline{1} \,-\, i\,\sigma_2\,\sin\theta_1\right) \,+\, \sinh\rho\left(\sigma_3\, \cos\theta_2\ + \ \sigma_1\, \sin\theta_2 \right)
\ , \label{global_ads3_1}
\eeq
where $0 \leq \rho< \infty$, $-\,\pi \leq \theta_{1,2} < \pi$.

The Schr\"odinger equation determines in general the relation
\beq
\underline{C}\left(z_m\right) \ = \ V \ \underline{C}\left(0\right) \ , \label{C0mV}
\eeq
where $V$ is an $SL(2,R)$ matrix, consistently with our definitions and the constancy of the Wronskian, and 
the eigenvalue equation is in general~\cite{selfadjoint}
\beq
\mathrm{Tr}\left[ {U}^{-1}\ V \right] \ = \ 2\,\cos\beta \ . \label{eigenveq}
\eeq

The large--$\rho$ limit of eq.~\eqref{P_constraint} yields independent boundary conditions at the ends depending on the two parameters $\theta_1$ and $\theta_2$, which can be cast in the form
\bea
&& \cos\left(\frac{\theta_1-\theta_2}{2}\right) C_1 \ - \ \sin\left(\frac{\theta_1-\theta_2}{2}\right)C_2  \ = \ 0 \ , \nonumber \\
&&  \sin\left(\frac{\theta_1+\theta_2}{2}\right) C_4 \ - \ \cos\left(\frac{\theta_1+\theta_2}{2}\right)C_3  \ = \ 0    \ \label{cond_sing1n1}
\eea
when $0 \leq \mu< 1$ and $0 \leq \tilde{\mu} < 1$.

\subsection{\sc Matrix Generalization of the Setup}
In some of the following sections we shall need a generalization of these results involving $n$-component states $\psi$ and $n \times n$ matrix potentials, so that near the $z=0$ boundary the Hamiltonian will have the limiting form
\beq
H \ = \ - \ \partial_z^2 \ + \ \frac{1}{z^2} \,V_0 \ , \label{HV0}
\eeq
with $V_0$ a Hermitian matrix with eigenvalues $\left[\mu_1^2-\frac{1}{4}, \ldots,\mu_n^2 - \frac{1}{4}\right]$. In a similar fashion, near the other end of the interval the Hamiltonian will have the limiting form
\beq
H \ = \ - \ \partial_z^2 \ + \ \frac{1}{\left(z_m - z\right)^2} \,V_m \ , \label{HVinf}
\eeq
with $V_m$ a Hermitian matrix with eigenvalues $\left[\tilde{\mu}_1^2-\frac{1}{4}, \ldots,\tilde{\mu}_n^2 - \frac{1}{4}\right]$. In general, the two matrices $V_0$ and $V_m$ are diagonalized into $D_0$ and $D_m$ by different unitary matrices $U_0$ and $U_m$, so that
\beq
V_0 \ = \ U_0\,D_0\,U_0^\dagger \ , \qquad V_m \ = \ U_m\,D_m\,U_m^\dagger \ .
\eeq
Consequently, if $0 < \mu_i < 1$ the limiting behavior of the wavefunction close to the left end of the interval has the general form
\beq
\psi \ \sim \ U_0 \left( \begin{array}{c}  \frac{C_{11}}{\sqrt{2\,\mu_1}} \,z^{\frac{1}{2}+\mu_1} \ + \  \frac{C_{12}}{\sqrt{2\,\mu_1}}  \,z^{\frac{1}{2} -\mu_1} \\ \ldots \\ \frac{C_{n1}}{\sqrt{2\,\mu_n}} \,z^{\frac{1}{2}+\mu_n} \ + \  \frac{C_{n2}}{\sqrt{2\,\mu_n}}  \,z^{\frac{1}{2} -\mu_n}\end{array}\right) \ .
\eeq
In complete analogy with the one--dimensional case, if some $\mu_i$ is larger than one the $L^2$ condition demands that the corresponding $C_{i2}$ vanish. Finally, if some $\mu_i=0$ the corresponding line becomes
\beq
C_{i1} \,z^{\frac{1}{2}}\, \log z \ + \ C_{i2}\, z^{\frac{1}{2}}  \ .
\eeq
The limiting behavior at the other end is similar, up to the replacement of $\mu_i$ with $\tilde{\mu}_i$, $z$ with $z_m-z$ and $C_{i\, 1,2}$ with new coefficients $C_{i\,3,4}$.

The Hamiltonian $H$ is self--adjoint if
\beq
\left[ \partial_z\,{\psi}^\dagger\, \chi \ - \ {\psi}^\dagger\, \partial_z\,\chi \right]_0^{z_m} \ = \ 0 \ ,
\eeq
and therefore, if $0 \leq \mu_i < 1$,
\beq
\sum_i \left[ C_{i1}^\star \, D_{i2} \ - \ C_{i2}^\star \, D_{i1} \ +\ C_{i3}^\star \, D_{i4} \ - \ C_{i4}^\star \, D_{i3} \right] \ = \ 0 \ , \label{san}
\eeq
where we have denoted by $C_{ij}$ and $D_{ij}$ the coefficients of $\psi$ and $\chi$. One can now define the $2n$-component vector $\underline{C}(0)_{ia}$, with $\underline{C}(0)_{i1}=C_{i1}$ and $\underline{C}(0)_{i2}=C_{i2}$  and  $\underline{C}(z_m)_{ia}$, with $\underline{C}(z_m)_{i1}=C_{i4}$ and $\underline{C}(z_m)_{i2}=C_{i3}$ and $i=1,\ldots,n$, and the preceding condition becomes
\beq
\underline{C}^\dagger(0) \ 1_n \otimes \sigma_2 \ \underline{D}(0) \ = \ \underline{C}^\dagger(z_m) \ 1_n \otimes \sigma_2 \ \underline{D}(z_m) \ . \label{sadjn}
\eeq
Consequently the self--adjoint boundary conditions are parametrized by elements ${\cal U}$ of $U(n,n)$ such that
\beq
C(z_m) \ = \ {\cal U} \,C(0) \ ,
\eeq
together with a similar relation for the $D$ coefficients. The independent boundary conditions are then obtained when both sides of eq.~\eqref{sadjn} vanish, which is the case if the linear conditions,
\beq
C(0) \ = \ \Lambda \, C(0) \ ,  \label{CUC}
\eeq
hold, where $\Lambda$ Hermitian, $\Lambda^2=1$, and
\beq
\left\{\Lambda, 1 \otimes \sigma_2\right\} \ = \ 0  \ .
\eeq
One can thus write
\beq
\Lambda \ = \ {\cal M}_1 \otimes \ \sigma_1 \ + \ {\cal M}_3 \otimes \ \sigma_3 \ ,
\eeq
with ${\cal M}_{1,3}$ Hermitian matrices such that
\beq
 {\cal M}_1{}^2 \ + \  {\cal M}_3{}^2 \ = \ 1 \ , \qquad  \left[ {\cal M}_1, {\cal M}_3\right] \ = \ 0 \ .
\eeq

The two matrices ${\cal M}_1$ and ${\cal M}_3$ can be simultaneously diagonalized, and can be cast in the form
\beq
 {\cal M}_1 \ = \ \Omega^\dagger \ \mathrm{diag}\Big( \sin\alpha_1 \,, \ldots ,\, \sin\alpha_n\Big) \ \Omega \ , \qquad  {\cal M}_3\ = \ \Omega^\dagger \ \mathrm{diag}\Big( \cos\alpha_1 \,, \ldots ,\,\cos\alpha_n\Big) \ \Omega \ ,
\eeq
with $\Omega$ a unitary $n \times n$ matrix of unit determinant. In detail, the boundary conditions~\eqref{CUC} read
\beq
C_{ia} \ = \ \left[ {\cal M}_1{}_i{}^j \ \sigma_1{}_a{}^b \ + \ {\cal M}_3{}_i{}^j \ \sigma_3{}_a{}^b \right] C_{jb} \ , 
\eeq
or regrouping the coefficients into a rectangular matrix
\beq
\Omega\,C \ = \left( D_1\,\sigma_1\ + \ D_3\,\sigma_3 \right)\Omega\, C \ , \label{omegaC}
\eeq
where
\beq
D_1 \ = \  \mathrm{diag}\Big( \sin\alpha_1 \,, \ldots ,\, \sin\alpha_n\Big) \qquad D_3 \ = \ \mathrm{diag}\Big( \cos\alpha_1 \,, \ldots ,\,\cos\alpha_n\Big)  \ .
\eeq

For an $n\times n$ second--order system, the general boundary conditions of this type thus involve $n(n+1)-1$ parameters.
Similar considerations apply at the other end, with $n$ angles $\tilde{\alpha}_i$ and another unitary $n \times n$ matrix $\widetilde{\Omega}$ of unit determinant. All in all, the independent boundary conditions are thus parametrized by $2n$ angles and a pair of special unitary $n \times n$ matrices. For $n=1$, these considerations recover the choices of self--adjoint boundary conditions discussed in the preceding pages. However, only $n$ of the $2n$ components of $C$ are arbitrary, due to the condition~\eqref{CUC}, so that effectively one ends up with $n$ parameters at each end. If some of the $\mu_i \geq 1$ (say, $m$ of them), the number $n$ is simply replaced by $n-m$ in the preceding considerations, and similarly for the $\tilde{\mu}_i$.

\subsection{\sc Exactly Solvable Hypergeometric Potentials}

Before exhibiting the actual potentials for the various sectors of our problem, is it convenient to introduce a family of exactly solvable Schr\"odinger systems that generalize those considered in~\cite{selfadjoint}, for which $\mu=\tilde{\mu}$, and share the same type of limiting behavior. These systems are characterized by the trigonometric potentials~\cite{math_literature_3}
\beq
V(\mu,\tilde{\mu},z) \ = \ \frac{\pi^2}{4\,z_m^2}\left[ \frac{\mu^2 \ - \ \frac{1}{4}}{\sin^2\left(\frac{\pi\,z}{2\,z_m}\right)} \ + \ \frac{\tilde{\mu}^2 \ - \ \frac{1}{4}}{\cos^2\left(\frac{\pi\,z}{2\,z_m}\right)} \right]  \ , \label{pot_hyp}
\eeq
where $0 < z< z_m$, which can be obtained starting from the hypergeometric equation~\cite{tables} and performing a change of independent variable followed by a redefinition of the function, in order to cast the result into the Schr\"odinger form
\beq
- \ \Psi''(z) \ + \ V\left(\mu,\tilde{\mu},z\right)\,\Psi(z) \ = \ \frac{\pi^2\,m^2}{z_m^2}\, \Psi(z) \ . \label{schrod_hyper}
\eeq
For $\mu=\tilde{\mu}$ the potential reduces to
\beq
V(\mu,z) \ = \ \frac{\pi^2}{z_m^2}\ \frac{\mu^2 \ - \ \frac{1}{4}}{\sin^2\left(\frac{\pi\,z}{z_m}\right)} \ ,
\eeq
and the resulting spectra were discussed in detail in~\cite{selfadjoint}.

For $\mu \neq 0$, which is always the case in Table~\ref{tab:tab_munu}, the general solution of eq.~\eqref{schrod_hyper} reads~\cite{tables}
\beq
\Psi(z) = \frac{A\ w_1(z) \,+\, B\ w_2(z)}{u(z)^{\mu-\frac{1}{2}} \ v(z)^{-\tilde{\mu}-\frac{1}{2}} } \ , \label{psihyp}
\eeq
where 
\bea
w_1(z) &=& {}_2F_1\left[a,b\,;c\,;u^2(z)\right]  \ , \nonumber \\
w_2(z) &=&  \ u(z)^{2(1-c)}\ {}_2F_1\left[a-c+1,b-c+1\,;2-c\,;u^2(z)\right] \ , 
\eea
and the ${}_2F_1$ are hypergeometric functions~\cite{tables}. Moreover
\bea
u(z) &=& \sin\left(\frac{\pi\,z}{2\,z_m}\right) \ , \qquad v(z) \ = \ \cos\left(\frac{\pi\,z}{2\,z_m}\right) \ , \nonumber \\
a &=& \frac{\tilde{\mu} \ - \ \mu \ + \ 1}{2} \ + \ m \ , \quad b \ = \  \frac{\tilde{\mu} \ - \ \mu \ + \ 1}{2} \ - \ m \ , \quad c \ = \ 1 \ - \ \mu \ ,
\eea
where, without loss of generality, one can assume that $\mu$ and $\tilde{\mu}$ be positive.
The two functions
\bea
w_3(z) &=& {}_2F_1\left[a,b\,;a+b-c+1\,;v^2(z)\right] \ , \nonumber \\
w_4(z) &=&  v(z)^{2(c-a-b)}\ {}_2F_1\left[c-a,c-b\,;c-a-b+1\,;v^2(z)\right]  \label{hypers12}
\eea
provide an alternative basis of solutions, and are related to previous pair according to~\cite{tables}
\bea
w_1(z) &=& \frac{\Gamma\left(c\right)\,\Gamma\left(c-a-b\right)}{\Gamma\left(c-a\right)\,\Gamma\left(c-b\right)}\ w_3(z) \ + \ \frac{\Gamma\left(c\right)\,\Gamma\left(a+b-c\right)}{\Gamma\left(a\right)\,\Gamma\left(b\right)} \ w_4(z) \ , \nonumber \\
w_2(z) &=& \frac{\Gamma\left(2-c\right)\,\Gamma\left(c-a-b\right)}{\Gamma\left(1-a\right)\,\Gamma\left(1-b\right)} \ w_3(z) \ + \ \frac{\Gamma\left(2-c\right)\,\Gamma\left(a+b-c\right)}{\Gamma\left(a-c+1\right)\,\Gamma\left(b-c+1\right)} \ w_4(z) \ . \label{connections_hyper}
\eea

Even in this more general setting, one can introduce first--order operators ${\cal A}_{\epsilon_1,\epsilon_2}$ and ${\cal A}_{\epsilon_1,\epsilon_2}^\dagger$, where
\beq
{\cal A}_{\epsilon_1,\epsilon_2} \ = \ \partial_z \ + \ \frac{\pi}{4\,z_m}\left(2\,\epsilon_1\,\mu\,+\,1\right) \ \cot\left(\frac{\pi\,z}{2\,z_m}\right) \ +\ \frac{\pi}{4\,z_m}\left(2\,\epsilon_2\,\tilde{\mu}\,-\,1\right) \ \tan\left(\frac{\pi\,z}{2\,z_m}\right) \ .
\eeq
which depend on the signs $\epsilon_1$ and $\epsilon_2$. One can then show that
\beq
{\cal A}_{\epsilon_1,\epsilon_2}\,{\cal A}_{\epsilon_1,\epsilon_2}^\dagger \ = \ - \ \partial_z^2 \ + \ V_{\epsilon_1,\epsilon_2}(z) \ ,
\eeq
where 
\beq
V_{\epsilon_1,\epsilon_2}(z) \ = \  V(z) \ - \ \frac{\pi^2}{4\,z_m^2}\Big(1\ + \ \epsilon_1\,\mu \ - \ \epsilon_2\,\tilde{\mu}\Big)^2 \ . \label{V12}
\eeq
The different Hamiltonians
\beq
H_{\epsilon_1,\epsilon_2} \ = \ - \ \partial_z^2 \ + \ V_{\epsilon_1,\epsilon_2}(z) 
\eeq
have the same eigenvectors as $H$ but have shifted eigenvalues, so that
\beq
m^2_{\epsilon_1,\epsilon_2} \ = \ m^2 \ - \ \frac{1}{4} \ \Big(1\ + \ \epsilon_1\,\mu \ - \ \epsilon_2\,\tilde{\mu}\Big)^2 \ .
\eeq
The solutions of
\beq{\cal A}_{\epsilon_1,\epsilon_2}^\dagger\ \Psi_{\epsilon_1,\epsilon_2} \ = \ 0 \ ,
\eeq
are
\beq
\Psi_{\epsilon_1,\epsilon_2}(z) \ = \ C \left[\sin\left(\frac{\pi\,z}{2\,z_m}\right)\right]^{\frac{1}{2} \,+\, \epsilon_1\,\mu}\ \left[\cos\left(\frac{\pi\,z}{2\,z_m}\right)\right]^{\frac{1}{2} \,-\, \epsilon_2\,\tilde{\mu}}  \label{psi12} \ .
\eeq
When they are normalizable, they are zero modes of $H_{\epsilon_1,\epsilon_2}$.

In order to discuss the possible self--adjoint boundary conditions for $V(z)$ in the different sectors of the spectrum, one must distinguish different ranges for $\tilde{\mu}$.
\begin{itemize}
\item The case $\mu > 1$ and $\tilde{\mu}>1 $ does not occur in Table~\ref{tab:tab_munu}, but it is simple and instructive. The $L^2$ condition at the origin implies that $A=0$ in eq.~\eqref{psihyp}, and the limiting behavior at the other end of the interval is determined by eqs.\eqref{connections_hyper}. The corresponding $L^2$ condition demands that the coefficient of $w_4(z)$ vanish, so that $a-c+1$ or $b-c+1$ must be negative integers, which determine altogether the stable spectrum
\beq
m^2 \ = \ \left(\frac{\mu \, + \, \tilde{\mu} \, + \, 1}{2} \ + \ n\right)^2 \ , \qquad n=0,1,\ldots \ .
\eeq
Consequently
\beq
m^2_{\epsilon_1,\epsilon_2} \ = \ \frac{1}{4}\left[\left(1+\epsilon_1\right)\mu \, + \, \left(1-\epsilon_2\right)\tilde{\mu} \ + \ 2\left(n+1\right)\right]\left[\left(1-\epsilon_1\right)\mu \, + \, \left(1+\epsilon_2\right)\tilde{\mu} \ + \ 2\,n\right] \ .
\eeq
In this case, among the zero mode wavefunctions~\eqref{psi12}, only $\Psi_{+-}$ is normalizable, and the corresponding zero--mass eigenvalue is recovered for $n=0$. 
\item If $0< \mu < 1$ and $\tilde{\mu} \geq 1$, which corresponds to case 2 in Table~\ref{tab:tab_munu}, both solutions in eq.~\eqref{psihyp} are normalizable, but one must again demand that the resulting contribution proportional to $w_4(z)$ vanish near the other end of the interval. In this case, the allowed self--adjoint boundary conditions are related to the ratio of the two coefficients $A$ and $B$ and, according to eq.~\eqref{C21}, they can be parametrized via an angle $\alpha$, so that
\beq
\frac{A}{B} \ = \ \tan\left(\frac{\alpha}{2}\right) \left(\frac{\pi}{2}\right)^{\,2\,\mu} \  .
\eeq
The resulting eigenvalue equation reads
\beq
\tan\left(\frac{\alpha}{2}\right) \ = \  \frac{C_2}{C_1} \ = \ - \ \left(\frac{\pi}{2}\right)^{\,-\,2\,\mu} \ \frac{\Gamma\left(1+\mu\right)\Gamma\left(\frac{\tilde{\mu}-\mu+1}{2} \,+\, m\right) \Gamma\left(\frac{\tilde{\mu}-\mu+1}{2} \,-\, m\right)}{\Gamma\left(1-\mu\right)\Gamma\left(\frac{\tilde{\mu}+\mu+1}{2} \,+\, m\right) \Gamma\left(\frac{\tilde{\mu}+\mu+1}{2} \,-\, m\right)} \ , \label{eigenvmutildelarger}
\eeq
and can be solved graphically, as in fig.~\ref{fig:hyperex} for both real values of $m$, which correspond to stable modes, and for imaginary ones, which correspond to tachyonic modes.
\begin{figure}[ht]
\centering
\begin{tabular}{cc}
\includegraphics[width=65mm]{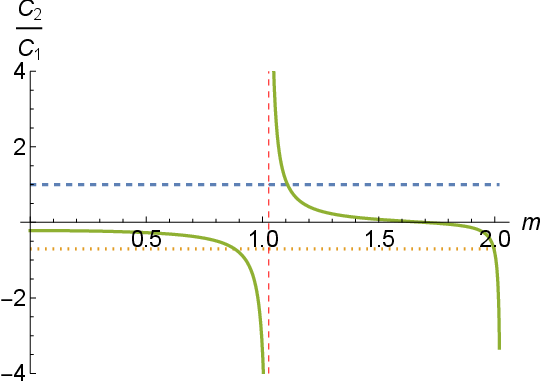} \qquad \qquad &
\includegraphics[width=65mm]{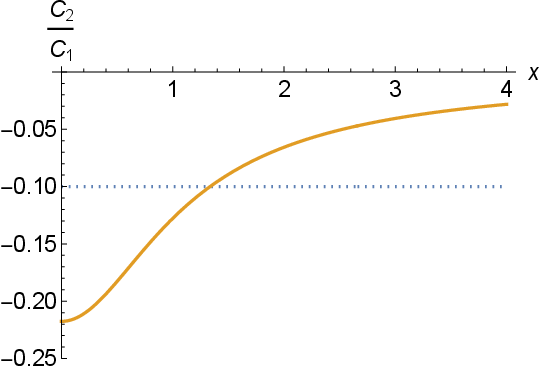} \\
\end{tabular}
\caption{\small The left panel illustrates the first stable eigenvalues of the potential~\eqref{pot_hyp} for $\mu=\frac{2}{3}$, $\tilde{\mu}=1.72$, $\tan\left(\frac{\alpha}{2}\right)=1$ (blue dashed line) and for $\tan\left(\frac{\alpha}{2}\right)=-0.21$ (orange dotted line). The right panel illustrates the presence of a tachyonic mode for the same values of $\mu$ and $\tilde{\mu}$ and $-0.21 <\tan\left(\frac{\alpha}{2}\right)<0$.}
\label{fig:hyperex}
\end{figure}

Some special cases are exactly solvable. 
\begin{itemize}
\item For $\alpha=0$ the denominator should have poles, so that the spectrum is given by
\beq
m^2 \ = \ \left(n \ + \ \frac{\tilde{\mu}+\mu+1}{2}\right)^2 \ ,
\eeq
which is stable, with no tachyons and no massless modes. 

\item For $\alpha = \pm \pi$ the numerator must have a pole, so that
\beq
m^2 \ = \ \left(n \ + \ \frac{\tilde{\mu}-\mu+1}{2}\right)^2 \ ,
\eeq
which is again stable, with no tachyons and no massless modes.
\item For general values of $\alpha$, one can solve the eigenvalue equation graphically, as illustrated in fig.~\ref{fig:hyperex}, and there is an infinite spectrum of real $m$ eigenvalues that correspond to stable modes with, in addition, at most one imaginary eigenvalue for $m$, which corresponds to a tachyonic mode. No tachyons are present if
\beq
\frac{C_2}{C_1} \ = \ \tan\left(\frac{\alpha}{2}\right)  \ < \ - \  \left(\frac{\pi}{2}\right)^{\,-\,2\,\mu}\frac{\Gamma(1+\mu)}{\Gamma(1-\mu)}\ \frac{\Gamma^2\left(\frac{\tilde{\mu}-\mu+1}{2}\right)}{\Gamma^2\left(\frac{\tilde{\mu}+\mu+1}{2}\right)} \quad \mathrm{or}\quad  \frac{C_2}{C_1}  \ > \ 0 \ .
\eeq
Two typical examples are displayed in fig.~\ref{fig:hyperex}
\end{itemize}
\item If $0< \mu < 1$ and $0<\tilde{\mu} <1$, which corresponds to case 3 in Table~\ref{tab:tab_munu}, one is free to use arbitrary combinations of the independent solutions at the two ends of the interval, and the self--adjoint boundary conditions relate them by a $U(1,1)$ matrix, according to eq.~\eqref{P_constraint}. Taking eqs.~\eqref{b.10}, \eqref{b.11} and \eqref{psihyp} into account, one can first conclude that
\beq
C_1 \ = \ B \, \sqrt{2\mu} \left(\frac{\pi}{2}\right)^{\frac{1}{2}+\mu} \ , \qquad C_2 \ = \ A \, \sqrt{2\mu} \left(\frac{\pi}{2}\right)^{\frac{1}{2}-\mu} \ , \label{C12AB}
\eeq
and then eqs.~\eqref{connections_hyper} determine $C_4$ and $C_3$ as
\bea
\frac{\left(\frac{\pi}{2}\right)^{\tilde{\mu}}\sqrt{\frac{\mu}{\tilde{\mu}}} \, C_4}{\Gamma\left(a+b-c\right)} \!\!&=&\!\!C_1\, \frac{\Gamma\left(2-c\right)\, \left(\frac{\pi}{2}\right)^{-\mu}}{\Gamma\left(a-c+1\right)\,\Gamma\left(b-c+1\right)} \ + \  C_2\,\frac{\Gamma\left(c\right)\, \left(\frac{\pi}{2}\right)^{\mu}}{\Gamma\left(a\right)\,\Gamma\left(b\right)} \ , \nonumber \\
\frac{\left(\frac{\pi}{2}\right)^{-\tilde{\mu}}\sqrt{\frac{\mu}{\tilde{\mu}}} \, C_3}{\Gamma\left(c-a-b\right)} \!\!&=&\!\!C_1\, \frac{\Gamma\left(2-c\right)\, \left(\frac{\pi}{2}\right)^{-\mu}}{\Gamma\left(1-a\right)\,\Gamma\left(1-b\right)} \ + \  C_2\,\frac{\Gamma\left(c\right)\, \left(\frac{\pi}{2}\right)^{\mu}}{\Gamma\left(c-a\right)\,\Gamma\left(c-b\right)} \ .
\eea
One can verify that the two pairs $(C_1,C_2)$ and $(C_4,C_3)$ are related by an $SL(2,R)$ transformation $V$, as in eq.~\eqref{C0mV}. The boundary conditions can now be parametrized via eq.~\eqref{global_ads3_1} and
an additional phase $\beta$, and the eigenvalue equation~\eqref{eigenveq}
reads
\bea
&& \xi\left(-\,\mu,\tilde{\mu},m\right) \left(\cos\theta_1 \cosh\rho-\cos\theta_2 \sinh\rho\right)
 \nonumber \\
 &&+ \ \xi\left(\mu,-\,\tilde{\mu},m\right)\left(\cos\theta_1\cosh\rho+\cos\theta_2 \sinh\rho\right) \nonumber \\
&&- \ \xi\left(\mu,\tilde{\mu},m\right) \left(\sin\theta_1 \cosh\rho + \sin\theta_2 \sinh\rho\right) \nonumber \\
&&+ \ \xi\left(-\,\mu,-\,\tilde{\mu},m\right) \left(\sin\theta_1 \cosh\rho-\sin\theta_2 \sinh\rho\right) \ = \ 2 \cos\beta \, ,
\eea
where
\beq
\xi\left(\mu,\tilde{\mu},m\right) \ = \  \frac{\left(\frac{\pi }{2}\right)^{\mu -\tilde{\mu} } \sqrt{\left|\frac{\tilde{\mu} }{\mu }\right|} \ \Gamma (1-\mu ) \ \Gamma (\tilde{\mu} ) }{\Gamma \left[\frac{1}{2} (-\mu +\tilde{\mu} +1)-m\right] \Gamma \left[m+\frac{1}{2} (-\mu +\tilde{\mu} +1)\right]} \ .
\eeq
In the $\rho \to \infty$ limit, which translates into independent boundary conditions at the ends of the interval, this expression reduces to
\bea
&& \xi\left(-\,\mu,\tilde{\mu},m\right) \left(\cos\theta_1 -\cos\theta_2 \right) \ + \ \xi\left(\mu,-\,\tilde{\mu},m\right)\left(\cos\theta_1+\cos\theta_2 \right) \nonumber \\
&&- \ \xi\left(\mu,\tilde{\mu},m\right) \left(\sin\theta_1 + \sin\theta_2 \right) \ + \ \xi\left(-\,\mu,-\,\tilde{\mu},m\right) \left(\sin\theta_1 -\sin\theta_2 \right) \ = \ 0 \, . \label{eigenvfinmumutilde}
\eea
Some special cases are exactly solvable. 
\begin{itemize}
\item If $\theta_1=\theta_2=0$, eq.~\eqref{eigenvfinmumutilde} reduces to $\xi\left(\mu,-\,\tilde{\mu},m\right)=0$, which is solved by
\beq
m^2 \ = \ \left(n\ - \ \frac{\mu \,+\,\tilde{\mu}\,-\,1}{2}\right)^2 \ , \qquad n=0,1,\ldots \ .
\eeq
There is a zero mode when $\mu+\tilde{\mu}=1$, which is in principle possible within the ranges that concern this case, but never occurs in Table~\ref{tab:tab_munu}; 
\item In a similar fashion, if $\theta_1=\theta_2=\frac{\pi}{2}$, eq.~\eqref{eigenvfinmumutilde} reduces to $\xi\left(\mu,\,\tilde{\mu},m\right)=0$, which is solved by
\beq
m^2 \ = \ \left(n\ - \ \frac{\mu \,-\,\tilde{\mu}\,-\,1}{2}\right)^2 \ , \qquad n=0,1,\ldots \ .
\eeq
\item If $\theta_1=-\,\theta_2=\frac{\pi}{2}$, eq.~\eqref{eigenvfinmumutilde} reduces to $\xi\left(-\,\mu,-\,\tilde{\mu},m\right)=0$, which is solved by
\beq
m^2 \ = \ \left(n\ + \ \frac{\mu \,-\,\tilde{\mu}\,+\,1}{2}\right)^2 \ , \qquad n=0,1,\ldots \ .
\eeq
\item Finally, if $\theta_1=0$ and $\theta_2=\pi$, eq.~\eqref{eigenvfinmumutilde} reduces to $\xi\left(-\,\mu,\tilde{\mu},m\right)=0$, which is solved by
\beq
m^2 \ = \ \left(n\ + \ \frac{\mu \,+\,\tilde{\mu}\,+\,1}{2}\right)^2 \ , \qquad n=0,1,\ldots \ .
\eeq
\end{itemize}
\item If $0< \mu < 1$ and $\tilde{\mu} = 0$, both solutions in eq.~\eqref{psihyp} are normalizable, and the asymptotic behavior at the left end defines again the two coefficients $C_1$ and $C_2$ according to eq.~\eqref{C12AB}, while $C_3$ and $C_4$ are defined according to 
\beq
\psi \ \sim \ \sqrt{1 - \ \frac{z}{z_m}}\left[C_4 \ + \  C_3\, \log\left(1 - \ \frac{z}{z_m}\right)\right] \ ,
\eeq

One can obtain the connection formulas as the $\tilde{\mu} \to 0$ limit of the preceding expressions in eqs.~\eqref{hypers12} and \eqref{connections_hyper}. Consequently, the behavior in the vicinity of the right end of the interval is now
\bea
w_1(z) &\sim& \xi_1(\mu,m) \ + \ \xi_2(\mu,m) \log\left(1\,-\,\frac{z}{z_m}\right) \ , \nonumber \\
w_2(z) &\sim& \xi_1(-\mu,m) \ + \ \xi_2(-\mu,m) \log\left(1\,-\,\frac{z}{z_m}\right) \ ,
\eea
where
\bea
\xi_1(\mu,m) &=& - \  \frac{\left(\frac{\pi}{2}\right)^\mu \Gamma(1-\mu) \left[\psi\left(-m-\frac{\mu }{2}+\frac{1}{2}\right)\,+\,\psi\left(m-\frac{\mu }{2}+\frac{1}{2}\right) \ - \ 2\,\psi\left(1\right)+ 2\log\left[\frac{\pi}{2}\right]\right]}{\sqrt{2\left|\mu\right|}\,\Gamma \left(-m-\frac{\mu }{2}+\frac{1}{2}\right) \Gamma \left(m-\frac{\mu }{2}+\frac{1}{2}\right) } \ \nonumber \\
\xi_2(\mu,m) &=& - \ \frac{ 2\,\left(\frac{\pi}{2}\right)^\mu \Gamma(1-\mu)}{\sqrt{2\left|\mu\right|}\,\Gamma \left(-m-\frac{\mu }{2}+\frac{1}{2}\right) \Gamma \left(m-\frac{\mu }{2}+\frac{1}{2}\right) } \ , \label{xi12def}
\eea
with 
\beq
\psi(z) \ = \frac{\Gamma'(z)}{\Gamma(z)} \ , \qquad \psi(1) = \ -  \ \gamma \ ,
\eeq
and $\gamma\sim 0.577$ the Euler--Mascheroni constant. The linear relations among the coefficients are now
\bea
C_4  &=& C_2 \  \xi_1(\mu,m) \ + \ C_1 \ \xi_1(-\,\mu,m) \ , \nonumber \\
C_3  &=& C_2 \  \xi_2(\mu,m) \ + \ C_1 \ \xi_2(-\,\mu,m)   \ . \label{eigenvmu0_rhoinf}
\eea
 
and consistently with~\cite{selfadjoint} they define an $SL(2,R)$ transformation $V$. The resulting eigenvalue equation is
\bea
&&\left(\cosh\rho\,\cos\theta_1 \,-\, \sinh\rho\,\cos\theta_2\right)  \xi_1(-\,\mu,m)\,+\,\left(\cosh\rho\,\sin\theta_1 \,-\, \sinh\rho\,\sin\theta_2\right)  \xi_2(-\,\mu,m) \nonumber \\
&&+\left(\cosh\rho\,\cos\theta_1 \,+\, \sinh\rho\,\cos\theta_2\right)  \xi_2(\mu,m)\,-\,\left(\cosh\rho\,\sin\theta_1 \,+\, \sinh\rho\,\sin\theta_2\right)  \xi_1(\mu,m) \nonumber \\
&&-\ 2 \,\cos \beta \ = \ 0\ .
\eea
In the $\rho \to \infty$ limit, which translates into independent boundary conditions at the ends of the interval, this expression reduces to
\bea
&&\left(\cos\theta_1 \,-\, \cos\theta_2\right)  \xi_1(-\,\mu,m)\,+\,\left(\sin\theta_1 \,-\, \sin\theta_2\right)  \xi_2(-\,\mu,m) \nonumber \\
&&+ \left(\cos\theta_1 \,+\, \cos\theta_2\right)  \xi_2(\mu,m)\,-\,\left(\sin\theta_1 \,+\, \sin\theta_2\right)  \xi_1(\mu,m)  \ = \ 0\ . \label{spec_mu0}
\eea

Some special cases are exactly solvable:
\begin{itemize}
\item if $\theta_1=\theta_2=0$, eq.~\eqref{spec_mu0} is solved by
\beq
m^2 \ = \ \left(n\ - \ \frac{\mu\,-\,1}{2}\right)^2 \ , \qquad n=0,1,\ldots \ ;
\eeq
\item if $\theta_1=0$ and $\theta_2=\pi$, the solution is
\beq
m^2 \ = \ \left(n\ + \ \frac{\mu\,+\,1}{2}\right)^2 \ , \qquad n=0,1,\ldots \ .
\eeq
\end{itemize}
\end{itemize}

Before turning to the different sectors of the actual spectrum, let us describe some features of the multi--dimensional case.

\subsection{\sc A Simple Matrix Generalization of the Hypergeometric Model} \label{sec:matrix_hyper}
One can generalize the preceding setup to cases when the wavefunction has $n$ components, replacing $\mu$ and $\tilde{\mu}$ in eq.~\eqref{pot_hyp} by real diagonal matrices, so that
\beq
\mu \ \rightarrow \ \mathrm{diag}\Big( \mu_1 \,, \ldots ,\, \mu_n\Big) \ ,
\eeq
and similarly for $\tilde{\mu}$. In this case there are $n$ decoupled hypergeometric equations, whose solutions, however, can be mixed by the boundary conditions. 

For simplicity, we can confine our attention to boundary conditions given independently at the two ends, and to the special cases of interest in this paper in Sections~\ref{sec:intvectexc} and \ref{sec:nonsingletscalrknot0}, where $n=2$, $0<\mu_{1,2}<1$, while $0<\tilde{\mu}_{1}<1$ and $\tilde{\mu}_2>1$. In these cases, the choice of self--adjoint boundary conditions involves five angles at the left end and one additional parameter at the right end.

Let us begin from the first end. Making use of eq.~\eqref{omegaC} and denoting the product $\Omega \,C$ by $\widetilde{C}$, the independent boundary conditions at the left end can be cast in the form
\bea
\widetilde{C}_{11} &=& \cos\alpha_1\, \widetilde{C}_{11} \ + \ \sin\alpha_1 \,\widetilde{C}_{12} \ , \nonumber \\
\widetilde{C}_{21} &=& \cos\alpha_2\, \widetilde{C}_{21} \ + \ \sin\alpha_2 \,\widetilde{C}_{22} \ ,
\eea
or equivalently
\beq
\widetilde{C}_{11}  \ = \  \cot\frac{\alpha_1}{2}\ \widetilde{C}_{12} \ , \qquad
\widetilde{C}_{21}  \ = \  \cot\frac{\alpha_2}{2}\ \widetilde{C}_{22} \ . \label{cot_definitions}
\eeq
Now the actual $C$ coefficients are related to the $\widetilde{C}$ by an $SU(2)$ matrix $\Omega^\dagger$, and one can parametrize $\Omega$ as
\beq
\Omega \ = \ \left( \begin{array}{cc} \cos\gamma\,e^{i \alpha} & \sin\gamma\,e^{i \beta} \\ - \, \sin\gamma\,e^{-\,i \beta} & \cos\gamma\,e^{- \,i \alpha} \end{array} \right) \ , 
\eeq
so that finally
\beq
C_{i 1} \ = \ \widetilde{E}_{ij}\ C_{j 2} \ , \label{Cj12}
\eeq
with
\beq
\widetilde{E} \ = \ \Omega^\dagger{} \, E \, \Omega \ , \qquad E \ = \ \left( \begin{array}{cc} \cot\frac{\alpha_1}{2} & 0  \\ 0 & \cot\frac{\alpha_2}{2}\end{array} \right) \ .
\eeq
For $\Omega=1$ one recovers the boundary conditions for independent equations that encompass all cases with $\mathbf{k}=0$.  When $\alpha_1=\alpha_2$, $\Omega$ commutes with $E$ and disappears altogether, and a similar simplification occurs, for a generic $E$, when $\Omega$ is diagonal.

If $\tilde{\mu}_2 \geq 1$, as will be the case in Section~\ref{sec:intvectexc}, one must demand that at the right end $C_{24}=0$, in order get $L^2$ solutions, and moreover the ratio
\beq
\frac{C_{13}}{C_{14}} \ = \ \cot \,\frac{\tilde{\alpha}_1}{2}  \label{C134}
\eeq
can be taken to parameterize the independent choices of self--adjoint boundary conditions there. The continuation to the right end of the interval of the hypergeometric solutions then yields the relations
\bea
\frac{\left(\frac{\pi}{2}\right)^{\tilde{\mu}_i}\sqrt{\frac{\mu_i}{\tilde{\mu}_i}} \, C_{i4}}{\Gamma\left(\tilde{\mu}_i\right)} &=& C_{i 1}\, \frac{\Gamma\left(1+\mu_i\right)\, \left(\frac{\pi}{2}\right)^{-\mu_i}}{\Gamma\left(\frac{\tilde{\mu}_i+\mu_i+1}{2}+m\right)\,\Gamma\left(\frac{\tilde{\mu}_i+\mu_i+1}{2}-m\right)} \nonumber \\  &+&  C_{i 2}\,\frac{\Gamma\left(1-\mu_i\right)\, \left(\frac{\pi}{2}\right)^{{\mu}_i}}{\Gamma\left(\frac{\tilde{\mu}_i-\mu_i+1}{2}+m\right)\,\Gamma\left(\frac{\tilde{\mu}_i-\mu_i+1}{2}-m\right)} \ , \nonumber \\
\frac{\left(\frac{\pi}{2}\right)^{-\tilde{\mu}_i}\sqrt{\frac{\mu_i}{\tilde{\mu}_i}} \, C_{i 3}}{\Gamma\left(-\,\tilde{\mu}_i\right)} &=& C_{i 1}\, \frac{\Gamma\left(1+\mu_i\right)\, \left(\frac{\pi}{2}\right)^{-\mu_i}}{\Gamma\left(\frac{\mu_i-\tilde{\mu}_i+1}{2}+m\right)\,\Gamma\left(\frac{\mu_i-\tilde{\mu}_i+1}{2}-m\right)} \nonumber \\ &+&  C_{i 2}\,\frac{\Gamma\left(1-\mu_i\right)\, \left(\frac{\pi}{2}\right)^{\mu_i}}{\Gamma\left(\frac{1- \tilde{\mu}_i-\mu_i}{2}+m\right)\,\Gamma\left(\frac{1- \tilde{\mu}_i-\mu_i}{2}-m\right)}  \ , \label{bc12}
\eea
and taking eq.~\eqref{Cj12} into account, the \emph{r.h.s} only involves the $C_{j2}$. These ingredients determine the eigenvalue equation as follows. One first expresses the $C_{i1}$ in terms of the $C_{i2}$ using eq.~\eqref{Cj12}, so that the preceding relations become
\beq
\sqrt{\frac{\mu_i}{\tilde{\mu}_i}} \ C_{i4}  \ = \ {\cal P}_4{}_{ij} \,C_{j2} \ , \qquad \sqrt{\frac{\mu_i}{\tilde{\mu}_i}}\ C_{i 3}\ = \ {\cal P}_3{}_{ij} \,C_{j2} 
\eeq
where
\bea
{\cal P}_4{}_{ij}\!\!\! &=&\!\!\!  \frac{\widetilde{E}_{ij}\,\Gamma\left(1+\mu_i\right)\, \Gamma\left(\tilde{\mu}_i\right)\left(\frac{\pi}{2}\right)^{-\mu_i-\tilde{\mu}_i}}{\Gamma\left(\frac{\tilde{\mu}_i+\mu_i+1}{2}+m\right)\,\Gamma\left(\frac{\tilde{\mu}_i+\mu_i+1}{2}-m\right)} \, +\,  \frac{\delta_{ij}\, \Gamma\left(1-\mu_i\right)\,\Gamma\left(\tilde{\mu}_i\right)\left(\frac{\pi}{2}\right)^{{\mu}_i-\tilde{\mu}_i}}{\Gamma\left(\frac{\tilde{\mu}_i-\mu_i+1}{2}+m\right)\,\Gamma\left(\frac{\tilde{\mu}_i-\mu_i+1}{2}-m\right)} \, , \nonumber \\
{\cal P}_3{}_{ij} \!\!\!&=&\!\!\!\frac{\widetilde{E}_{ij}\,\Gamma\left(1+\mu_i\right)\, \Gamma\left(-\,\tilde{\mu}_i\right)\left(\frac{\pi}{2}\right)^{-\mu_i+\tilde{\mu}_i}}{\Gamma\left(\frac{\mu_i-\tilde{\mu}_i+1}{2}+m\right)\,\Gamma\left(\frac{\mu_i-\tilde{\mu}_i+1}{2}-m\right)} \, + \, \frac{\delta_{ij}\, \Gamma\left(1-\mu_i\right)\, \Gamma\left(-\,\tilde{\mu}_i\right)\left(\frac{\pi}{2}\right)^{\mu_i+\tilde{\mu}_i}}{\Gamma\left(\frac{1- \tilde{\mu}_i-\mu_i}{2}+m\right)\,\Gamma\left(\frac{1- \tilde{\mu}_i-\mu_i}{2}-m\right)} \, .
\eea

One must now demand that $C_{24}=0$, while also relating $C_{13}$ and $C_{14}$ according to eq.~\eqref{C134}, and the resulting eigenvalue equation
\beq
\cot \,\frac{\tilde{\alpha}_1}{2} \ = \ \frac{P_{422} \,P_{311} \ - \ P_{421} \,P_{312}}{P_{422} \,P_{411} \ - \ P_{421} \,P_{412}}  \label{eigenv_munot0}
\eeq
depends on six parameter, the five parameters contained in $\widetilde{E}$ and $\tilde{\alpha}_1$.

The other case of interest, $\tilde{\mu}_1=0$ and $\widetilde{\mu}_2>1$, will present itself in Section~\ref{sec:nonsingletscalrknot0}.
Now the preceding equations~\eqref{bc12} for $i=1$ are replaced by
\bea
C_{14}  &=& C_{12} \  \xi_1(\mu_1,m) \ + \ C_{11} \ \xi_1(-\,\mu_1,m) \ , \nonumber \\
C_{13}  &=& C_{12} \  \xi_2(\mu_1,m) \ + \ C_{11} \ \xi_2(-\,\mu_1,m)   \ , \label{eigenvmu0_rhoinf2}
\eea
where $\xi_1$ and $\xi_2$ are defined in eqs.~\eqref{xi12def},
while they still hold for $i=2$. The final form of the eigenvalue equation is then
\beq
\cot \,\frac{\tilde{\alpha}_1}{2} \ = \ \frac{P_{422}\left(\xi_2(\mu_1,m) \,-\, \xi_2(-\mu_1,m)\,\widetilde{E}_{11}\right)  \ - \ \xi_2(-\mu_1,m)\,\widetilde{E}_{12}\,P_{421}}{P_{422}\left(\xi_1(\mu_1,m) \,+\, \xi_1(-\mu_1,m)\,\widetilde{E}_{11}\right)  \ - \ \xi_1(-\mu_1,m)\,\widetilde{E}_{12}\,P_{421}} \ , \label{eigenv_mu0}
\eeq
and the number of free parameters remains the same.

\section{\sc The Perturbed Dilaton--Axion System} \label{sec:dilatonaxion}

In a generic metric of the form
\beq
ds^2 \ = \ e^{\,2\,A(r)}\, dx^2 \ + \ e^{\,2\,B(r)}\, dr^2 \ + \ e^{\,2\,C(r)}\, dy^2 \ , \label{metricABC}
\eeq
and in the harmonic gauge
\beq{}{}{}{}{}{}{}{}{}{}{}{}{}{}{}{}{}{}{}{}{}{}{}{}{}{}{}{}{}
B \ = \ 4\, A \ + \ 5\, C \ , \label{harm_gauge}
\eeq
the perturbed dilaton equation follows from the quadratic terms in the type--IIB effective action, which for manifolds without boundaries contains the terms
\bea{}{}{}{}{}{}{}{}{}{}{}{}{}{}{}{}{}{}{}{}
{\cal S} \ = \ - \ \frac{1}{4\,k_{10}^2}\, \int d^{10}\, x \left\{ e^{2(B-A)}\,\eta^{\mu\nu}\, \partial_\mu\,\varphi\,\partial_\nu\,\varphi \ + \e^{2(B-C)}\,\delta^{ij}\, \partial_i\,\varphi\,\partial_j\,\varphi  \ + \ \left(\partial_r\,\varphi\right)^2  \right\}\ , \label{3.1}
\eea
and can be cast in the form
\beq
\Box\,\varphi \ +\ e^{2(A-C)}\,\nabla^2\, \varphi  \ + \ e^{2(A-B)}\, \partial_r^2\,\varphi \ = \ 0 \ .
\eeq
Here $\Box$ and $\nabla^2$ denote the d'Alembertian operator for four--dimensional Minkowski space and the Laplace operator for the internal torus.
In the analysis of the different sectors we shall rely implicitly on the self--adjoint action discussed in~\cite{ms_23_3}, which differs from eq.~\eqref{3.1} by the addition of boundary terms, in order to explore general 
self-adjoint boundary conditions. In this sector this action reduces to
\bea{}{}{}{}{}{}{}{}{}{}{}{}{}{}{}{}{}{}{}{}
{\cal S}_{s.a.} \ = \ \frac{1}{4\,k_{10}^2}\, \int d^{10}\, x \left\{ e^{2(B-A)}\ \varphi\,\Box\,\varphi \ + \ e^{2(B-C)}\,\varphi\,\nabla^2\,\varphi  \ + \ \varphi\, \partial_r{}^2\,\varphi  \right\}\ , \label{3.11}
\eea
and one can now separate variables letting
\beq
\varphi(x,r,y) \ = \ \varphi(x)\,f(r)\,e^{i\mathbf{k}\cdot \mathbf{y}} \, ,
\eeq
while also defining the four--dimensional squared mass via
\beq
\Box\,\varphi(x) = m^2 \, \varphi(x) \ ,
\eeq
so that the perturbed dilaton equation becomes
\beq
f''(r) \ - \ \mathbf{k}^2 \,e^{2(B-C)}\,f(r) \ + \ m^2 \,e^{2(B-A)}\,f(r) \ = \ 0 \ . \label{secondorderfdil}
\eeq

This equation determines the allowed values of $m^2$, and in fact one can recast it in a form where these become eigenvalues of a Hermitian second--order operator.  We can now describe the procedure in detail, since it will recur in the following sections. The first step consists in trading $r$ for the conformal variable $z$, defined via
\beq
dz \ = \ e^{B-A}\, dr \ , \label{dzdr}
\eeq
with $z(0)=0$. Note that $z$ has a finite range for the background of eqs.~\eqref{back_epos_fin2}, $0\leq z \leq z_m$, with $z_m$ given by
\beq{}{}{}{}{}{}{}{}{}{}{}{}{}
z_m \ = \ \int\, e^{B-A}\ dr \ \simeq \ 2.24 \, z_0 \ , \label{zm_app}
\eeq
and
\beq{}{}{}{}{}{}{}{}{}{}{}{}{}{}{}{}{}{}{}{}
z_0  \ = \ \rho\, h^\frac{1}{2} \ = \ \left(2 H \rho^3\right)^\frac{1}{2} \ . \label{z0}
\eeq
Here and in the rest of the paper $z$-derivatives will be often denoted by a subscript, so that, for instance
\beq{}{}{}{}{}{}{}{}{}{}{}{}{}{}{}{}{}{}{}{}
A_z \ \equiv \ \frac{dA}{dz}\ = \ e^{A-B}\, A'(r)  \ ,
\eeq
where $A'(r)$ denotes the derivative with respect to $r$.
Performing this change of variable, the original equation~\eqref{secondorderfdil} becomes
\beq
\frac{d^2\,f}{dz^2} \ + \ \left(B_z\,-\,A_z\right)\,\frac{df}{dz} \ + \ \left[m^2 \ - \ \mathbf{k}^2\,e^{2(A-C)} \right]f \ = \ 0  \ .
\eeq
The second step entails, in this case, the field redefinition
\beq
f \ = \ e^{\,-\,\frac{1}{2}\left(B-A\right)} g \ ,
\eeq
which finally leads to the manifestly Hermitian Schr\"odinger--like equation
\beq
-\, \frac{d^2\,g}{dz^2} \ + \ V \,g \ = \ m^2\,g  \ . \label{schrodg}
\eeq
The potential is
\beq
V \ = \  \frac{1}{4}\left(B_z\,-\,A_z\right)^2\ + \ \frac{1}{2}\left(B_{zz}\,-\,A_{zz}\right) \ + \ \mathbf{k}^2 \,e^{2\left(A-C\right)} \ , \label{pot_dilaton}
\eeq
and its detailed expression as a function of $r$, which can be obtained using results in Appendix~\ref{app:background}, reads
\beq
V \ = \ -\,\frac{e^{\frac{r}{\rho}\,\sqrt{\frac{5}{2}}}}{4\,z_0^2\,\sinh\left(\frac{r}{\rho}\right)^3}\left[1+\frac{1}{4}\left(\cosh\left(\frac{r}{\rho}\right)-\sqrt{\frac{5}{2}}\,\sinh\left(\frac{r}{\rho}\right)\right)^2 \right]\ + \ \mathbf{k}^2 \,e^{2\left(A-C\right)}\ . \label{vdilgen}
\eeq
Note that
\beq{}{}{}{}{}{}{}
V \ = \ \frac{1}{z_0^2}\left[ f_1\left(\frac{z}{z_0}\right) \ + \ \left(k \rho\right)^2 \, f_2\left(\frac{z}{z_0}\right) \right] \ ,
\eeq
where
\beq{}{}{}{}{}{}{}
k \ = \ \frac{\left|\mathbf{n}\right|}{R} \ ,
\eeq
with $\left|\mathbf{n}\right|$ an integer--valued vector independent of $R$, and consequently for $\mathbf{k}=0$ all squared masses are proportional to $\frac{1}{z_0^2}$, while in general they also depend on the ratio $\frac{\rho}{R}$.
\begin{figure}[ht]
\centering
\begin{tabular}{cc}
\includegraphics[width=65mm]{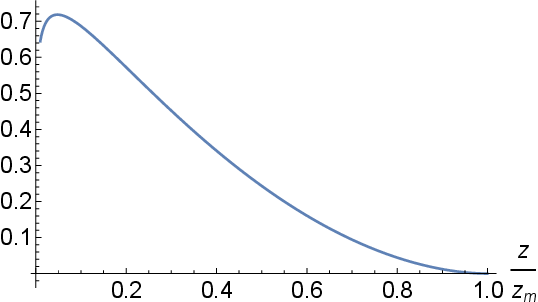} \qquad \qquad &
\includegraphics[width=65mm]{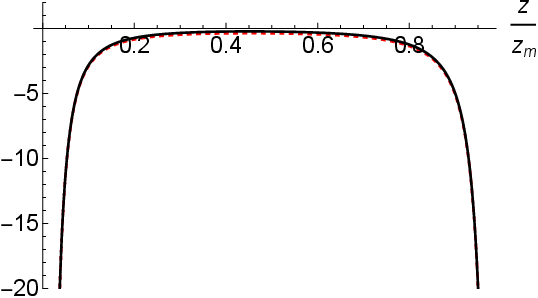} \\
\end{tabular}
\caption{\small The left panel shows the normalized zero--mode wavefunction~\eqref{dil_zero_mode}. The right panel compares the actual dilaton potential (black, solid) with its approximation~\eqref{pot_hyp} with $(\mu,\tilde{\mu})=\left(\frac{1}{3},0\right)$ (red, dashed), which is almost superposed to it. They are in units of $\frac{1}{{z_0}^2}$, with $z_m$ and $z_0$ defined in eqs.~\eqref{zm_app} and \eqref{z0}.}
\label{fig:dil_pot}
\end{figure}

The $\mathbf{k}$-independent portion of the potential is displayed in fig.~\ref{fig:dil_pot} as a function of $\frac{z}{z_0}$: it has the form of an inverted well, with singularities at the two ends $z=0$ and $z=z_m$, where
\beq{}{}{}{}{}
V \ \sim \ \frac{\mu^2\,-\,\frac{1}{4}}{z^2} \ ,
\qquad
V \ \sim \ \frac{\tilde{\mu}^2\,-\,\frac{1}{4}}{\left(z-z_m\right)^2} \ , \label{lim_dil_axion2}
\eeq
with $\mu=\frac{1}{3}$ and $\tilde{\mu}=0$, so that it belongs to case 1 in Section~\ref{sec:exactschrod}.
The supersymmetric limit is recovered as $\rho \to \infty$, where the interval becomes of infinite length, and then $V$ approaches
\beq{}{}{}{}{}{}{}{}{}{}{}{}{}{}{}{}{}{}{}{}{}{}{}{}{}{}{}{}{}
V_{susy} \ = \ - \ \frac{5}{36\, z^2} \ + \ \frac{\mathbf{n}^2}{R^2} \left(\frac{1}{3\left|H\right| z}\right)^\frac{2}{3} \ .
\eeq

One can actually recast eq.~\eqref{schrodg} in the form
\beq
\left[{\cal A}\,{\cal A}^\dagger \ + \ \mathbf{k}^2\,e^{2(A-C)}\right] g \ = \ m^2 \, g \ , \label{aadaggerdilaton}
\eeq
where the operators ${\cal A}$ and ${\cal A}^\dagger$ are
\beq
{\cal A} \ = \ \partial_z \ + \ \frac{1}{2}\left(3\,A_z\,+\,5\,C_z\right)  \ , \qquad {\cal A}^\dagger \ = \ - \ \partial_z \ + \ \frac{1}{2}\left(3\,A_z\,+\,5\,C_z\right)  \ . \label{aadaggerdilform}
\eeq

\begin{figure}[ht]
\centering
\includegraphics[width=65mm]{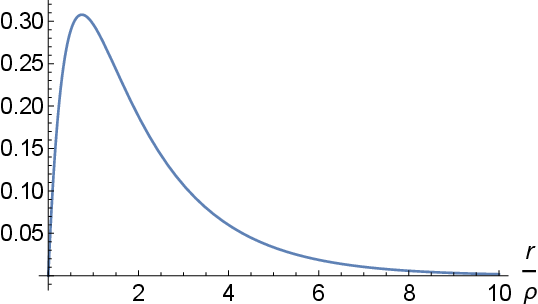}
\caption{\small The distribution function $\Pi_\phi\left(\frac{r}{\rho}\right)$ for the dilaton zero mode, in units of $\frac{1}{\rho}$. The average value of $r$ is about ${2.1}\,{\rho}$.}
\label{fig:distr_dil}
\end{figure}
\noindent One can thus find a massless mode for the dilaton, solving the first--order equation
\beq
{\cal A}^\dagger \, g \ = \ 0  \  . \label{zeromoddil}
\eeq
The solution reads
\beq{}{}
g \ = \ g_0 \ e^\frac{3 A + 5 C}{2} \ , \label{dil_zero_mode}
\eeq
where $g_0$ a constant, and this wavefunction is normalizable, since
\beq
\int_0^{z_m} dz\, g^2 \ = \ \int_0^\infty dr \,e^{2(B-A)}\, g_0^2 \label{distr_G}
\eeq
is clearly finite. The corresponding normalized $r$--distribution
\beq{}{}
\Pi_\phi(r) \ \simeq \ \frac{3}{2\,\rho}\, \sinh\left(\frac{r}{\rho}\right)\,e^{-\frac{r}{\rho}\,\sqrt{\frac{5}{2}}} \ , \label{dil_profile}
\eeq
which is localized in the vicinity of the effective BPS orientifold, is displayed in fig.~\ref{fig:distr_dil}.
Note that the actual dilaton zero--mode wavefunction
\beq{}{}{}{}{}{}{}{}{}{}{}{}{}{}{}{}{}{}{}{}{}{}{}{}{}{}{}{}{}{}{}{}{}{}{}{}{}{}{}{}{}{}{}{}{}{}{}{}{}{}{}{}{}{}{}{}{}{}{}{}{}{}{}{}{}{}{}{}{}{}
\varphi(x,r,y) \ = \ \varphi(x)\,f_0
\eeq
has a constant $r$ profile.

The behavior of the zero--mode wavefunction~\eqref{dil_profile} near the right end of the interval is
\beq
g \sim \left(1 \ - \ \frac{z}{z_m}\right)^\frac{1}{2} \ , \label{no_log}
\eeq
without a $\log$ term, so that, in the notation of the previous section, $C_3=0$. In a similar fashion, the dominant behavior of the actual zero mode near the left end of the interval is
\beq
g \ \simeq \ \left(\frac{z}{z_m}\right)^\frac{1}{6} \ - \ 0.71 \left(\frac{z}{z_m}\right)^\frac{5}{6} \ . \label{gmodel_leading}
\eeq
The two limiting behaviors define the self--adjoint boundary conditions characterizing the zero mode.
The argument presented in~\cite{selfadjoint} shows that, with this choice of boundary conditions, no instabilities are present in this sector.
\begin{figure}[ht]
\centering
\begin{tabular}{cc}
\includegraphics[width=65mm]{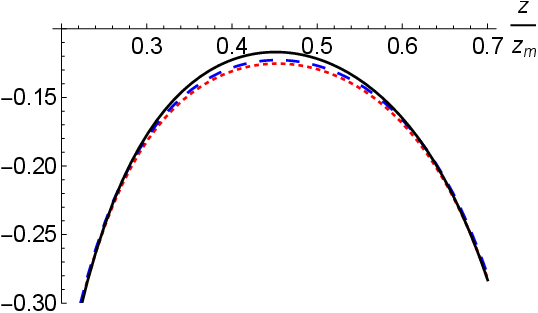} \qquad \qquad &
\includegraphics[width=65mm]{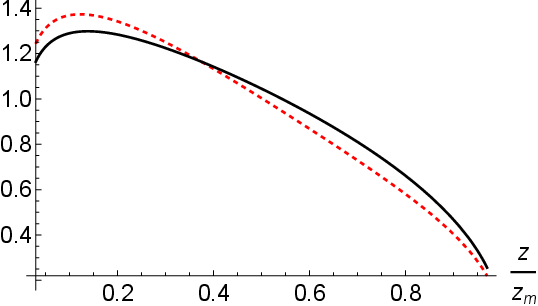} \\
\end{tabular}
\caption{\small The left panel shows how shifting slightly eq.~\eqref{pot_hyp} can optimize the correspondence with the dilaton potential in the middle region. The potentials are multiplied by the ratio $\frac{z_m^2}{\pi^2}$ in order to enhance their differences. The dashed curve corresponds to a shift $a \simeq 0.26$, which is also suggested by perturbation theory, while the dotted curve corresponds to $a \simeq 0.255$. The right panel compares the normalized ground--state wavefunction~\eqref{dil_zero_mode} (black, solid) with its approximation obtained combining the two zero modes in eq.~\eqref{psi12}, in such a way that the leading behavior corresponds to eq.~\eqref{gmodel_leading}.}
\label{fig:dil_pot_shift}
\end{figure}

\begin{figure}[ht]
\centering
\begin{tabular}{cc}
\includegraphics[width=65mm]{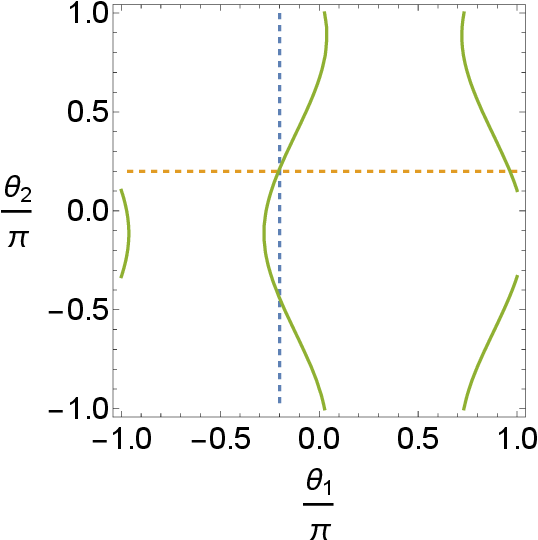} \qquad\qquad &
\includegraphics[width=65mm]{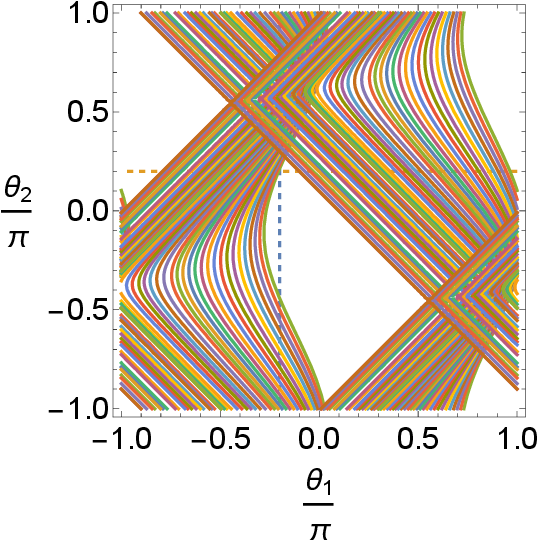} \\
\end{tabular}
\caption{\small The curves in the left panel identify the boundary conditions leading to a massless mode. The zero mode~\eqref{dil_zero_mode} lies at the intersection of the vertical and horizontal dashed lines for $(\theta_1,\theta_2) \simeq (-0.2,0.2)\pi$, on the line $\theta_1+\theta_2=0$ that characterizes boundary conditions leading to $C_3=0$.  The shaded regions in the right panel identify the boundary conditions leading to instabilities. }
\label{fig:dil_instabilities}
\end{figure}

More general boundary conditions can be explored relying on the model potential~\eqref{pot_hyp} with $\left(\mu,\tilde{\mu}\right)=\left(\frac{1}{3},0\right)$
\beq
V \ \simeq \ \frac{\pi^2}{4\,z_m^2}\left[ \frac{- \ \frac{5}{36}}{\sin^2\left(\frac{\pi\,z}{2\,z_m}\right)} \ - \ \frac{\frac{1}{4}}{\cos^2\left(\frac{\pi\,z}{2\,z_m}\right)} \right] \ + \ \frac{\pi^2}{z_m^2} \ a^2\ ,
\eeq
where we have allowed for an overall shift determined by $a$. The preferred values are $a=0.255$ (dashed curve) and $a=0.26$ (dotted curve), as shown in fig.~\ref{fig:dil_pot_shift}, and the latter is also suggested by perturbation theory. The resulting spectrum is now determined by eq.~\eqref{spec_mu0}, but the actual masses $m_{dil}$ are related to $m$ according to
\beq
m_{dil}^2 \ = \ m^2 \ + \ a^2 \ ,
\eeq
so that the massless mode corresponds to $m=i a$. The large--$\rho$ boundary conditions leading to a massless mode are displayed in the left panel of fig.~\ref{fig:dil_instabilities}. Those corresponding to $C_3=0$ lie on the diagonal $\theta_1=-\,\theta_2$, on account of the second of eqs.~\eqref{cond_sing1n1}, while the second of eqs.~\eqref{eigenvmu0_rhoinf}, with $m\simeq 0.26\,i$ gives
\beq
\frac{C_1}{C_2} \ = \tan \theta_1 \ = \ - \ \frac{\xi_2\left(\frac{1}{3},i\,a\right)}{\xi_2\left(-\,\frac{1}{3},i\,a\right)} \ \simeq\ - \ 0.74 \ ,
\eeq
to be compared with the coefficient 0.71 that enters eq.~\eqref{gmodel_leading}, which is thus captured up to an error of about 3\%.
The resulting stability region is displayed in the right panel of fig.~\ref{fig:dil_instabilities}. A large portion of the moduli space is excluded, but there is nonetheless a wide range of boundary conditions that can make the dilaton massive.  An alternative procedure to determine the shift $a$, which we shall favor in the following, is to determine it so that the ratio of $C_1$ and $C_2$ coincides with the result that can be deduced from the exact zero mode, which would be $-\,0.71$ in this case. This would lead to $a\simeq 0.3$, and to a very similar stability region.

Summarizing, the requirement of stability removes a large portion of the moduli space of self--adjoint boundary conditions for the dilaton, and massless modes are only present on the curve in the left panel of fig.~\ref{fig:dil_instabilities}. There are two equivalent special points on this curve, which correspond to the solution of eq.~\eqref{zeromoddil}. Using the same boundary conditions for the axion, which solves an identical equation, can thus lead to a pair of massless scalars. With this choice, no additional instabilities can emerge from the modes with $\mathbf{k} \neq 0$, since the corresponding spectrum is lifted in mass, as is manifest in eq.~\eqref{vdilgen}.

\section{\sc The Perturbed Type--IIB Three--Forms} \label{sec:IIB3forms}

We can now turn to the modes of the two type--IIB three-forms. This is a more intricate sector, with some unfamiliar features, as we are about to see.
In the following, it will be convenient to work with the complex combination
\beq{}{}{}{}{}{}{}
{\cal H} \ \equiv \ d{\cal B} \ = \ d\, {{\cal B}}_2^1 \ + \ i\, d\, {{\cal B}}_2^2
\eeq
of the type--IIB three--form field strengths,
so that ${\cal B}$ will now denote the corresponding complex two--form gauge field. In this notation the self-adjoint action describing the quadratic fluctuations in this sector~\cite{IIB} becomes
\beq{}{}{}{}{}{}{}{}{}{}{}{}{}{}{}{}{}{}{}{}{}{}{}
{\cal S} \ = \ \frac{1}{4\,k_{10}^2}\, \int_{\cal M} \Big[ - \ \frac{1}{2} \ \overline{\cal B} \,\wedge \, d\,\star\, {\cal H}  \ - \ \frac{1}{2} \ {\cal B} \,\wedge \, d\,\star\, \overline{\cal H} \ - \ i\left( \overline{\cal B}\,\wedge {\cal H}\ - \  {\cal B}\,\wedge \overline{\cal H} \right)\,\wedge\,{\cal H}_5^{(0)}\Big] \ , \label{action_cs}
\eeq
and the corresponding field equations read~\footnote{The factor 2 in this equation and the previous ones is consistent with the original work in~\cite{IIB}, and the differences with respect to the excellent work in~\cite{bkrvp} reflect our different definition of ${\cal H}_5$ and our identical normalization for the three-forms.}
    \beq
    d \star {\cal H} \ = \ - \ 2\,{i} \, {\cal H} \wedge  {\cal H}_5^{(0)} \ , \label{form_eq}
    \eeq
or, in components
\beq{}{}{}{}{}{}{}{}{}{}{}{}{}{}{}{}{}{}{}{}{}{}
D^M\,{\cal H}_{MNP} \ = \ - \ \frac{2\,i}{3}\, {\cal H}_5^{(0)}{}_{NPQRS}\ {\cal H}^{QRS} \ .
\eeq
These equations are clearly invariant under
\beq{}{}{}{}{}{}{}{}{}{}{}{}{}{}{}{}{}{}{}{}{}{}{}{}{}
\delta\,{\cal B} \ = \ d\,\Lambda \ ,
\eeq
while the boundary contributions to the variation of the action,
\bea
\delta\,{\cal S} \ = \ \frac{1}{4\,k_{10}^2}\, \int_{\partial\,{\cal M}} \Big[ - \ \frac{1}{2} \ \overline{\Lambda} \,\wedge \left( d\,\star\, {\cal H}  \ +\ 2\,{i} \, {\cal H} \wedge  {\cal H}_5^{(0)}\right)  \ + \ \mathrm{c.c} \ \Big] \ , \label{action_cs2}
\eea
 only vanish on shell. Note that, even if one started from the conventional first--order action, the Chern--Simons term would still yield a non--vanishing contribution. There are two options to deal with this kind of problem: one can constrain the gauge parameter to vanish on the boundary, or alternatively one can introduce Stueckelberg fields living there that compensate the variation in eq.~\eqref{action_cs2}. In this case the second option would rest on a complex nine--dimensional one--form $A$ living on the boundary such that
 \beq
\delta\,A \ = \ - \ \left. \Lambda \, \right|_{\partial{\cal M}} \ , \label{gauge_A}
 \eeq
and on the modified action
\beq{}{}{}{}{}{}{}{}{}{}{}{}{}{}{}{}{}{}{}{}{}{}{}
{\cal S}_{\mathrm{tot}} \ = \ {\cal S} \ - \ \frac{1}{4\,k_{10}^2}\, \int_{\partial\,{\cal M}} \Big[ \frac{1}{2} \ \overline{A} \,\wedge \left( d\,\star\, {\cal H}  \ +\ 2\,{i} \, {\cal H} \wedge  {\cal H}_5^{(0)}\right)  \ + \ \mathrm{c.c} \ \Big]
\ . \label{action_cs3}
\eeq
Varying $\overline{A}$ and its complex conjugate $A$ one simply obtains eqs.~\eqref{form_eq} reduced to the boundary. Removing $A$ via eq.~\eqref{gauge_A} is possible, but at the cost of limiting the residual gauge transformations to those vanishing on the boundary, as we have said. When $A$ is retained, one does not modify the bulk equations provided the boundary term is stationary under variations of ${\cal B}$. 
The complete variation of the action~\eqref{action_cs3} reads
\beq
\delta\,{\cal S}_{\mathrm{tot}} \ = \  \frac{1}{4\,k_{10}^2}\, \int_{\cal M} \left[-\ \frac{1}{2}\ \delta\,\overline{\cal B} \left(  d \star {\cal H}\  + \ 2\,{i} \, {\cal H} \wedge  {\cal H}_5^{(0)}\right) \ + \ \mathrm{c.c.} \right] \ + \ 
\left. \delta\,{\cal S}_{\mathrm{tot}}\right|_{\partial_{\cal M}} \, , \label{delta_tot}
\eeq
and includes a boundary term, whose form we can now spell out after introducing the $9+1$ dimensional decomposition
\beq
{\cal B} \ = \ {\cal C} \ + \ {\cal D}\ dr \ , \qquad d \ = \ d_9 \ + \ dr \ \partial_r \ , \label{Bd9}
\eeq
where ${\cal C}$ is a nine--dimensional two-form and ${\cal D}$ is nine--dimensional one-form. In terms of $A$
and of these two quantities, one can see the total boundary term in eq.~\eqref{delta_tot} is
\bea
\left. \delta\,{\cal S}_{\mathrm{tot}}\right|_{\partial_{\cal M}} &=&  \frac{1}{4\,k_{10}^2}\, \int_{\partial{\cal M}} \Bigg\{\ \delta\,{\cal C} \left[- \ \frac{1}{2}\ e^{-B} \star_9\left( d_9 \,\overline{\cal D} \,+\,\partial_r\,\overline{\cal C} \,\right) \ + \ i\left(\overline{\cal C} \,+\, d_9\,\overline{A} \right) \wedge {\cal H}_5^{(0)} \right] \nonumber \\
&+& \frac{1}{2}\ e^{-B}\ \partial_r\,\delta\,{\cal C} \, \star_9\left(\overline{\cal C} \,+\, d_9\,\overline{A} \right) \ + \ \frac{1}{2}\, e^{-B} \delta\,{\cal D}\ d_9 \star_9 \left(\overline{\cal C} \,+\, d_9\,\overline{A} \right) \nonumber \\
&-& \delta\,\overline{A} \left[ i\ d_9\,{\cal C} \wedge {\cal H}_5^{(0)} \ + \ \frac{1}{2} \ e^{-B} \, d_9 \star_9 \left( d_9 \,{\cal D} \,+\,\partial_r\,{\cal C} \,\right) \right] \ + \ \mathrm{c.c} \Bigg\} \ ,
\eea
where $\star_9$ denotes the nine--dimensional \emph{curved} ($r$-dependent) Hodge dual.

The contribution involving $\delta\,{\cal D}$ vanishes provided
\beq
d_9 \star_9 \left({\cal C} \,+\, d_9\,{A} \right) \ = \ 0  \ .  \label{412}
\eeq
This is an equation of motion for ${A}$ that describes a complex vector, massless in four dimensions, for which the divergence of $C$ is a source. Moreover, the contribution proportional to $\delta A$ is the induced equation of motion on the boundary for the original bulk fields, which is tantamount to eq.~\eqref{form_eq} if considered together with its $r$-component
\beq
e^B \, d_9 \star_9 d_9\, {\cal C} \ + \ \partial_r\left( e^{-B}\, \star_9\, \partial_r\,{\cal C}\right) \ + \ 2\,i\, \left(d_9\,{\cal D} \,+\, \partial_r\,{\cal C} \right) \wedge {\cal H}_5^{(0)} \ = \ 0 \ .
\eeq
Finally, the vanishing of the remaining terms,  
\bea
\delta\,{\cal C} \left[- \ \frac{1}{2}\ e^{-B} \star_9\left( d_9 \,\overline{\cal D} \,+\,\partial_r\,\overline{\cal C} \,\right) \, + \,i\left(\overline{\cal C} \,+\, d_9\,\overline{A} \right) \wedge {\cal H}_5^{(0)} \right] + \frac{1}{2}\ \partial_r\,\delta\,{\cal C} \, \star_9\left(\overline{\cal C} \,+\, d_9\,\overline{A} \right) ,
\eea
is a gauge--invariant counterpart of the self--adjoint condition
\beq
\left. \Big( \partial_z \, \Psi \, \delta\,\Psi \ - \ \Psi \, \partial_z\,\delta\,\Psi \Big) \right|_{\partial{\cal M}} \ = \ 0  \ ,
\eeq
discussed at length in~\cite{ms_23_3} for a scalar field, together with the corresponding expression for gravity.

Let us now address the gauge fixing of this formulation, taking into account that the gauge parameter $\Lambda$ can be decomposed according to
\beq
\Lambda \ = \ \Lambda_2 \ + \ \Lambda_1 \ dr \ ,
\eeq
where we have distinguished in it a nine--dimensional two-form $\Lambda_2$ and a nine--dimensional one-form $\Lambda_1$.
Then, taking eq.~\eqref{Bd9} into account, one can conclude that
\beq
\delta\,{\cal C} \ = \ d_9\,\Lambda_2 \ , \qquad \delta\,{\cal D} \ = \ d_9\,\Lambda_1 \ + \ \partial_r\,\Lambda_2 \ , \qquad \delta\,A \ = \ - \ \left. \Lambda_2 \, \right|_{\partial{\cal M}} \ .
\eeq
There are two options at this point.
\begin{itemize}
\item A first option is removing ${\cal D}$ altogether, which leaves a residual gauge symmetry associated to $\Lambda_2$ gauge parameters satisfying the condition
\beq
\partial_r\,d_9\,\Lambda_2 \ = \ 0 \ .
\eeq
The bulk equations of motion then become
\bea
&& e^B \, d_9 \star_9 d_9\, {\cal C} \ + \ \partial_r\left( e^{-B}\, \star_9\, \partial_r\,{\cal C}\right) \ + \ 2\,i\, \partial_r\,{\cal C} \, \wedge {\cal H}_5^{(0)} \ = \ 0 \ , \nonumber \\
&& \ d_9\left[i\,{\cal C} \wedge {\cal H}_5^{(0)} \ + \ \frac{1}{2} \ e^{-B} \, \star_9 \,\partial_r\,{\cal C}\right] \ = \ 0 \ .
\eea
A complex massless vector field $A$ lives on the boundary, and the solution for ${\cal C}$ is a source for it. Strictly speaking, it might seem that two different $A$ fields are left at the two ends of the interval, but the residual gauge symmetry with an $r$--independent $\Lambda_2$ can be used to remove one of them.
\item There is a second option, which is less convenient. It consists in enforcing the Lorentz gauge condition $d \star {\cal B} =0$, which becomes
\beq
d_9 \, \star_9\,{\cal D} \ = \ 0 \ , \qquad \star_9\,\partial_r\,{\cal D} \ - \ d_9\,\star_9\,{\cal C} \ = \ 0 
\eeq
in the 9+1 decomposition. This turns eq.~\eqref{form_eq} into
\beq
\Box_{10}\, {\cal B} \ - \ 2 \,i \,\star\left(d\,{\cal B} \wedge {\cal H}_5^{(0)}\right) \ = \ 0 \ ,
\eeq
so that the kinetic contribution becomes simpler, but the equations still involve both ${\cal C}$ and ${\cal D}$.
\end{itemize}

In order to make progress, it is now important to take into account the detailed form of the background.
To this end, it is useful to distinguish the spacetime and toroidal coordinates among the nine residual ones. In this fashion, one can identify in ${\cal B}$ a four--dimensional two-form, two types of one--forms and scalars, according to
\bea
&& b \ = \ \frac{1}{2}\,{\cal B}_{\mu\nu}\, dx^\mu\,dx^\nu \ , \quad a \ = \ {\cal B}_{\mu r}\, dx^\mu \ , \quad a_i \ = \ {\cal B}_{\mu\,i} \, dx^\mu \ , \quad {\cal B}_{ri} \ , \quad {\cal B}_{ij} \ ,
\eea
and one can similarly decompose $\Lambda$ into a four--dimensional one-form $\lambda$ and scalars $\Lambda_r$ and $\Lambda_i$, so that
\bea{}{}{}{}{}{}{}{}{}{}{}{}{}{}{}{}{}{}{}{}{}{}{}{}{}
&& \delta\,b \ = \ d_4\,\lambda \ , \qquad \delta\, a \ = \ d_4\,\Lambda_r \ -\ \partial_r\,\lambda \ , \qquad  \delta\,{\cal B}_{ij} \ =  \ i\left(k_i\,\Lambda_j \ - \ k_j\,\Lambda_i\right) \ , \nonumber \\  &&\delta\, a_i \ = \ d_4\,\Lambda_i  \ - \ i\,k_i\,\lambda \ , \qquad \delta\,{\cal B}_{ri} \ = \ \partial_r\,\Lambda_i \ - \ i\,k_i\,\Lambda_r \ . \label{gauge_transf_Lambda}
\eea

For $\mathbf{k}\neq 0$, in the spirit of what we said in the Introduction, one can eliminate the longitudinal parts of ${\cal B}_{ij}$, ${\cal B}_{ri}$ and $a_i$. Both for $\mathbf{k}\neq 0$ and for $\mathbf{k}=0$, in view of the preceding discussion, $a$ and ${\cal B}_{ri}$ could be eliminated using the gauge parameters $\lambda$ and $\Lambda_i$. However, in all cases one should retain on the boundary the $A$ field, within which one can distinguish a complex four--dimensional one-form $A_4$ and five complex scalars $A_i$. There is however a further simplification: the Chern--Simons term plays a role only in the equations for ${\cal B}_{\mu\nu}$ and ${\cal B}_{\mu r}$, on account of the special form of the five--form field strength present in the background of eqs.~\eqref{back_epos_fin2}.

We can now examine the available modes, treating separately those corresponding to different $SO(5)$ representations since, as we have stressed in the Introduction, this is an internal symmetry for the $\mathbf{k}=0$ sector, on which our analysis is largely focused. 
\subsection{\sc The Modes Originating from ${\cal B}_{\mu\nu}$ and ${\cal B}_{\mu r}$} \label{sec:Bmn and Bmr}

The modes originating from  ${\cal B}_{\mu\nu}$ and ${\cal B}_{\mu r}$ entail some complications, due to the role played by the Chern--Simons term that, as we have seen, contributes to the small fluctuations due to the five--form background profile~\eqref{back_epos_fin2}. Let us now focus on the modes of this type with $\mathbf{k}=0$.
In this case there are only two types of curvature components, ${\cal H}_{\mu\nu\rho}$ and ${\cal H}_{\mu\nu r}$, so that the decomposition
    \beq
    {\cal B} = \ \,b \ + \ a\, dr \ ,
    \eeq
    where $b$ is a four--dimensional complex two-form and $a$ is a four-dimensional complex one-form, translates into
    \beq{}{}{}{}{}
    {\cal H} \ = \ d_4 \,b\ + \ \left(\partial_r\,b + d_4 \,a\right) dr \ , \label{GBmunu}
    \eeq
    with $d_4$ the four--dimensional exterior derivative. The gauge transformations act on $a$ and $b$ as
    \bea{}{}{}{}{}{}{}{}{}{}{}
    \delta \,a \ = \ d_4\,\Lambda_r \ - \ \partial_r\,\Lambda \ , \qquad \delta\,b \ = \ d_4\,\Lambda \ ,
    \eea
    and allow one to remove $a$ altogether, but at the expense of introducing an $A$ field on the boundary, as we have stressed. One is thus left with the system
\bea
&& d_4\Big[ e^{-4A}\,\star_4 \,\partial_r\,b   \ - \ \frac{i\,h}{\rho}\,b \Big] \ = \ 0 \ , \nonumber \\
&& e^{2A+10C}\,d_4\,\star_4 \,d_4\,b \,+\, \partial_r \left[e^{-4A}\,\star_4 \,\partial_r\,b \ - \  \frac{i\, h}{\rho} \,b \right]\ = \ 0  \label{eqsbm1}
\eea
in the bulk, and on the boundary
\beq
e^{5C}\ \Big[\,d_4 \,\star_4\,d_4 \,A \ + \ d_4\,\star_4\,b \Big] \ = \ 0 \ . \label{eq_boundary}
\eeq
Here $\star_4$ denotes the four--dimensional \emph{flat} Hodge dual, and we have used the harmonic gauge condition~\eqref{harm_gauge} for the background. As we have seen, a residual $r$--independent gauge transformation can remove $A$ from one of the two boundaries, where the last equation still sets to zero the divergence of $b$. 

It is now convenient to define
\beq{}{}{}{}{}{}{}{}{}{}{}{}{}{}{}{}{}{}{}{}{}{}
u \ = \ e^{-4A}\,\star_4 \,\partial_r\,b   \ , \label{mass-shell-u}
\eeq
so that eqs.~\eqref{eqsbm1} become
\bea
&& d_4\,u  \ - \ \frac{i\,h}{\rho}\,d_4\,b \ = \ 0 \ , \label{eqsbm2} \\
&& e^{2A+10C}\,d_4\,\star_4 \,d_4\,b \,+\, \partial_r \, u \ + \  \frac{i\, h}{\rho} \,\star_4\, e^{4A}\,u\ = \ 0 \ , \nonumber
\eea
and combining them leads to an equation for $u$ that is of first order in $r$:
\beq{}{}{}{}{}{}{}{}{}{}{}{}{}{}{}{}{}{}{}{}{}{}
\partial_r \, u \  - \ \frac{i\,\rho}{h}\,e^{2A+10C}\, d_4 \,\star_4\,d_4\,u\ + \frac{i\, h}{\rho} \,e^{4A}\,\star_4\, u\ = \ 0 \ . \label{equ}
\eeq
Note that, using eqs.~\eqref{GBmunu}, \eqref{mass-shell-u} and the first of eqs.~\eqref{eqsbm2}, one can link ${\cal H}$ to $u$ according to
\beq
{\cal H} \ = \ - \ \frac{i\,\rho}{h} \ d_4\,u \ - \ e^{4A}\,\star_4 u \ dr \ ,  \label{HdU1}
\eeq
or
\beq
{\cal H} \ = \ - \ \frac{i \rho}{h} \left( \partial_r \,u\ dr \ + \ d_4\,u\right) \ - \ \left(\frac{\rho}{h} \right)^2 \ e^{2A+10C} \ d_4 \,\star_4\,d_4\,u \ dr \ . \label{HdU}
\eeq

One can now exhibit different modes within $u$, making use of the key identity
\beq{}{}{}{}{}{}{}{}{}{}{}{}{}{}{}{}{}{}{}{}{}{}
\star_4 d_4 \star_4 d_4 \ + \ d_4 \star_4 d_4\, \star_4 \ = \ - \ \Box  \ = \ - \ m^2 \ , \label{massshellu}
\eeq
which defines the four--dimensional mass--shell condition.
To this end, one can apply $d_4$ and then $d_4\,\star_4$ to eq.~\eqref{equ}, and combining the result with the $d_4$ of  eq.~\eqref{massshellu} leads to the system
\bea{}{}{}{}{}{}{}{}{}{}{}{}{}{}{}{}{}{}
&& \partial_r \, d_4\,u \ + \ \frac{i\, h}{\rho} \,e^{4A}\ d_4\,\star_4\, u\ = \ 0 \ , \nonumber \\
&& \partial_r \, d_4\,\star_4\, u \ + \  \frac{i\,m^2\,\rho}{h}\,e^{2A+10C}\, d_4 \,u\ - \frac{i\, h}{\rho} \,e^{4A}\, d_4\,u\ = \ 0 \ . \label{eqs4du}
\eea
The first equation can now be solved for $d_4\,\star_4\,u$, 
and substituting the result into the second gives a second--order equation in $r$ involving only $d_4\,u$ and not $d_4\star_4 u$:
\beq{}{}{}{}{}{}{}{}{}{}{}{}{}{}{}{}{}{}{}{}{}{}
\partial_r\left(e^{-4A}\,\partial_r\,d_4\,u\right) \ + \ m^2\, e^{2A+10C}\,d_4\,u \ - \ \left(\frac{h}{\rho}\right)^2\, e^{4A}\,d_4\,u \ = \ 0 \ . \label{secordu}
\eeq

One can now separate variables in eq.~\eqref{secordu}, letting
\beq{}{}{}{}{}{}{}{}{}{}{}{}{}{}{}{}{}{}{}{}{}{}{}{}{}
d_4\,u(x,r) \ = \ d_4\,U(x)\, f(r) \ ,
\eeq
which is equivalent to
\beq
u(x,r) \ = \ U(x)\, f(r) \ + \ d_4\,\Lambda(x,r) \ , \label{uUf}
\eeq
with $\Lambda(x,r)$ a four--dimensional one-form also depending on $r$, which is not necessarily separable. This decomposition should be consistent with the initial equation~\eqref{equ}, and substituting in it eq.~\eqref{uUf} leads to
\bea{}{}{}{}{}{}{}{}{}{}{}{}{}{}{}{}{}{}{}{}{}{}{}{}{}
\partial_r\, d_4\,\Lambda(x,r) &+& \frac{i\,h}{\rho}\, e^{4A} \, \star_4\,d_4\,\Lambda(x,r) \, = \, - \, \frac{i\,h}{\rho}\, e^{4A}\,\star_4\,U(x)\,f(r) \,-\,  U(x)\,f'(r) \nonumber \\ &+& \frac{i\,\rho}{h}\ e^{2A+10C}\, f(r) d_4\,\star_4\,d_4\,U(x) \ . \label{consistency1}
\eea
The parameter $\Lambda$ of eq.~\eqref{uUf} is defined up to an exact form, and consequently it can be chosen to be transverse. The exterior derivative of this equation then gives
\beq{}{}{}{}{}{}{}{}{}{}{}{}{}{}{}{}{}{}{}{}{}{}{}{}{}{}{}
d_4\, \star_4\,U(x)\,f(r) \ - \ m^2\,\star_4\,\Lambda(x,r) \ = \ \frac{i\,\rho}{h}\, e^{-4A}\, d_4 U\, f'\ , \label{constr_L}
\eeq
which determines $\Lambda$ algebraically for the modes with $m^2 \neq 0$. One can verify that eq.~\eqref{consistency1} is identically satisfied if one makes use of this solution for $\Lambda(x,r)$. Therefore, the massive spectrum can be analyzed referring solely to a Schr\"odinger system, as in Section~\ref{sec:dilatonaxion}, starting from eq.~\eqref{secordu}. 

\subsection{\sc The Massive Sector} \label{sec:massive_bmn}

As we have seen, all the preceding equations are consistent with the separation of variables~\eqref{uUf} in the massive case, where $\Lambda$ is determined algebraically by eq.~\eqref{constr_L}. One can then focus on eq.~\eqref{secordu}, assuming that $d_4\,u\neq 0$. In fact, if $d_4\,u=0$, eqs.~\eqref{eqs4du} imply that $d_4\,\star_4\,u=0$, and taking these results into account eq.~\eqref{massshellu} implies that $m=0$. We shall return to this case later.

\begin{figure}[ht]
\centering
\begin{tabular}{cc}
\includegraphics[width=65mm]{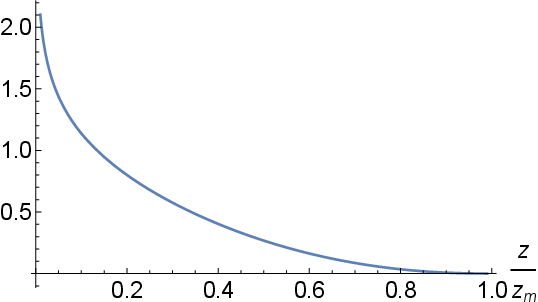} \qquad \qquad &
\includegraphics[width=65mm]{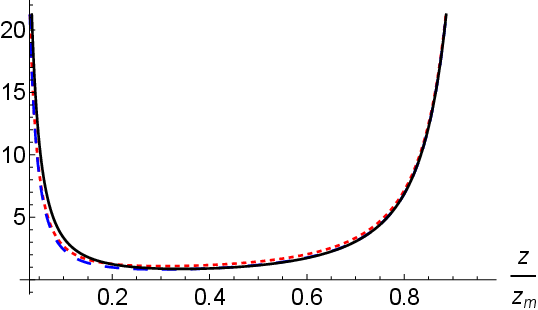} \\
\end{tabular}
\caption{\small The left panel shows the normalized zero--mode wavefunction~\eqref{zeromode_Bmn}. The right panel compares the actual $B_{\mu\nu}$ potential (black, solid), its approximation~\eqref{pot_hyp} with $(\mu,\tilde{\mu})=\left(\frac{2}{3},1.72\right)$ (red, dotted) and the improvement of the latter obtained with a slight negative shift, all in units of $\frac{1}{{z_0}^2}$. $z_m$ and $z_0$ are defined in eqs.~\eqref{zm_app} and \eqref{z0}.}
\label{fig:pot_Bmn}
\end{figure}

For the time being, let us thus concentrate on modes with $d_4u \neq 0$, which are complex massive two-forms and are dual to complex massive vectors in the resulting four--dimensional spacetime. In view of eq.~\eqref{secordu} $f$ satisfies
\beq
\partial_r\left(e^{-4A}\,\partial_r\,f\right) \ + \ m^2\, e^{2A+10C}\,f \ - \ \left(\frac{h}{\rho}\right)^2\, e^{4A}\,f \ = \ 0 \ .
\eeq
In terms of the conformal variable $z$ defined in eq.~\eqref{dzdr}, this equation takes the form
\beq{}{}{}{}{}{}{}{}{}{}{}{}{}{}{}{}{}{}{}{}{}{}
- \left(\partial_z\,+\,B_z\,-5\,A_z\right)\partial_z\,f \ + \ \left(\frac{h}{\rho}\right)^2\, e^{2\left(A-5C\right)}\,f \ = \ m^2\,f \ ,
\eeq
and the further redefinition
\beq{}{}{}{}{}{}{}{}{}{}{}{}{}{}{}{}{}{}{}{}{}{}
f(r) \ = \ e^{\frac{5\,A\,-\,B}{2}}\, \Psi(z) \label{fpsi}
\eeq
finally yields the Schr\"odinger--like equation
\beq
\widetilde{\cal A}\,\widetilde{\cal A}{}^\dagger\,\Psi \ = \ m^2 \,\Psi \label{schrodpsi} \ .
\eeq
The two operators
\beq
\widetilde{\cal A} \ = \ \partial_z \ + \ \frac{7}{2}\,A_z \ + \ \frac{5}{2}\,C_z \ , \qquad \widetilde{\cal A}{}^\dagger \ = \  - \ \partial_z \ + \ \frac{7}{2}\,A_z \ + \ \frac{5}{2}\,C_z \ ,
\eeq
can be identified after making use of the identities for the background in eqs.~\eqref{hamiltonian_F},
and the resulting Schr\"odinger potential,
\beq
V \ = \ \frac{1}{64\,z_0^2} \ \frac{e^{\sqrt{\frac{5}{2}} \frac{r}{\rho}}}{\sinh\left(\frac{r}{\rho}\right)^3} \left[2 \sqrt{10} \, \sinh\left(\frac{2\,r}{\rho}\right)\ + \ \cosh \left(\frac{2\,r}{\rho}\right)+27
\right] \ , \label{schroV}
\eeq
is displayed in fig.~\ref{fig:pot_Bmn}.

Eq.~\eqref{schrodpsi} now implies that
\beq
\Psi \ = \ C \ e^\frac{7\,A\,+\,5\,C}{2} \label{zeromode_Bmn}
\eeq
is an exact normalizable zero mode of the Schr\"odinger system. Note that this result translates into
\beq
f(r) \ = \ C\, e^{4 A} \ =  \ \frac{C}{h\,\sinh\left(\frac{r}{\rho}\right)} \ , \label{Fdist}
\eeq
in view of eq.~\eqref{fpsi}, consistently with eq.~\eqref{zeromode_u}. Taking the measure into account, the zero mode~\eqref{Fdist} corresponds to the normalized $r$-distribution
\beq
\Pi_f(r) \ = \ \frac{1}{\rho}\ \sqrt{\frac{5}{2}} \ e^{\,-\,\frac{r}{\rho}\,\sqrt{\frac{5}{2}}}  \ .
\eeq
The arguments presented in~\cite{selfadjoint} show that the self--adjoint boundary conditions satisfied by this zero mode would identify a complete spectrum of excitations for the Schr\"odinger operator~\eqref{schrodpsi} that is free from tachyonic instabilities. However, we originally assumed that $m \neq 0$, and in fact this expression for $f(r)$ \emph{does not} satisfy eq.~\eqref{constr_L} if $m =0$, unless the two conditions
\beq
d_4\, \star_4\,U(x) \ = \ 0 \ , \qquad d_4\,U(x) \ = \ 0 
\eeq
hold. Moreover, even if these conditions hold, one should make sure that the original equation~\eqref{equ} be satisfied. This condition will be spelled out in detail in the next two sections, where the actual nature of the zero modes will emerge. In contrast, when $m \neq 0$ the Schr\"odinger problem is \emph{equivalent} to the original equation. 

Summarizing, from the Schr\"odinger system one can retain, without further ado, the whole spectrum aside from the zero mode that can possibly be present. When a zero mode is present, it must be handled with care, as we shall see. Nonetheless, the zero mode profile~\eqref{Fdist} will now prove useful in characterizing the optimal shift of the hypergeometric approximation of the actual potential~\eqref{schroV}.

At the two ends the potential is dominated once more by the behavior in eq.~\eqref{lim_dil_axion_intro},
with $\mu=\frac{2}{3}$ and $\tilde{\mu}=1.72$. Consequently, we are now in case 2 of Table~\ref{tab:tab_munu}, so that the limiting behavior near the left end is given by the first of eqs.~\eqref{b.10}, while near $z_m$ it is given by the last of eqs.~\eqref{b.11}, and is actually fixed to be
\beq
\psi \ \sim \left(1 \ - \ \frac{z}{z_m}\right)^{2.22} \ . \label{bcbmnright}
\eeq
Self--adjoint boundary conditions thus depend on a single parameter characterizing, close to the origin, the relative weight of the two independent contributions
\beq
\psi \ \sim \ C_1\, \left(\frac{z}{z_m}\right)^\frac{7}{6} \ + \  C_2\, \left(\frac{z}{z_m}\right)^{\,-\,\frac{1}{6}} \ , \label{psic1c2}
\eeq
which can be identified with the ratio between $C_2$ and $C_1$. For the zero mode~\eqref{zeromode_Bmn} the self--adjoint boundary conditions are determined by
\beq
\frac{C_2}{C_1} \ \sim \ - \ 2.85 \ . \label{masslessB}
\eeq
This result can determine an optimal shift of the hypergeometric approximation~\eqref{pot_hyp} of the potential. One can see from fig.~\ref{fig:hyperex} that, in the hypergeometric model, the ground state should correspond to $m \simeq 0.99$, so that the optimal potential for this sector is
\beq
V \ \simeq \ V\left(\frac{2}{3},1.72,z\right) \ - \ \frac{\pi^2}{z_m^2}\ (0.99)^2 \ .
\eeq
Taking the shift into account, the hypergeometric eigenvalue equation becomes in general
\beq
\frac{C_2}{C_1} \ \simeq \ - \left(\frac{2}{\pi}\right)^{\frac{4}{3}} \ \frac{\Gamma\left(\frac{5}{3}\right)\Gamma\left(1.03\,+\,\sqrt{m_{B_{\mu\nu}}^2+(0.99)^2}\right) \Gamma\left(1.03\,-\, \sqrt{m_{B_{\mu\nu}}^2+(0.99)^2}\right)}{\Gamma\left(\frac{1}{3}\right)\Gamma\left(1.69\,+\,\sqrt{m_{B_{\mu\nu}}^2+(0.99)^2}\right) \Gamma\left(1.69\,-\,\sqrt{m_{B_{\mu\nu}}^2+(0.99)^2}\right)} \ . \label{massiveB}
\eeq
This approximation reveals the presence of an instability region corresponding to
\beq
- \ 2.85 \ < \ \frac{C_2}{C_1} \ < \ 0  \ .
\eeq
On the other hand, outside this region, if $\frac{C_2}{C_1}$ differs from the ``critical'' value $ -\, 2.85$, the spectrum consists of purely massive complex two-forms, which are dual to complex massive vectors, as we have stressed. 

For $\mathbf{k} \neq 0$, one can gauge away all excitations that are longitudinal in $\mathbf{k}$, and then these perturbations continue not to mix with others. The resulting masses increase, as dictated by the additional positive potential $\mathbf{k}^2 \e^{2(A-C)}$.

We can now turn to a detailed analysis of the zero modes that can be contained in $u$.

\subsection{\sc The Massless Sector}

When the boundary conditions allow a massless sector, the resulting setup is rather peculiar. In analyzing it, we shall explore two solutions of the massless equations of increasing complexity.
\subsubsection{\sc the Case $d_4\,U(x)=0$} \label{sec:d4u0}
    If $d_4\,U(x)=0$, eq.~\eqref{constr_L} implies that $d_4\,\star_4\,U(x)=0$, and then $f(r)$ is arbitrary. Letting
    \beq
d_4\,\Lambda \ = \ - \ U(x) \, f(r) \ +\ d_4\,g(x,r) \ ,
\eeq
so that
\beq
d_4\,g(x,r) \ =  u(x,r) \ .
\eeq
Eq.~\eqref{consistency1}, or equivalently~\eqref{equ}, leads to
\beq{}{}{}{}{}{}{}{}{}{}{}{}{}{}{}{}{}{}{}{}{}{}
\partial_r \, d_4\,g(x,r) \  + \  \frac{i\, h}{\rho} \,e^{4A}\,\star_4\, d_4\,g(x,r) \ = \ 0  \ .\label{equ2}
\eeq
Decomposing now the complex two-form $d_4\,g$ into selfdual and anti-selfdual parts according to
\beq{}{}{}{}{}{}{}{}{}{}{}{}{}{}{}{}{}{}{}{}{}{}
d_4\,g \ = \ \left(d_4\,g\right)^+ \ + \ \left(d_4\,g\right)^- \ ,
\eeq
with
\beq
\left(d_4\,g\right)^\pm \ = \ \frac{1}{2} \left(1\ \mp \ i\, \star_4\right)d_4\,g \ ,
\eeq
leads to
\beq
\left(d_4\,g\right)^\pm \ = \ \left(d_4\,g\right)^\pm(x) \left[\tanh\left(\frac{r}{2\,\rho}\right)\right]^{\pm\,1} \ , \label{upm}
\eeq
and finally to
\beq
u \ = \ \left(d_4\,g\right)^+(x) \ \tanh\left(\frac{r}{2\,\rho}\right)\ + \ \left(d_4\,g\right)^-(x) \coth\left(\frac{r}{2\,\rho}\right)
\eeq
or
\beq
u \ = \ d_4\,g \ = \  d_4\,g(x) \ \coth\left(\frac{r}{\rho}\right)  + \ \frac{i\, \star_4\,d_4\,g(x) }{\sinh\left(\frac{r}{\rho}\right)}  \ . \label{ud4u0}
\eeq
The condition $d_4\,u=0$, which was our starting point, demands that
\beq
d_4\,\star_4\,d_4\,g(x) \ = \ 0  \ . \label{maxwell}
\eeq
Making use of eq.~\eqref{ud4u0} gives
\beq
b \ = \ - \ \frac{i\,\rho}{h} \ u  \ ,
\eeq
up to an $r$-independent two-form that is pure gauge, and using eqs.~\eqref{HdU1},~\eqref{ud4u0} and \eqref{maxwell} gives
\beq
{\cal H} \ = \ \frac{i}{h\,\sinh\left(\frac{r}{\rho}\right)}\left[ i\, \star_4\,d_4\,g(x) \ \coth\left(\frac{r}{\rho}\right)  \ + \ \frac{d_4\,g(x) }{\sinh\left(\frac{r}{\rho}\right)}\right] dr \ .
\eeq

The norm is determined by the sum of the two contributions
\bea
\int\, \overline{\cal H} \, \wedge\,\star\, {\cal H} &=&  V_5 \int \, \frac{dr \ d_4\,\bar{g}\,\wedge\,\star_4\,d_4\,g}{h \ \sinh\left(\frac{r}{\rho}\right)} \ , \nonumber \\
\int\, 2 \, \mathrm{Im} \left(\overline{b} \,\wedge \,{\cal H}\right) \,\wedge\, {\cal H}_5^{(0)} &=& - \ V_5 \int\, \frac{dr\,dV_5 \ d_4\,\bar{g}\,\wedge\,\star_4\,d_4\,g}{h \ \sinh\left(\frac{r}{\rho}\right)}  \ .
\eea
These terms cancel, and actually they are both total derivatives if eq.~\eqref{maxwell}, which is a constraint in this case, is used. Therefore, the modes obtained in this fashion have vanishing norm, and must be rejected. In the next section we shall recover eq.~\eqref{maxwell}, but as an equation of motion, not as a consistency condition. 
\subsubsection{\sc the Case $d_4\,U(x) \neq 0$} \label{sec:d4unot0}
If $d_4\, U(x) \neq 0$, eq.~\eqref{constr_L} with $m=0$ implies that
\beq
f'(r) \ = \  \gamma\,e^{4 A} \, f(r) \ , \qquad
d_4\, \star_4\,U(x) \ = \  \frac{i\,\rho \,\gamma}{h}\  d_4 U(x) \ ,
\eeq
where $\gamma$ is a constant. Substituting in eq.~\eqref{secordu} now leads to
\beq
d_4\,U \left[ \gamma^2 \ - \ \left(\frac{h}{\rho}\right)^2\right] e^{4A}\, f \ = \ 0 \ ,
\eeq
so that
\beq
\gamma  \ = \ \pm \ \frac{h}{\rho} \ .
\eeq
There are thus two solutions of this type, 
\bea
&& f^+(r) \ = \ \tanh\left(\frac{r}{2\,\rho} \right) \ , \qquad d_4\left( 1\ + \ i\, \star_4\right)U^+(x) \ = \ 0 \ , \nonumber \\
&& f^-(r) \ = \ \coth\left(\frac{r}{2\,\rho} \right) \ , \qquad d_4\left( 1\ - \ i\, \star_4\right)U^-(x) \ = \ 0  \ , \label{m0sols}
\eea
where the overall constants are included in $U^\pm(x)$, which are both consistent with the separation of variables~\eqref{uUf}. 
Consequently one can conclude that
\beq
U^+(x) \ = \  U^{++}(x) \ + \ A^{+-}(x) \ , \qquad
U^-(x) \ = \  U^{--}(x) \ + \ A^{-+}(x) \ , 
\eeq
where $U^{++}$ and $A^{-+}$ are selfdual two-forms, $U^{--}$ and $A^{+-}$ are anti-selfdual two-forms, and furthermore
\beq
d_4\,A^{+-} \ = \ 0 \ , \qquad d_4\,A^{-+} \ = \ 0 \ .
\eeq

In general, a complex (anti-)selfdual two-form ${\cal G}_2^\pm$, such that
\beq
\star_4\,{\cal G}_2^\pm \ = \ \pm \,i \, {\cal G}_2^\pm
\eeq
can be expressed in terms of a real two-form $G_2$ according to
\beq
{\cal G}_2^\pm \ = \ \left(1 \  \mp \ i\, \star_4\right)G_2 \ .
\eeq
Consequently one can write
\bea
U^{++} &=&  \left(1 \ - \ i\,\star_4\right)u^+ \ , \qquad A^{+-} \ = \  \left(1 \ + \ i\,\star_4\right)a^+ \ , \nonumber \\
U^{--} &=&  \left(1 \ + \ i\,\star_4\right)u^-\ , \qquad A^{-+} \ = \  \left(1 \ - \ i\,\star_4\right)a^- \ , \label{Upm}
\eea
where $u^\pm$ and $a^\pm$ are real two-forms, with
\beq
d_4 \,a^\pm \ = \ 0 \ , \qquad d_4 \,\star_4\,a^\pm \ = \ 0 \ . \label{aclosed}
\eeq
Collecting all these contributions, the general massless $u$ profile of this type, obtained combining the two separable solutions that we started from, reads
\bea
u(x,r) &=& \left[ a^+ \,+\, u^+\ +\ i \,\star_4\left(a^+ -u^+\right)\right] \, \tanh\left(\frac{r}{2\,\rho}\right) \nonumber \\
&+& \left[a^- \,+\, u^-\ -\ i \,\star_4\left(a^- -u^-\right)\right] \, \coth\left(\frac{r}{2\,\rho}\right) \ + \  d_4 \Lambda(x,r) \ , \label{u_complete}
\eea
where $d_4\,\Lambda$ is determined by eq.~\eqref{equ}. One can thus conclude that
\beq
d_4\,u(x,r) \,=\,d_4\left(1\,-\,i\,\star_4\right)u^+\, \tanh\left(\frac{r}{2\,\rho}\right) \,+\, d_4\left(1\,+\,i\,\star_4\right)u^-\, \, \coth\left(\frac{r}{2\,\rho}\right) \ . \label{d4udec}
\eeq
In particular, for a divergence--free $u$ with $u^+=-u^-$ this expression recovers the zero mode of the Schr\"odinger system.

In order to determine $\Lambda$, taking eqs.~\eqref{aclosed} into account, one can let
\beq
d_4\,\Lambda(x,r) \ = \ - \ \left(1+ i\,\star_4\right) a^+ \, \tanh\left(\frac{r}{2\,\rho}\right) \ - \ \left(1- i\,\star_4\right) a^- \, \coth\left(\frac{r}{2\,\rho}\right) \ + \ d_4\,\lambda(x,r) \ ,
\eeq
thus eliminating all terms involving $a^\pm$ from $u(x,r)$, which becomes
\beq
u(x,r) \,=\,\left(1\,-\,i\,\star_4\right)u^+\, \tanh\left(\frac{r}{2\,\rho}\right) \,+\, \left(1\,+\,i\,\star_4\right)u^-\, \, \coth\left(\frac{r}{2\,\rho}\right) \ + \  d_4 \lambda(x,r) \ , \label{u_complete2}
\eeq
and now $d_4\,\lambda$ can be determined by eq.~\eqref{equ}.
Decomposing it into selfdual and anti-selfdual portions according to
\beq
d_4\,\lambda(x,r)  \ = \ \left(d_4\,\lambda(x,r)\right)^+ \ + \  \left(d_4\,\lambda(x,r)\right)^- \ , \label{lambdapm}
\eeq
and using the identity
\beq
\left(1 \ - \ i\,\star_4\right) d_4 \star_4 d_4 \left(1 \ - \ i\,\star_4\right) \ = \ 0 
\eeq
for massless modes, one can see that $\left(d_4\,\lambda\right)^\pm(x,r)$ satisfy the decoupled first--order equations
\beq
\partial_r\, \left(d_4\,\lambda(x,r)\right)^\pm \ \mp \  \frac{h}{\rho}\, e^{4A} \, \left(d_4\,\lambda(x,r)\right)^\pm \  = \ \frac{i\,\rho}{h}\ e^{2A+10C}\, f^\mp(r) d_4\,\star_4\,d_4\,U^{\mp\mp}(x) \ . \label{consistency13}
\eeq
These are solved by
\beq
\left(d_4\,\lambda\right)^\pm \ = \ \left[ \tanh\left(\frac{r}{2\,\rho}\right)\right]^{\pm\,1} \left[C^\pm(x) \ + \ \frac{i g h}{4} \int_{r_0}^r ds \ e^{-\,\frac{5 s}{\rho \sqrt{10}}} \left( e^\frac{s}{2\rho} \pm e^{-\,\frac{s}{2\rho}}\right)^4 \xi^\mp(x) \ \right]\ , 
\eeq
where the $C^\pm(x)$ are proportional to $\left(d_4\,\lambda\right)^\pm$ at $r=r_0$, and
\beq
\xi^\mp(x) \ = \  d_4 \star_4 d_4 U^{\mp\mp}(x) \ = \ d_4 \star_4 d_4 \left(1 \ \pm i\,\star_4\right)u^\mp(x)\ . \label{xipm}
\eeq
In fact, given the different $r$ dependence of $\left(d_4\,\lambda\right)^\pm$, the decomposition~\eqref{lambdapm} is only consistent if
\beq
d_4\, C^\pm \ = \ 0 \ ,
\eeq
and consequently
\beq
d_4 \star_4 d_4 \,\lambda \ = \ 0 \ , \label{constr_lambda}
\eeq
so that $d_4\,\lambda$ is also co-closed.
One finally obtains
\bea
u(x,r) \!\!\!\!&=&\!\!\!\! \left(1\,-\,i\,\star_4\right)u^+\, \tanh\left(\frac{r}{2\,\rho}\right) \,+\, \left(1\,+\,i\,\star_4\right)u^-\, \, \coth\left(\frac{r}{2\,\rho}\right)  \label{u_completef} \\
\!\!\!\!&+&\!\!\!\! \left[ \tanh\left(\frac{r}{2\,\rho}\right)\right] \left[C^+(x) \, + \, \frac{i g h}{4} \int_{r_0}^r ds \ e^{-\,\frac{5 s}{\rho \sqrt{10}}} \left( e^\frac{s}{2\rho} \,+\, e^{-\,\frac{s}{2\rho}}\right)^4 \xi^-(x) \ \right] \nonumber \\
\!\!\!\!&+&\!\!\!\! \left[ \coth\left(\frac{r}{2\,\rho}\right)\right] \left[C^-(x) \, + \, \frac{i g h}{4} \int_{r_0}^r ds \ e^{-\,\frac{5 s}{\rho \sqrt{10}}} \left( e^\frac{s}{2\rho} \,-\, e^{-\,\frac{s}{2\rho}}\right)^4 \xi^+(x) \ \right]
, \nonumber
\eea
which is consistent with the decomposition~\eqref{d4udec}.

We can now compute the gauge--invariant field strength ${\cal H}$, whose expression in terms of $u(x,r)$ is given in eq.~\eqref{HdU1}. One thus finds
\bea
{\cal H} &=& - \ \frac{i\,\rho}{h}d_4 \left( 1\ -\ i \,\star_4\right)\,u^+ \, \tanh\left(\frac{r}{2\,\rho}\right) \ - \  \frac{i\,\rho}{h} d_4\left(1\ +\ i \,\star_4\right)u^- \, \coth\left(\frac{r}{2\,\rho}\right) \nonumber \\
&-& e^{4 A}\,i\,dr \left(1 \ - \ i\,\star_4\right)u^+ \, \tanh\left(\frac{r}{2\,\rho}\right) \, +\, e^{4A}\,i\, dr\left(1 \ + \ i\,\star_4\right)u^- \, \coth\left(\frac{r}{2\,\rho}\right) \nonumber \\
&-& e^{4 A} \, dr\, \star_4\, d_4 \,\lambda \ , 
\eea
and
\bea
b &=& - \ \frac{i\,\rho}{h} \left(1 \ - \ i\,\star_4\right)u^+ \, \tanh\left(\frac{r}{2\,\rho}\right) \ - \ \frac{i\,\rho}{h}  \left(1 \ + \ i\,\star_4\right)u^-\, \coth\left(\frac{r}{2\,\rho}\right) \nonumber \\
&-& \star_4\,d_4\,\int e^{4 A}\, \lambda(x,r) \ .
\eea

Let us begin by analyzing the asymptotic behavior of these quantities as $r \to \infty$. To this end, it is important to note that in this limit the second contribution on the left--hand side of eqs.~\eqref{consistency13} can be neglected, and consequently
\beq
d_4\,\lambda \ \sim \ \frac{i\,\rho^2\,h}{4\left(2 \ - \ \sqrt{\frac{5}{2}}\right)}\ e^{\frac{r}{\rho}\left(2\,-\,\sqrt{\frac{5}{2}}\right)}\ d_4 \,\star_4\,d_4 \left[\left(1\,+\,i\,\star_4\right) u^-\,+\,\left(1\,-\,i\,\star_4\right) u^+\right] \ ,
\eeq
while 
\beq
{\cal H} \ \sim \ \ - \ \frac{i\,\rho}{h} \left[ d_4 \left( 1\ -\ i \,\star_4\right)\,u^+  \ + \   d_4\left(1\ + \ i \,\star_4\right)u^- \right] \ , 
\eeq
or
\beq
{\cal H} \ \sim \ \ - \ \frac{i\,\rho}{h} \ d_4 \, \mu \ ,
\eeq
where
\beq
\mu \ = \ u^+ \ + \ u^- \ - \ i \ \star_4\left(u^+ \ - \ u^- \right) \ ,
\eeq
since the other contributions are sub--dominant in the limit.

Near the right end of the interval the leading behavior of $b$ is
\beq
b \ \sim  \ - \ \frac{i\,\rho}{h} \ u \  ,
\eeq
and the only two--derivative contribution to the kinetic action integral thus originates from the first term in eq.~\eqref{action_cs}, and is proportional to
\beq
\int\, e^{\left(2-\sqrt{\frac{5}{2}}\right)\frac{r}{\rho}} \ d_4 \,\bar{\mu} \ \wedge \star_4 \ d_4\,\mu \ .
\eeq
Consequently, a finite result only obtains if the two conditions
\beq
d_4\left( u^+ \ + \ u^- \right) \ = \ 0 \ , \qquad d_4\ \star_4 \left( u^+ \ - \ u^- \right) \ = \ 0  \ . \label{5.49}
\eeq
hold for normalizable massless modes. These conditions are solved letting
\beq
u^\pm \ = \ d_4\,\gamma \ \pm \, \star_4\,d_4\,\delta , \label{gamma_delta}
\eeq
where $\gamma$ and $\delta$ are two real one-forms, which could be taken to be divergence-free. Consequently,
the overall content of $u^\pm$ corresponds at most to a pair of real vectors, and 
\beq
\xi^\mp(x) \ = \  \pm \ d_4 \star_4 d_4 \star_4 d_4 \left(\delta - i \gamma\right) \ .
\eeq

Making use of eq.~\eqref{gamma_delta}, one can conclude that
\bea
{\cal H} &=&  \frac{2\, \rho}{h}\ \frac{d_4\,\star_4\,d_4\left(\gamma\,+\,i\,\delta\right) }{\sinh\left(\frac{r}{\rho}\right)} \ - \   \frac{2\,dr \left[\cosh\left(\frac{r}{\rho}\right) \ \star_4 \ - \ i \right]}{h\,\sinh^2\left(\frac{r}{\rho}\right)} \ d_4\left(\gamma \,+\,i\,\delta\right) { \ - \ e^{4 A} \, dr\, \star_4\, d_4 \,\lambda} \ , \nonumber \\
\star {\cal H} &=& {2\, \rho\,h} \ dr \, dV_5 \,   \star_4\,d_4\,\star_4\,d_4\left(\gamma\,+\,i\,\delta\right) \  \sinh\left(\frac{r}{\rho}\right) e^{\,-\,\frac{r}{\rho}\sqrt{\frac{5}{2}}}  \nonumber \\
&-& \frac{2\,d\,V_5 \left[\cosh\left(\frac{r}{\rho}\right) \star_4 \ - \ i \right]}{\sinh\left(\frac{r}{\rho}\right)} \ \star_4\,d_4\left(\gamma \,+\,i\,\delta\right) \ { + \ d_4\,\lambda}\ , \\
b &=& -\ \frac{2\,i\,\rho}{h\,\sinh\left(\frac{r}{\rho}\right)} \left[1 \ + \ i\,\star_4\,\cosh\left(\frac{r}{\rho}\right)\right]\left(d_4\,\gamma\,+\,\star_4\,d_4\,\delta\right) { \ - \ \star_4\,d_4\,\int e^{4 A}\, \lambda(x,r)} \  , \nonumber
\eea
but taking eq.~\eqref{constr_lambda} into account one can see all terms involving $\lambda$ do not contribute to the action~\eqref{action_cs}. Consequently
\bea
\int \ \overline{\cal H} \,\wedge\,\star \,{\cal H} &=& 4\,\rho^2\, \int dr\,dV_5\, d_4\,\star_4\,d_4\left(\gamma\,-\,i\,\delta\right) \,\wedge\,\star_4\,d_4\,\star_4\,d_4\left(\gamma\,+\,i\,\delta\right) \, e^{\,-\,\frac{r}{\rho}\sqrt{\frac{5}{2}}}\nonumber \\ &-& \int \ \frac{4\,dr\,dV_5}{h\,\sinh\left(\frac{r}{\rho}\right)} \ \Big( d_4\,\gamma\,\wedge\,\star_4\,d_4\,\gamma \ + \  d_4\,\delta\,\wedge\,\star_4\,d_4\,\delta \Big) \ , \nonumber \\
- \ 2\ \int \ \mathrm{Im}\left(b \wedge \overline{\cal H}\right)\,\wedge\,{\cal H}_5^{(0)} &=& \int \ \frac{4\,dr\,dV_5}{h\,\sinh\left(\frac{r}{\rho}\right)} \ d_4\,\delta\,\wedge\,\star_4\,d_4\,\delta \ .
\eea

The terms involving $\delta$ that are singular at the origin cancel among the two contributions above, so that finiteness only demands that $\gamma=0$. One is thus left with a real vector $\delta$, whose contribution to the action, in the first term, is finite but contains higher derivatives. 

The corresponding measure is precisely the one captured by the Schr\"odinger system for the massive modes. Keeping only $\delta$ one finds indeed
\bea
{\cal H} &=&  \frac{2\,i\, \rho}{h}\ \frac{d_4\,\star_4\,d_4\,\delta}{\sinh\left(\frac{r}{\rho}\right)} \
 - \   2\,\frac{dr \left[1 \ + \ i \, \star_4\,\cosh\left(\frac{r}{\rho}\right) \right]}{h\,\sinh^2\left(\frac{r}{\rho}\right)} \ d_4\,\delta \ , \nonumber \\
u &=& 2\,\frac{i\,\cosh\left(\frac{r}{\rho}\right)\ - \ \star_4}{\sinh\left(\frac{r}{\rho}\right)} \ d_4\,\delta \ , \nonumber \\
b &=& \frac{2\,\rho}{h\,\sinh\left(\frac{r}{\rho}\right)} \left[\star_4 \ - \ i\,\cosh\left(\frac{r}{\rho}\right)\right] d_4\,\delta \ , \label{u_complete3}
\eea
so that
\beq
d_4\,u \ = \  - \ 2\,\frac{d_4\,\star_4\,d_4\,\delta}{\sinh\left(\frac{r}{\rho}\right)} \ . \label{zeromode_u}
\eeq
We have thus recovered for $d_4\,u$ the $r$ profile of the massless mode of the Schr\"dinger system in eq.~\eqref{Fdist}. However, the $x$ dependence of $u$ is determined by a real vector $\delta$. The results of the preceding section are recovered if 
\beq
d_4\,\star_4\,d_4\,\delta = 0 \ ,
\eeq
so that $\delta$ reduces to the field $g$ introduced there. The novelty here is the absence of this constraint.

The equation for $\delta$ in four dimensions, 
\beq
d_4\,\star_4\,d_4\,\star_4\,d_4\,\delta \ = \ 0 \ , \label{fineq_delta}
\eeq
follows if one substitutes eqs.~\eqref{u_complete3} in the ten--dimensional equations~\eqref{eqsbm1}. It has a peculiar form, and contains an odd number of derivatives. However, it is consistent with the four-derivative equation that follows from the effective action
\beq
{\cal S}_4 \ = \ \int d_4\,\star_4\,d_4\,\delta\,\wedge\,\star_4\,d_4\,\star_4\,d_4\,\delta \ .
\eeq
Note also that only $d_4\,u$ has a separable form in $x$ and $r$.

Denoting $d_4\,\delta$ by $F$, eq.~\eqref{fineq_delta} can be cast in the form
\beq
\partial_{[\mu}\,\partial^\rho\,F_{\nu]\,\rho} \ =  \ 0 \ ,
\eeq
which is equivalent to
\beq
\partial^\rho\,F_{\nu\,\rho} \ = \ \partial_\nu\,\sigma \ , \qquad  \Box\,\sigma \ = \ 0 \ , \label{eqsbmn}
\eeq
where $\sigma$ is a massless scalar field. All in all, one is thus left with a real massless vector and a real massless scalar in this sector. To these massless modes one must add a complex massless vector that lives generically in the boundary.

Summarizing, as we have seen in Section~\ref{sec:massive_bmn}, the massive modes of this sector are complex two-forms, which in four dimensions are dual to complex massive vectors. Naively, when massless modes are present,  what happens in circle compactification would suggest the presence of a complex scalar from ${\cal B}_{\mu\nu}$ and a complex vector from ${\cal B}_{\mu r}$. The surprise is that the peculiar system~\eqref{equ} actually leads to a third--order equation, which results in the halving of the massless modes, in a way that resonates with chiral projections of Fermi systems. In addition, in an interval, as we have seen, another vector mode lives on the boundary, in a way reminiscent of what is familiar for twisted sectors in orbifolds~\cite{orbifolds} or orientifolds~\cite{orientifolds} in String Theory, or from the Horava--Witten construction~\cite{HW}. 

\begin{figure}[ht]
\centering
\includegraphics[width=65mm]{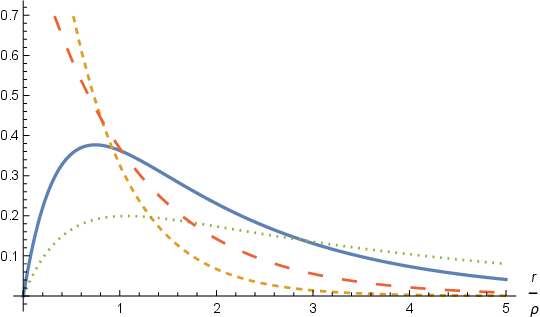}
\caption{\small The $r$--distributions $\Pi_{\phi}\left(\frac{r}{\rho}\right)$ of eq.~\eqref{dil_profile} (blue, solid),
$\Pi_{f}\left(\frac{r}{\rho}\right)$ of eq.~\eqref{Fdist} (yellow, dashed),
$\Pi_{B_{\mu i}}\left(\frac{r}{\rho}\right)$ of eq.~\eqref{pibmui} (green, dotted), $\Pi_{B_{ij}}\left(\frac{r}{\rho}\right)$ of eq.~\eqref{pibij} (red, long-dashed), in units of $\frac{1}{\rho}$. The corresponding mean values of $\frac{r}{\rho}$ are about 2.1, 0.6, 4.2 and 1.1.}
\label{fig:distr_Gdil}
\end{figure}

\subsection{The Modes Originating from ${\cal B}_{\mu i}$ and ${\cal B}_{r i}$}
\begin{figure}[ht]
\centering
\begin{tabular}{cc}
\includegraphics[width=65mm]{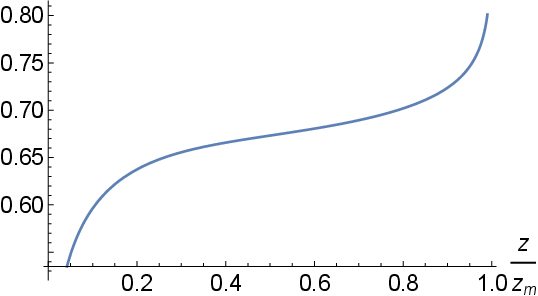} \qquad \qquad &
\includegraphics[width=65mm]{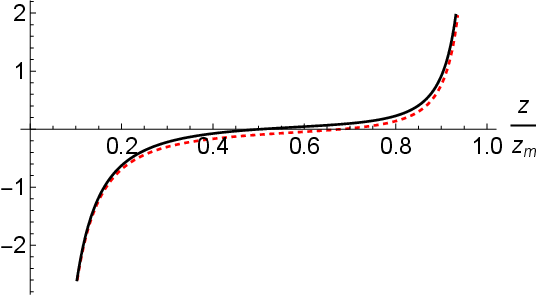} \\
\end{tabular}
\caption{\small The left panel shows the normalized zero--mode wavefunction~\eqref{bmui_prof}. The right panel compares the actual $B_{\mu i}$ potential (black, solid) with its approximation~\eqref{pot_hyp} with $(\mu,\tilde{\mu})=\left(\frac{1}{3},0.54\right)$ (red, dashed), both in units of $\frac{1}{{z_0}^2}$. $z_m$ and $z_0$ are defined in eqs.~\eqref{zm_app} and \eqref{z0}.}
\label{fig:Bmi_pot}
\end{figure}

For this mode sector, one can start from
\beq
{\cal B} \ =\ a_i(x) f(r) \,dy^i  \ ,
\eeq
where the $a_i$ are five 4D complex one-forms since. As we already stressed after eq.~\eqref{gauge_transf_Lambda}, $\Lambda_i$ can be used to remove the ${\cal B}_{ri}$, but as in the previous section this is at the cost of introducing a set of complex scalars $A_i$ living in the boundary.
The field strength corresponding to ${\cal B}$ is in this case
\beq
{\cal H}\ =\ f\,d_4\,a_i\ dy^i +\ f'\,a_i\,dy^i\,dr  \ ,
\eeq
and the equations of motion give
\beq
f'd_4*_4\,a_i\ = \ 0\ ,\qquad m^2fe^{B+3C}+(f'e^{-2C-2A})'\ = \ 0 \ .
\eeq
Letting
\beq
f\ =\ g\,e^{\,-\,\frac{A+3C}{2}} \ ,
\eeq
one obtains a manifestly Hermitian Schr\"odinger--like equation
\beq
\left(\partial_z+\frac{(A_z+3C_z)}{2}\right)\left(-\partial_z+\frac{(A_z+3C_z)}{2}\right)g \ = \ m^2\, g \ , \label{schrod_bmi}
\eeq
of the familiar ${\cal A}\,{\cal A}^\dagger$ form, with the potential
\beq
V \,=\, -\ \frac{1}{4} \left({A_z}\,+\,3 \,{C_z}\right) \left(5 \,{A_z}\,+\,7 \,{C_z}\right)\ -\ 4\, {{\cal W}_5}^2\ . \label{potBmi}
\eeq
This is displayed in fig.~\ref{fig:Bmi_pot} as a function of $z$, together with the corresponding hypergeometric approximation. In terms of $r$ its detailed form is
\beq
V \ = \ \frac{e^{\sqrt{\frac{5}{2}} \frac{r}{\rho}}}{32 \,z_0^2\,\sinh\left(\frac{r}{\rho}\right)}  \left[\sqrt{10}\, \sinh \left(\frac{2 r}{\rho}\right)\ -\ \frac{31}{10}\, \cosh\left(\frac{2 r}{\rho}\right) \ -\ \frac{69}{10}\right] \ .
\eeq

At the two ends of the interval, this potential has the singular limiting behavior of eq.~\eqref{lim_dil_axion_intro}, with $\mu=\frac{1}{3}$ and $\tilde{\mu}=0.54$, and thus belongs to case 3 of Section~\ref{sec:exactschrod}. As for the dilaton-axion pair, there are boundary conditions given independently at the two ends that are parametrized by a pair on angles $(\theta_1,\theta_2)$. 

Before examining the possible choices of boundary conditions, let us remark that the ${\cal A}\,{\cal A}^\dagger$ form of the Schr\"odinger system implies the presence of the zero mode
\beq{}{}{}
g \ = \ g_0 \ e^{\frac{A+3C}{2}} \ ,  \label{bmui_prof}
\eeq
which corresponds to
\beq
f \ = \ g_0 \ ,
\eeq
where $g_0$ is a constant. This zero mode is normalizable, since
\beq
\int_0^{z_m} dz \ e^{A+3C}\ = \ \int_0^\infty dr\  e^{4(A+2C)} <\infty \ ,
\eeq
and the corresponding normalizable $r$-distribution is
\beq{}{}
\Pi_{B_{\mu i}} \ = \  \frac{3}{5\,\rho}\, \left[\sinh\left(\frac{r}{\rho}\right)\right]\ e^{-\frac{4r}{\rho\sqrt{10}}} \ .
\label{pibmui}
\eeq
Note that the zero--mode wavefunction has the dominant behaviors
\beq
g \ \sim \ \left(\frac{z}{z_m}\right)^\frac{1}{6} \ - \ 0.35 \  \left(\frac{z}{z_m}\right)^\frac{5}{6} 
\eeq
close to $z=0$, and
\beq
g \ \sim \ \left(1\ - \ \frac{z}{z_m}\right)^{-\,0.04} \ - \ 0.14 \  \left(1\ - \ \frac{z}{z_m}\right)^{1.04} 
\eeq
close to $z=z_m$. 

\begin{figure}[ht]
\centering
\includegraphics[width=65mm]{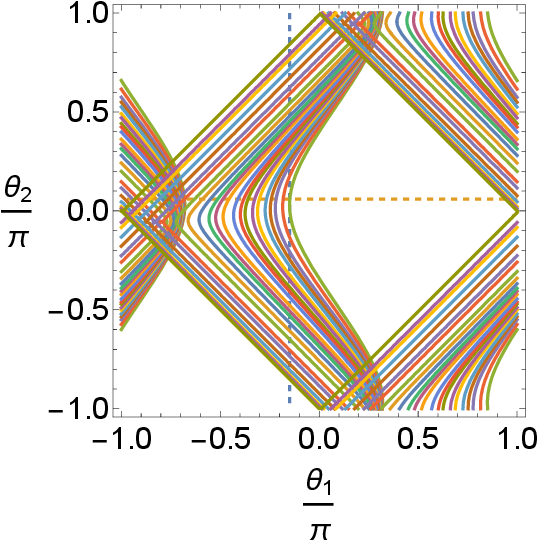}
\caption{\small The point $\left({\theta_1},{\theta_2}\right)=\left(-0.15,0.06\right)\pi$ identifies the special boundary conditions for ${\cal B}_{\mu i}$  corresponding to the zero mode~\eqref{bmui_prof}. The shaded regions identify the boundary conditions leading to instabilities. }
\label{fig:Bmi_instabilities}
\end{figure}

The arguments of~\cite{selfadjoint} lead one to conclude that this whole sector is stable with the boundary conditions of this zero mode, which correspond to $\left({\theta_1},{\theta_2}\right)=\pi\left(-0.15,0.06\right)$, in the notation of Section~\ref{sec:exactschrod}, in view of eqs.~\eqref{cond_sing1n1}. The comparison with the hypergeometric potentials of eq.~\eqref{pot_hyp} requires a slight shift of the latter, as can be seen from fig.~\ref{fig:Bmi_pot}. The precise shift can be determined demanding that the exact zero mode lie on the resulting massless curve, and amounts to adding to the hypergeometric potential the constant
\beq
\Delta\,V \ \simeq \ (0.22)^2\, \frac{\pi^2}{z_m^2} \ .
\eeq
Once this is done, one can rely on the hypergeometric approximation, and in particular on the corresponding exact eigenvalue equation~\eqref{eigenvfinmumutilde}, to identify the boundary conditions resulting in stable modes, which correspond to the unshaded region displayed in fig.~\ref{fig:Bmi_instabilities}

In conclusion, this sector can yield altogether, \emph{ten massless real Abelian vectors} in the bulk, with suitable boundary conditions, which are accompanied by ten real massless scalars living in the boundary.
When allowing for nonzero values of $\mathbf{k}$, it is convenient not to eliminate $B_{ri}$ but rather to impose the transversality of all fields to $\mathbf{k}$. In this fashion, one can see that $B_{ri}$ is set to zero by the equations of motion while $m^2$, as usual, is replaced by $m^2 - e^{2(A-C)}\,\mathbf{k}^2$, and the whole spectrum is lifted in this sector, which thus contains no unstable modes for suitable boundary conditions.

\subsection{The Modes Originating from ${\cal B}_{ij}$}
\begin{figure}[ht]
\centering
\begin{tabular}{cc}
\includegraphics[width=65mm]{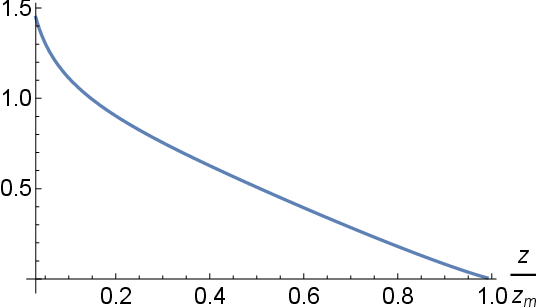} \qquad \qquad &
\includegraphics[width=65mm]{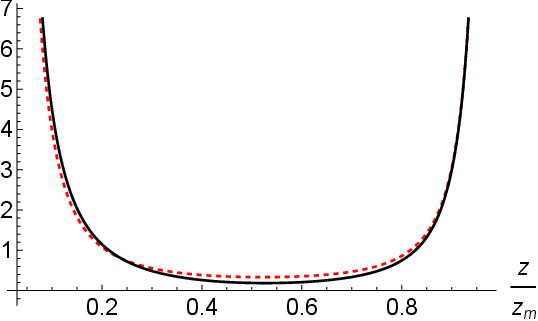} \\
\end{tabular}
\caption{\small The left panel shows the normalized zero--mode wavefunction~\eqref{bij_prof_g}. The right panel compares the actual $B_{ij}$ potential (black, solid) with its approximation~\eqref{pot_hyp} with $(\mu,\tilde{\mu})=\left(\frac{2}{3},0.63\right)$ (red, dashed), both in units of $\frac{1}{{z_0}^2}$. $z_m$ and $z_0$ are defined in eqs.~\eqref{zm_app} and \eqref{z0}.}
\label{fig:Bij_pot}
\end{figure}

Turning now to the modes that are scalar fields valued in the antisymmetric of $SO(5)$, let us first note that the two-form is in this case
\beq
{\cal B} \ = \ \frac{1}{2}\,{\cal B}_{ij}\,dy^idy^j\ ,
\eeq
while the corresponding field strength is
\beq
{\cal H}\ = \ \frac{1}{2}(d_4 {\cal B}_{ij}+\partial_r\,{\cal B}_{ij}dr)dy^idy^j 
\eeq
for $\mathbf{k}=0$. The non--trivial equation is in this case
\beq{}{}{}
m^2e^{6A+6C} \, {\cal B}_{ij} \ + \ \partial_r\left(e^{-4C}\partial_r\, {\cal B}_{ij}\right)\ =\ 0 \ ,
\eeq
and in terms of the $z$ variable it becomes
\beq{}{}{}
m^2\, {\cal B}_{ij} \ + \ \left(\partial_z \,+\,3\,A_z\,+\,C_z\right)\partial_z {\cal B}_{ij}\ = \ 0 \ .
\eeq
The redefinition
\beq{}{}{}
{\cal B}_{ij} \ = \ e^{-\frac{1}{2}\left(3A+C\right)} \, g(z) \, b_{ij}(x) \label{bij_prof}
\eeq
leads once more to a manifestly Hermitian Schr\"odinger--like equation of the ${\cal A}\,{\cal A}^\dagger$ form,
\beq
\left(\partial_z+\frac{(3A_z+C_z)}{2}\right)\left(-\partial_z+\frac{(3A_z+C_z)}{2}\right)g \ = \ m^2\, g \ , \label{schrod_bij}
\eeq
with the potential
\beq{}{}{}
V \,=\,\frac{1}{4}\left(3 A_z + C_z\right)^2 \ + \ \frac{1}{2}\,\partial_z\left(3 A_z +  C_z\right) \  , \label{potBij}
\eeq
which is displayed in fig.~\ref{fig:Bij_pot} as a function of $z$. In terms of $r$ it detailed form is
\beq
V \ = \ \frac{e^{\sqrt{\frac{5}{2}} \frac{r}{\rho}}}{320 \,z_0^2\,\sinh\left(\frac{r}{\rho}\right)}  \left[- \,6\,\sqrt{10}\, \sinh \left(\frac{2 r}{\rho}\right)\ + \ 21\, \cosh\left(\frac{2 r}{\rho}\right) \ +\ 119\right] \ . \label{pot_bij}
\eeq

Close to $z=0$ the potential has the singular limiting behaviors of eq.~\eqref{lim_dil_axion_intro}, with $\mu=\frac{2}{3}$ and $\tilde{\mu}=0.63$, so that it corresponds to case 3 in Table~\ref{tab:tab_munu}, as for the previous sector.
There is a normalizable ground state, which corresponds to
\beq{}{}{}
g \ = \ g_0\ e^{\frac{3A+C}{2}} \ , \label{bij_prof_g}
\eeq
with a constant $g_0$, and thus to ${\cal B}_{ij}$ independent of $r$:
\beq{}{}{}
{\cal B}_{ij} \ = \ g_0 \, b_{ij}(x) \ . \label{bij_prof1}
\eeq

\begin{figure}[ht]
\centering
\includegraphics[width=65mm]{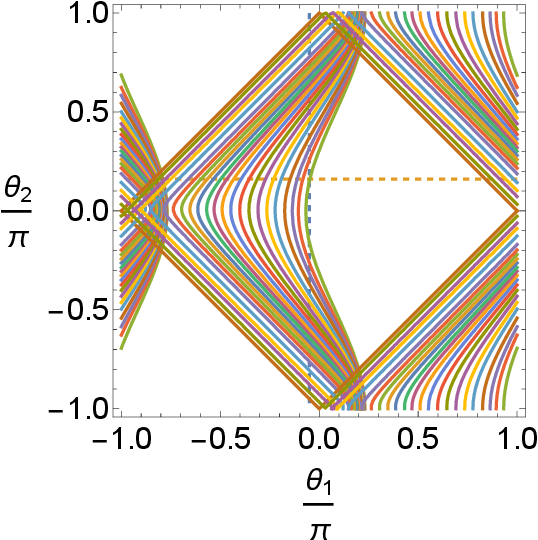}
\caption{\small The point $\left({\theta_1},{\theta_2}\right)=\pi\left(-0.05,0.16\right)$ identifies the special boundary conditions corresponding to the zero mode~\eqref{bmui_prof}. The shaded regions identify the boundary conditions leading to instabilities. }
\label{fig:Bij_instabilities}
\end{figure}

Note that the zero--mode wavefunction has the dominant behaviors
\beq
g \ \sim \ \left(\frac{z}{z_m}\right)^{\,-\,\frac{1}{6}} \ - \ 0.35 \  \left(\frac{z}{z_m}\right)^\frac{7}{6} 
\eeq
close to $z=0$, and
\beq
g \ \sim \ \left(1\ - \ \frac{z}{z_m}\right)^{-\,0.13} \ + \ 0.18 \  \left(1\ - \ \frac{z}{z_m}\right)^{1.13} 
\eeq
close to $z=z_m$. The arguments of~\cite{selfadjoint} lead one to conclude that this whole sector is stable with the boundary conditions of this zero mode, which correspond to $\left({\theta_1},{\theta_2}\right)=\pi\left(-0.16,0.05\right)$, in the notation of Section~\ref{sec:exactschrod}. The comparison with the hypergeometric potentials of eq.~\eqref{pot_hyp} requires a slight shift of the latter. The shift can be determined demanding that the exact zero mode lie on the resulting massless curve, and amounts to adding to the hypergeometric potential the constant
\beq
\Delta\,V \ \simeq \ - \ (0.08)^2\, \frac{\pi^2}{z_m^2} \ .
\eeq
Once this is done, one can rely on the hypergeometric approximation, and in particular on the corresponding exact eigenvalue equation~\eqref{eigenvfinmumutilde}, to identify the boundary conditions for ${\cal B}_{ij}$ resulting in stable modes, which correspond to the unshaded region displayed in fig.~\ref{fig:Bij_instabilities}

The Sch\"odinger system also indicates that the norm would be proportional to
\beq{}{}{}
\int dz\ e^{3A+C} \ = \ \int_0^\infty dr \ e^{6(A+C)} \ ,
\eeq
which is finite, as we have anticipated. This expression identifies the $r$-distribution
\beq
\Pi_{B_{ij}}(r) \ = \ \frac{1}{\rho}\, \frac{3}{\sqrt{10}}\ e^{\,-\,\frac{r}{\rho}\frac{3}{\sqrt{10}}} \ . \label{pibij}
\eeq
Only ${\cal H}_{\mu i j}$ is present for these zero modes, and therefore the no-flow conditions of~\cite{ms_20} are identically satisfied.

With non--vanishing internal momenta $\mathbf{k}$, the Chern--Simons term still does not contribute while, once more, $m^2$ is replaced by $m^2 \,-\, e^{2(A-C)} \,\mathbf{k}^2$. All the preceding choices of boundary conditions remain possible, once the wavefunctions are modified by the addition of suitable $\mathbf{k}$-dependent corrections. Once this restriction is taken into account, the $\mathbf{k}^2$ term lifts the mass spectrum further, and no instabilities emerge with suitable boundary conditions.

\section{\sc The Perturbed Einstein - Five-Form System} \label{sec:perturbed_tensor}

A peculiar feature of type--IIB supergravity~\cite{IIB} is the presence of a four--form gauge field whose field strength satisfies the self--duality condition
\beq
{\cal H}_5 \ = \ \star {\cal H}_5 \ . \label{HHdual}
\eeq
We have already seen its crucial role in the class of backgrounds at stake, and now we want to explore its perturbations. Eq.~\eqref{HHdual} is indeed the complete four--form equation of motion, together with the Bianchi identity for ${\cal H}_5$, once the field strength is properly dressed with fermionic terms and Chern--Simons forms, which are however irrelevant for the linearized analysis.

Linear perturbations of the four--form gauge field in eq.~\eqref{HHdual} mix with metric perturbations, and eq.~\eqref{HHdual} is to be combined with the linearized Einstein equations, which can be deduced starting from
\beq
R_{MN} \ = \  \frac{1}{24} \left({\cal H}_5^2\right)_{MN}  \ .\label{einstein_complete}
\eeq
The type of system one is confronted with is rather unfamiliar and, as we shall see, is somewhat complicated. In organizing the analysis, it is always useful to distinguish various sectors of modes relying on an internal symmetry, whenever this is present. The internal toroidal directions are helpful in this respect, and it will be important, as in the preceding sections, to distinguish matters according to whether or not the toroidal momentum $\mathbf{k}$ vanishes.

As we have already seen, the modes with $\mathbf{k}=0$ have an SO(5) internal symmetry inherited from the internal space. In flat space, after a IIB compactification on $S^1 \times T^5$, the original 35 modes of the 10D self-dual five-form field strength would translate, within this sector, into 35 massless modes independent of the $r$ coordinate. They would build a multiplet of Abelian vectors in the $\underline{10}$ of $SO(5)$, together with two scalar multiplets in the $\underline{10}$ and $\underline{5}$ representations of $SO(5)$. This can be seen in the four--dimensional light--cone gauge, leaving aside the components along $r$, which are linked to the others by the self--duality relations~\eqref{HHdual}. In this fashion, one must deal with three types of four--dimensional modes, $B_{ijkl}$, $B_{ijka}$ and $B_{ijab}$. Here $(i,j,k)=1,..,5$ correspond to the internal directions and $(a,b)=1,2$ correspond the two four--dimensional directions transverse to the light cone. If massless in four dimensions, these three sets of fields, up to internal and spacetime dualities, would indeed describe a $\underline{5}$ of scalars, a $\underline{10}$ of vectors and an additional $\underline{10}$ of two--tensors, which are dual to scalars. These modes would add to 35 other modes originating from the metric field. These are the two graviton polarizations from $h_{\mu\nu}$, a multiplet of vectors $h_{\mu i}$ in the $\underline{5}$ of $SO(5)$, an additional singlet vector $h_{\mu r}$, a $\underline{5}$ of scalars $h_{ri}$, a singlet, $h_{rr}$, and finally a $\underline{14}$ and a singlet scalar from $h_{ij}$. In the present context, one expects to find, in general, a subset of these massless modes emerging in the bulk.

Although some of these massless modes will disappear in our background, the comparison with this benchmark will often prove a convenient tool in the following. When modes become massive due to the behavior along $r$ but still correspond to $\mathbf{k}=0$, the $SO(5)$ remains manifest, but the ten scalars from the five form can be eaten by the corresponding 10 vectors, or equivalently the vectors can be eaten by the ten two-forms. On the gravity side, the pattern is more familiar from Kaluza--Klein theory.

For any given choice of $\mathbf{k} \neq 0$, the relevant symmetry is the $SO(4)$ associated to the four internal directions orthogonal to it, as we have already stated. While so far we have largely confined our attention to $\mathbf{k}=0$ modes, reserving only brief comments to sectors with non--vanishing internal momenta, here it will be important to also scrutinize perturbations with $\mathbf{k}\neq 0$, since they will entail special subtleties and, as we saw in detail in~\cite{bms}, can host instabilities as a result of mixings.

\subsection{\sc Perturbing the Tensor Equations}

Perturbing eqs.~\eqref{HHdual} around the background leads to
\bea
{\delta\, {\cal H}}_{P_1\ldots P_5}
&=& {h_{[P_1}}^{M_1} \ {\cal H}^{(0)}_{5\,P_2 \ldots P_5] M_1} \ - \ \frac{1}{2}\ {h_M}^M \ {\cal H}^{(0)}_{5\,P_1 \ldots P_5} \nonumber \\
&+& \frac{g_{P_1 M_1}^{(0)} \ldots g_{P_5 M_5}^{(0)}}{5!\ \sqrt{-g^{(0)}}}\, \epsilon^{M_1 \ldots M_5 N_1 \ldots N_5}\, \delta {\cal H}_{N_1 \ldots N_5} \ , \label{selfdual_lin}
\eea
where the antisymmetrizations have unit coefficients,
quantities bearing the superscript $(0)$ refer to the background, ${\delta\, {\cal H}}$ denotes perturbations of the tensor field strength and $h$ denotes perturbations of the metric, defined in eq.~\eqref{field_pert}.
Moreover, ${\cal H}_5^{(0)}$ is defined in eq.~\eqref{back_epos_fin2}, and here the Levi--Civita tensor, such that $\epsilon^{01 \ldots 9} = -1$, will be lowered with the Minkowski metric.

The perturbation $\delta {\cal H}$ is to be expressed in terms of the four--form gauge field, whose independent components can be conveniently parametrized as follows:
\bea
&& \delta B_{\mu\nu\rho\sigma} \ = \ \epsilon_{\mu\nu\rho\sigma} \, b \ , \ \qquad \delta B_{\mu \nu \rho r} \ = \ {\epsilon_{\mu\nu\rho}}^{\sigma}\,b_{\sigma}  , \qquad  \delta B_{\mu\nu\rho i} \ = \ {\epsilon_{\mu\nu\rho}}^{\sigma} \, b_{\sigma i} \ , \nonumber \\
&& \delta B_{\mu \nu r i} \ = \ b_{\mu \nu i}  \ , \qquad \delta B_{\mu \nu i j} \ = \ b_{\mu \nu i j}    \ , \qquad \delta B_{\mu r i j} \ = \ b_{\mu i j}^{(1)}   \ , \nonumber \\
&& \delta B_{\mu i j k} \ = \ \frac{1}{2}\ \epsilon_{i j k l m}\, b_\mu^{(2)l m} \ , \qquad  \delta B_{r i j k}  \ = \ \frac{1}{2}\ \epsilon_{i j k l m}\, b^{l m}  \ , \qquad  \delta B_{i j k l}  \ = \ \epsilon_{i j k l m} \, b^m\, . \label{tensor_linear_H2}
\eea
Using eqs.~\eqref{selfdual_lin} and the results collected in Appendix~\ref{app:fiveform}, and in particular eqs.~\eqref{deltaH5}, one can see that the independent self--duality conditions reduce to
\bea
\partial_{[\mu}\,b^{(2)}{{}_{\nu]}}^{l m} \ + \ \frac{1}{2}\,\epsilon^{p q r l m}\,\partial_p\,b_{\mu \nu q r}   &=& - \,  \frac{e^{-4A-4C}}{2}\, \epsilon_{\mu\nu\rho\sigma} \nonumber \\ &\times& \left( \partial^{[\rho}\,b^{(1) \,\sigma] l m} \, + \, \partial_r\, b^{\rho \sigma l m }
\, - \, \partial^{[l}\,b^{\rho \sigma | m]}\right) \, ,\nonumber \\
\partial_\mu\,b^{l m} \ - \ \partial_r\,{b^{(2)}}{{}_\mu}^{l m}\ + \ \frac{1}{2}\,\epsilon^{p q s l m}\,\partial_p\, {b^{(1)}}_{\mu q s} &=& - \  {e^{2A+6C}}\left(\partial^{[l}\,{b_\mu}^{m]}\,+\,\frac{1}{2}\,{\epsilon^{\alpha\beta\gamma}}_\mu\,\partial_\alpha\,{b_{\beta\gamma}}^{lm} \right)\, ,
\nonumber \\
\partial_\mu\,b^m \ - \ \partial_n\,{{b^{(2)}}_\mu}{}^{m n}  &=& e^{-2C}\left[\frac{h}{2\,\rho} \ {h_\mu}^m \right. \nonumber \\
&+& \left. {e^{-6A}}\left(\partial^m\,b_\mu\,-\, \partial_r\,{b_\mu}^m\,-\,\frac{1}{2}\,{\epsilon^{\alpha\beta\gamma}}_\mu\,\partial_\alpha\,{b_{\beta\gamma}}^m\right)\right]\, ,
\nonumber \\
\partial_r\,b^m \ - \ \partial_n\,b^{m n}  &=& e^{-2C}\left[\frac{h}{2\,\rho}\,{h_r}^m\ - \ {e^{10C}}\left( \partial^m\,b\,-\,\partial_\mu\,b^{\mu m} \right)\right]\, ,
\nonumber \\
\partial_p\,b^p  &=& \left[ \frac{h}{4\,\rho}  \left( - \ e^{-2A}\,{h_{\alpha}}^\alpha\ - \  e^{-2B}\,h_{r r} \ + \ e^{-2C}\,{h_{i}}^i \right) \right. \nonumber \\ &+& \left.  e^{-8A} \left(\partial_r\,b \ - \ \partial_\tau\,b^\tau \right) \right] \ , \label{eqs_sd}
\eea
where we expressed the background metric in terms of the three functions $A(r)$, $B(r)$ and $C(r)$, as in eq.~\eqref{metricABC}.

\subsection{\sc Tensor Gauge Fixing of the Einstein--Five form System} \label{sec:tensorgfix}

In order to analyze the modes arising from the Einstein--five-form sector, one must perform a gauge fixing of the resulting equations.

Diffeomorphisms act on metric fluctuations $h_{MN}$ as
\beq{}{}{}
\delta_\xi\,h_{MN} \,=\, \nabla_M\,\xi_N \ + \ \nabla_N\, \xi_M \ , \nonumber \\
\eeq
and consequently the Ricci curvature perturbations transform according to
\beq
\delta_\xi \left( \delta\,R_{N R}\right) \ = \ \nabla_N\, \xi^P\, R^{(0)}{}_{P R} \ + \ \nabla_R\, \xi^P\, R^{(0)}{}_{N P} \ ,
\eeq
where $R^{(0)}{}_{MN}$ is the background Ricci curvature. They are thus covariant under diffeomorphisms, albeit not invariant as would be the case when working around flat space.

The four--form gauge potential is also affected by tensor gauge transformations, and a convenient presentation of their combined action with diffeomorphisms is
\bea
\delta_{\xi,\Lambda}\left(\delta \,B_{MNPQ}\right) &=&  \xi^R\,{\cal H}^{(0)}_{5\,RMNPQ} \ + \ \partial_{[M}\,\Lambda_{NPQ]}\ , \nonumber \\
\delta_\xi\left(\delta{\cal H}_{S M N P Q}\right) &=& \left(\nabla_{[S}\xi^R\right) {\cal H}^{(0)}_{5\,M N P Q ] R} \ + \ 5\,\xi^R \nabla_R \,{\cal H}^{(0)}_{5\,S M N P Q} \ .
\eea
Here $\Lambda$ is a three--form gauge parameter and, as in other portions of this paper, square brackets denote antisymmetrizations without overall factors.
The independent components of $\Lambda_{MNP}$ can be conveniently parametrized as
\bea
\Lambda_{\nu \rho \sigma} &=& \epsilon_{\nu\rho\sigma\tau}\,\Lambda^\tau \ , \quad
\Lambda_{\nu \rho r} \ = \ \Lambda_{\nu\rho} \ , \quad
\Lambda_{\nu \rho i} \ , \qquad
\Lambda_{\nu r i} \ = \ \Lambda_{\nu i} \ , \nonumber \\
\Lambda_{r i j } &=& {\Lambda^{(1)}}_{i j} \ , \qquad
\Lambda_{\mu i j} \ , \qquad
\Lambda_{ijk} \ = \  \frac{1}{2}\,\epsilon_{i j k l m}\,\Lambda^{(2)\, l m} \ .
\eea
Diffeomorphisms and tensor gauge transformation thus act on the independent fields according to
\bea
\delta\,h_{\mu\nu} &=& \partial_\mu\,\xi_\nu \ + \ \partial_\nu\, \xi_\mu\,+\,2\,\eta_{\mu\nu}\,A'\,e^{2(A-B)}\,\xi_r \ ,\nonumber \\
\delta\,h_{\mu r} &=& \partial_\mu\,\xi_r \ + \ \left(\partial_r \,-\,2\,A'\right) \xi_\mu \ , \qquad
\delta\,h_{\mu i} \ = \  \partial_\mu\,\xi_i \ + \ \partial_i\, \xi_\mu  \ , \nonumber \\
\delta\, h_{r r} &=& 2 \left(\partial_r \,-\, B' \right)\xi_r \ , \qquad
\delta\,h_{r i} \ = \  \left(\partial_r \,-\, 2\, C'\right)\xi_i \ + \ \partial_i\, \xi_r \ , \nonumber \\
\delta\,h_{i j} &=& \partial_i\,\xi_j \ + \ \partial_j\, \xi_i\,+\,2\,\delta_{i j}\,C'\,e^{2(C-B)}\,\xi_r \ , \nonumber \\
\delta\,b &=& \frac{h}{2\,\rho}\,e^{-10C}\,\xi_r - \ \partial_\mu\,\Lambda^\mu \ , \qquad
\delta\,b_i \ = \  \frac{h}{2\,\rho}\,e^{-2C}\,\xi_i \ +\ \partial^j\,\Lambda^{(2)}_{i j} \ , \nonumber \\
\delta\,b_\mu &=& - \ \frac{h}{2\,\rho}\,e^{6A}\,\xi_\mu \ - \ \frac{1}{2}\,\epsilon_{\mu\nu\rho\sigma}\,\partial^\nu\,\Lambda^{\rho\sigma} \ - \ \partial_r\,\Lambda_\mu\ , \label{xilambda_gauge}
\eea
while the remaining tensor components are invariant under diffeomorphisms and have the tensor gauge transformations
\bea
\delta\,\,b_{\mu\,i} &=& - \ \frac{1}{2}\,\epsilon_{\mu\nu\rho\sigma}\,\partial^\nu\,\Lambda^{\rho\sigma\,i} \ - \ \partial_i\,\Lambda_\mu \ , \qquad
\delta\,\,{b_{\mu\nu i}} \ = \ \partial_{[\mu}\,\Lambda_{\nu] i} \ + \ \partial_r\,\Lambda_{\mu\nu i} \ - \ \partial_i \, \Lambda_{\mu\nu}\ , \nonumber \\
\delta\,\,b_{\mu\nu i j} &=& \partial_{[\mu}\,\Lambda_{\nu] i j}\ + \ \partial_{[i}\,\Lambda_{ \mu\nu |j]}\ , \qquad
\delta\,\,b_{\mu i j}^{(1)} \ = \  \partial_{\mu}\,{\Lambda^{(1)}}_{ij} \ - \ \partial_r\,\Lambda_{\mu i j} \ + \ \partial_{[i}\,\Lambda_{ \mu |j]}\ ,  \label{gauge_transf_tensork}\\
 \delta\,\,{{b^{(2)}}_{\mu}}{}_{i j} &=& \partial_\mu\,{\Lambda^{(2)}}_{ij} \ - \ \frac{1}{2}\,\epsilon_{i j k l m}\,\partial^{k}\,{\Lambda_\mu}^{l m} \ ,\qquad
 \delta\,\,b_{i j} \ = \  \partial_r\,{\Lambda^{(2)}}_{ij} \ - \ \frac{1}{2}\,\epsilon_{i j k l m}\,\partial^{k}\,\Lambda^{(1)\,l m}  \ . \nonumber
\eea

Using the tensor gauge transformations one can now set
\beq
B_{rMNP} \ = \ 0  \ ,\label{4formgauge}
\eeq
for all choices of $M$, $N$ and $P$, thus removing all fields whose gauge transformations involve the radial derivative of a parameter with the same Lorentz structure. In analogy with what we said for the two-forms, this gauge fixing brings along other modes living in the boundary, which can be associated to a nine--dimensional three form. 

The gauge condition~\eqref{4formgauge} translates into
\beq
b_\mu \,=\,0 \ , \qquad  b_{\mu\nu i} \,=\,0 \ , \qquad {b^{(1)}}_{\mu i j}\,=\,0 \ , \qquad  b_{ij}\,=\,0 \ ,
\eeq
and reduces the system of tensor equations to
\bea
\partial_{[\mu}\,b^{(2)}{{}_{\nu]}}^{l m} \ + \ \frac{1}{2}\,\epsilon^{l m n p q}\,\partial_n\,b_{\mu \nu p q}   &=& - \,  \frac{e^{-4A-4C}}{2}\, \epsilon_{\mu\nu\rho\sigma}\, \partial_r\, b^{\rho \sigma l m } \, ,\nonumber \\
 \partial_r\,{b^{(2)}}{{}_\mu}^{l m} &=&  {e^{2A+6C}}\left[\partial^{[l}\,{b_\mu}^{m]}\,+\,\frac{1}{2}\,{\epsilon^{\alpha\beta\gamma}}_\mu\,\partial_\alpha\,{b_{\beta\gamma}}^{lm} \right]\, ,
\nonumber \\
\partial_\mu\,b^m \ - \ \partial_n\,{{b^{(2)}}_\mu}{}^{m n}  &=& e^{-2C}\left[\frac{h}{2\,\rho} \ {h_\mu}^m  \,-\, {e^{-6A}}\, \partial_r\,{b_\mu}^m \right]\, ,
\nonumber \\
\partial_r\,b^m   &=& e^{-2C}\left[\frac{h}{2\,\rho}\,{h_r}^m\ - \ {e^{10C}}\left( \partial^m\,b\,-\,\partial_\mu\,b^{\mu m} \right)\right]\, ,
\label{first_ord_5form} \\
\partial_p\,b^p  &=&  \frac{h}{4\,\rho}  \Big[ - \ e^{-2A}\,{h_{\alpha}}^\alpha\, - \,  e^{-2B}\,h_{r r} \, + \, e^{-2C}\,{h_{i}}^i \Big] \,+\,  e^{-8A} \,\partial_r\, b  \, .  \nonumber
\eea
This simpler system is still invariant under some residual gauge transformations, with arbitrary parameters $\Lambda_{\rho\sigma}$, $\Lambda_{\mu i}$ and $\Lambda^{(1)}{}_{i j}$, while the radial dependence of the others is determined by
\bea
\partial_r\,\Lambda_\mu &=& - \ \frac{1}{2}\,\epsilon_{\mu\nu\rho\sigma}\,\partial^\nu\,\Lambda^{\rho\sigma} \,-\,\frac{h}{2\,\rho}\,e^{6A}\,\xi_\mu \ , \qquad
\partial_r\,\Lambda_{\mu\nu i} \ = \  \partial_i \, \Lambda_{\mu\nu} \ -\ \partial_{[\mu}\,\Lambda_{\nu] i} \ , \nonumber \\
\partial_r\,\Lambda_{\mu i j} &=& \partial_{\mu}\,{\Lambda^{(1)}}_{ij} \ + \ \partial_{[i}\,\Lambda_{ \mu |j]}\ , \qquad
\partial_r\,{\Lambda^{(2)}}_{ij} \ = \  \frac{1}{2}\,\epsilon_{i j k l m}\,\partial^{k}\,\Lambda^{(1)\,l m} \ . \label{res_tensor_sym}
\eea

These residual symmetries will be instrumental to exhibit the modes actually emerging from this unfamiliar sector. Indeed, even in the flat--space limit, where the metric functions $A$, $B$, $C$ and the parameter $h$ are all vanishing, the recovery of the expected modes that we have listed at the beginning of this section is not evident. There is actually a subtlety here, when comparing with the case of circle compactification, whose internal zero modes, which would be independent of $r$ in the present notation, cannot be gauged away. That would require gauge parameters linear in $r$, which are not allowed in order to maintain periodicity, consistently with the global translational symmetry on the circle. In our case the internal $r$-space is an interval, and the requirement of periodicity is replaced by proper boundary conditions. Therefore, the parameters on the right--hand side of eqs.~\eqref{res_tensor_sym} are arbitrary, which justifies our choice.
We shall see this explicitly in the following sections, where we shall also fix diffeomorphism invariance on a case-by-case basis. As we have already stressed, however, other modes appear generically at the ends of the interval.

\subsection{\sc Perturbing the Einstein Equations}

When the complete Einstein equations~\eqref{einstein_complete} are linearized around the background, their left--hand side becomes
\bea
R_{MN}&=& R^{(0)}_{MN}\ + \ \delta\,R_{MN} \ , \label{pert_RMN_0}
\eea
where
\beq
- 2\,\delta\,R_{N R}\ = \  \Box_{10}\,h_{N R} \ +\ \nabla_N\,\nabla_R\,{h_S}^S \ - \ \nabla^P\left(\nabla_N\,h_{P R}\,+\,\nabla_R\,h_{P N} \right) \ ,
\eeq
where the derivatives and the d'Alembertian are covariant with respect to the background. The components of the first--order correction to the energy--momentum tensor can be obtained in a similar fashion, and these steps lead to the linearized Einstein equations
\bea
&&\Box_{10}\,h_{N R} \ +\ \nabla_N\,\nabla_R\,{h_S}^S \ - \ \nabla^P\left(\nabla_N\,h_{P R}\,+\,\nabla_R\,h_{P N} \right) \nonumber \\
&&=\ -\, \frac{1}{12}\left( \delta{\cal H}_{5 (N} \cdot {\cal H}_{5 R)}^{(0)} \ - \ 4 \, {\cal H}_{5 NK}^{(0)} \cdot {\cal H}_{5 RL}^{(0)} \ h^{KL}\right) \ . \label{EinstMN}
\eea

The spacetime components of eqs.~\eqref{EinstMN} are
\bea
\alpha\beta&:& \left(e^{-2A} \, \Box\,+\,e^{-2C}\,\Delta\right)h_{\alpha\beta} \ - \ e^{-2B}\left(\partial_r\,-\,2\,A'\right)\partial_{(\alpha}\,h_{\beta)r}\nonumber \\
&+&\eta_{\alpha\beta}\,A'\,e^{2(A-B)}\big[e^{-2A}\left(\partial_r-2 A'\right){h^\mu}_\mu - e^{-2B}\left(\partial_r-\,2 B'\right)h_{rr} + e^{-2C}\left(\partial_r-2 C'\right){h^i}_i\big] \nonumber\\
&-& 2\,\eta_{\alpha\beta}\,A'\,e^{2(A-B)}\big[e^{-2A}\,\partial^\mu\,h_{\mu r}\,+\, e^{-2C}\,\partial^i\,h_{i r}\big] \nonumber \\
&+&e^{-2B}\big[\left(\partial_r-4A'\right)\partial_r\,+\,4\,(A')^2\big]h_{\alpha\beta} \nonumber \\
&+& \partial_\alpha \, \partial_\beta\left( e^{-2A}\, {h^\mu}_\mu \,+\, e^{-2B}\,{h^r}_r\,+\,e^{-2C}\, {h^i}_i  \right) \,-\,e^{-2A}\,\partial^\mu\,\partial_{(\alpha}\,h_{\beta)\mu}\,-\, e^{-2C}\, \partial^i\,\partial_{(\alpha} \,h_{\beta)i} \nonumber \\
&=&  \,\frac{h}{2\,\rho} \left\{ 4\,\,e^{2(A-B)}\,\partial_r\,b \, \eta_{\alpha\beta} \, - \,\frac{h}{\rho} \,e^{- 10 C} \left( \eta_{\alpha\beta}\,{h_\rho}^\rho - h_{\alpha\beta} \right) \right\} \ , \label{Einstein_alphabeta}
\eea
while the $\alpha r$ components are
\bea
\alpha r&: & \left(e^{-2A} \, \Box\,+\,e^{-2C} \, \Delta \right) h_{\alpha r} \,-\,e^{-2A}\left(\partial_r\,-\,2\,A' \right)\partial^\rho\,h_{\alpha \rho}\nonumber \\
&+& \left(A'-B'\right)e^{-2B}\,\partial_\alpha\,h_{rr}\,-\,e^{-2C}\left(\partial_r\,-\,2 A'\right)\partial^i\,h_{i\alpha} \nonumber \\&+& e^{-2A}\,\left(\partial_r\,-\,2\,A' \right)\partial_\alpha\,{h^\mu}_\mu \,+\, e^{-2C}\left( \partial_r\,-\,A'\,-\,C'\right)\partial_\alpha\,{h_i}^i\nonumber \\
&-& e^{-2A}\,\partial_\alpha\,\partial^\rho\,h_{\rho r}\,-\,e^{-2C}\,\partial_\alpha\,\partial^i\,h_{r i} \ = \  0 \ , \label{Einstein_alphar}
\eea
and the $\alpha i$ components are
\bea
\alpha i&:& e^{-2A}\,\Box\,h_{\alpha i} \,+\,e^{-2C}\,\Delta\,h_{\alpha i}\,+\,e^{-2B}\left[\left(\partial_r\,-\,A'\,-\,C' \right)^2 \ - \ \left(A'\,-\,C'\right)^2\right]h_{\alpha i} \nonumber \\
&-& e^{-2B}\left(\partial_r\,-\,2\,A' \right)\partial_\alpha\,h_{ir} \ - \ e^{-2B}\left(\partial_r\,-\,2\,C' \right)\partial_i\,h_{\alpha r} \nonumber \\
&+& \partial_\alpha\,\partial_i \left( e^{-2A}\,{h^\mu}_\mu \,+\, e^{-2B}\,{h^r}_r\,+\,e^{-2C}\, {h^k}_k  \right) \nonumber \\
&-& e^{-2A}\partial^\rho\left(\partial_\alpha\,h_{\rho i}\,+\, \partial_i\,h_{\rho \alpha}\right) \ - \ e^{-2C}\,\partial^j\left( \partial_\alpha \,h_{i j} \,+\, \partial_i \,h_{\alpha j}\right) \nonumber \\ &=& - \ \frac{h\,e^{-8C}}{2\,\rho}\left[ 4\left( \partial_\alpha\,b_i \ - \ \partial_n \, b^{(2)}{{}_{\alpha i}}^{n}\right) \ - \ \frac{h}{\rho}\, e^{-2C}\,h_{\alpha i} \right]\ . \label{Einstein_alphai}
\eea
Moreover, the $rr$ component is
\bea
rr &: &\left[e^{-2A}\, \Box\,-\,e^{-2B}\,B'\left(\partial_r\,-\,2\,B'\right) \,+\,e^{-2C}\,\Delta \right]h_{rr} \nonumber \\
&-& 2\,e^{-2A}\,\left(\partial_r\,-\,B'\right)\partial^\mu\,h_{\mu r} \,-\, 2\,e^{-2C}\,\left(\partial_r\,-\,B'\right)\partial^i\,h_{r i}\nonumber \\ &+&e^{-2A}\left(\partial_r\,-\,B'\right)\left(\partial_r\,-\,2\,A'\right){h^\mu}_\mu \,+\, e^{-2C}\left(\partial_r\,-\,B'\right)\left(\partial_r\,-\,2\,C'\right){h^i}_i
\nonumber \\ &=&  \frac{h}{2\,\rho} \Big[ 4\,\partial_r\,b  \,-\,  \frac{h}{\rho}\, e^{6A}\, {h_\rho}^\rho \Big] \ , \label{Einstein_rr}
\eea
while the $r i$ components are
\bea
r i&: & \left(e^{-2A} \, \Box\,+\,e^{-2C} \, \Delta \right) h_{i r} \,-\, e^{-2C}\left(\partial_r\,-\,2\,C'\right)\partial^k\,h_{k i} \nonumber \\
&+&\partial_i\Big\{ e^{-2A}\left[ \left(\partial_r-A'-C'\right){h^\alpha}_\alpha \,-\,\partial^\alpha\,h_{\alpha r}\right]\,+\,e^{-2B}\left(C'-B'\right)h_{r r}  \nonumber \\&+&  e^{-2C}\left[\left(\partial_r-2C'\right){h^k}_k\,-\,\partial^k\,h_{k r} \right]\Big\} \ - \ e^{-2A}\left(\partial_r\,-\,2\,C'\right)\partial^\alpha\,h_{\alpha i} \nonumber \\
&=&  \frac{2\,h}{\rho}\left(\partial_i\,b\,-\,\partial_\mu\,{b^\mu}_i \right)\ .  \label{Einstein_ir}
\eea
Finally, the internal $ij$ components are
\bea{}{}{}{}{}
i j &: & \left(e^{-2A}\,\Box \,+\, e^{-2C} \, \Delta\right)h_{i j}\,-\, e^{-2B}\left(\partial_r\,-\,2\,C'\right)\partial_{(i}\,h_{j)r}\,-\,2\,\delta_{i j}\,C''\,e^{2(C-2B)}\,h_{rr}\nonumber \\
&+&\delta_{i j}\,C'\,e^{2(C-B)}\big[e^{-2C}\left(\partial_r-2 C'\right){h^k}_k - e^{-2B}\left(\partial_r-\,2 B'\right)h_{rr} + e^{-2A}\left(\partial_r-2 A'\right){h^\mu}_\mu\big] \nonumber\\
&-& 2\,\delta_{i j}\,C'\,e^{2(C-B)}\big[e^{-2C}\,\partial^k\,h_{k r}\,+\, e^{-2A}\,\partial^\mu\,h_{\mu r}\big] \,+\,e^{-2B}\big[\left(\partial_r-4C'\right)\partial_r\,+\,4\,(C')^2\big]h_{i j} \nonumber \\
&+& \partial_i \, \partial_j\left( e^{-2C}\, {h^k}_k \,+\, e^{-2B}\,h_{rr}\,+\,e^{-2A}\, {h^\mu}_\mu  \right) \nonumber \,-\,e^{-2C}\,\partial^k\,\partial_{(i}\,h_{j)k}\,-\, e^{-2A}\, \partial^\mu\,\partial_{(i} \,h_{j)\mu} \nonumber \\
&=& -\ \frac{h}{2\,\rho} \left[ 4\,\delta_{ij}\,e^{-8C}\,\partial_p\,b^p \,-\, \frac{h}{\rho}\,e^{-10C}\left( \delta_{ij}\,{h_k}^k \ - \ h_{ij} \right) \right] \ . \label{Einstein_ij}
\eea

Summarizing, the equations of motion for the coupled Einstein--five form system are eqs.~\eqref{first_ord_5form}, together with eqs.~\eqref{Einstein_alphabeta}--\eqref{Einstein_ij}. We can now analyze their modes, starting from some sectors where the contributions from gravity and the five-form are decoupled, whose dynamics is thus simpler. These are fields in the antisymmetric of $SO(5)$, which originate solely from the four--form gauge field, spin-two modes and scalar modes in the symmetric traceless $SO(5)$ representation, which originate solely from the gravity sector. All other modes are mixed, which makes their analysis more involved and will treated in Sections~\ref{sec:singlet_vector} -- \ref{sec:singletscalarmodes}.

\section{\sc Tensor Modes Decoupled from Gravity Perturbations} \label{sec:tensor_no_grav}

According to eqs.~\eqref{first_ord_5form}, this sector involves the two fields $b^{(2)}{}_\mu{}^{lm}$ and $b_{\mu\nu}{}^{lm}$, which are both valued, for $\mathbf{k}=0$, in the antisymmetric of $SO(5)$, and no gravity contributions. We begin our analysis from the modes with $\mathbf{k}=0$, which are somewhat simpler.

\subsection{\sc $\mathbf{k}=0$ Tensor Modes}

For the modes with $\mathbf{k}=0$, eqs.~\eqref{first_ord_5form} reduce to
\bea
\partial_{[\mu}\,b^{(2)}{{}_{\nu]}}^{l m}   &=& - \  \frac{e^{-4A-4C}}{2}\, \epsilon_{\mu\nu\rho\sigma} \, \partial_r\, b^{\rho \sigma l m } \, ,\nonumber \\
 \partial_r\,{b^{(2)}}{{}_\mu}^{l m} &=&   {e^{2A+6C}}\,\frac{1}{2}\,{\epsilon^{\alpha\beta\gamma}}_\mu\,\partial_\alpha\,{b_{\beta\gamma}}^{lm} \, , \label{eqsb2b}
\eea
and it is now clearly convenient to let
\beq
{\beta_{\mu\nu}}^{lm} \ = \ \frac{1}{2}\,\epsilon_{\mu\nu\rho\sigma}\,b^{\rho\sigma l m} \ ,
\eeq
so that the equations simplify, and become
\beq
\partial_{[\mu}\,b^{(2)}{{}_{\nu]}}^{l m}   \ = \  - \  {e^{-4A-4C}}\, \partial_r\, {\beta_{\mu\nu}}^{l m } \, ,\qquad
 \partial_r\,{b^{(2)}}{{}_\mu}^{l m} \ = \    {e^{2A+6C}}\,\,\partial^\alpha\,{\beta_{\alpha\mu}}^{l m} \, .
\eeq
These equations link quantities that are invariant under residual gauge transformations, with parameters independent of $r$:
\beq
\delta\,\,\beta_{\mu\nu i j} \ = \  \epsilon_{\mu\nu\rho\sigma}\,\partial^{\rho}\,{\Lambda^{\sigma}}_{i j}\ , \qquad
 \delta\,\,{{b^{(2)}}_{\mu}}{}_{i j} \ = \  \partial_\mu\,{\Lambda^{(2)}}_{ij}  \ . \label{gauge_zero_modes}
\eeq

One can now separate the radial dependence introducing a factor $f(r)$ for $\beta_{\mu\nu i j}$ and a factor $f^{(2)}(r)$ for ${b^{(2)}}_{\mu i j}$, according to
\beq
\beta_{\mu\nu i j}(x,r) \ = \  f(r)\,\beta_{\mu\nu i j}(x) \ , \qquad
{{b^{(2)}}_{\mu}}{}_{i j}(x,r) \ = \ f^{(2)}(r)\,{{b^{(2)}}_{\mu}}{}_{i j}(x) \ .
\eeq
This leads to the system
\beq
f'\ = \  a_1\,e^{4A+4C}\,f^{(2)} \ , \qquad
f^{(2)}{}' \ = \  a_2 \, e^{2A+6C}\,f \ , \label{system_ff2}
\eeq
where $a_1$ and $a_2$ are two real constants,
and the resulting space--time modes satisfy
\bea
\partial_{[\mu}\,{b^{(2)}{}_{\nu]}}{}^{l m}   &=& - \,  a_1\, {\beta_{\mu\nu}}^{l m} \, ,\nonumber \\
a_2\,{{b^{(2)}}_\mu}^{l m} &=&   \partial^\alpha\,{\beta_{\alpha\mu}}^{lm} \, . \label{4tensor_zeromodes}
\eea
From these equations, if $a_1 a_2 \neq 0$ one obtains a second--order Proca equation
\bea
\Box\,{b^{(2)}}_\nu{}^{lm}\ - \ \partial_\nu\,\partial^\rho\, {b^{(2)}}_\rho{}^{lm}\ + \ a_1\,a_2\,{b^{(2)}}_\nu{}^{lm} &=& 0 \ ,  \label{Proca}
\eea
with a squared mass
\beq
m^2 \ = \ - \ a_1\,a_2  \label{ma1a2} \ ,
\eeq
and $\beta$ is completely determined in terms of $b^{(2)}$.

On the other hand, if $a_1=a_2=0$, $f$ and $f^{(2)}$ are independent of $r$, and then all non--trivial curvature components arising from the two potentials $\beta_{\mu\nu i j}$ and ${b^{(2)}}_{\mu i j}$,
\bea{}{}{}{}{}{}{}{}{}{}{}{}{}{}{}{}{}{}{}{}{}{}{}{}{}{}{}
H_{\mu\nu\rho i j} &=& - \ \epsilon_{\mu\nu\rho\sigma}\, \partial_\lambda\,\beta^{\lambda\sigma}{}_{ij} \ , \qquad H_{\mu\nu r i j} \ = \  \partial_r\, b_{\mu\nu i j} \ , \nonumber \\
H_{\mu\nu i j k} &=& \frac{1}{2}\,\epsilon_{i jklm}\, \partial_{[\mu}\,b_{\nu]}^{(2)\, lm} \ , \qquad H_{\mu r i jk} \ = \ - \ \frac{1}{2}\,\epsilon_{i jklm}\, \partial_r\,b_\mu^{(2)\,lm} \ , \label{curvs_bb2}
\eea
vanish, on account of eqs.~\eqref{eqsb2b}, so that this type of solution is pure gauge.

The actual values of $m^2$ are determined by eqs.~\eqref{system_ff2}, and as in other sectors it is convenient to
recast the system into a manifestly Hermitian form and then attain self--adjointness by a proper choice of boundary conditions. To this end, let us define the two functions $g^+$ and $g^-$ via
\beq
f(r) \ = \  g^-(r)\ e^\frac{A-C}{2} \ , \qquad
f^{(2)}(r) \ = \  g^+(r)\ e^{\,-\,\frac{A-C}{2}} \ ,
\eeq
while also introducing the variable $z$ of eq.~\eqref{dzdr}. These steps lead to
\beq
{\mathcal A}\, g^- \ = \ a_1\,g^+ \ , \qquad
{\mathcal A}^\dagger\, g^+ \ = \ - \ a_2 \,g^- \ , \label{sysg1g2}
\eeq
where
\beq{}{}{}
{\mathcal A} \ = \ \partial_z \,+\, \frac{A_z-C_z}{2} \ , \qquad  {\mathcal A}^\dagger \ = \ - \ \partial_z \,+\, \frac{A_z-C_z}{2} \ . \label{AAdagger_4tensor}
\eeq
Combining the two first--order equations and making use of eq.~\eqref{ma1a2} leads to either of the two manifestly Hermitian Schr\"odinger--like equations
\beq{}{}{}
{\mathcal A}^\dagger\,{\mathcal A} \,g^- \ = \  m^2 \, g^- \ , \qquad
{\mathcal A}\,{\mathcal A}^\dagger \,g^+ \ = \  m^2 \, g^+ \ ,
\label{schrod_so5tens}
\eeq
which include the potentials
\beq{}{}{}
V^\mp \ = \  \frac{1}{4}\left(A_z-C_z\right)^2\ \mp \ \frac{1}{2}\partial_z\left(A_z-C_z\right)  \ , \label{v4tensor}
\eeq
or, in detail,
\bea
V^-(r) &=& \frac{e^{\sqrt{\frac{5}{2}} \frac{r}{\rho}}}{320\, z_0^2\,\sinh^3\left(\frac{r}{\rho}\right)}  \left[10 \sqrt{10} \ \sinh \left(\frac{2\,r}{\rho}\right) \ - \ 19 \cosh\left(\frac{2\,r}{\rho}\right)\ - \ 81\right] \ , \nonumber \\
V^+(r) &=& \frac{e^{\sqrt{\frac{5}{2}} \frac{r}{\rho}}}{320\, z_0^2\,\sinh^3\left(\frac{r}{\rho}\right)}  \left[ - \ 14 \sqrt{10} \ \sinh \left(\frac{2\,r}{\rho}\right) \ + \ 41 \cosh\left(\frac{2\,r}{\rho}\right)\ + \ 99\right] \,. \label{pots12}
\eea

Their limiting behaviors at the two ends are once more of the form~\eqref{lim_dil_axion_intro}, with  $\left(\mu,\tilde{\mu}\right)=\left(\frac{1}{3},1.09\right)$ for $g^-$ and $\left(\mu,\tilde{\mu}\right)=\left(\frac{2}{3},0.09\right)$ for $g^+$. However, only a subset of the solutions of the two Schr\"odinger--like equations solves the original first--order system~\eqref{sysg1g2}, as we now explain.

Let us first note that the solution of ${\cal A}\,g^-=0$ is not normalizable. Consequently, $a_1$ cannot vanish, and it is thus convenient to absorb it in $g^+$, turning the system~\eqref{sysg1g2} into
\beq
{\mathcal A}\, g^- \ = \ g^+ \ , \qquad
{\mathcal A}^\dagger\, g^+ \ = \ m^2 \,g^- \ . \label{sysg1g2n}
\eeq
The second equation yields, for $m=0$, a zero mode for $g^+$, which solves
\beq
{\cal A}^\dagger\, g^+ \ = \ 0
\eeq
and reads
\beq
g^+ \ = \ g_{0} \ e^{\,\frac{1}{2}\left(A-C\right)} \ . \label{g2zm}
\eeq
This corresponds to a constant $f^{(2)}$ and behaves as $z^{\,-\,\frac{1}{6}}$ close to the origin. 
However, one must also solve the first equation in~\eqref{sysg1g2n}, and a normalizable solution reads
\beq
g^- \ = \ - \ e^\frac{C-A}{2}\ g_0 \ \int_z^{z_m} dz'\ e^{A(z')-C(z')}  \ \sim \ e^{B\,-\,\frac{A+C}{2}} \ = \ e^\frac{7\, A\,+\, 9\, C}{2}\ . \label{g1}
\eeq
Identifying the upper end of the integral with $z_m$ is crucial in order to obtain a normalizable mode. In fact, the integral in eq.~\eqref{g1} can be simply computed, and the result is
\beq
g^- \ = \ - \ \frac{\sqrt{10}\,\rho}{2}\ g_0 \ \left[h\,\sinh\left(\frac{r}{\rho}\right)\right]^\frac{1}{4} \ e^{\,-\,\frac{9}{4\sqrt{10}}\,\frac{r}{\rho}}\ , \label{gmzm}
\eeq
which behaves as $r^\frac{1}{4} \sim z^\frac{1}{6}$ at the origin. These zero modes describe ten massless vectors, as dictated by eqs.~\eqref{4tensor_zeromodes}, and have a constant internal profile $f^{(2)}$.

Consequently
\beq
\int_0^{z_m} \ dz \ \left( g^- \right)^2 \ \sim \ \int_0^\infty dr \ e^{\,3 B- 2 A - C} 
\eeq
is clearly convergent, since the integrand vanishes as $r$ at the origin and as $e^{\,-\,\frac{r}{\rho} \, \frac{10-7 \sqrt{10}}{10}}$ at the other end. All in all, one can work with the Schr\"odinger equation
\beq
{\cal A}^\dagger\,{\cal A}\, g^- \ = \ m^2\, g^- \ ,  \label{2ordgm}
\eeq
determining $g^+$ via $g^+={\cal A}\,g^-$. The proper normalization integral is also determined by $g^-$, according to
\beq
\int \ dz \left[ \left(g^+\right)^2 \ + \ \left(g^-\right)^2 \right] \ = \ \left(1 \,+\, m^2\right) \int \ dz \ \left(g^-\right)^2 \ .
\eeq
The main subtlety, in this case, is that the proper zero mode is not the solution of ${\cal A}\,g^-=0$, but the other independent massless solution of the second--order equation~\eqref{2ordgm}, which is given in eq.~\eqref{gmzm}.

\begin{figure}[ht]
\centering
\begin{tabular}{cc}
\includegraphics[width=65mm]{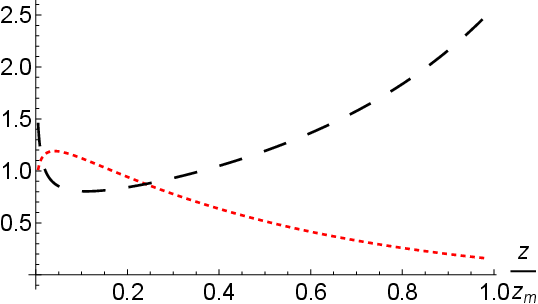} \qquad \qquad &
\includegraphics[width=65mm]{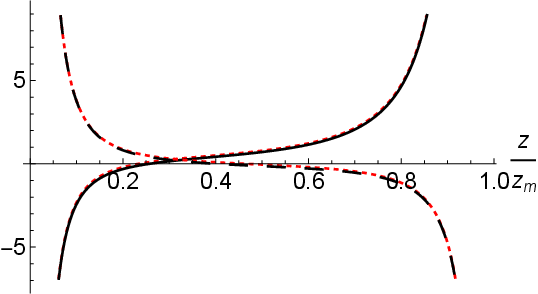} \\
\end{tabular}
\caption{\small The left panel shows the normalized zero--mode wavefunctions of eqs.~\eqref{g1} (black, dashed) and~\eqref{g2zm} (red, dotted). The right panel compares the corresponding potentials $V^-$ and $V^+$ of eq.~\eqref{pots12} (black, long-dashed and black, solid) with their approximations~\eqref{pot_hyp} with $(\mu,\tilde{\mu})=\left(\frac{1}{3},1.09\right)$ and $(\mu,\tilde{\mu})=\left(\frac{2}{3},0.09\right)$ (red,dash-dotted and red, dashed), both in units of $\frac{1}{{z_0}^2}$. $z_m$ and $z_0$ are defined in eqs.~\eqref{zm_app} and \eqref{z0}.}
\label{fig:pot_g1}
\end{figure}

We can now identify the stable self--adjoint boundary conditions for this sector referring to $g^-$, which corresponds to case 2 in Section~\ref{sec:exactschrod}. The self--adjoint boundary conditions are thus determined by the behavior at the origin,
\beq
g^- \ \sim \  C_{1} \left(\frac{z}{z_m}\right)^\frac{5}{6} \ + \ C_{2} \left(\frac{z}{z_m}\right)^{\frac{1}{6}} , \label{lim_g1g20}
\eeq
and can be parametrized by the ratio $\frac{C_2}{C_1}$, and for the zero mode~\eqref{gmzm}
\beq
\frac{C_2}{C_1} \ \simeq \ - \ 0.7 \ . \label{c21_tensor}
\eeq

\begin{figure}[ht]
\centering
\begin{tabular}{cc}
\includegraphics[width=65mm]{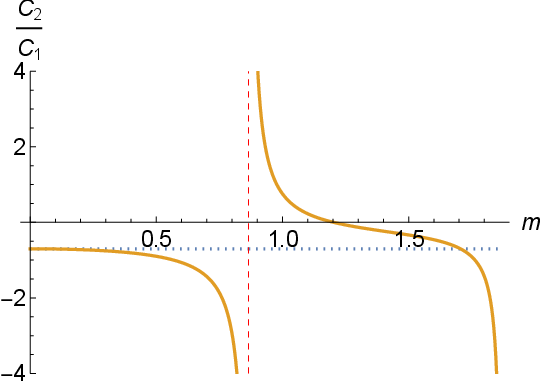} \qquad \qquad &
\includegraphics[width=65mm]{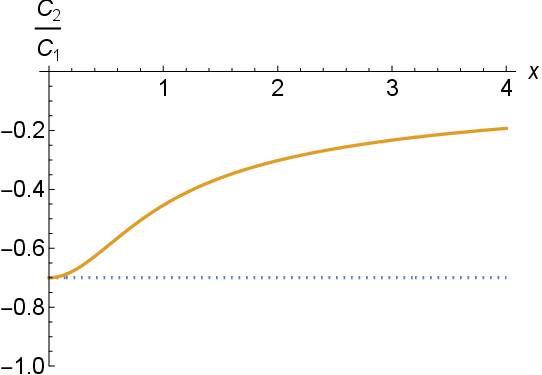} \\
\end{tabular}
\caption{\small The left panel illustrates the stable eigenvalues of the potential $V^-$ of ~\eqref{pots12} for $\mu=\frac{1}{3}$, $\tilde{\mu}=1.09$, while the red dashed line corresponds to the value $\frac{C_2}{C_1}=-0.7$ that leads to a massless mode. The right panel illustrates the presence of a tachyonic mode, with $m^2=-x^2$, for the same values of $\mu$ and $\tilde{\mu}$ and $-0.7 <\frac{C_2}{C_1}<0$.}
\label{fig:hyperex_tensor}
\end{figure}

As in previous cases, we now approximate $V^-$ of eq.~\eqref{pots12} with a hypergeometric potential~\eqref{pot_hyp} with $(\mu,\tilde{\mu})=\left(\frac{1}{3},1.09\right)$ and a suitable shift $\Delta\,V$, which we determine demanding that the boundary condition~\eqref{c21_tensor} correspond to a massless solution for the shifted potential. In this fashion one finds
\beq
\Delta\,V \ \simeq \ - \ (0.15)^2\, \frac{\pi^2}{z_m^2} \ ,
\eeq
and the resulting hypergeometric eigenvalue equation becomes
\beq
\frac{C_2}{C_1} \ = \ - \ \left(\frac{\pi}{2}\right)^{\,-\,\frac{2}{3}} \ \frac{\Gamma\left(\frac{5}{3}\right)\Gamma\left(0.88 \,+\, \sqrt{m^2+0.15^2}\right) \Gamma\left(0.88 \,-\, \sqrt{m^2+0.15^2}\right)}{\Gamma\left(1.21 \,+\, \sqrt{m^2+0.15^2}\right) \Gamma\left(1.21 \,-\, \sqrt{m^2+0.15^2}\right)} \ . \label{eigenvmutildelarger4}
\eeq
The solutions are illustrated in fig.~\ref{fig:hyperex_tensor}, and comprise and infinite number of massive modes and at most a single tachyonic mode.
These results are qualitatively similar to what we displayed in fig.~\ref{fig:hyperex}, and the instability region corresponds in this case to the range
\beq
- 0.7 \ < \ \frac{C_2}{C_1} \ < \ 0 \ .  \label{unst_tensor}
\eeq
Note that the tachyon mass grows indefinitely, in absolute value, as the ratio $\frac{C_2}{C_1}$ approaches zero from negative values.

We can now examine how the choice of self--adjoint boundary conditions affects the flow of the energy--momentum tensor across the boundary. In this sector, the relevant component of the energy--momentum tensor behaves as
\beq
\sqrt{-g}\ T^r{}_\mu \ \sim \ g^+ \, g^- \ ,
\eeq
so that the no--flow conditions~\eqref{bc_bose_interval_mui} are violated, at the origin, whenever $C_2 \neq 0$, since $g^+\,g^- \sim C_2{}^2$ at $z=0$, and in particular by the boundary condition~\eqref{c21_tensor} allowing the zero mode. 

Summarizing, within the stable region $\frac{C_2}{C_1} \geq 0$ or $\frac{C_2}{C_1} \leq - 0.7$ there are three classes of boundary conditions: there is a unique choice leading to a massless spectrum for ten vectors in the antisymmetric of $SO(5)$ corresponding to eq.~\eqref{c21_tensor}, while all other choices lead to purely massive spectra. Among them, a special choice, obtained as $\frac{C_2}{C_1}$ approaches zero from above, respects the no-flow condition.

\subsection{\sc $\mathbf{k} \neq 0$ Tensor Modes} \label{sec:tensor_no_grav_knot0}

We can now discuss the modes with a non--vanishing internal momentum $\mathbf{k}$.
The starting point is now the system in eqs.~\eqref{first_ord_5form}, and including (complex) Fourier modes $e^{i \mathbf{k}\cdot\mathbf{y}}$ in the internal torus leads to the first--order equations
\bea
\partial_{[\mu}\,b^{(2)}{{}_{\nu]}}^{l m} \ + \ \frac{i}{2}\,\epsilon^{p q r l m}\,k_p\,b_{\mu\nu q r}   &=& - \,  \frac{e^{-4A-4C}}{2}\, \epsilon_{\mu\nu\rho\sigma} \ \partial_r\, b^{\rho \sigma l m } \, ,\nonumber \\
 - \ \partial_r\,{b^{(2)}}{{}_\mu}^{l m}&=& - \  {e^{2A+6C}}\,\frac{1}{2}\,{\epsilon^{\alpha\beta\gamma}}_\mu\,\partial_\alpha\,{b_{\beta\gamma}}^{lm} \, ,
\nonumber \\
 k_n\,{{b^{(2)}}_\mu}{}^{m n}  &=& 0 \ , \label{sys_internal}
\eea
where we omit the $\mathbf{k}$ suffix for brevity. Taking the divergence with respect to the internal coordinates gives
\beq
k_m\,\partial_r\, b^{\rho \sigma l m } \ = \  0 \ , \qquad
k_m\,\partial_{[\alpha}\,{b_{\beta\gamma]}}^{lm} \ = \ 0 \ .
\eeq
Using the residual gauge symmetries of eqs.~\eqref{res_tensor_sym}, one can demand that $k_m\,b^{\rho\sigma l m}=0$, because the first equation grants that is independent of $r$, while all preceding gauge choices only involved the $r$-dependent portions of the gauge parameters. As a result, one can work with fields that are transverse in the internal space.

Now the momentum $\mathbf{k}$ picks a direction in the internal space, so that for $\mathbf{k} \neq 0$ the two fields in eqs.~\eqref{sys_internal} are in the antisymmetric $\underline{6}$ of the residual internal $SO(4)$ that is transverse to it. One can thus decompose them further into portions that are self-dual and antiself-dual with respect to this $SO(4)$ group, which we denote by $b^{(\epsilon)}$ and $b^{(2,\epsilon)}$, with $\epsilon=\pm 1$, such that
\beq
b^{(\epsilon) \,\rho \sigma\, l m} \ = \ \frac{\epsilon}{2} \, \epsilon^{l m p q s} \,\frac{k_s}{\left|\mathbf{k}\right|} \ {b^{(\epsilon) \,\rho \sigma}}_{p q}\ , \qquad b^{(2,\epsilon) \,\rho \,l m} \ = \ \frac{\epsilon}{2} \, \epsilon^{l m p q s} \,\frac{k_s}{\left|\mathbf{k}\right|} \ {b^{(2,\epsilon) \,\rho }}_{p q} \ .
\eeq
As a result, for example, $b^{(\epsilon) \,\rho \sigma}$ is $SO(4)$--valued along directions transverse to $\mathbf{k}$, and the system becomes
\bea
\partial_{[\mu}\,b^{(2,\epsilon)}{{}_{\nu]}}^{l m} \ + \ {i\,\left|\mathbf{k}\right| \,\epsilon}\,{b^{(\epsilon)}{}_{\mu \nu}}^{l m}   &=& - \,  \frac{e^{-4A-4C}}{2}\, \epsilon_{\mu\nu\rho\sigma} \ \partial_r\, b^{(\epsilon) \,\rho \sigma l m } \, ,\nonumber \\
 - \ \partial_r\,{b^{(2,\epsilon)}}{{}_\mu}^{l m}&=& - \  {e^{2A+6C}}\,\frac{1}{2}\,{\epsilon^{\alpha\beta\gamma}}_\mu\,\partial_\alpha\,{{b^{(\epsilon)}}_{\beta\gamma}}^{l m} \, . \label{tensorknot0}
 \eea
One can now separate variables, letting
\beq{}{}{}{}{}{}{}{}{}{}{}{}{}{}{}{}{}{}{}{}{}{}{}{}{}{}{}{}
b^{(2,\epsilon)}{}_{\nu}{}^{l m}(x,r) \ = \ f_2^\epsilon(r) \, b^{(2,\epsilon)}{}_{\nu}{}^{l m}(x) \ , \qquad b^{(\epsilon)}{}_{\mu\nu}{}^{l m}(x,r) \ = \ f^{(\epsilon)}(r) \, b^{(\epsilon)}{}_{\mu\nu}{}^{l m}(x) \ ,
\eeq
and the first of eqs.~\eqref{tensorknot0} implies that this is only possible if $b^{(\epsilon)}{}_{\mu\nu}{}^{lm}$ is also self--dual or antiself-dual in spacetime, so that
\beq
{b^{(\epsilon,\zeta)}{}_{\mu \nu}}^{l m} \ = \ i\, \frac{\zeta}{2}\,\epsilon_{\mu\nu\rho\sigma}\, b^{(\epsilon,\zeta)\rho\sigma\,l m} \ , \label{bmunuzeta}
\eeq
with $\zeta=\pm 1$. As a result, $f$ and $f_{2}$ satisfy
\bea
&& i\zeta\,e^{-4(A+C)}\,\frac{df{}^{(\epsilon,\zeta)}}{dr} \ - \  i\,\left|\mathbf{k}\right| \epsilon \, f^{(\epsilon,\zeta)} \ = \ a_1\,f_2{}^{(\epsilon,\zeta)} \ , \nonumber \\
&& \frac{df_2^{(\epsilon,\zeta)}}{dr} \ = \ i \zeta a_2\,e^{2A+6C}\,f^{(\epsilon,\zeta)} \ , \label{sysff2p2}
\eea
where $a_1$ and $a_2$ are two constants, not necessarily real anymore since we are dealing with Fourier modes, while the spacetime equations become
\bea
\partial_{[\mu}\,b^{(2,\epsilon)}{{}_{\nu]}}^{l m} &=& a_1\, {b^{(\epsilon,\zeta)}{}_{\mu \nu}}^{l m} \ , \nonumber \\
a_2\,b^{(2,\epsilon)}{{}_{\nu}}^{l m} &=& - \ \partial^\alpha\,{{b^{(\epsilon,\zeta)}}_{\alpha\mu}}^{l m} \ .
\eea
If $a_1 \neq 0$, the first implies that the field strength of $b_\mu{}^{(2,\epsilon) lm}$ must also satisfy eq.~\eqref{bmunuzeta}, and
combining them yields, as before, a second--order Proca equation for ${b^{(2)}}{}_\mu{}^{lm}$,
\beq
\Box\,{b^{(2,\epsilon)}}{}_\mu{}^{lm} \ - \ \partial_\mu\,\partial^\nu\,{b^{(2,\epsilon)}}{}_\nu{}^{lm} \ - \ m^2\,{b^{(2,\epsilon)}}{}_\mu{}^{lm} \ = \ 0  \ ,  \label{tensormnotzero}
\eeq
with
\beq
m^2 \ =\ - \ a_1\,a_2 \ , \label{a12m}
\eeq
where $a_1$ and $a_2$ are the constants entering eqs.~\eqref{sysff2p2}.

In order to recast the system in a manifestly Hermitian form, we turn once more to the $z$ variable of eq.~\eqref{dzdr} and redefine the two functions $f$ and $f_2$ according to
\beq{}{}{}
f^{(\epsilon,\zeta)} \ = \ g^- \ e^{\frac{A-C}{2}\,+\,\chi} \ , \qquad f_{2}^{(\epsilon,\zeta)} \ = \ g^+ \ e^{\frac{C-A}{2}\,+\,\chi}  \ ,
\eeq
where
\beq{}{}{}
\partial_z\,\chi \ = \ \frac{\left|\mathbf{k}\right|}{2}\,\epsilon\,\zeta\, e^{A-C} \ , \label{chi_eq}
\eeq
and for brevity we leave the two signs $\epsilon$ and $\zeta$ implicit in $g^\pm$.
The end result is
\beq
{\mathcal A}\, g^- \ = \ - \ i\,\zeta\,a_1\,g^+ \ , \qquad
{\mathcal A}^\dagger\, g^+ \ = \ - \ i\,\zeta\,a_2\,g^- \label{systknot0}
\eeq
where now
\beq
{\mathcal A} \ = \ \partial_z \,+\, \frac{A_z-C_z}{2} \ - \ \partial_z\,\chi , \qquad  {\mathcal A}^\dagger \ = \ - \ \partial_z \,+\, \frac{A_z-C_z}{2}\ - \ \partial_z\,\chi\ ,
\eeq
which modify the combinations in eqs.~\eqref{AAdagger_4tensor}. The solution to ${\cal A}\,g^-=0$ is
\beq
g^- \ = \  c_1 \ e^{\frac{C-A}{2}\,+\, \chi} \ ,
\eeq
and, as for $\mathbf{k}=0$, is not normalizable. As a result, $a_1$ must be different from zero and can again absorbed in $g^+$. The system~\eqref{systknot0} can thus be replaced with
\beq
{\mathcal A}\, g^- \ = \ g^+ \ , \qquad
{\mathcal A}^\dagger\,{\cal A}\, g^- \ = \ m^2 \,g^- \ , \label{systknot2}
\eeq
where the Schr\"odinger--like equation is now
    \beq
\left[ - \partial_z^2 \ + \ V^- \ + \ \frac{\left|\mathbf{k}\right|^2}{4} \  e^{2(A-C)}\right] g^- \ = \ m^2\ g^- \ ,
    \eeq
with $V^-$ the first potential in eq.~\eqref{pots12}, and the additional $\left|\mathbf{k}\right|^2$ term is subdominant, at both ends, with respect to $V^-$. Therefore the leading singularities remain the same as for $\mathbf{k}=0$, and the allowed self--adjoint boundary conditions are still determined by the ratio $\frac{C_2}{C_1}$. Since one is adding to the potential of the preceding section a positive contribution, all modes corresponding to stable boundary conditions for $\mathbf{k}=0$ are simply lifted in mass by the internal momentum.  Moreover, the tachyonic modes, which have a continuous spectrum of masses determined by the boundary conditions, can be lifted to zero mass. In fact, for all values of $\mathbf{k}$ there are massless modes, which are obtained solving
\beq
{\cal A}^\dagger\,g^+ \ = \ 0 \ , \qquad {\mathcal A}\, g^- \ = \ g^+ \ ,
\eeq
whose internal wavefunctions are
\beq
g^+ \ = \  c_2 \ e^{\frac{A-C}{2}\,-\,\chi} \ , \qquad g^- \ \sim  \ e^{\frac{C-A}{2}\,+\,\chi}\ \int_{z_m}^z dz'\ e^{A(z')-C(z') - 2 \chi(z')} \ .
\eeq
    Here $\chi$ can determined exactly as a function of $r$ solving eq.~\eqref{chi_eq}, and reads
        \beq
\chi \ =  \ -  \ \sqrt{10}\,\epsilon\,\zeta\,\rho\,\left|\mathbf{k}\right| \,e^{\,-\,\frac{r}{2\,\rho\sqrt{10}}} \ .
    \eeq
    It approaches a constant at both ends of the interval, but nonetheless it changes the value of the ratio $\frac{C_2}{C_1}$ by $\mathbf{k}$-dependent terms. Note that the emergence of additional massless modes occurs in all sectors, when starting from tachyonic boundary conditions. The peculiar feature of this sector is that, even for $\mathbf{k} \neq 0$, the $r$-profile of these modes can be still deduced from the first--order equation~\eqref{systknot2}.
    
    Summarizing, for all values of $\frac{C_2}{C_1}$ in the range~\eqref{unst_tensor} there is a boundary condition lifting a $\mathbf{k}=0$ tachyon to a massless mode carrying an internal momentum. However, for all boundary conditions that yield only stable modes for $\mathbf{k}=0$ the complete spectrum for $\mathbf{k} \neq 0$ is also stable.

\section{\sc Spin--2 Modes from $h_{\mu\nu}$} \label{sec:4Dgrav}

We can now turn to modes involving perturbations of the metric field. The simplest case is obtained considering a traceless and divergence--free $h_{\mu\nu}$, whose massless modes can describe long--range gravity in lower dimensions. This spin--2 portion cannot mix with anything else, and therefore one can set to zero all other perturbations when addressing it. The dynamical $\alpha\beta$ Einstein equation~\eqref{Einstein_alphabeta} then reduces to
\beq{}{}{}{}{}{}{}{}{}{}{}{}{}{}{}{}{}{}{}{}{}{}{}{}{}{}{}{}{}{}{}{}{}{}{}{}{}{}{}{}{}{}{}{}{}{}{}{}{}{}{}{}{}{}{}{}{}{}{}{}{}{}{}{}{}{}{}{}{}{}{}
\left(e^{-2A} \, \Box\,-\,e^{-2C}\,\mathbf{k}^2\right)h_{\alpha\beta} \, + \, e^{-2B}\big[\left(\partial_r-4A'\right)\partial_r\,+\,4\,(A')^2\big]h_{\alpha\beta} \, - \,  \frac{h^2}{2\,\rho^2} \ e^{- 10 C} \  h_{\alpha\beta} \, = \, 0 \, ,
\eeq
while the other equations are identically satisfied.
As usual, the d'Alembertian operator defines $m^2$, so that the preceding result is equivalent to
\beq
\left[ \left(\partial_r\,-\,2\,A'\right)^2 \ + \ \left(m^2\,e^{-2A}\,-\,\mathbf{k}^2\,e^{-2C}\right)e^{2B} \right] h_{\alpha\beta} \ = \ 0 \ ,
\eeq
where we used some properties of the background, and in particular eqs.~\eqref{A2C2}.
It is now convenient to introduce once more the variable $z$ of eq.~\eqref{dzdr}, and redefining the field according to
\beq
h_{\alpha\beta} \ = \ e^{-\frac{1}{2}\,B\,+\,\frac{5}{2}\,A}\ \tilde{h}_{\alpha\beta} \ , \label{hhtilde}
\eeq
leads finally to the Schr\"odinger--like equation
\beq
\left[\partial_z^2 \ + \ m^2\ - \ \mathbf{k}^2\,e^{2(A-C)}\ - \ \frac{1}{4}\left(B_z\,-\,A_z\right)^2\,-\,\frac{1}{2}\left(B_{zz}\,-\,A_{zz}\right)\right] \tilde{h}_{\alpha\beta} \ = \ 0 \ . \label{htilde_modes}
\eeq

This equation is identical to the one obtained for the dilaton--axion system in~\eqref{aadaggerdilaton}, and leads to the same spectra for the self--adjoint boundary conditions discussed there.  The same considerations apply, and in particular there are boundary conditions granting spectra with $m^2 \geq0$, as in fig.~\ref{fig:dil_instabilities}. For $\mathbf{k}=0$ there is the zero mode~\eqref{dil_zero_mode}, which translates in this case into the graviton wavefunction
\beq{}{}{}{}{}
{h}_{\alpha\beta}(r) \ = \ e^{2A}\, \tilde{h}^{(0)}{}_{\alpha\beta} \ , \label{h0_nottilde}
\eeq
which satisfies the boundary condition
\beq
\partial_r\left( r^\frac{1}{2}\ h_{\mu\nu}\right) \ \sim 0  \ .
\eeq
The finite normalization integral for this ground--state wavefunction,
\beq{}{}{}{}{}
\int_0^\infty dr \ e^{2B-6A}\,h_{\alpha\beta}\,h^{\alpha\beta} \ \sim \ \int_0^\infty dr \ e^{2(B-A)} \ ,
\eeq
also determines a finite value of the four--dimensional Planck mass, as discussed in~\cite{ms22_1}. This zero mode is crucial, since it grants the existence of a long--range four--dimensional effective gravity. It corresponds to the same self--adjoint boundary conditions identified in Section~\ref{sec:dilatonaxion}. As for the dilaton, for any given boundary condition the spectrum for $\mathbf{k} \neq 0$ is lifted upwards in mass, and therefore all stable choices remain stable also in the presence of an internal momentum.

\section{\sc Scalar Modes from $h_{i j}$} \label{sec:hij}

Metric perturbations arising from $h_{ij}$ are also simple to analyze. These are symmetric traceless $SO(5)$ two-tensors for $\mathbf{k}=0$, which are also transverse in the internal space if $\mathbf{k}\neq 0$ and cannot mix with other modes. All Einstein equations are then identically satisfied, aside from the $ij$ equation~\eqref{Einstein_ij}, which reduces to
\bea
(ij)&:& \left(e^{-2A}\,\Box \,+\, e^{-2C} \, \Delta\right)h_{i j} \,+\,e^{-2B}\big[\left(\partial_r-4C'\right)\partial_r\,+\,4\,(C')^2\big]h_{i j}  \nonumber \\ &=&  -\ \frac{h^2}{2\,\rho^2}\,e^{-10C}\,h_{ij} \ .
\eea
In terms of Fourier modes and of the $z$ variable, this becomes
\beq
 \left(m^2 \,-\, e^{2(A-C)} \, \mathbf{k}^2 \right)h_{i j} \,+\,\Big(\partial_z\,+\,3\left(A_z+C_z\right)\Big)\Big(\partial_z\,-\,2C_z\Big) h_{i j} \ = \ 0 \ ,
\eeq
after using some identities for the background collected in Appendix~\ref{app:background}. One can now turn this equation into a manifestly Hermitian form, letting
\beq{}{}{}{}{}
h_{i j} \ = \ \tilde{h}_{ij}\ e^{\,-\,\frac{3A+C}{2}}  \ ,
\eeq
and the result is identical to eq.~\eqref{htilde_modes} or eq.~\eqref{aadaggerdilaton}. Therefore, the arguments of the preceding section apply almost verbatim, the only difference being that the spacetime profile of the massless mode, which is now valued in the $\underline{14}$ of $SO(5)$, is in this case
\beq{}{}{}{}{}{}{}{}{}{}{}{}{}{}{}{}{}{}{}{}{}{}{}{}{}{}{}{}{}{}{}{}{}
h_{ij} \ \sim \ e^{B-A} \ . \label{limitinghij}
\eeq
As a result $h_{ij} \ \sim r^\frac{1}{2}$ as $r \to 0$, or equivalently as $z^\frac{1}{6}$ as $z\to 0$, s. Moreover, this wavefunction behaves as $\left(z_m-z\right)^{-\,\frac{2}{\sqrt{10}}}$ as $z \to z_m$.

\section{\sc Singlet Vector Modes} \label{sec:singlet_vector}

We can now turn to the four--dimensional vector modes that are invariant under the effective internal symmetry group ($SO(5)$ or $SO(4)$, as we have seen, depending on the values of $\mathbf{k}$) . These vector modes originate partly from the tensor, with
\beq{}{}{}{}{}
b_{\mu i} \ = \ \frac{1}{\Sigma}\,\partial_i\, V_{1\,\mu} \ ,
\eeq
and partly from gravity, with the relevant fluctuations parametrized as
\beq
h_{\mu\nu} \,=\, \frac{1}{\Sigma}\,\partial_{(\mu}\,V_{2\,\nu)}  \ , \qquad
h_{\mu r} \,=\, V_{3\,\mu} \ , \qquad
h_{\mu i} \,=\, \frac{1}{\Sigma}\,\partial_i \, V_{4\,\mu} \ ,
\eeq
where $\Sigma$ is a scale that grants the different fields the standard dimension. Moreover, the four vector fields $V_1$, $V_2$, $V_3$ and $V_4$ are all assumed to be divergence-free,
\beq{}{}{}{}{}
\partial^\mu\,V_{a\,\mu} \ = \ 0 \qquad \quad (a=1,\ldots, 4) \ ,
\eeq
in order not to include scalar components in them.
As a result, the tensor equations~\eqref{first_ord_5form} become
\bea
\partial^{[l}\,{b_\mu}^{m]} &=& 0 \ ,
\qquad
\frac{h}{2\,\rho} \ {h_\mu}^m  \,-\, {e^{-6A}}\, \partial_r\,{b_\mu}^m \, =\,  0 \ ,
\nonumber \\
\partial_\mu\,b^{\mu m} &=& 0 \ , \qquad
{h^\mu}_\mu \,=\, 0 \ ,\label{eqs_sd2_2}
\eea
and the only non--trivial one is the second, which leads to
\beq{}{}{}{}{}
\partial^m \left[\frac{h}{2\,\rho}\,V_{4\,\mu} \ - \ e^{-6A}\,\partial_r\,V_{1\,\mu} \right] \ = \ 0 \ , \label{V14}
\eeq
while the others are identically satisfied. Now $V_{2\,\mu}$ can be gauged away, and the Einstein equations~\eqref{Einstein_alphabeta} -- \eqref{Einstein_alphai} reduce to
\bea
(\alpha\beta)&:&  \partial_{(\alpha} \left[e^{-2B}\left(\partial_r\,-\,2\,A'\right)V_{3\,\beta)}\,+\, \frac{\mathbf{k}^2}{\Sigma}\,e^{-2C}V_{4\,\beta)} \right]\ = \  0 \ , \nonumber \\
(\alpha r)&:& \left(e^{-2A} \, m^2\,-\,e^{-2C} \, \mathbf{k}^2 \right) V_{3\,\alpha} \,+\, \frac{\mathbf{k}^2}{\Sigma}\, e^{-2C}\left(\partial_r\,-\,2 A'\right)\,V_{4\,\alpha}  \ = \  0 \ ,
\nonumber \\
(\alpha i)&:&  \partial_i\Big\{ \frac{e^{-2A}}{\Sigma}\,m^2\,V_{4\,\alpha}\ - \ e^{-2B}\left(\partial_r\,-\,2\,C' \right)V_{3\,\alpha}  \label{eqs_singlet_vector} \\
&+&\frac{1}{\Sigma}\, e^{-2B}\left[\left(\partial_r\,-\,A'\,-\,C' \right)^2 \ - \ \left(A'\,-\,C'\right)^2\right]V_{4\,\alpha}   \, - \, \frac{e^{-10C}}{2\,\rho^2\,{\Sigma}}\,h^2\,V_{4\,\alpha}\Big\} = 0\ ,\nonumber
\eea
while the rest,
the $(r r)$, $(r i)$ and $(i j)$ equations, are all identically satisfied. One can also remove the overall $\partial_\alpha$ from the first group of equations, while the $(\alpha i)$ equation, together with the $V_{1\,\mu}$ and $V_{4\,\mu}$ fields, is only present for nonzero momenta.

\subsection{\sc $\mathbf{k}=0$ Modes}

For $\mathbf{k}=0$ only the first two of eqs.~\eqref{eqs_singlet_vector} are left, with only $V_3$, and give
\bea
(\alpha\beta)&:& \left(\partial_r\,-\,2\,A'\right)V_{3\,\beta} \ = \  0 \ , \nonumber \\
(\alpha r)&:& m^2\, V_{3\,\alpha}  \ = \  0 \ . \label{vector_zerom}
\eea
Therefore, there is in principle a massless vector with a wavefunction in the radial direction
\beq
V_{3\,\mu} \ = \ h_{\mu r} \ = \ e^{\,2A}\, V^{(0)}{}_{3\,\mu}(x) \ . \label{firstzeromode}
\eeq
This, however, is not a normalizable mode, since the norm inherited from the Einstein--Hilbert action,
\beq
\int_0^\infty dr \ \sqrt{-g}\,\partial_\rho\,h_{\mu r}\,\partial_\sigma h_{\nu r}\, \eta^{\mu\nu}\,\eta^{\rho\sigma}\, e^{-2(2A+B)} \ = \ - \,\int_0^\infty dr \ V^{(0)}{}_{3\,\mu}\,\Box\,V^{(0)}{}_{3\,\nu}\, \eta^{\mu\nu}\ ,
\eeq
diverges. In conclusion, this sector describes no modes altogether.

\subsection{\sc $\mathbf{k} \neq 0$ Modes}

For $\mathbf{k}\neq 0$, the system becomes
\bea
(\alpha\beta)&:&  e^{-2B}\left(\partial_r\,-\,2\,A'\right)V_{3\,\beta}\,+\, \frac{\mathbf{k}^2}{\Sigma}\,e^{-2C}V_{4\,\beta} \ = \  0 \ , \nonumber \\
(\alpha r)&:& \left(e^{-2A} \, m^2\,-\,e^{-2C} \, \mathbf{k}^2 \right) V_{3\,\alpha} \,+\, \frac{\mathbf{k}^2}{\Sigma}\, e^{-2C}\left(\partial_r\,-\,2 A'\right)\,V_{4\,\alpha}  \ = \  0 \ , \nonumber 
\\
(\alpha i)&:&  \frac{e^{-2A}}{\Sigma}\,m^2\,V_{4\,\alpha}\ - \ e^{-2B}\left(\partial_r\,-\,2\,C' \right)V_{3\,\alpha} \nonumber \\
&+&\frac{1}{\Sigma}\, e^{-2B}\left[\left(\partial_r\,-\,A'\,-\,C' \right)^2 \ - \ \left(A'\,-\,C'\right)^2\right]V_{4\,\alpha}   \, - \, \frac{e^{-10C}}{2\,\rho^2\,{\Sigma}}\,h^2\,V_{4\,\alpha} = 0\ . \label{eqs_singlet_vector_k}
\eea
The first of these equations determines $V_4$ in terms of $V_3$, and this relation can be used in the second, which becomes
\beq
e^{2(A-B)}\left(\partial_r\,-\,2\,A'\,-2\,B'\,+\,2\,C'
\right)\left(\partial_r\,-\,2\,A'\right) V_{3\,\alpha} \ + \ \left(m^2 \, -\, \mathbf{k}^2\,e^{2(A-C)}\right)V_{3\,\alpha} \ = \ 0 \ , \label{eigenvm}
\eeq
while the third equation follows from the first two. Notice that the scale $\Sigma$ does not enter this eigenvalue equation, which determines the mass spectrum.

In terms of $z$--derivatives, making use of eq.~\eqref{dzdr}, eq.~\eqref{eigenvm}  becomes
\beq
\left(\partial_z\,-\,3\,A_z\,-\,B_z\,+\,2\,C_z
\right)\left(\partial_z\,-\,2\,A_z\right) V_{3\,\alpha} \ + \ \left(m^2 \, -\, \mathbf{k}^2\,e^{2(A-C)}\right)V_{3\,\alpha} \ = \ 0 \ ,
\eeq
and letting
\beq
\omega \ = \ \frac{5 A_z\,+\,3\,C_z}{2} \ , \qquad
V_{3\,\alpha} \ = \ e^{\frac{9\,A\,+\,3\,C}{2}}\,\widetilde{V}_\alpha(x) \, f(z) \ ,
\eeq
leads to the manifestly Hermitian equation
\beq
\left(\partial_z\,-\,\omega\right)\left(\partial_z\,+\,\omega\right) f \ + \ \left(m^2 \, -\, \mathbf{k}^2\,e^{2(A-C)}\right) f \ = \ 0 \ . \label{eq_singlet_vector_k}
\eeq
\begin{figure}[ht]
\centering
\includegraphics[width=65mm]{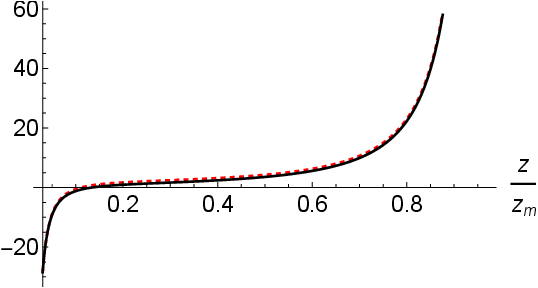}
\caption{\small The $\mathbf{k}=0$ portion of the Schr\"odinger potential for $f$ in eq.~\eqref{pot_vm} (black, solid) and its approximation~\eqref{pot_hyp} (red, dashed) with $(\mu,\tilde{\mu})=\left(\frac{1}{3},2.18\right)$, in units of $\frac{1}{{z_0}^2}$. $z_m$ and $z_0$ are defined in eqs.~\eqref{zm_app} and \eqref{z0}.}
\label{fig:pot_vm}
\end{figure}

We have thus reached once more a familiar form
\beq{}{}{}{}{}{}{}
\left[ {\cal A}^\dagger\,{\cal A} \ + \ \mathbf{k}^2\,e^{2(A-C)}\right] f \ = \ m^2\, f \ , \label{schrod_vm}
\eeq
where now
\beq{}{}{}{}{}
{\cal A} \ = \ \partial_z \ + \ \omega \ , \qquad {\cal A}^\dagger \ = \ - \ \partial_z \ +\ \omega \ ,
\eeq
so that the Schr\"odinger potential is
\beq
V \ = \ \frac{e^{\sqrt{\frac{5}{2}} \,\frac{r}{\rho}}}{320\,z_0^2\,\sinh^3\left(\frac{r}{\rho}\right)} \left[10 \sqrt{10}\, \sinh\left(\frac{2\,r}{\rho}\right) \ + \ 29 \cosh\left(\frac{2\,r}{\rho}\right)\ -\ 129\right]\ . \label{pot_vm}
\eeq

The limiting behavior of the potential at the two ends is as in eq.~\eqref{lim_dil_axion_intro}, with $\mu=\frac{1}{3}$ and $\tilde{\mu}=\frac{1}{15} \left(4 \sqrt{10}+20\right) \simeq 2.18$. Since $\tilde{\mu}>1$, the only limiting behavior allowed at the right end is
\beq
f \ \sim \ \left( 1 \ - \ \frac{z}{z_m}\right)^\frac{55 \ + \ 8 \sqrt{10}}{30} \ ,
\eeq
while there are different options at the left end, with
\beq
f \ \sim \ C_1 \ \left(\frac{z}{z_m}\right)^\frac{5}{6} \ + \ C_2\ \left(\frac{z}{z_m}\right)^\frac{1}{6} \ ,
\eeq
and the self--adjoint boundary conditions are determined by the ratio $\frac{C_2}{C_1}$. Note that the solution of ${\cal A}f=0$ that we found in the previous section is not normalizable, due to its behavior as $r \to \infty$, but there is another independent massless solution of the Schr\"odinger equation~\eqref{schrod_vm} with $\mathbf{k}=0$. It can be determined by the Wronskian method, and is given by
\beq
f \ = \ e^\frac{11 A+ 13 C}{2} \ .
\eeq
It is normalizable and is characterized by
\beq
\frac{C_2}{C_1} \ \simeq \ - \ 0.47 \ .
\eeq
Note that this zero mode is not a solution of eqs.~\eqref{eqs_singlet_vector_k} for $\mathbf{k}=0$, but nonetheless it can be used to determine the shift for the hypergeometric approximation to the $\mathbf{k}$-independent portion of the potential,
\beq
\Delta\,V \ \simeq \ - \ \frac{\pi^2}{z_m^2} (0.39)^2 \ .
\eeq
The resulting eigenvalue equation reads
\beq
\frac{C_2}{C_1} \ = \ - \ \left(\frac{\pi}{2}\right)^{\,-\,\frac{2}{3}} \ \frac{\Gamma\left(\frac{4}{3}\right)\Gamma\left(1.42 \,+\, \sqrt{m^2+(0.39)^2}\right) \Gamma\left(1.42 \,-\, \sqrt{m^2+(0.39)^2}\right)}{\Gamma\left(\frac{2}{3}\right)\Gamma\left(1.76 \,+\, \sqrt{m^2+(0.39)^2}\right) \Gamma\left(1.76 \,-\, \sqrt{m^2+(0.39)^2}\right)} \ , \label{eigenvmutildelarger2}
\eeq
so that the corresponding instability region for the $\mathbf{k}=0$ portion of the potential is
\beq
- 0.47 \ < \ \frac{C_2}{C_1} \ < \ 0 \ .
\eeq
In conclusion, if $\frac{C_2}{C_1}$ lies outside this range, only massive modes emerge from this sector, when one takes into account the complete $\mathbf{k}$-dependent potential in eq.~\eqref{schrod_vm}. However, fine--tuning the ratio as a function of the toroidal radius it is possible to lift a tachyonic mode of the $\mathbf{k}$--independent potential to zero mass, whenever it is present.

\section{\sc Non--Singlet Vector Modes} \label{sec:nonsingletvectors}
These modes also originate partly from the self--dual tensor and partly from gravity, and can be exhibited letting
\bea
b^{(2)}{}_\mu{}^{i j} &=& \frac{1}{\Sigma}\ \partial^{[i}\,W_{1}{}_\mu{}^{j]} \ , \qquad
\frac{1}{2}\, \epsilon^{\alpha\beta\mu\nu}\,b_{\mu\nu}{}^{i j} \,=\, \frac{1}{{\Sigma}^2}\ \partial^{[i}\partial^{[\alpha}W_{2}{}^{\beta]}{}^{j]}\,+\, \frac{1}{2\,{\Sigma}^2}\,\epsilon^{\alpha\beta\mu\nu}\,\partial_{[\mu}\partial^{[i}\widetilde{W}_2^{j]}{}_{\nu]} \ , \nonumber \\
b_\mu{}^i &=& W_{3\,\mu}{}^i \ , \qquad
h_\mu{}^i \,=\, W_{4\,\mu}{}^i \ ,
\eea
where $\Sigma$ is again a scale that grants the different fields standard dimensions.
All other perturbations are set to zero.
Moreover, in order to leave aside scalar and singlet vector modes, the $W$ fields thus defined are chosen to be \emph{divergence-free} both in spacetime and in the internal space, where the second condition is only relevant for $\mathbf{k} \neq 0$. In this fashion, the tensor equations reduce to
\bea
\partial^{[i}\partial_{[\mu}\,W_{1\,\nu]}{}^{j]}    &=& - \,  \frac{e^{-4A-4C}}{\Sigma}\, \partial_r\, \partial^{[i}\left( \partial_{[\mu} W_2{}_{\nu]}{}^{j]} \,+\, \frac{1}{2}\,\epsilon_{\mu\nu}{}^{\alpha\beta}\, \partial_{[\alpha} \widetilde{W}_2{}_{\beta]}{}^{j]}\right)\, ,\nonumber \\
\frac{1}{\Sigma}\  \partial_r\,\partial^{[i}\,W_{1}{}_\mu{}^{j]} &=&  {e^{2A+6C}}\left(\partial^{[i}\,{W_{3\,\mu}}^{j]} \ + \ \frac{m^2}{{\Sigma}^2}\, \partial^{[i}\,W_2{}_{\mu}{}^{j]}\ \right) \, ,
\nonumber \\
 - \ \frac{\mathbf{k}^2}{\Sigma}\ W_{1}{}_\mu{}^{i}  &=& e^{-2C}\left[\frac{h}{2\,\rho} \ {W_{4\,\mu}}^i  \,-\, {e^{-6A}}\, \partial_r\,{W_{3\,\mu}}^i \right]\, .
\label{eqs_sd2_3}
\eea
Moreover, only the $\alpha i$ Einstein equation~\eqref{Einstein_alphai} is left, and reads
\bea
\alpha i &: & e^{-2A}\,m^2\,W_{4\,\alpha i} \,-\,e^{-2C}\,\mathbf{k}^2\,W_{4\,\alpha i}\, + \, e^{-2B}\left[\left(\partial_r-A'-C' \right)^2 \, - \, \left(A'\,-\,C'\right)^2\right]W_{4\,\alpha i} \nonumber \\
&=& e^{-8C} \left[ \frac{2\,\mathbf{k}^2\,h}{\rho\,{\Sigma}}\ W_{1 \,\alpha \,i} \, + \, \frac{h^2}{2\,\rho^2}\, e^{-2C}\,W_{4\, \alpha \,i} \right] ,  \label{w4_dyn}
\eea
after taking into account that the fields are divergence-free, and after using, as in previous cases, the conditions $\Box=m^2$ and $\Delta=-\mathbf{k}^2$.

\subsection{\sc $\mathbf{k}=0$ Modes} \label{sec:nonsingletk0vectors}

We can now solve these equations, starting from the toroidal zero modes. In this case $W_1$, $W_2$ and $\widetilde{W}_2$ are absent, while $W_3$ is linked to $W_4$ according to
\beq
\partial_r\,W_{3\,\mu}{}^i \ = \ \frac{h}{2\,\rho}\, e^{6A}\,W_{4\,\mu}{}^i \ . \label{w3w4}
\eeq
This leaves in principle an $r$--independent contribution to $W_{3\,\mu}^i$, which does not affect the tensor field strength of Appendix~\ref{app:fiveform} and therefore can be gauged away.
Only $W_4$ is thus left as an independent field in this sector. It is divergence-free, as we have stated, and satisfies
\beq
e^{-2A}\,m^2\,W_{4\,\alpha i} \,+\, e^{-2B}\left[\left(\partial_r\,-\,A'\,-\,C' \right)^2 \, - \, \left(A'\,-\,C'\right)^2\right]W_{4\,\alpha i} \ = \ \frac{h^2}{2\,\rho^2}\, e^{-10C}\, W_{4\,\alpha i} \ ,
\eeq
or, using the results for the background metric in Appendix~\ref{app:background},
\beq
m^2\,W_{4\,\alpha i} \,+\, e^{2(A-B)}\left(\partial_r-2C'\right)\left(\partial_r-2A'\right)W_{4\,\alpha i} \ = \ 0 \ .
\eeq

\begin{figure}[ht]
\centering
\includegraphics[width=65mm]{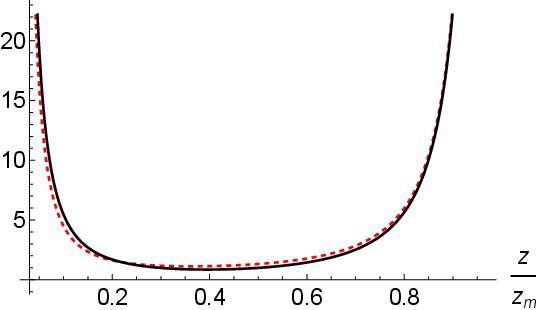} \\
\caption{\small The potential $V$ of eq.~\eqref{hmi_potential} (black, solid) with its approximation~\eqref{pot_hyp} (red, dashed) with $(\mu,\tilde{\mu})=\left(\frac{2}{3},1.18\right)$, in units of $\frac{1}{{z_0}^2}$. $z_m$ and $z_0$ are defined in eqs.~\eqref{zm_app} and \eqref{z0}.}
\label{fig:hmi_pot}
\end{figure}

Turning to the $z$ variable of eq.~\eqref{dzdr} and performing the redefinition
\beq
W_{4\,\alpha i} \ = \ e^{\,-\,\frac{A+3C}{2}} \ Z_{\alpha i}(x) f(z)
\eeq
leads finally to the manifestly Hermitian Schr\"odinger--like equation for $f$,
\beq
 \left(-\,\partial_z \, + \, \beta\right)\left(\partial_z\,+\,\beta\right)f \ = \ m^2\,f \ , \label{schrod_hmi}
\eeq
with
\beq
\beta \ = \ - \ \frac{1}{2} \left(5\,A_z \ + \ 3\,C_z\right)\ .
\eeq
The Schr\"odinger potential is now
\beq
V \ = \ \frac{e^{\sqrt{\frac{5}{2}} \frac{r}{\rho}}}{320\, z_0^2\,\sinh^3\left(\frac{r}{\rho}\right)} \left[2 \sqrt{10} \sinh\left(\frac{2\,r}{\rho}\right)\ +\ 9\, \cosh\left(\frac{2\,r}{\rho}\right) \ +\ 131\right] \ , \label{hmi_potential}
\eeq
and is displayed as a function of $z$ in fig.~\ref{fig:hmi_pot}.
Note that the Schr\"odinger--like equation is once more of the form
\beq
{\cal A}\,{\cal A}^\dagger \, f\ = \ m^2\, f \ ,
\eeq
with
\beq
{\cal A} \ = \ -\  \partial_z\ + \ \beta \ , \qquad {\cal A}^\dagger \ = \ - \ \partial_z\ + \ \beta \ .
\eeq

Close to $z=0$ the potential is as in eq.~\eqref{lim_dil_axion_intro}, with $\mu=\frac{2}{3}$, while $\tilde\mu=\frac{1}{15} \left(4 \sqrt{10}+5\right) \simeq 1.18$.

The allowed wavefunctions thus behave as
\beq
f \ \sim \left( 1 \ - \ \frac{z}{z_m}\right)^{1.68}
\eeq
near the right end of the interval, while there are different options at the left end that are compatible with the leading behavior of the potential, with
\beq
f \ \sim \ C_1 \ \left(\frac{z}{z_m}\right)^\frac{7}{6} \ + \ C_2\ \left(\frac{z}{z_m}\right)^{\,-\,\frac{1}{6}} \ , \label{nonsingvk0}
\eeq
and self--adjoint boundary conditions are characterized by fixed values of the ratio between these two coefficients.

There is a normalizable ground-state wavefunction,
\beq
f(r) \ = \ f_0\,e^{\frac{5 A+3 C}{2}} \ , \label{zero_mode_hmi}
\eeq
with $f_0$ a constant, which solves ${\cal A}^\dagger\,f=0$ and corresponds to the special choice of boundary conditions 
\beq
\frac{C_2}{C_1} \ \simeq \ - \ 2.42 \ .
\eeq
This result translates into the spacetime wavefunction
\beq
W_{4\,\alpha i}(x,r) \ = \ f_0\,Z_{\alpha i}(x)\,e^{2A} \ . \label{W4zeromode}
\eeq

Comparing with the hypergeometric potential~\eqref{pot_hyp} determines the corresponding shift,
\beq
\Delta\,V \ \simeq \ - \ \frac{\pi^2}{z_m^2} (0.71)^2 \ ,
\eeq
so that the eigenvalue equation for this sector reads
\beq
\frac{C_2}{C_1} \ = \ - \ \left(\frac{\pi}{2}\right)^{\,-\,\frac{4}{3}} \ \frac{\Gamma\left(\frac{4}{3}\right)\Gamma\left(0.76 \,+\, \sqrt{m^2+(0.71)^2}\right) \Gamma\left(0.76 \,-\, \sqrt{m^2+(0.71)^2}\right)}{\Gamma\left(\frac{1}{3}\right)\Gamma\left(1.42 \,+\, \sqrt{m^2+(0.71)^2}\right) \Gamma\left(1.42 \,-\, \sqrt{m^2+(0.71)^2}\right)} \ .\label{eigenvmutildelarger3}
\eeq
A tachyonic instability is present, in this sector, within the range
\beq
- \ 2.42 \ < \  \frac{C_2}{C_1} \ < \ 0  \ .
\eeq

The limiting form of $W_{4}$ in eq.~\eqref{W4zeromode}, which diverges proportionally to $r^{- \frac{1}{2}}$ as $r \to 0$, violates the no-flow conditions of~\cite{ms_20}. This can be seen relatively simply starting from the Kaluza-Klein toroidal reduction to five dimensions, so that the gravity field components of interest, $h_{\mu i}$, behave as Abelian vector fields. The further reduction on the interval can then be analyzed referring to their Maxwell energy momentum tensor, and this leads simply to the preceding conclusion.

\subsection{\sc $\mathbf{k}\neq 0$ Modes} \label{sec:intvectexc}

For nonzero internal momenta, one must consider the full system of eqs.~\eqref{eqs_sd2_3}
and~\eqref{w4_dyn}, but the transversality conditions that the fields satisfy allow one to remove the overall $\partial_i$. We can now describe how one can build from them a manifestly Hermitian system of second--order equations for these modes.

\subsubsection{\sc The Schr\"odinger--like System}

Taking the exterior derivative of the first equation and making use of the transversality conditions implies that $\widetilde{W}_2$ vanishes identically for $m^2\neq 0$. On the other hand, for $m^2=0$ it can be regarded as a mere redefinition of $W_2$, which does not enter the other equations. In the following we shall thus remove $\widetilde{W}_2$ altogether, and as a result one can also remove one spacetime derivative from the first of eqs.~\eqref{eqs_sd2_3}, so that the system reduces further to
\bea
W_{1\,\mu}{}^{i}    &=& - \  \frac{e^{-4A-4C}}{\Sigma}\, \partial_r\,W_2{}_{\mu}{}^{i}\, ,\nonumber \\
\frac{1}{\Sigma}\  \partial_r\,W_{1}{}_\mu{}^{i} &=& {e^{2A+6C}}\left({W_{3\,\mu}}^{i} \ + \ \frac{m^2}{{\Sigma}^2}\, W_2{}_{\mu}{}^{i}\ \right) \, ,
\nonumber \\
 - \ \frac{\mathbf{k}^2}{\Sigma}\, W_{1}{}_\mu{}^{i}  &=& e^{-2C}\left[\frac{h}{2\,\rho} \ {W_{4\,\mu}}^i  \,-\, {e^{-6A}}\, \partial_r\,{W_{3\,\mu}}^i \right]\, .
\label{eqs_sd2_5}
\eea
Note that the second--order equation~\eqref{w4_dyn} relates $W_1$ to $W_4$, so that it is natural to deduce from this system another second--order equation linking these two fields. In order to do this, one can start from the second of eqs.~\eqref{eqs_sd2_5}, which  implies
\beq
\partial_r\,{W_{3\,\mu}}^{i} \ = \ \frac{1}{\Sigma}\,e^{-2A-6C}\left(\partial_r\,-\,2\,A'\,-\,6\,C'\right)\partial_r\,{W_{1\,\mu}}^{i} \ - \ \frac{m^2}{{\Sigma}^2}\, \partial_r\,W_2{}_{\mu}{}^{i} \ .
\eeq
Combining it with the first gives
\beq
\partial_r\,{W_{3\,\mu}}^{i} \ = \ \frac{1}{\Sigma}\,e^{-2A-6C}\left(\partial_r\,-\,2\,A'\,-\,6\,C'\right)\partial_r\,{W_{1\,\mu}}^{i} \ + \ e^{4A+4C}\, \frac{m^2}{\Sigma}\, W_1{}_{\mu}{}^{i} \ ,
\eeq
and substituting these results into the last of eqs.~\eqref{eqs_sd2_5} gives
\bea
 - \ \frac{\mathbf{k}^2}{\Sigma}\, W_{1}{}_\mu{}^{i}  &=& \frac{h}{2\,\rho}\,e^{-2C}\,W_{4}{}_\mu{}^{i} \ - \ \frac{e^{-8(A+C)}}{\Sigma}\,\left(\partial_r\,-\,2\,A'\,-\,6\,C'\right)\partial_r\,{W_{1\,\mu}}^{i}\nonumber \\ &-& \frac{e^{2(C-A)}}{\Sigma} \,m^2\,{W_{1\,\mu}}^{i} \ .
\eea
This is the second equation we were after, to be considered in combination with eq.~\eqref{w4_dyn}. One can now separate
variables, letting
\beq
W_{a}{}_\mu{}^{i}(x,z) \ = \ w_\mu{}^{i}(x) \ W_a(z) \ , \qquad (a=1,2)
\eeq
which leads to the matrix form
\beq
{\cal M} \, Y \ = \ m^2\, Y \ ,
\eeq
where $Y$ is a column vector containing the two fields $W_1$ and $W_4$, 
\beq
Y \ = \  \left(\begin{array}{c} W_1 \\ W_4 \end{array}\right) \ ,
\eeq
and
\beq
{\cal M} =  \left(\begin{array}{cc} {\cal K}^2- e^{2(A-B)}\left(\partial_r - 2 A' - 6 C'\right)\partial_r  \!\!\!\!\!\!& \frac{h\,{\Sigma}}{2\,\rho}\, e^{2A-4C} \\ \!\!\!\!\!\!\!\!\!\!\!\!\frac{2\,h \,{\cal K}^2}{\rho\,{\Sigma}}\,e^{-6C} \!\!\!\!\!\!\!\!& {\cal K}^2 - e^{2(A-B)}\left(\partial_r - 2 C'\right)\left(\partial_r - 2\,A'\right) \end{array}\right) \ ,
\eeq
with
\beq{}{}{}{}{}
{\cal K} \ = \ \left|\mathbf{k}\right|\,e^{A-C}  \ . \label{calK}
\eeq

The mass spectrum is determined by this system, and therefore it is convenient to try to reduce it to a manifestly Hermitian Schr\"odinger--like form. To this end, we turn once more to the $z$ variable of eq.~\eqref{dzdr}, so that $M$ becomes
\bea
{\cal M} \!\!\!\!\!&=&\!\!\!\!\! \left(\begin{array}{cc} {\cal K}^2 - \left(\partial_z + A_z - C_z\right)\partial_z  & \!\!\!\!\!\!\!\!\!\frac{h\,{\Sigma}}{2\,\rho}\, e^{2A-4C} \\ \!\!\!\!\!\!\!\!\!\!\!\!\!\!\!\!\!\frac{2\, h\,{\cal K}^2}{{\Sigma}\,\rho}\,e^{-6C} \!\!\!\!\!\!\!\!\!& {\cal K}^2 - \left(\partial_z +3 \,A_z + 3\,C_z\right)\left(\partial_z - 2\,A_z\right) \end{array}\right) \ . \label{M_non_hermitian}
\eea
This operator is still not manifestly Hermitian, but after redefining the fields according to
\beq
Y \ = \ \Lambda\,Z \ ,
\eeq
with
\beq
\Lambda \ = \  \left(\begin{array}{cc}\sqrt{\frac{\Sigma}{{2}\,\left|\mathbf{k}\right|}}\ e^\frac{C\,-\,A}{2}  & 0 \\ 0 & \sqrt{\frac{{2}\,\left|\mathbf{k}\right|}{\Sigma}}\ e^{\,- \,\frac{A\,+\,3 C}{2}}  \end{array}\right) \, , \qquad Z \ = \ \left(\begin{array}{c} Z_1 \\ Z_2 \end{array}\right) \ , \label{LAMBDA_vector_non_singlet}
\eeq
the system finally takes the desired form,
\beq
\widetilde{\cal M} \, Z \ = \ m^2\, Z \ . \label{schrod_Z}
\eeq
Now the matrix is
\beq
\widetilde{\cal M} \ = \ \left(\begin{array}{cc} {\cal K}^2\ +\  \left(-\,\partial_z + \alpha_z\right)\left(\partial_z + \alpha_z\right)  & \frac{\left|\mathbf{k}\right| \,h}{\rho}\, e^{2(A-3C)} \\ \frac{\left|\mathbf{k}\right| \,h}{\rho}\,e^{2(A-3C)} & {\cal K}^2 \ +\ \left(-\partial_z +\beta_z\right)\left(\partial_z + \beta_z\right) \end{array}\right) \, ,  \label{mtilde2}
\eeq
where
\beq
\alpha_z \ = \ \frac{C_z - A_z}{2} \ , \qquad \beta_z \ = \ - \ \frac{5A_z+3C_z}{2} \ , \label{alpha_beta_vec}
\eeq
and is manifestly Hermitian and independent of $\Sigma$. The scalar product
\beq
\int dz \left(\left|Z_1\right|^2 \ + \ \left|Z_2\right|^2 \right) \ = \ \frac{\Sigma}{2\left|\mathbf{k}\right|}\ \int dz \left[\left(\frac{2\left|\mathbf{k}\right|}{\Sigma}\right)^2 e^{A-C} \left|W_1\right|^2 \ + \ e^{A+3 C} \left|W_4\right|^2 \right]
\eeq
can also be deduced from these results, and the contribution within square brackets is precisely what the effective action would yield. Note that, although $\widetilde{\cal M}$ is well defined in the $\mathbf{k}\to 0$ limit, the scalar product is singular when expressed in terms of the original variables. The behavior of $\widetilde{\cal M}$ in this limit and its implications for the boundary conditions are examined in detail in Appendix~\ref{app:Mtildek0}.

\subsubsection{\sc Boundary Conditions and Stability Analysis}

We can now address the possible emergence of instabilities for nonzero values of $\mathbf{k}$. This very problem jeopardizes~\cite{bms} the non--supersymmetric $AdS \times S$ vacua of~\cite{gubsermitra,ms_16}, but here there is an important novelty, since the toroidal radius $R$ is a free parameter, insofar as it is large enough with respect to the Planck scale to make the present low--energy setup reliable. 
\begin{figure}[ht]
\centering
\includegraphics[width=65mm]{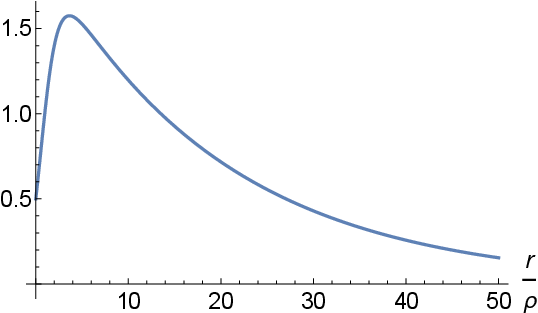}
\caption{\small The bounded function $u$ of eq.~\eqref{function_u}, in units of $\frac{\left|\mathbf{k}\right|\rho}{z_0^2}$}
\label{fig:bounded_u}
\end{figure}

To begin with, let us note that letting
\beq
Q \ = \ \left(\begin{array}{cc} \partial_z\,+\alpha_z & {\cal K}   \\ {\cal K}   & \partial_z\,+\beta_z \end{array}\right) \ ,
\eeq
where ${\cal K}$ was defined in eq.~\eqref{calK}, $\widetilde{\cal M}$ can be cast in the form
\beq
\widetilde{\cal M} \ = \ Q^\dagger\,Q \ - \ u\, \sigma_1 \ , \label{MQQu}
\eeq
with $\sigma_1$ the familiar Pauli matrix and
\beq
u \ = \ - \ \left|\mathbf{k}\right| \left( 4\,A_z\,e^{A-C}\,+\,\frac{h}{\rho}\,e^{2(A-3C)}\right) \ = \ \frac{\left|\mathbf{k}\right| \rho}{2\,z_0^2}\ \frac{e^{\,\frac{3\,r}{\rho\,\sqrt{10}}}}{\cosh^2\left(\frac{r}{2\,\rho}\right)}\ , \label{function_u}
\eeq
where $z_0$ is the scale first defined in eq.~\eqref{z0_i}.

Note that $u$ is a positive and bounded function of $r$, which is displayed in fig.~\ref{fig:bounded_u}, but nonetheless the presence of the Pauli matrix $\sigma_1$ in eq.~\eqref{MQQu} yields negative contributions to the $m^2$ eigenvalues.
As a result, $\widetilde{\cal M}$ is \emph{not} a manifestly positive operator, but for large values of $\mathbf{k}$ the contribution of $u$ is subdominant with respect to the ${\cal K}^2$ terms in $Q^\dagger\,Q$. The internal momentum $\mathbf{k}$ is quantized in units of $\frac{1}{R}$, with $R$ the radius of the internal torus, and enters all these expressions via the dimensionless combination
\beq
\xi \ = \  \mathbf{k}\,\rho \ \sim \ \frac{\left|\mathbf{k}\,z_m\right|} {\left(3 \,H\,z_m\right)^\frac{1}{3}} \ . \label{xidef}
\eeq
Therefore, there is a minimum nonzero value for $\xi$ in this sector, corresponding to the minimal non--vanishing internal momenta. It can be expressed in several equivalent ways, as
\beq
\xi_0 \ = \ \left(\frac{z_m{}^2}{3\,H\,R^3}\right)^\frac{1}{3} \ \simeq \frac{\rho}{R} \ \simeq 2\,\pi\left(\frac{\ell}{\Phi^\frac{1}{4}}\right)^\frac{4}{5} \ , \label{xi0}
\eeq
where the flux $\Phi$ and the length $\ell$ of the interval were defined in the Introduction.
For $\xi_0$ larger than a critical value $\xi_c$, and thus for $\frac{R}{\rho}$ below a corresponding critical value, no unstable modes can be present, since as we have stressed the terms depending on ${\cal K}$ eventually dominate. However, special choices of boundary conditions may remove the tachyons altogether, and in order to address this question and to obtain an estimate for $\xi_c$ one needs to characterize how $m^2$ depends on the boundary conditions, while also taking into account the contributions depending on the internal momentum $\mathbf{k}$.

Our next goal is to characterize the independent self--adjoint boundary conditions at the two ends of the interval that grant the positivity of $Q^\dagger\,Q$. Let us begin by considering the behavior at the origin, which is more intricate.
At the $z=0$ end, the eigenvalue equation~\eqref{schrod_Z} reduces to
\beq
Q_0^\dagger\,Q_0 \, Z \ = \ 0 \ , 
\eeq
where $Q_0$, the dominant term in $Q$ near $z=0$, is given by
\beq
Q_0 \ = \   \partial_z \ + \ \frac{1}{6\,z}\, \ + \ \xi \left(\frac{z_m}{z}\right)^\frac{1}{3}\ \sigma_1 \ . \label{Q0}
\eeq
The contributions involving $u$ and $m^2$ are negligible with respect to these singular terms, which include some $\mathbf{k}$-dependent contributions at this end. As a result, the analysis can be split into a pair of steps. One first solves
\beq
Q_0^\dagger\, \Psi \  = \ 0 \ , \label{Qvector}
\eeq
obtaining
\beq
\Psi \ = \ \left(\frac{z}{z_m}\right)^{\frac{1}{6}}\ \left\{\cosh\left[\frac{3\,\xi}{2}\,\left(\frac{z}{z_m}\right)^\frac{2}{3}\right] \ + \ \sigma_1\ \sinh\left[\frac{3\,\xi}{2}\,\left(\frac{z}{z_m}\right)^\frac{2}{3}\right] \right\} \Psi_0 \ ,
\eeq
with $\Psi_0$ a constant vector, and the complete solution of $Q_0^\dagger\,Q_0\,Z = 0$ can then be obtained solving the inhomogeneous equation
\beq
Q_0\,Z \ = \ \Psi \ .
\eeq
It reads
\bea
Z &=& \left(\frac{z_m}{z}\right)^{\frac{1}{6}}\left\{\cosh\left[\frac{3\,\xi}{2}\,\left(\frac{z}{z_m}\right)^\frac{2}{3}\right] \ - \ \sigma_1\ \sinh\left[\frac{3\,\xi}{2}\,\left(\frac{z}{z_m}\right)^\frac{2}{3}\right] \right\} \nonumber \\
&\times& \left\{ \chi_0 \ + \ \int^{\frac{z}{z_m}} \ d y \ y^\frac{1}{3}\left[\cosh\left(3\,\xi\, y^\frac{2}{3}\right) \ + \ \sigma_1\ \sinh\left(3\,\xi\, y^\frac{2}{3}\right)\right] \Psi_0 \right\} \ ,
\eea
where $\chi_0$ ia another constant vector~\footnote{In this section, for simplicity, we are not rescaling the $C_{ij}$ by the $\mu_i$, as in Section~\ref{sec:intvectexc}, since $\mu_1=\mu_2$, and therefore this would only introduce an overall factor in the following equations.}. We now let
\beq
\chi_0 \ = \ \left(\begin{array}{c} C_{12} \\ C_{22} \end{array} \right) \ , \qquad \Psi_0 \ = \ \frac{4}{3} \left(\begin{array}{c} C_{11} \\ C_{21} \end{array} \right) \ ,
\eeq
and the limiting behavior of the preceding expression close to $z=0$ then yields
\bea{}{}{}{}{}{}{}
Z_1 &\sim&  C_{11}  \, \left(\frac{z}{z_m}\right)^{\frac{7}{6}}  + \,  C_{12}\, \left(\frac{z}{z_m}\right)^{-\,\frac{1}{6}} + \  \frac{\xi}{2}\left[C_{21} \, \left(\frac{z}{z_m}\right)^\frac{11}{6} - \ 3 \,C_{22} \, \left(\frac{z}{z_m}\right)^\frac{1}{2} \right] , \nonumber \\
Z_2 &\sim&  C_{21}  \, \left(\frac{z}{z_m}\right)^{\frac{7}{6}} \, + \, C_{22} \left(\frac{z}{z_m}\right)^{-\,\frac{1}{6}} + \  \frac{\xi}{2}\left[C_{11} \, \left(\frac{z}{z_m}\right)^\frac{11}{6} - \ {3}\,C_{12} \left(\frac{z}{z_m}\right)^\frac{1}{2}\right] .
\label{Z_origin}
\eea

This limiting behavior will be important in the following, since it characterizes generic wavefunctions $Z$ such that $\widetilde{\cal M}\,Z$ is also in $L^2$, insofar as the left end of the interval is concerned. Note that the structure of eqs.~\eqref{Z_origin} is fully determined by the $\xi$--independent terms, which rest on the indicial exponents for the limiting diagonal system discussed in Appendix~\ref{app:Mtildek0}. Yet, the additional terms that we have identified are instrumental to guarantee that the $\mathbf{k}$-dependent terms contained in $\widetilde{\cal M}$ do not give rise to divergent contributions from the $z=0$ end.

As discussed in Section~\ref{sec:exactschrod}, the boundary terms that ought to vanish in order to grant self--adjointness are of the form
\beq
\left[ {\widetilde{Z}}^\dagger \, \partial_z {Z} \ - \  \left(\partial_z\,{\widetilde{Z}}^\dagger\right) \, Z \right]_{z=0}^{z=z_m} \ , \label{sadjcd}
\eeq
with $Z$ and ${\widetilde{Z}}$ a pair of two--component vectors whose asymptotic behavior at the origin is as in eq.~\eqref{Z_origin}, with coefficients $C$ and $\tilde{C}$. We shall focus on the natural choice of conditions given independently at the two ends.At the origin the coefficients must thus satisfy
\beq
 \tilde{C}_{12}^*\,C_{11}\ - \ \tilde{C}_{11}^*\,C_{12} \ + \ \tilde{C}_{22}^*\,C_{21} - \,\tilde{C}_{21}^*\,C_{22} \ = \ 0  \label{sa4}
\ ,
\eeq
which is independent of $\xi$, as expected.

The linear relations granting that eq.~\eqref{sa4} holds were discussed in Section~\ref{sec:exactschrod} for systems involving $n$-component vectors. However, in the present $n=2$ case there is an alternative, more convenient formulation that we can now describe. Let us therefore start by considering the two--component vectors
\beq
\underline{X}_1 \ = \left(\begin{array}{c} C_{11} \\ C_{12} \end{array} \right) \ , \qquad \underline{X}_2 \ = \left(\begin{array}{c} C_{22} \\ C_{21} \end{array}\right) \ ,
\eeq
and their counterparts $\underline{\widetilde{X}}_{1,2}$, while also recasting eq.~\eqref{sa4} as the invariance condition for a quadratic form resting on $\sigma_2$:
\beq
\underline{\widetilde{X}}_1^\dagger \ \sigma_2 \ \underline{X}_1 \ = \ \underline{\widetilde{X}}_2^\dagger \ \sigma_2 \ \underline{X}_2 \ .
\eeq
The self--adjoint boundary conditions given independently at $z=0$ rest on a $U(1,1)$ matrix $U$,
such that
\beq
U^\dagger\, \sigma_2\, U \ = \ \sigma_2 \ ,
\eeq
which links the two vectors $\underline{X}_{1,2}$ according to
\beq
\underline{X}_2 \ = \ U \ \underline{X}_1 \ . \label{bcUbeta}
\eeq
The $U$ matrix is conveniently parametrized as
\beq
U\left(\rho,\theta_1,\theta_2,\beta\right) \ = \ e^{i\beta}\left[\cosh\rho\left(\cos\theta_1 \, \underline{1} \,-\, i\,\sigma_2\,\sin\theta_1\right) \,+\, \sinh\rho\left(\sigma_3\, \cos\theta_2\ + \ \sigma_1\, \sin\theta_2 \right)\right] 
\ , \label{global_ads3_12}
\eeq
and the self--adjoint boundary conditions at $z=0$ thus depend, in general, on four real numbers.

Moreover, an integration by parts leads to
\beq
\int_0^{z_m} dz \ Z^\dagger\,Q^\dagger\,Q\, Z \ = \ \int_0^{z_m} dz \left| Q\,Z\right|^2 \ - \ \left[ Z^\dagger\,Q\, Z\ \right]_{z=0}^{z=z_m} \ , \label{ham_boundary}
\eeq
where the first term on the \emph{r.h.s.} is manifestly positive. Consequently, the
contribution from the $z=0$ end is not negative if the limiting contributions from $Z^\dagger\,Q\, Z$ at both ends are positive. In particular, at $z=0$ one is led to
\beq
\mathrm{Re}\left[C_{12}^* \ C_{11}  \ +\ C_{22}^* \ C_{21} \right] \ \geq \ 0\ , \label{pos_co}
\eeq
where the self--adjointness condition was taken into account. Making use of eq.~\eqref{bcUbeta}, this positivity condition at $z=0$ is equivalent to demanding the positivity of the symmetric matrix 
\beq
{\cal S} \ = \ \sigma_1 \ + \ U^\dagger\,\sigma_1 \, U \ ,
\eeq
as shown in~\cite{selfadjoint}. In terms of the global parametrization~\eqref{global_ads3_1}, 
the restrictions thus enforced on the parameters read
\beq
\sin\left(\theta_1 \,+\,\theta_2 \right) \ \geq \ 0 \ , \qquad \tanh^2\rho \,\cos^2\theta_2 \ - \ \cos^2\theta_1 \ \leq \ 0 \ . \label{positiveU}
\eeq
Identical conditions emerged, in a similar context, in Section~3.2.5 of~\cite{selfadjoint}.

Our next task is to identify the possible independent self--adjoint boundary conditions at $z=z_m$. These depend on the values $(\tilde{\mu_1},\tilde{\mu_2}) = (0.09,1.1)$ that were given after eq.~\eqref{mtilde20}.
This analysis is simpler, since the $\mathbf{k}$-dependent terms vanish in the limit, and consequently the dominant terms in $\widetilde{\cal M}$ are captured by the diagonal matrix
\beq
Q_{\infty}^\dagger\ Q_\infty \ ,
\eeq
with
\beq
Q_\infty \ = \ \left(\begin{array}{cc} \partial_z + \frac{\frac{1}{2}+\tilde{\mu}_1}{z_m-z}  & 0 \\ 0 & \partial_z + \frac{\frac{1}{2}+\tilde{\mu}_2}{z_m-z} \end{array} \right) \ .
\eeq
The allowed limiting behaviors are
\beq
Z_1 \ \sim  \ C_{13} \left(\frac{z_m-z}{z_m}\right)^{\frac{1}{2}+\tilde{\mu}_1} \!\!+ \ C_{14} \left(\frac{z_m-z}{z_m}\right)^{\frac{1}{2}-\tilde{\mu}_1} \ , \qquad
Z_2\  \sim \  C_{23} \left(\frac{z_m-z}{z_m}\right)^{\frac{1}{2}+\tilde{\mu}_2} \ , \label{Z_infinity}
\eeq
since $\tilde{\mu}_2>1$, and therefore only one of the two possible options for $Z_2$ leads to normalizable solutions. The contribution from the upper end to the boundary term in eq.~\eqref{sadjcd} yields the condition
\beq
 \tilde{C}_{14}^*\,C_{13}\ - \ \tilde{C}_{13}^*\,C_{14} \ = \ 0 \ ,
\eeq
and consequently the ratio between $C_{13}$ and $C_{14}$ must be a real number, and we let
\beq
\cot\left(\frac{\tilde{\alpha}}{2}\right) \ = \ \frac{C_{13}}{C_{14}} \ . \label{BCright}
\eeq
Turning now to the positivity condition, let us note that there is a difference with respect to the other end, since $Q_\infty$ annihilates the least singular contribution to $Z_1$, while $Q_0$ was suppressing the most singular one. As a result, the boundary contribution from $z_m$ to eq.~\eqref{ham_boundary} is now dominated by
\beq
- \ \frac{1\,-\,\tilde{\mu}_1}{z_m} \ \left| C_{14} \right|^2 \left(1 \ - \ \frac{z}{z_m}\right)^{\,-\,2\,\tilde{\mu}_1} \ ,
\eeq
which is singular and negative. Although this term can be canceled by the bulk contribution in eq.~\eqref{ham_boundary}, positivity is clearly guaranteed if $C_{14}=0$, or $\tilde{\alpha}=0$, and we shall abide to this choice in the following.

We can now explore whether, for the self--adjoint boundary conditions granting positivity of $Q^\dagger\,Q$, which in this case are characterized by $\tilde{\alpha}=0$ and by the restrictions on the $U(1,1)$ parameters in eq.~\eqref{positiveU}, instabilities can arise when the $\mathbf{k}$--dependent terms in $\widetilde{\cal M}$ are fully taken into account.
Finding this out exactly is difficult, but one can rely on the variational principle, according to which the ground--state energy of the Schr\"odinger system is given by
\beq{}{}{}{}{}{}{}{}{}{}{}{}{}{}{}{}{}{}{}{}{}{}{}{}{}{}{}{}
m_0^2\left(U,\tilde{\alpha}\right) \ = \ \underset{{\Psi \in \mathcal{S}}} {\mathrm{Inf}} \left[\frac{\langle \Psi | \widetilde{M} | \Psi \rangle}{\langle \Psi | \Psi \rangle}\right] \ ,
\eeq
where the \emph{infimum} is over the set $\mathcal{S}$ of all normalizable wavefunctions $\Psi$ for which $\widetilde{\cal M}\,\Psi$ is also normalizable, and which satisfy the boundary conditions~\eqref{bcUbeta} and~\eqref{BCright}. In practice, working within a subset $\mathcal{S}_0 \subset \mathcal{S}$ determined by a finite number of parameters and minimizing over them can yield results that approximate closely from above the actual value of $m_0^2$:
\beq
m^2\left(\mathcal{S}_0\right)\ = \ \underset{{\Psi \in \mathcal{S}_0}} {\mathrm{Inf}} \left[\frac{\langle \Psi | \widetilde{M} | \Psi \rangle}{\langle \Psi | \Psi \rangle}\right] \ \geq \ m_0^2  \ . \label{m2_est}
\eeq
Negative values of $m^2\left(\mathcal{S}_0\right)$ arising from these estimates thus signal the presence of tachyonic instabilities.

\begin{figure}[ht]
\centering
\begin{tabular}{cc}
\includegraphics[width=65mm]{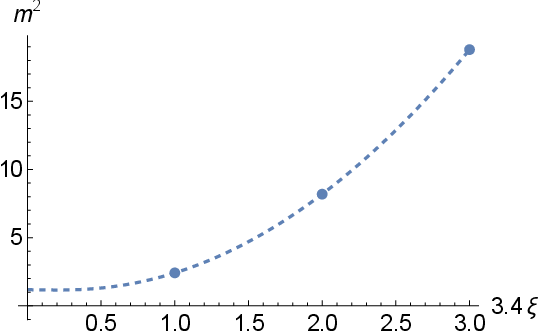} \qquad \qquad &
\includegraphics[width=65mm]{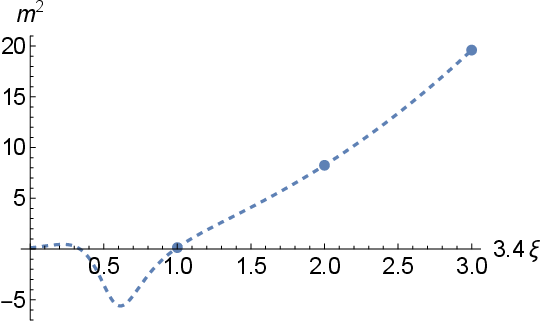} \\
\end{tabular}
\caption{\small $m^2$ as a function of $\xi$, with $\left(\rho,\theta_1,\theta_2,\beta\right)=\left(2,-\frac{\pi}{2},\frac{\pi}{2},0\right)$ (left panel) and with $\left(\rho,\theta_1,\theta_2,\beta\right)=\left(0,0,0,0\right)$ (right panel). The dots are examples of possible allowed values for $\xi$, if the quantization of $\mathbf{k}$ is taken into account.}
\label{fig:xi_quantized}
\end{figure}

Some details on our variational tests are collected in Appendix~\ref{app:variationa2}, and their indications can be summarized as follows:
\begin{itemize}
    \item self--adjoint boundary conditions exist for which \emph{no tachyons are present for all values of $\xi_0$}. An example of this type, with $\left(\rho,\theta_1,\theta_2,\beta\right)=\left(2,-\frac{\pi}{2},\frac{\pi}{2},0\right)$, is presented in the left panel of fig.~\ref{fig:xi_quantized};

\item other boundary conditions lead to the absence of tachyons provided $\xi_0$ is larger than a critical value $\xi_c$. An example of this type, with $\left(\rho,\theta_1,\theta_2,\beta\right)=\left(0,0,0,0\right)$ determines $\xi_c\simeq 3.4$, and is presented in the right panel of fig.~\ref{fig:xi_quantized}.
\end{itemize}

In conclusion, no tachyonic modes are present in this sector with proper choices of boundary conditions. The parameter space characterizing them depends on the background through the combination $\xi_0$ of eq.~\eqref{xi0}, and its measure grows as $\xi_0 \sim \frac{\rho}{R}$ increases.

\section{\sc Non--Singlet Scalar Modes} \label{sec:nonsingletscalars}
We can now turn to scalar perturbations, and we begin by considering the modes that for $\mathbf{k}=0$ are valued in the fundamental representation of $SO(5)$. These originate from the field profiles
\bea
&&b^i\ = \  \phi_1^i\ ,\qquad b_\mu{}^i\ =\ \frac{1}{\Sigma}\,\partial_\mu\phi_2^i\ ,\qquad  b^{(2)}_\mu{}^{ij}\ =\ \frac{1}{{\Sigma}^2}\, \partial_\mu\partial^{[i}\phi_3{}^{j]}\,,\nonumber\\
&&h^{ij}\ =\ \frac{1}{\Sigma}\,\partial^{(i}\phi_4{}^{j)} \ , \qquad h_{\mu}{}^i \ = \ \frac{1}{\Sigma}\,\partial_\mu\phi_5^i\ ,\qquad h_{r}{}^i\ =\
\,\phi_6^{i} \ ,
\eea
which involve the independent scalar fields $\phi_a{}^i$, with $a=1,\ldots ,6$.
A convenient choice of gauge fixing, using the $\xi^i$ of eqs.~\eqref{xilambda_gauge}, is in this case
\beq
\phi_5^i \ = \ 0 \label{gauge_scalar_fund} \ ,
\eeq
and, as for the preceding sectors, we begin our analysis from the modes with $\mathbf{k}=0$.

\subsection{\sc $\mathbf{k}=0$ Modes} \label{sec:nonsingletscalarsk0}
In this case the fields $\phi_3$ and $\phi_4$ disappear, and in the gauge of eq.~\eqref{gauge_scalar_fund} the tensor equations~\eqref{first_ord_5form} reduce to
\bea
\phi_1^i   &=& - \ \frac{1}{\Sigma}\ e^{-2C-6A} \partial_r\,\phi_2^i  ,
\nonumber \\
\partial_r\,\phi_1^i   &=& e^{-2C}\left[\frac{h}{2\,\rho}\,{\phi_6}^i\ + \frac{m^2}{\Sigma}\ {e^{10C}}\phi_2^i \right]\, .
 \label{tenss1}
\eea
In addition, the $\alpha i$ and $\alpha r$ Einstein equations~\eqref{Einstein_alphai} and \eqref{Einstein_ir} reduce to
\bea
 && e^{-2B}\left(\partial_r\,-2A'\right)\phi_6^i\
= \ \frac{2\,h}{\rho}\ e^{-8C} \,\phi_1^i \ ,  \nonumber  \\
&&  m^2\,{\Sigma}\,e^{-2A} \phi_6^i
  \ =\  -\ m^2\,\frac{2\,h}{\rho}\, \phi_2^i \ . \label{alphai_scalar_fund}
\eea
When $m\neq 0$, making use of this last equation, one can express $\phi_6$ in terms of $\phi_2$, and then the first of eqs.~\eqref{alphai_scalar_fund} becomes identical to the first of eqs.~\eqref{tenss1}, while the second becomes
\beq
e^{-2B+8C}(\partial_r-2B'+8C')(\partial_r-2A')
\phi_6^i  \ = \ e^{-2C}\left[\frac{h^2}{\rho^2}\,{\phi_6}^i\ - m^2\ {e^{10C-2A}}\phi_6^ i \right]\, .
 \label{tenss2}
\eeq
Note, however, that the end result also applies if $m=0$, as can be seen retracing the preceding steps.

\begin{figure}[ht]
\centering
\begin{tabular}{cc}
\includegraphics[width=65mm]{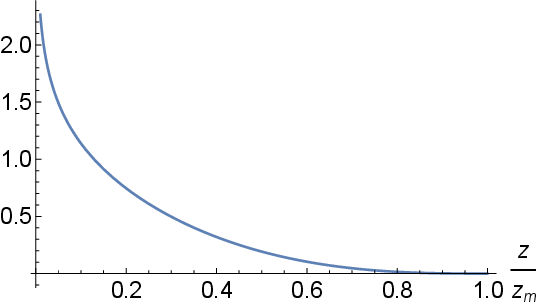} \qquad \qquad &
\includegraphics[width=65mm]{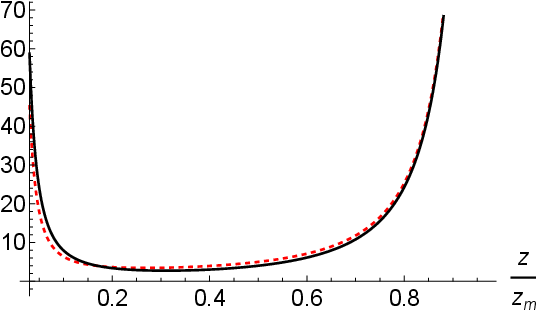} \\
\end{tabular}
\caption{\small The left panel shows the normalized zero--mode wavefunctions of eqs.~\eqref{zeromode_chi6i} . The right panel compares the corresponding potential $V$ of eq.~\eqref{chi6i_potential} (black, solid) with its approximation~\eqref{pot_hyp} (red,dashed)  with $(\mu,\tilde{\mu})=\left(\frac{2}{3},2.27\right)$, in units of $\frac{1}{{z_0}^2}$. $z_m$ and $z_0$ are defined in eqs.~\eqref{zm_app} and \eqref{z0}.}
\label{fig:pot_chi6i}
\end{figure}

 Working in terms of the variable $z$ of eq.~\eqref{dzdr} and performing the field redefinition and the separation of variables
\beq
\phi_6^i(x,z)=e^{\frac{7A-3C}{2}}\,\chi_6^i(x)\ f(z) \ , \label{zeromode_chi6i}
\eeq
leads to the manifestly Hermitian Schrodinger--like equation
\beq
 \left(\partial_z+\frac{3}{2}(C_z-A_z)\right)
 \left(\partial_z-\frac{3}{2}(C_z-A_z)\right)f(z)\ = \
-m^2\chi_6^i+\frac{h^2}{\rho^2}e^{2A-10C}f(z) \ . \label{schrod_vector_scalar_k0}
\eeq
The corresponding potential 
\beq
V(r) \ = \ \frac{e^{\sqrt{\frac{5}{2}} \frac{r}{\rho}}}{320\, z_0^2\,\sinh^3\left(\frac{r}{\rho}\right)} \left[18 \sqrt{10}\, \sinh\left(\frac{2\,r}{\rho}\right)\ +\ 9\, \cosh\left(\frac{2\,r}{\rho}\right) \ +\ 131\right] \label{chi6i_potential}
\eeq
is displayed in fig.~\ref{fig:pot_chi6i}.

Using eqs.~\eqref{hamiltonian_F}, eq.~\eqref{schrod_vector_scalar_k0} can be surprisingly recast in the form
\beq
{\cal A}{}^\dagger\,{\cal A}\,f(z) \ = \ m^2 \,f(z) \ , \label{AAtilde}
\eeq
 where
\beq
{\cal A} \ = \ \partial_z \ - \ \frac{9}{2}\,A_z \ - \ \frac{7}{2}\,C_z \ , \qquad {\cal A}{}^\dagger \ = \  - \ \partial_z \ - \ \frac{9}{2}\,A_z \ - \ \frac{7}{2}\,C_z \ ,
\eeq
using the identities involving $A_z$ and $C_z$ collected in Appendix~\ref{app:background}.

Close to $z=0$ the potential is as in eq.~\eqref{lim_dil_axion_intro}, with $\mu=\frac{2}{3}$, while $\tilde{\mu}=\frac{1}{5} \left(2 \sqrt{10}+5\right) \simeq 2.27$. Consequently, a single limiting behavior is allowed at $z_m$, with $f(z) \sim (z_m-z)^{2.77}$, but close to $z=0$ there is more freedom, and in general
\beq
f(z) \ \sim \ C_1{} \left(\frac{z}{z_m}\right)^\frac{7}{6} \ + \ C_2{} \left(\frac{z}{z_m}\right)^{-\,\frac{1}{6}}
\eeq
so that self--adjoint boundary conditions exist for arbitrary fixed ratios $\frac{C_2}{C_1}$. A particular choice solves ${\cal A} f(z) = 0$,
\beq
f(z) \ = \ C \ e^\frac{9\,A\,+\,7\,C}{2} \ . \label{zero_mode_nonsinglet_scalar}
\eeq
This wavefunction is normalizable, as the reader can simply verify, and corresponds to the choice
\beq
\frac{C_2}{C_1} \ = \ - \  6.2 \ , \label{C21ss}
\eeq
and thus to the profile
\beq
\phi_6^i(x,z)=e^{8A+2C}\,\chi_6^i(x) \  \label{zeromode_chi6i_final}
\eeq
for $h_r{}^{i}$.

As in other sectors, one can approximate the potential by the hypergeometric form of eq.~\eqref{pot_hyp}, up to a shift that can be determined demanding that the resulting massless mode correspond to the ratio between $C_2$ and $C_1$ in eq.~\eqref{C21ss}. In this case one finds
\beq
\Delta\,V \ = \ - \ \frac{\pi^2}{z_m^2}\ (1.23)^2 \ ,
\eeq
so that the instability region for this sector is
\beq
- \ 6.2 \ < \ \frac{C_2}{C_1} \ < \ 0 \ . \label{bcrange}
\eeq

\subsection{\sc $\mathbf{k} \neq 0$ Modes} \label{sec:nonsingletscalrknot0}
Let us now turn to modes with a nonzero internal momentum $\mathbf{k}$.
As in previous cases, the scalars are valued in the fundamental representation of the $SO(4)$ transverse to $\mathbf{k}$, and the fields satisfy the conditions
\beq
k_i\,\phi_a^i\ =\ 0,\qquad a\ =\ 1,\dots 6 \ .
\eeq
The longitudinal excitations will contribute to the singlet scalar spectrum.

\subsubsection{\sc The Schr\"odinger--like System}

In the gauge of eq.~\eqref{gauge_scalar_fund}, the tensor equations become
\bea
 \partial_r\,\phi_3^{j} &=&   {e^{2A+6C}}\,{\Sigma}\,\phi_2^{j}\ ,
\nonumber \\
\phi_1^i \, -\,\frac{\mathbf{k}^2}{{\Sigma}^2}  \ \phi_3{}^{i}  &=& -\ \frac{1}{\Sigma}\ e^{-6A-2C} \,\partial_r\,\phi_2^i\  ,
\nonumber \\
\partial_r\,\phi_1{}^i   &=& e^{-2C}\left[\frac{h}{2\,\rho}\,\phi_6{}^i + \frac{m^2}{\Sigma}
{e^{10C}} \phi_2^i \right]\ .
\label{eqs_sd2_3i}
\eea
The  $\alpha i$ Einstein equation~\eqref{Einstein_alphai} reduces to
\beq
 e^{-2B}\left(\partial_r\,-\,2\,A' \right)\phi_6^i
  -\ e^{-2C}\,\frac{\mathbf{k}^2}{\Sigma}\, \phi_4^i  \  =  \ e^{-8C}\,\frac{2\,h}{\rho}\left(\phi_1^i \ - \ \frac{\mathbf{k}^2}{{\Sigma}^2}\, \phi_3^i\right) \ , \label{eqai2}
\eeq
while the $ri$ and $ij$ Einstein equations~\eqref{Einstein_ir} and ~\eqref{Einstein_ij} reduce to
\bea
&& \left(e^{-2A}m^2\, -\,e^{-2C} \, \mathbf{k}^2 \right) \phi_6^i \,+\, \frac{1}{\Sigma}\, e^{-2C}\left(\partial_r\,-\,2\,C'\right)\mathbf{k}^2 \phi_4^i
 \ =  \ -\ \frac{2\,m^2\,h}{\rho\,{\Sigma}}\, \phi_2^i  \ , \label{riij} \\
&& m^2e^{-2A} \phi_4^i -  {\Sigma}\,e^{-2B}\left(\partial_r - 2\,C'\right)\phi_6^i + e^{-2B}\big[\left(\partial_r-4C'\right)\partial_r + 4\,(C')^2\big]\phi_4^i  = -\, \frac{h^2}{2\,\rho^2} e^{-10C}  \phi_4^i  \ , \nonumber
\eea
which can also be cast in the form
\bea
&& \mathbf{k}^2e^{-2C} \left[{\Sigma}\,\phi_6^i - \left(\partial_r - 2\,C'\right)\phi_4^i\right]   \ = \ \frac{2 m^2\,h}{\rho}\left[\phi_2^i \ + \ \frac{\rho\,{\Sigma}}{2\,h}\,e^{-2A}\,\phi_6^i \right] \ , \nonumber \\
 && m^2\,\phi_4^i -  {\Sigma}\,e^{2(A-B)}\left(\partial_r - 2\,C'\right)\left[{\Sigma}\,\phi_6^i - \left(\partial_r - 2\,C'\right)\phi_4^i\right]   \ = \ 0 \ .
\eea

This is a complicated--looking system for five different fields, which comprises one second--order equation and five first--order ones. Note that there is still a local symmetry that allows one to shift $\phi_1^i$ and $\phi_3^i$ by $r$--independent amounts in such a way that $\left(\phi_1^i \ - \ \frac{\mathbf{k}^2}{{\Sigma}^2}\, \phi_3^i\right)$ is unaffected. Up to this gauge symmetry, the system determines uniquely $\phi_1^i$, $\phi_3^i$ and $\phi_4^i$ in terms of $\phi_2^i$ and $\phi_6^i$.

We can now obtain a system of second--order equations for $\phi_2^i$ and $\phi_6^i$. To this end, one can use the second of eqs.~\eqref{eqs_sd2_3i}  in eq.~\eqref{eqai2}, obtaining
\beq
 e^{-2B}\,{\Sigma}\,\left(\partial_r\,-\,2\,A' \right)\phi_6^i
  - \ e^{-2C}\,\mathbf{k}^2\, \phi_4^i  \  =
  \ - \ \frac{2\,h}{\rho}\,e^{-6A-10C}\,\partial_r\,\phi_2^i \label{eqai3} \ .
\eeq
Next, one can combine the derivative of the second of eqs.~\eqref{eqs_sd2_3i} with the first and the third, obtaining the first equation we were after
\beq
e^{-6A-2C}\left(\partial_r-6\,A'-2\, C'\right)\partial_r\phi_2^i \ = \ - \frac{{\Sigma}\,h}{2\,\rho}\,e^{-2C}\,\phi_6^i\,-\,\left(m^2\,e^{8C}\,-\,\mathbf{k}^2\,{ e^{2A+6C}}\right)\phi_2^i \ . \label{eqphi2}
\eeq
In order to obtain the second, one can differentiate eq.~\eqref{eqai3}, and adding it to the first of eqs.~\eqref{riij} gives
\bea
&& e^{-2B}\left(\partial_r-2B'\right)\left(\partial_r-2A'\right)\phi_6^i + \left(e^{-2A}m^2-e^{-2C}\,\mathbf{k}^2\right)\phi_6^i \nonumber \\
&&=- \ \frac{2\,h}{{\Sigma}\,\rho}\left[e^{-6A-10C}\left(\partial_r-6 A'-10 C'\right)\partial_r\ + \ m^2 \right]\phi_2^i \ . \label{eqphi6}
\eea
One can also combine eqs.~\eqref{eqphi2} and \eqref{eqphi6} in order to eliminate the second derivative of $\phi_2$ from the second equation, and the system can be presented in the more convenient form
\bea
&-& \left[ e^{2(A-B)}\left(\partial_r- 6A'-2C'\right)\partial_r - \mathbf{k}^2 e^{2(A-C)}\right] \phi_2^i \,-\,\frac{{\Sigma}\,h}{2\,\rho}\,e^{-10C}\,\phi_6^i \ = \ m^2\,\phi_2^i \ , \nonumber \\
&-& \left[e^{2(A-B)}\left(\partial_r-2B'\right)\left(\partial_r - 2 A' \right) - \mathbf{k}^2\,e^{2(A-C)}\, - \frac{h^2}{\rho^2}\,e^{2A-10C}\right]\phi_6^i \nonumber \\
&+& \frac{2\,h}{{\Sigma}\,\rho}\left(8 C'e^{-4A-10C}\partial_r-\mathbf{k}^2 {e^{4A-2C}} \right)\phi_2^i
\ = \ m^2\, \phi_6^i \ .
\eea

Now, as a first step toward attaining a manifestly Hermitian form, we turn to the independent variable $z$ of eq.~\eqref{dzdr}, and the system becomes
\bea
&-& \left[\left(\partial_z- 3A_z+3C_z\right)\partial_z - \mathbf{k}^2 \,{ e^{2(A-C)}}\right] \phi_2^i \,-\,\frac{{\Sigma}\,h}{2\,\rho}\,e^{-10C}\,\phi_6^i \ = \ m^2\,\phi_2^i\ , \nonumber \\
&-& \left[\left(\partial_z-5(A_z+C_z)\right)\left(\partial_z - 2 A_z \right) - e^{2(A-C)}\,\mathbf{k}^2 - \frac{h^2}{\rho^2}\,e^{2A-10C}\right]\phi_6^i \nonumber \\
&+& \frac{2\,h}{{\Sigma}\,\rho}\,e^{2A}\left(8 C_z\partial_z-\mathbf{k}^2 e^{2(A-C)} \right)\phi_2^i \nonumber \ = \ m^2\, \phi_6^i \ .
\eea

Separating variables letting
\beq
\phi_a^i(x,z) \ = \ \phi^i(x) \ f_a(z) \qquad (a=2,6) \ ,
\eeq
in matrix notation the system becomes
\beq{}{}{}{}{}{}
{\cal M}\, \Psi \ = \ m^2 \, \Psi \ ,
\qquad
\mathrm{with} \qquad
\Psi \ = \ \left(\begin{array}{c} f_2 \\ f_6 \end{array} \right)
\eeq
and
\beq
{\cal M}= \left(\begin{array}{cc} - \left[\left(\partial_z- 3A_z+3C_z\right)\partial_z - {\cal K}^2\right] & \,-\,\frac{{\Sigma}\,h}{2\,\rho}\,e^{-10C} \\ \!\!\!\!\!\!\!\!\!\!\!\!\!\frac{2\,h}{{\Sigma}\,\rho}\,e^{2A}\left[8 C_z\partial_z-{\cal K}^2\right]  & \!\!\!\!\!\!\!\!\!\!\!\! - \left[\left(\partial_z-5(A_z+C_z)\right)\left(\partial_z - 2 A_z \right) - {\cal K}^2 - 16 {\cal W}_5^2\right]
\end{array} \right)\, . \label{matrixPhi}
\eeq

The two quantities ${\cal K}^2$ and ${\cal W}_5$ that enter this expression were defined in eqs.~\eqref{calK} and \eqref{W5K}, and a number of related properties can be found in Appendix~\ref{app:background}. Note that the differential operator ${\cal M}$ is, once more, not in a manifestly Hermitian form. This form can reached, as in Section~\ref{sec:intvectexc}, by a convenient change of basis, which is now slightly more involved.

In general, letting
\beq{}{}{}{}{}{}
\Psi \ = \ \Lambda \, \widetilde{\Psi} \ ,
\eeq
with $\Lambda$ an invertible matrix, one obtains for $\widetilde{\Psi}$ the new system
\beq{}{}{}{}{}{}
\widetilde{\cal M}\,\widetilde{\Psi} \ = \ m^2 \, \widetilde{\Psi} \ ,
\eeq
with
\beq{}{}{}{}{}{}
\widetilde{\cal M} \ = \ \Lambda^{-1}\,{\cal M}\,\Lambda \ .
\eeq
In this sector the matrix $\Lambda$ and its inverse are not diagonal, and read
\beq
\Lambda \ = \ \left(\begin{array}{cc} e^{\xi_1} & 0 \\ -\,e^{\xi_3} & e^{\xi_2} \end{array} \right)\ , \qquad \Lambda^{-1} \ = \ \left(\begin{array}{cc} e^{-\xi_1} & 0 \\ e^{\xi_3-\xi_1-\xi_2} & e^{-\xi_2} \end{array} \right) \ , \label{Lambda123}
\eeq
where
\beq
e^{\xi_1} \ = \ e^{\frac{3}{2}\,\left(A\,-\,C\right)}\,\sqrt{\frac{\Sigma}{2\,\left|\mathbf{k}\right|}} \ , \quad e^{\xi_2} \ = \ e^{\frac{1}{2}\,\left(7\,A\,+\,5\,C\right)}\,\sqrt{\frac{2\,\left|\mathbf{k}\right|}{\Sigma}} \ , \quad e^{\xi_3} \ = \ e^{\frac{1}{2}\,\left(7\,A\,-\,3\,C\right) }\,\frac{h}{\rho\,{\Sigma}}\,\sqrt{\frac{2{\Sigma}}{\left|\mathbf{k}\right|}} \ .
\eeq
These redefinitions lead to the manifestly Hermitian operator
\beq
\widetilde{\cal M} \ = \ \left(\begin{array}{cc} \left(-\partial_z +\alpha_z\right)\left(\partial_z+ \alpha_z\right)+ {\cal K}^2 +  16\,{\cal W}_5^2 &  - \, 4\, {\cal W}_5\,{\cal K} \\ - \, 4\, {\cal W}_5\,{\cal K} &   \!\!\!\!\!\!\!\!\!\!\!\!\left(-\partial_z+ \beta_z\right)\left(\partial_z + \beta_z\right) + {\cal K}^2  \end{array} \right) \ ,
\eeq
where
\beq
\alpha_z \ = \ \frac{3}{2}\,\left(A_z\,-\,C_z\right)  \ , \qquad \beta_z \ = \ - \ \frac{1}{2}\,\left(3\,A_z\,+\,5\,C_z\right) \ ,
\eeq
and the scalar product, after separating variables, takes the form
\bea{}{}{}{}{}{}{}{}{}{}{}{}{}{}{}{}{}{}{}{}{}{}{}{}{}{}{}{} &\int dz& \!\!\!\left[ \left|\widetilde{\Psi}_1\right|^2 \ + \ \left|\widetilde{\Psi}_2\right|^2 \right]\nonumber \\ &=& \frac{\Sigma}{2\left|\mathbf{k}\right|} \int dz \left[ \left(\frac{2\left|\mathbf{k}\right|}{\Sigma}\right)^2 e^{3(C-A)} \left|f_2\right|^2 \ +\ e^{-(3A+5C)} \left| f_2 \, + \, e^{-2A}\,\frac{\rho\,{\Sigma}}{2\,h}\,f_6\right|^2\right] \, . \label{scalar_scalfund}
\eea
The correspondence with the scalar product implied by the Schr\"odinger system~\eqref{AAtilde} of the previous section can be exhibited dividing this expression, which becomes singular for vanishing $\mathbf{k}$, by an overall factor proportional to $\left|\mathbf{k}\right|$, and focusing on the first term.

Actually, using the result in Section~\ref{sec:nonsingletscalarsk0}, $\widetilde{\cal M}$ can be recast in a form that is very similar to eq.~\eqref{mtilde2},
\beq
\widetilde{\cal M} \ = \ \left(\begin{array}{cc} \left(-\partial_z +\widetilde{\alpha}_z\right)\left(\partial_z+ \widetilde{\alpha}_z\right)+ {\cal K}^2 &  - \, \frac{\left|\mathbf{k}\right| \,h}{\rho}\, e^{2(A-3C)}  \\ - \, \frac{\left|\mathbf{k}\right| \,h}{\rho}\, e^{2(A-3C)}  &   \!\!\!\!\!\!\!\!\!\!\!\!\left(-\partial_z+ \beta_z\right)\left(\partial_z + \beta_z\right) + {\cal K}^2  \end{array} \right) \ ,
\eeq
where
\beq
\widetilde{\alpha}_z \ = \ - \ \frac{9}{2}\,A_z \ - \ \frac{7}{2}\,C_z \ , \qquad \beta_z \ = \ - \ \frac{1}{2}\,\left(3\,A_z\,+\,5\,C_z\right) \ ,
\eeq
and the sign of the off--diagonal terms is not significant, since it could be flipped conjugating by $\sigma_3$, or equivalently redefining one of the two wavefunctions by an overall sign.

\subsubsection{\sc Boundary Conditions and Stability Analysis}

As before, self--adjoint boundary conditions are determined by the leading behavior of the wavefunctions at the two ends, and thus by the indices $\mu_i$ and $\tilde{\mu}_1$ that first emerged in eq.~\eqref{lim_dil_axion_intro}. In this case $\mu_1=\frac{2}{3}$ and $\mu_2=\frac{1}{3}$, so that close to the origin, proceeding as in Section~\ref{sec:intvectexc}, one can identify the limiting behavior~\footnote{For simplicity, all these $C_{ij}$ coefficients were redefined by an overall factor $\sqrt{3}$ with respect to Section~\ref{sec:exactschrod}.} 
\bea
\widetilde{\Psi}{}_1 &\sim& \frac{C_{11}}{\sqrt{2}} \left(\frac{z}{z_m}\right)^\frac{7}{6} + \frac{C_{12}}{\sqrt{2}} \left(\frac{z}{z_m}\right)^{\,-\,\frac{1}{6}} \ + \ 2\,\xi\left[- \ \frac{3}{5}\ C_{21} \left(\frac{z}{z_m}\right)^\frac{3}{2} \ + \  C_{22} \left(\frac{z}{z_m}\right)^\frac{5}{6}\right]\nonumber \\
&+& \frac{3}{4 \sqrt{2}}\, \xi^2\left[ \frac{19}{40}\ C_{11} \left(\frac{z}{z_m}\right)^\frac{5}{2} \ - \ 3\,C_{12}\left(\frac{z}{z_m}\right)^\frac{7}{6}\, \log\left(\frac{z}{z_m}\right)\right] \ , \nonumber \\
\widetilde{\Psi}{}_2 &\sim& C_{21} \left(\frac{z}{z_m}\right)^\frac{5}{6} + C_{22} \left(\frac{z}{z_m}\right)^\frac{1}{6} \ +\ \sqrt{2}\,\xi\left[-\, \frac{1}{5}\,C_{11}  \left(\frac{z}{z_m}\right)^\frac{11}{6} \ + 3\,C_{12}  \left(\frac{z}{z_m}\right)^\frac{1}{2}\right]\nonumber \\
&+& \frac{1}{8}\,\xi^2\left[\frac{27}{5}\,C_{21} \left(\frac{z}{z_m}\right)^\frac{13}{6} \ -\ C_{22} \left(\frac{z}{z_m}\right)^\frac{3}{2}\right] , \label{scalarnearzero}
\eea
where
\beq
\xi \ = \  \frac{\left|\mathbf{k}\,z_m\right|} {\left(3 \,H\,z_m\right)^\frac{1}{3}} 
\eeq
is the dimensionless quantity that already emerged in eq.~\eqref{xidef}.
These expressions are more complicated than those obtained in Section~\ref{sec:intvectexc}, but the additional terms proportional to $\mathbf{k}^2$ are needed to grant that $\widetilde{\cal M}\,\widetilde{\Psi}$ be in $L^2$.

The behavior at the other end of the interval is simpler, since $\widetilde{\cal M}$ is dominated by the diagonal matrix
\beq
\widetilde{\cal M} \ = \ \left(\begin{array}{cc} \left(-\partial_z +\frac{\frac{1}{2}+\tilde{\mu}_1}{z_m-z}\right)\left(\partial_z +\frac{\frac{1}{2}+\tilde{\mu}_1}{z_m-z}\right) &  0  \\ 0  &   \!\!\!\!\!\!\!\!\!\!\!\!\left(-\partial_z +\frac{\frac{1}{2}}{z_m-z}\right)\left(\partial_z +\frac{\frac{1}{2}}{z_m-z}\right)  \end{array} \right) \ ,
\eeq
with the two indices $\tilde{\mu}_1 \simeq 2.27$ and $\tilde{\mu}_2=0$. Consequently, the limiting behavior of the wavefunctions close to $z_m$ is captured by
\beq
\widetilde{\Psi}{}_1 \ \sim \  \left(1\ - \ \frac{z}{z_m}\right)^{2.77}  \ , \qquad
\widetilde{\Psi}{}_2  \ \sim \   \left(1\ - \ \frac{z}{z_m}\right)^\frac{1}{2} \left[C_{23} \ \log\left(1\ - \ \frac{z}{z_m}\right) \ + \ C_{24} \right] \ ,
\eeq
since the other possible exponent for $\widetilde{\Psi}{}_1$ would be incompatible with the $L^2$ condition.

As in Section~\ref{sec:intvectexc}, the limiting behaviors at the origin  that grant self-adjointness are parame\-trized by a $U(1,1)$ matrix such that
\beq
\left(\begin{array}{c} C_{22} \\ C_{21} \end{array}\right) \ =\ U \left(\begin{array}{c} C_{11} \\ C_{12} \end{array}\right) \ ,
\eeq
while the allowed choices at the other end are parametrized by
\beq
\cot\left(\frac{\tilde{\alpha}}{2} \right) \ = \ \frac{C_{23}}{C_{24}} \ .
\eeq

At the upper end positivity is guaranteed if $C_{23}=0$, while at the lower end one is led again to eq.~\eqref{pos_co}, and thus positivity is surely guaranteed if the conditions in eq.~\eqref{positiveU} hold. We shall focus on these options in the following.

Before ending the section, let us note that in this sector $\widetilde{\cal M}$ admits the decomposition
\beq{}{}{}{}{}{}{}
\widetilde{\cal M} \ = \ \widehat{Q}^\dagger\,\widehat{Q} \ + \ \Delta \ ,  \label{MQQD}
\eeq
where
\beq{}{}{}{}{}{}{}
\widehat{Q} \ = \ \left(\begin{array}{cc} \partial_z+ \alpha_z & 0 \\ 0 &   \partial_z + \beta_z \end{array} \right)
\eeq
and
\beq{}{}{}{}{}{}{}
\Delta \ = \ \left(\begin{array}{cc} {\cal K}^2 +  16\,{\cal W}_5^2 &  - \, 4\, {\cal W}_5\,{\cal K}\\ - \, 4\, {\cal W}_5\,{\cal K} & {\cal K}^2  \end{array} \right) \ ,
\eeq
is a manifestly positive-definite matrix, since its trace and determinant are both positive, while $\widehat{Q}^\dagger\,\widehat{Q}$ can be positive definite with self--adjoint boundary conditions determined by $\alpha_z$ and $\beta_z$. However, $\Delta$ contains contributions proportional to $\frac{1}{z^2}$, and therefore the self--adjoint boundary conditions for $\widetilde{\cal M}$ are different in general, so that eq.~\eqref{MQQD} does not suffice to imply its positivity.

With general self--adjoint boundary conditions appropriate to $\widetilde{\cal M}$, our variational tests summarized in Appendix~\ref{app:variationa2} provide evidence that stability is generally granted, in this sector, by values of $\xi_0$ of eq.~\eqref{xi0} beyond a few units, while special choices of boundary conditions lead to spectra that are fully stable without any conditions on $\xi_0$.

\section{\sc Singlet Scalar Modes} \label{sec:singletscalarmodes}

We can now conclude our analysis with a discussion of the most intricate sector of the spectrum, which concerns scalar singlets. These originate from a number of different tensor and metric perturbations, which can be parametrized as follows:
\bea
b&=&\phi_1 \ ,\qquad b^i=\frac{1}{{\Sigma}}\,\partial^i\phi_2\ ,\qquad b_\mu^i \ = \ \frac{1}{{\Sigma}^2}\,\partial_\mu\partial^i\phi_3\ ,\nonumber\\
h_{\mu\nu}&=&\eta_{\mu\nu}\phi_4+\frac{1}{{\Sigma}^2}\,\partial_\mu\partial_\nu\phi_5 \ ,
\qquad h_{\mu r}\ = \ \frac{1}{{\Sigma}}\,\partial_\mu\phi_6\ ,\qquad h_{\mu i}=\frac{1}{{\Sigma}^2}\,\partial_\mu\partial_i\phi_7\ ,\nonumber\\
h_{rr}&=&\phi_8\ ,\qquad h_{ri}\ = \ \frac{1}{{\Sigma}}\,\partial_i\phi_9\ ,\qquad h_{ij}\ = \ \delta_{ij}\phi_{10}+\frac{1}{{\Sigma}^2}\,\partial_i\partial_j\phi_{11}\ . \label{scalar_singlet_fields}
\eea

Since the background values of the dilaton--axion pair are constant, their perturbations decouple from these other scalar modes. For this reason we could treat them separately in Section~\ref{sec:dilatonaxion}, and the same happened in Section~\ref{sec:IIB3forms} for the perturbations arising from the two-forms $B_{MN}^i$, so that here we can set all of them to zero. One can also fix diffeomorphism invariance, using the three parameters $\xi_\mu$, $\xi_r$ and $\xi_i$, making the convenient gauge choice
\beq
\phi_5\ = \ \phi_6 \ = \ \phi_7=0 \ ,
\eeq
but these steps leave nonetheless, in general, eight fields within this sector. They also introduce boundary fields, as in other sectors. For brevity we shall not discuss them, although they can change the massless spectrum, since they cannot give rise to instabilities. Moreover, here we confine our attention to the $\mathbf{k}=0$ modes, where some simplifications occur, leaving the general case, which involves a number of novel features, to a future publication~\cite{ms_23_3}.

For $\mathbf{k}=0$, only the four fields $\phi_1,\phi_4,\phi_8$ and $\phi_{10}$ are left.
The tensor equations reduce to
\beq
\frac{h}{4\,\rho}  \left( - \ 4\,e^{-2A}\,\phi_4\, - \,  e^{-2B}\,\phi_8 \, + \, 5\,e^{-2C}\,\phi_{10} \right) \,+\,  e^{-8A} \,\partial_r\, \phi_1 \ = \ 0 \ , \label{tensorsingletscalar}
\eeq
and determine $\phi_1$ in terms of the different metric perturbations,
while the $\alpha\beta$ Einstein equation involves two different structures, associated to $\partial_\alpha\partial_\beta$ and $\eta_{\alpha\beta}$, and the corresponding terms are to vanish separately. This gives for the modes belonging to this sector the two additional equations
\bea
\alpha\beta_1&: &  2\,e^{-2A}\, \phi_4 \,+\, e^{-2B}\,\phi_8\,+\,5\,e^{-2C}\,\phi_{10}  \ = \ 0  \ , \label{alphabeta1n} \\
\alpha\beta_2&: & e^{-2A} \, m^2\,\phi_4 \ + \ A'\,e^{2(A-B)}\big[4\,e^{-2A}\left(\partial_r-2 A'\right)\phi_4 \nonumber \\ &-& e^{-2B}\left(\partial_r-\,2 B'\right)\phi_8 + 5\,e^{-2C}\left(\partial_r-2 C'\right)\phi_{10}\big] \nonumber\\
&+&e^{-2B}\big[\left(\partial_r-4A'\right)\partial_r\,+\,4\,(A')^2\big]\,\phi_4 \,=\,  2\,\frac{h}{\rho}\,e^{2(A-B)}\,\partial_r\,\phi_1\, - \,\frac{3}{2}\,\frac{h^2}{\rho^2}\, e^{- 10 C} \,\phi_4 \ . \label{alphabeta2n}
\eea
The remaining Einstein equations become
\bea
\alpha r &:&  3\,e^{-2A}\left(\partial_r\,-\,2\,A' \right)\phi_4 \,+\, \left(A'-B'\right)e^{-2B}\,\phi_8\,+\, 5\,e^{-2C}\left( \partial_r\,-\,A'\,-\,C'\right)\phi_{10} \ = \  0 \, , \nonumber\\
r r &:& e^{-2A}\, m^2\,\phi_8\,-\,B'\,\partial_r\left(e^{-2B}\,\phi_8\right) \nonumber \\
&+& 4\,e^{-2A}\left(\partial_r\,-\,B'\right)\left(\partial_r\,-\,2\,A'\right)\phi_4 \,+\, 5\,e^{-2C}\left(\partial_r\,-\,B'\right)\left(\partial_r\,-\,2\,C'\right)\phi_{10}
\nonumber \\ &=&  2\,\frac{h}{\rho}\,\,\partial_r\,\phi_1 \,-\,  2\,\frac{h^2}{\rho^2} \, e^{6A}\, \phi_4  \ , \nonumber \\
ij&:& e^{-2A}\,m^2 \,\phi_{10}\,+\, 4\,C'\,e^{2(C-A-B)}\left(\partial_r-2 A'\right)\phi_4\,-\,e^{2(C-2B)}\left[2\,C''\,+\,C' \left(\partial_r-\,2 B'\right)\right]\phi_8\nonumber \\
&+&e^{-2B}\big[\partial_r^2\,+\,C'\,\partial_r\,-\,6\,(C')^2\big]\phi_{10} \ = \  2\,\frac{h^2}{\rho^2}\,e^{-10C}\,\phi_{10} \ . \label{rreinsteinsingletscalij}
\eea

One can now eliminate $\phi_8$ using eq.~\eqref{alphabeta1n}, and then eq.~\eqref{tensorsingletscalar}  reduces to
\beq{}{}{}{}{}{}
\partial_r\,\phi_1 \ = \ \frac{h}{2\,\rho} \left(e^{6A}\,\phi_4 \ - \ 5 \,e^{8A-2C}\,\phi_{10} \right) \ ,
\eeq
and determines $\phi_1$
up to an $r$--independent contribution that is pure gauge, so that only $\phi_4$ and $\phi_{10}$ are left. After the convenient redefinitions
\beq
e^{-2A}\,\phi_4 \ = \ \chi_4 \ , \qquad e^{-2B}\,\phi_{8} \ = \ \chi_{8}  \ , \qquad e^{-2C}\,\phi_{10} \ = \ \chi_{10} \ ,
\eeq
the $\alpha\beta_1$ Einstein equation~\eqref{alphabeta1n} reduces to the simple algebraic constraint
\beq{}{}{}{}{}{}
2 \chi_4 \ + \ \chi_8 \ + \ 5\, \chi_{10} \ = \ 0 \ , \label{alphabeta1}
\eeq
and eliminating $\chi_8$ now leads to a system of four equations for the two fields $\chi_4$ and $\chi_{10}$
\bea
\alpha\beta_2&:& m^2\,\chi_4 \,+\,  e^{2(A-B)}\left[ 10\,A'\,\partial_r\,+\,\frac{5\,h^2}{\rho^2}\,e^{8A}\right]\chi_{10} \nonumber\\
&&+e^{2(A-B)}\left[\partial_r^2\,+\,6\,A'\,\partial_r\,+\,\frac{h^2}{\rho^2}\,e^{8A}\right]\chi_4  \, = \,  0 \ , \nonumber
\\
rr&:& - m^2\left(2\chi_4+5\chi_{10}\right) \,+\, e^{2(A-B)}\left[ \left(4 \partial_r\,+\,8 A'\,-\,2 B' \right)\partial_r +\frac{h^2}{\rho^2}\,e^{8A} \right]\chi_4 \nonumber \\ &&+ \ e^{2(A-B)}\left[5 \left(\partial_r\,+\,2\,C'\right)\,\partial_r + \frac{5\,h^2}{\rho^2}\,e^{8A}\right]\chi_{10}\, = \, 0 \ ,  \nonumber \\
ij&:& m^2 \,\chi_{10}\,+\, e^{2(A-B)}\left[ 6\,C'\,\partial_r - \frac{h^2}{\rho^2}\,e^{8A} \right]\chi_4 \nonumber \\
&&+ \ e^{2(A-B)}\left[\partial_r^2\,+\,10\,C'\,\partial_r\,-\,\frac{5\,h^2}{\rho^2}\,e^{8A}\right]\chi_{10} \, = \,  0 \ , \nonumber \\
\alpha r&:&  \left(3\,\partial_r+6 A'+ 10\,C'\right)\chi_4 \,+\, 5\,\left(\partial_r+2A'+6 C'\right)\chi_{10}\,=\,0 \ . \label{eqsgravss}
\eea

In terms of the $z$ variable of eq.~\eqref{dzdr}, this system takes the more compact form
\bea{}{}{}{}{}{}{}{}{}{}{}{}{}{}{}{}{}{}{}{}{}{}{}{}{}{}{}{}{}{}{}{}{}
&& 10\left(A_z\,\partial_z\,+\,8\,{\cal W}_5^2\right)\chi_{10} +  \left[\left(\partial_z+9 A_z+5 C_z\right)\partial_z + m^2 + 16\,{\cal W}_5^2 \right]\chi_4 \, = \, 0 \,,\nonumber \\
&& 2\left[\left(2 \partial_z+6 A_z+5 C_z\right)\partial_z - m^2 + 8\,{\cal W}_5^2\right]\chi_4  +  5\left[\left(\partial_z+3 A_z+7C_z\right)\partial_z-m^2+ 16\,{\cal W}_5^2\right]\chi_{10} =  0 \ ,\nonumber \\
&& 2\left(3 C_z\,\partial_z\,-\,8\,{\cal W}_5^2\right)\chi_{4}+\left[\left(\partial_z+3 A_z+15 C_z\right)\partial_z+m^2- 80\,{\cal W}_5^2\right]\chi_{10} = 0 \ ,  \nonumber \\
&& \left(\partial_z+2 A_z\right)\left(3\,\chi_4\,+\,5\,\chi_{10}\right) \ + \ 10\,C_z\left(\chi_4\,+\,3\,\chi_{10}\right) \,=\,0 \ . 
\eea
However, the $rr$ equation is a linear combination of the $\alpha\beta$ and $i j$ ones, once the $\alpha r$ equation is used. Moreover, the three second--order equations can be combined into another first--order one. This is in fact the $rr$ Einstein equation, which we often refer to as ``Hamiltonian constraint'',
\bea{}{}{}{}{}
&-& m^2 \left(3 \chi_4 + 5 \chi_{10}\right) \,+\,e^{2(A-B)}\left[ -20(A'+C')\partial_r + \frac{5\,h^2}{\rho^2}\,e^{8A}\right] \chi_{10}\nonumber \\
&+& e^{2(A-B)}\left[ -4(3A'+5 C')\partial_r + \frac{h^2}{\rho^2}\,e^{8A}\right] \chi_4 \ = \ 0 \ , \label{grav_constr}
\eea
or, in terms of the $z$ variable,
\beq{}{}{}{}{}{}{}{}{}{}{}{}{}{}{}{}{}{}{}{}{}{}{}{}{}{}{}{}{}{}{}{}{}
 m^2 \left(3 \chi_4 + 5 \chi_{10}\right) \, +\,20\left[\left(A_z+C_z\right)\,\partial_z \, - \, 4\,{\cal W}_5^2\right]\chi_{10}\,+\, 4\left[\left(3A_z+5 C_z\right)\partial_z\,-\, 4 \,{\cal W}_5^2\right] \chi_4 \ = \ 0 \ .
\eeq
In this fashion, one ends up with two first--order equations linking $\chi_4$ and $\chi_{10}$, and all second--order equations follow from them. Therefore, the system only contains one independent scalar degree of freedom, from which all other wavefunctions can be deduced.

In flat space, with the internal $T^5$ and an $S_1$ along the $r$-direction, these equations would allow \emph{two} $r$-independent wavefunctions for $m=0$, and actually a pair of four--dimensional massless scalars $\chi_4(x)$ and $\chi_{10}(x)$. The constant parts of $\chi_8$ and $\chi_{10}$ would be associated to independent deformations of the radius of the $S_1$ and the overall size of the internal torus. In our case the length $\ell$ of the $r$-interval is fixed by the background, and indeed a single independent constant part is left by the preceding conditions, since the first three of eqs.~\eqref{eqsgravss} demand that $r$--independent quantities satisfy
\beq{}{}{}{}{}{}
\chi_4 \ + \ 5\, \chi_{10} \ = \ 0 \ .
\eeq
Equivalently, using eq.~\eqref{alphabeta1}, constant shifts of $B$ and $C$, which are captured by $\chi_8$ and $\chi_{10},$ are surprisingly not independent, but
\beq{}{}{}{}{}{}
\chi_8 \ = \ 5\, \chi_{10} \ .
\eeq
The reason behind this result is explained in detail in~\cite{ms22_1}: the equal radii $R$ of the internal torus can be scaled out, so that the background only depends on two parameters, the conserved flux $\Phi$ and the length $\ell$ of the $r$-interval, whose perturbation is described by $\chi_8$. As a result, the size of the internal torus is effectively determined by the combination $\left(\Phi\,\ell\right)^\frac{1}{5}$, as can be seen in eqs.~\eqref{4d_PhiEpos2}. Note, however, that there is no massless field in four dimensions associated to this deformation, which is just a constant shift.

In flat space, for $m^2 \neq 0$ eqs.~\eqref{eqsgravss} or, equivalently, eq.~\eqref{grav_constr}, demand that
\beq
\chi \ \equiv \ 3 \chi_4 + 5 \chi_{10} \ = \ 0 \ .
\eeq
This is a familiar result in Kaluza--Klein theory: massive scalar excitations are eaten by corresponding massive vectors.
In our background, however, it is actually convenient to obtain a second--order equation for this very field,
\beq
\chi \ \equiv \ 3 \chi_4 + 5 \chi_{10} \ ,
\eeq
which will be singular in the flat limit.
To this end, we also let
\beq
\psi = \chi_4 + 3 \chi_{10} \ ,
\eeq
and then the $\alpha r$ Einstein equation becomes
\beq
\left(\partial_r + 2 A'\right)\chi \ + \ 10 \,C'\,\psi \ = \  0 \ , \label{chipsialg}
\eeq
and determines $\psi$ algebraically, while eq.~\eqref{grav_constr} takes the form
\bea{}{}{}{}{}{}
0&=& - \ m^2 \chi\,-\, 2\,e^{2(A-B)}\left[\left(2 A'+5 C'\right)\partial_r\,+\, \frac{h^2}{4\,\rho^2} \,e^{8A} \right]\chi \nonumber \\ &+& 10\,e^{2(A-B)}\left[C'\,\partial_r\,+\, \frac{h^2}{4\,\rho^2}\,e^{8A} \right]\psi \, . \label{grav_constr_chipsi}
\eea

In terms of the $z$-variable of eq.~\eqref{dzdr}, these equations become
\bea{}{}{}{}{}{}{}
&& \left(\partial_z + 2 A_z\right)\chi \ + \ 10 \,C_z\,\psi \ = \  0 \ , \nonumber \\
&& m^2 \chi\,+\, 2 \left[\left(2 A_z+5 C_z\right)\partial_z\,+\, 4\,{\cal W}_5^2\right]\chi\,-\, 10 \left[C_z\,\partial_z\,+\, 4\,{\cal W}_5^2 \right]\psi \ = \ 0 \ , \label{1221}
\eea
where ${\cal W}_5$ was defined in eq.~\eqref{W5K}.
The two equations~\eqref{1221} can now be combined into a second--order one,
\bea{}{}{}{}{}{}{}
\partial_z\left(\partial_z+2 A_z\right)\chi &+&  \left(\frac{4\,{\cal W}_5^2}{C_z} - \frac{C_{zz}}{C_z}\right)\left(\partial_z+2 A_z\right)\chi \ + \  m^2 \chi \nonumber \\
&+& 2 \left[\left(2 A_z+5 C_z\right)\partial_z\,+\, 4\,{\cal W}_5^2\right]\chi \ = \ 0 \ ,
 \eea
that, using the results collected in Appendix~\ref{app:background}, can be cast in the form
  \bea{}{}{}{}{}{}{}
 m^2 \chi\,+\, \partial_z^2\,\chi \,+\, \left[3(3\,A_z\,+\,5\,C_z)\,+\,\frac{8\,{\cal W}_5^2}{C_z}\right]\partial_z\,\chi\,+\,16\,{\cal W}_5^2\left(1\,+\,\frac{A_z}{C_z}\right)\chi \ = \ 0 \ ,
 \eea
\begin{figure}[ht]
\centering
\includegraphics[width=65mm]{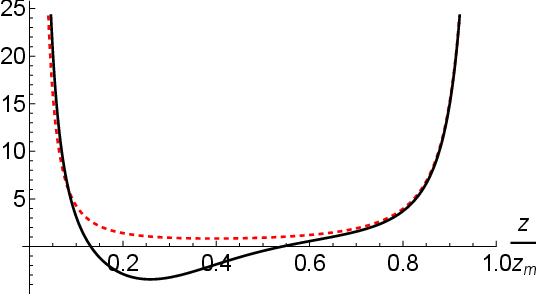}
\caption{\small The potential $V$ of eq.~\eqref{pot_phi} (black, solid) with its approximation~\eqref{pot_hyp} (red,dashed)  with $(\mu,\tilde{\mu})=\left(\frac{2}{3},1\right)$, in units of $\frac{1}{{z_0}^2}$. $z_m$ and $z_0$ are defined in eqs.~\eqref{zm_app} and \eqref{z0}.}
\label{fig:pot_singlet}
\end{figure}

The redefinition
 \beq{}{}{}{}{}{}{}
 \chi \ = \ Y \, C_z\,e^{\,-\,\frac{1}{2}\left(3 A + 5 C\right)} \ ,  \label{chiYk0}
 \eeq
 leads finally to the Schr\"odinger--like equation
 \beq{}{}{}{}{}{}{}
 m^2\,Y \ = \ - \ \frac{d^2\,Y}{dz^2} \ + \ V\, Y \ , \label{schrod_scalarsinglet}
 \eeq
and letting
 \beq{}{}{}{}{}{}{}
 \alpha \ = \ \frac{4\,{\cal W}_5^2}{C_z} \ +\ \frac{3}{2}\left( 3 A_z + 5C_z\right) \ ,
 \eeq
 the potential takes the form
 \beq{}{}{}{}{}{}{}
 V \ = \  - \ 16\,{\cal W}_5^2\left(1\,+\,\frac{A_z}{C_z}\right) \ +\ \alpha^2 \ + \ \frac{d\,\alpha}{d\,z} \ ,
 \eeq
 or in detail
 \bea
V &=& \frac{5 e^{\sqrt{\frac{5}{2}} r}}{256 \,z_0^2\,\sinh^5\left(\frac{r}{\rho}\right)\left[\sqrt{10}-5 \coth\left(\frac{r}{\rho}\right)\right]^2} \Big[-\,84 \sqrt{10} \sinh\left(\frac{4\,r}{\rho}\right)\,+\,128 \sqrt{10} \sinh\left(\frac{2\,r}{\rho}\right) \nonumber \\ &{}& \nonumber \\ &-& 928 \cosh\left(\frac{2\,r}{\rho}\right)+267 \cosh\left(\frac{4\,r}{\rho}\right)+1221\Big] \ . \label{pot_phi}
 \eea
 Consequently, near the two ends $V$ behaves as in eq.~\eqref{lim_dil_axion_intro}, with $\mu = \frac{2}{3}$ and $\tilde{\mu}=1$. As a result, the allowed wavefunctions have the limiting behaviors
 \beq
Y \ \sim \ C_1 \left(\frac{z}{z_m}\right)^\frac{7}{6} \ + \ C_2 \left(\frac{z_m}{z}\right)^\frac{1}{6} \qquad \mathrm{and} \qquad Y \ \sim \ \left(1 \ - \ \frac{z}{z_m}\right)^\frac{3}{2} \ ,
 \eeq
 and self--adjoint boundary conditions are determined by the ratio $x=\frac{C_2}{C_1}$. In fact, eq.~\eqref{schrod_scalarsinglet} can be cast in the formally positive form
\beq{}{}{}{}{}{}{}
 m^2\,Y \ = \ {\cal A}^\dagger\, {\cal A} \ Y \ + \ {\cal V} \, Y \ , \label{scalar_Schrod_complete}
 \eeq
 where now
 \beq{}{}{}{}{}{}{}
 {\cal A} \ = \ \frac{d}{dz} \ - \ \alpha , \qquad
 {\cal A}^\dagger \ = \ - \ \frac{d}{dz} \ - \ \alpha \ ,
 \eeq
 and
 \beq{}{}{}{}{}{}{}
 {\cal V} \ = \ - \ 16\,{\cal W}_5^2\left(1\,+\,\frac{A_z}{C_z}\right) \ = \ \frac{16\,{\cal W}_5^2}{\frac{\sqrt{10}}{2}\,\coth\left(\frac{r}{\rho}\right) \ - \ 1} \ > \ 0  \ .
 \eeq

The $ {\cal A}^\dagger\, {\cal A}$ portion of the potential has a zero mode,
  \beq
Y \ \sim \ e^{\,\int\alpha\, dz} \ ,
\eeq
which is normalizable and behaves as
\beq
Y \ \sim \ \left(\frac{z}{z_m}\right)^\frac{7}{6} 
\eeq
close to the left end of the interval, where $\alpha \sim \ \frac{7}{6\,z}$. For this zero mode the absence of the $z^{-\frac{1}{6}}$ contribution indicates that
\beq
\frac{C_2}{C_1} \ = \ 0 \ . \label{bc_scalar_pert}
\eeq
\begin{figure}[ht]
\centering
\begin{tabular}{cc}
\includegraphics[width=65mm]{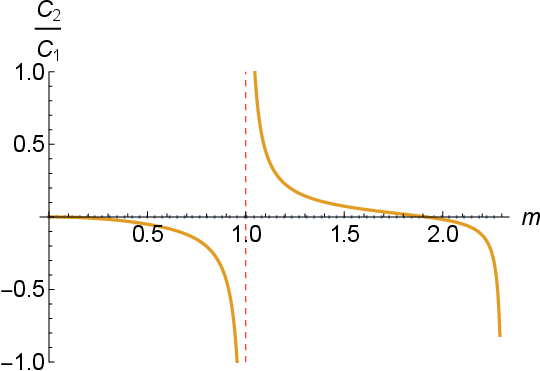} \qquad \qquad &
\includegraphics[width=65mm]{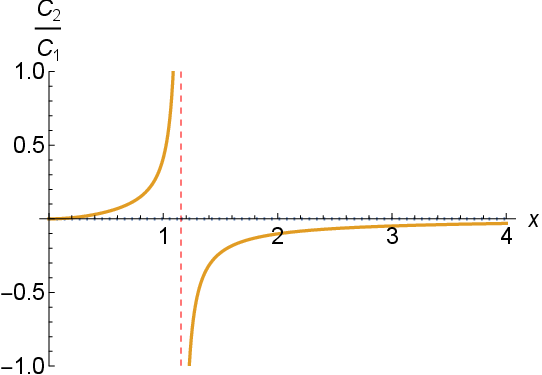} \\
\end{tabular}
\caption{\small The left panel illustrates how the stable eigenvalues of the shifted hypergeometric potential $V_{1,-1}$ of eq.~\eqref{V12} with $\mu=\frac{2}{3}$, $\tilde{\mu}=1$ depend on the boundary conditions. The right panel illustrates the corresponding tachyonic eigenvalues, with $m^2=-x^2$, for the same values of $\mu$ and $\tilde{\mu}$.}
\label{fig:singletscalar}
\end{figure}

The ${\cal A}^\dagger{\cal A}$ portion of the potential can be approximated by the shifted hypergeometric potential of eq.~\eqref{V12} with $\epsilon_1=1$ and $\epsilon_2=-1$, whose zero mode in eq.~\eqref{psi12} has the same boundary condition~\eqref{bc_scalar_pert}. The corresponding shift,
\beq
\Delta\,V \ = \ - \ \left(\frac{4}{3}\right)^2 \frac{\pi^2}{z_m^2} \ ,
\eeq
can be read from eq.~\eqref{V12}. By the general argument of~\cite{selfadjoint}, the ${\cal A}{\cal A}^\dagger$ potential has a stable spectrum if $C_2=0$, and a closer look at the eigenvalue equation shows that every other boundary condition gives rise to an unstable mode (see fig.~\ref{fig:singletscalar}). Therefore, in this sector there is at least one stable boundary condition for the full potential in eq.~\eqref{scalar_Schrod_complete}, where the would be massless mode becomes slightly massive, due to the positive contribution ${\cal V}$. As a result, a small range of stable boundary conditions exists for the complete potential, where stability holds for $\frac{C_2}{C_1} \geq 0$.

\section{\sc Conclusions and Open Issues} \label{sec:conclusions} 
In this paper we have explored in detail the bosonic modes that can emerge in a family of four--dimensional IIB Randall--Sundrum--like~\cite{randallsundrum} non-supersymmetric vacua, and the resulting indications for vacuum stability. These backgrounds are supported by a flux of the self--dual five-form field strength that is homogeneous in spacetime and in an internal five--torus, and depend on a parameter $\rho$ that determines the size of an internal interval and the corresponding scale of supersymmetry breaking~\footnote{The presence of an interval with singularities at its ends establishes a connection between this work and the ongoing activity devoted to ``dynamical cobordism''~\cite{DC}. The link will be made more precise in~\cite{ms_23_3}, where we shall compute tensions and charges at the ends of the interval.}. The vacua thus obtained have a constant dilaton profile, and thus overcome the strong string--coupling problem that typically plagues broken supersymmetry, but the interval introduces some complications related to the choices of boundary conditions at its ends. The $\alpha'$ corrections can be bounded, away from the singularities~\cite{ms22_1} at the ends of the interval, within wide portions of the internal space. All these reasons, together with the well--known stability problems of string vacua in the absence of supersymmetry, motivated us to undertake the detailed analysis presented in the preceding sections, combining a number of different techniques. 

Our analysis relied, all along, on the most general self--adjoint boundary conditions that are in principle available for the different fields, and thus, implicitly, on second--order actions for the internal profiles. For example, for a free complex scalar field $\phi$ in $D$ dimensions the second--order action
\beq
{\cal S}_2\ = \ \int d^{D}x \ \overline{\phi}\ \Box\,\phi \ , \label{scalar_2}
\eeq
allows the most general self--adjoint boundary conditions, while only a subset of them eliminate the boundary terms that emerge when varying the standard first--order action
\beq
{\cal S}_2\ = - \ \int d^{D}x \ \partial^\mu\,\overline{\phi} \ 
\partial_\mu\,\phi \ . \label{scalar_1}
\eeq
In this fashion, we also ran across a peculiar property of gauge fields in the presence of an internal interval: additional modes localized on the boundary can emerge, in general, unless one confines the attention to gauge transformations that vanish there. We could thus classify the stable boundary conditions for the different sectors of the spectrum, and we could also tackle, in rather general terms, the possible instabilities of their Kaluza--Klein excitations. These represent an essentially insurmountable problem~\cite{bms} for the non--supersymmetric AdS vacua of the ten--dimensional strings of~\cite{so1616,susy95,bsb}, since the length scales of the internal sphere and of the AdS spacetimes are correlated by the Einstein equations. In our setting, which includes an internal torus, or more generally with a Ricci--flat internal manifold, there are boundary conditions compatible with a stable spectrum, although the available choices can depend on the background when mixings involving Kaluza--Klein modes occur. For the Kaluza--Klein excitations, the boundary conditions that yield massless modes with $\mathbf{k}=0$ can bring along a finite number of tachyons, as we saw in Sections~\ref{sec:intvectexc} and~\ref{sec:nonsingletscalrknot0}). However, this pathology can be eluded if the parameter $\rho$ in eqs.~\eqref{back_epos_fin2} lies one or two orders of magnitude above the radius $R$ of the internal torus. Equivalently, this condition sets on the scale of supersymmetry breaking
\beq{}{}{}{}{}{}{}{}{}{}{}{}{}{}{}{}{}{}{}{}{}{}{}{}{}{}{}{}{}{}
\mu_S \ = \ \frac{1}{\ell\,h^\frac{1}{4}} \ ,
\eeq
 which we identified in~\cite{ms22_1}, the upper bound
\beq{}{}{}{}{}{}{}{}{}{}{}{}{}{}{}{}{}{}{}{}{}{}{}{}{}{}{}{}{}{}
\mu_S \left(\Phi\,h\right)^\frac{1}{4} \ < \ {\cal O}\left(10^{-2}\right) \ . \label{stability_window}
\eeq
Alternatively, one can select boundary conditions granting the absence of tachyons for all values of $\mathbf{k}$, but these typically eliminate the massless modes with $\mathbf{k}=0$.

\begin{table}[h!]
\centering
\begin{tabular}{||c c c c ||}
 \hline
$4D\,\mathrm{hel.} \times SO(5)$ & 4D $m=0$ Content & $10D$ origin & Equation \\ [0.5ex]
 \hline\hline
 $\left(0,1\right)$ & {1 dilaton} & $\phi$  & \eqref{dil_zero_mode} \\
  $\left(0,1\right)$ & {1 axion} &  $a$  & \eqref{dil_zero_mode} \\
  $\left(\pm 1,0\right)$ & {1 real vector,  1 real scalar}   & $B_{\mu \nu}^{1,2}$,$B_{\mu r}^{1,2}$  & \eqref{eqsbmn} \\
    $2\left(\pm 1,5\right)$ & 10 real vectors  & $B_{\mu i}^{1,2}$, $B_{r i}^{1,2}$  & \eqref{bmui_prof} \\
    $2\left(0,15\right)$ & 30 real scalars    & $B^{1,2}_{ij}$  &  \eqref{bij_prof1}  \\
$\left(\pm 1,10\right)$ & 10 real vectors    & $B_{\mu\nu i j}$  &  \eqref{g2zm} \\
$\left(\pm\,{2},1\right)$ & { 1  graviton}   & $h_{\mu\nu}$ & \eqref{h0_nottilde} \\
$\left(0,14\right)$ & 14 real scalars    & $h_{ij}$  &  \eqref{limitinghij} \\
$\left(\pm 1,5\right)$ & 5 real vectors    & $h_{\mu i},B_{\mu\nu\rho i}$  &  \eqref{W4zeromode} \\
$\left(0,5\right)$ & 5 real scalars   & $h_{r i}, \ B_{\mu\nu\rho i}, \ B_{ijkl}$  &  \eqref{zeromode_chi6i_final} \\
$\left(0,1\right)$ & 1 real scalar${}^\star$ & $b,b^i, b_\mu{}^i,h_{\mu\n}, h_{r r}$ &
\\[1ex]
 \hline
\end{tabular}
\caption{The maximum numbers of four--dimensional real massless bosonic modes that can arise from the bulk, for generic values of $R$, within the stability window of eq.~\eqref{stability_window}. The scalar singlet in the last line, accompanied by the $(\star)$ symbol, could be fine-tuned to zero mass. However, we do not have an analytic form for its wavefunction, which originates from the second--order equation~\eqref{schrod_scalarsinglet}. There are at most 26 vectors and 53 scalars, for a total of 107 massless bosonic degrees of freedom after including the graviton.}
\label{table:tab_B}
\end{table}

While the preceding results are clearly encouraging, our analysis is still incomplete, since we left out the Kaluza--Klein excitations of singlet scalars. These appear resilient to our approach, since they lead to a three--component Schr\"odinger system where the potential cannot be put in a symmetric form with the techniques used in the other cases. We are thus leaving to a future work a proper identification of the norm of these perturbations, and of its correspondence with the convenient Henneaux--Teitelboim action~\cite{henneaux}, together with a final statement on the stability region determined by this sector. 

The first two columns in Table~\ref{table:tab_B} collect the maximum numbers of massless modes found explicitly in the previous sections. In four dimensions, these correspond to a graviton, 26 real vectors and 53 real scalars, which are a large fraction of the modes that would emerge, from the type--IIB theory, after a toroidal compactification to four dimensions. These numbers are purely indicative, since we are focusing on the quadratic terms, and interactions and/or quantum corrections could lift in mass many of these modes. In addition, their number could be reduced by choices of boundary conditions dictated by symmetry requirements. For example, some of these modes lead to the flow~\cite{ms_20}, across the boundary, of charges that would be conserved in its absence. This was the case for the ten vector modes of section~\ref{sec:tensor_no_grav} arising from the four-form gauge field, and for the five vector modes from $h_{\mu i}$ of Section~\ref{sec:nonsingletscalarsk0} arising from the metric field, and self--adjoint boundary conditions eliminating the flow can make all these modes massive. Table~\ref{table:tab_C} collects the additional massless modes that could be present on the boundary.

\begin{table}[h!]
\centering
\begin{tabular}{||c c c ||}
 \hline
$4D\,\mathrm{hel.} \times SO(5)$ & 4D $m=0$ Content & corresponding gauge parameters \\ [0.5ex]
 \hline\hline
  $\left(\pm 1,0\right)$ & {2 real vectors}   & $\Lambda_{\mu}^{1,2}$ \\
    $\left(0,5\right)$ & {10 real scalars}   & $\Lambda_{i}^{1,2}$ \\
$\left(0,5\right)$ & {5 real scalars}    & $\Lambda_{\mu\nu i}$   \\
$\left(\pm 1,10\right)$ & {10 real vectors}    & $\Lambda_{\mu i j}$   \\
$\left(0,10\right)$ & {10 real scalars}    & $\Lambda_{i j k}$   \\
$\left(\pm\,{1},0\right)$ & { 1 real vector}   & $\xi_{\mu}$  \\
$\left(\pm\,{0},5\right)$ & { 5 real scalars}   & $\xi_{i}$ 
\\[1ex]
 \hline
\end{tabular}
\caption{The maximum numbers of four--dimensional real massless bosonic modes that can arise from the boundary of the internal interval. There are in principle 56 degrees of freedom of this type, if one concentrates them on one of the two boundaries. The resulting vector equations are gauge invariant, in view of the discussion presented in~\cite{ms_23_3}.}
\label{table:tab_C}
\end{table}

Summarizing, the bosonic spectrum of the vacua of eqs.~\eqref{back_epos_fin2} confronted us with a number of technical difficulties, revealing some novelties and bringing along some surprises. The main novelty was the indication that \emph{stable vacua may be attained in non--supersymmetric compactifications to four--dimensional Minkowski space}. 
The main surprises were the emergence of additional moduli related to boundary conditions and of corresponding boundary modes, and two technical findings. In Section~\ref{sec:d4unot0} massless modes of the type--IIB three-forms led to dynamical equations with three derivatives, and in Section~\ref {sec:tensor_no_grav_knot0} massless modes emerged, for all internal momenta, from the first--order equations of tensor modes. This last result is not pathological, since these modes were excitations of tachyonic ground states with different $\mathbf{k}$-dependent boundary conditions.

Identifying the
effective four-dimensional theory resulting from this type  of compactifications would be clearly an interesting further step, for which the present work can provide some indications. Local supersymmetry is a key requirement, and boundary terms will be needed to grant it, as we saw for bosonic gauge symmetries.  The $T^5$ reduction to five dimensions of the type--IIB theory, which would yield the corresponding maximal supergravity, could provide some useful guidance~\cite{cremmer}.
Finally, the widespread activity devoted, over the years, to supersymmetric flux compactifications (for reviews, see~\cite{grana}) rests on different geometrical setups and has typically addressed supersymmetric vacua. Generalizing the geometric approach to non--supersymmetric settings should elicit detailed links with the present work, and is likely to lead to further progress.

\section*{\sc Acknowledgments}
\vskip 12pt
We are grateful to C.~Bachas, G.~Dall'Agata, E.~Dudas and A.~Tomasiello for stimulating discussions. AS was supported in part by Scuola Normale, by INFN (IS GSS-Pi) and by the MIUR-PRIN contract 2017CC72MK\_003. JM is grateful to Scuola Normale Superiore for the kind hospitality while this work was in progress. AS is grateful to Universit\'e de Paris Cit\'e and DESY--Hamburg for the kind hospitality, and to the Alexander von Humboldt foundation for the kind and generous support, while this work was in progress. Finally, we are both very grateful to Dr.~M.~Nardelli, who kindly helped us to retrieve some mathematical literature.

\newpage
\begin{appendices}
\section{\sc Conventions and Properties of the Background} \label{app:background}
In this Appendix we collect some useful properties of the background described in Section~\ref{sec:intro}.
Our main conventions are the following. Capital Latin labels like $M$ denote curved ten--dimensional indices, and Greek or Latin labels like $(\mu,r,i)$ denote their spacetime or internal portions. 
Moreover, when we need to distinguish the curved radial index $r$ from the remaining nine--dimensional ones, we denote them collectively by $m$, while primed labels like $a'$ denote the flat internal indices $(r,a)$ and others like $i'$ will denote the corresponding curved ones. 
We use a ``mostly--plus'' signature, defining the Riemann curvature tensor via~\footnote{These conventions are as in~\cite{ms21_1} and \cite{ms21_2}.}
\beq
[ \nabla_M, \nabla_N] V_P \ = \ {R_{MNP}}^{Q}\, V_Q \ ,
\eeq
so that
\beq
{R_{MNP}}^{Q} \ = \ \partial_N\,{\Gamma^Q}_{MP} \ - \ \partial_M\,{\Gamma^Q}_{NP}  \ + \ {\Gamma^Q}_{NR}\, {\Gamma^R}_{MP} \ - \ {\Gamma^Q}_{MR}\, {\Gamma^R}_{NP} \ .
\eeq
We also define the Ricci tensor as
\beq
R_{MP} \ = \ {R_{MNP}}^{N} \ .
\eeq

For backgrounds of the type
\beq
ds^2 \ = \ e^{\,2\,A(r)}\, dx^2 \ + \ e^{\,2\,B(r)}\, dr^2 \ + \ e^{\,2\,C(r)}\, dy^2 \ , \label{metricABCapp}
\eeq
where the $x^\mu$--coordinates, with $\mu=0,\ldots,3$, refer to the four--dimensional spacetime, while the $y^i$--coordinates, with $i=1,\ldots,5$ refer to the internal torus,
the Christoffel symbols are
\bea
&& {\Gamma^\mu}_{\nu r} \ = \ A' \, {\delta^\mu}_\nu  \ , \quad
{\Gamma^i}_{j r} \ = \ C' \, {\delta^i}_j \ , \nonumber \\
&& {\Gamma^r}_{r r} \ = \ B'\ , \quad {\Gamma^r}_{\mu\nu} \ = \ - \ \eta_{\mu\nu}\, A'\, e^{2(A-B)}\ , \quad {\Gamma^r}_{ij} \ = \ - \ \delta_{ij}\, C'\, e^{2(C-B)} \ .\label{gamma0}
\eea

The components of the Ricci tensor read
\bea
R^{(0)}{}_{\mu\nu} &=& - \ \eta_{\mu\nu}e^{2A}\left[3\left(A'\right)^2 \,e^{-2B} \ +\  5\,A'\,C'e^{-2B}\ +\ \left(A'e^{A-B}\right)'e^{-A-B} \right] \nonumber \\
&=& - \ \eta_{\mu\nu}\,e^{2(A-B)} \left[A'\left(4 A'+5 C'- B'\right) \ + \ A'' \right]\ , \nonumber \\
R^{(0)}{}_{rr} &=& - \left[ 4 \left(A'e^{A-B}\right)'e^{B-A}\ + \ 5\left(C'e^{C-B}\right)'e^{B-C}\right] \ \nonumber \\
&=& -\left[4 A'(A'-B')+5 C'(C'-B') + 4 A'' + 5 C''\right] \nonumber \ , \\
R^{(0)}{}_{ij} &=&  -\ \delta_{ij}e^{2C}\left[4\left(C'\right)^2 \,e^{-2B}\ + \ 4\,A'\,C'e^{-2B} \ +\ \left(C'e^{C-B}\right)'e^{-B-C} \right] \nonumber \\
&=& -\ \delta_{ij}e^{2(C-B)}\left[C'\left(4 A'+5 C'- B'\right) \ + \ C'' \right]\ ,
\eea
and in the ``harmonic'' gauge, where
\beq
B\ =\ 4\,A \ +\ 5\,C \ , \label{harmonicapp}
\eeq
they reduce to
\bea
R^{(0)}{}_{\mu\nu} &=& - \eta_{\mu\nu}\,e^{2(A-B)}\,A'' \ , \nonumber \\
R^{(0)}{}_{rr} &=& - \left[ 4 \left(A'e^{A-B}\right)'e^{B-A}\ + \ 5\left(C'e^{C-B}\right)'e^{B-C}\right] \ \nonumber \\
&=& \left[ 4 A'(3A'+5 C') + 20 C'(A'+C') - 4 A'' - 5 C''\right]  \ , \nonumber \\
R^{(0)}{}_{ij} &=&  -\ \delta_{ij}\,e^{2(C-B)}\, C''\ .
\eea
Consequently the scalar curvature takes the form
\bea{}
R^{(0)}&=& e^{-2B} \left[ - 8 A'' - 10 C'' + 4 A'(3A'+5 C') + 20 C'(A'+C')\right] \ ,
\eea
and the components of the Einstein tensor read
\bea{}{}{}{}{}{}{}{}{}{}{}{}{}{}{}{}{}{}
G^{(0)}{}_{\mu\nu} &=& \eta_{\mu\nu}\,e^{2(A-B)}\left[3\,A'' \,+\, 5\,C''\,-\,2\left(3\,(A')^2 \ + \ 10 A'C' \ + \ 5 (C')^2\right)  \right]\ , \nonumber \\
G^{(0)}{}_{rr} &=& 2 \left[ 3\,(A')^2 \ + \ 10 A'C' \ + \ 5 (C')^2 \right] \ , \nonumber \\
G^{(0)}{}_{ij} &=& \delta_{ij}\,e^{2(C-B)}\left[4\left(A''+C''\right) \,-\,2\left(3\,(A')^2 \ + \ 10 A'C' \ + \ 5 (C')^2\right) \right] \ . \label{eqs_delta}
\eea

For the vacua of interest the Einstein equations
\beq
{G^{(0)}}_{MN}  =  \frac{1}{4!}\ \left({\cal H}_5^{(0)}{}^2\right){}_{M N} = \frac{1}{24} \ g^{(0) PP'}\,g^{(0) QQ'}\,g^{(0) RR'}\,g^{(0) SS'}\ {\cal H}_{5\,M P Q R S}^{(0)}\ {\cal H}_{5\,N P' Q' R' S'}^{(0)}
\eeq
reduce to
\bea
{G^{(0)}}_{\mu\nu}  &=& {R^{(0)}}_{\mu\nu} \ = \  - \ \frac{h^2}{4\,\rho^2}\  e^{2A-10C}\, \eta_{\mu\nu} \ , \quad
{G^{(0)}}_{r r}  \ = \ {R^{(0)}}_{r r} \ = \  - \  \frac{h^2}{4\,\rho^2}\  e^{2B-10C} \ ,  \nonumber \\
{G^{(0)}}_{i j}  &=&  {R^{(0)}}_{i j} \ = \  \frac{h^2}{4\,\rho^2}\  e^{-8C}\, \delta_{ij} \ . \label{einst_conf}
\eea
In particular, the $rr$ equation is the ``Hamiltonian constraint''
\beq{}{}{}{}{}{}{}{}{}{}{}{}{}{}{}{}{}{}{}{}{}{}{}{}{}{}{}{}{}{}{}{}{}{}{}{}{}{}{}{}{}{}{}{}{}{}{}{}{}{}{}{}{}{}{}{}{}{}{}{}{}{}{}{}{}{}{}{}{}{}{}{}{}{}{}{}
3\left(A'\right)^2 \ + \ 10\, A'\, C' \ + \ 5 \left(C'\right)^2 \ = \ - \ \frac{h^2}{8\,\rho^2} \ e^{8 A} \ = \ - \ \frac{H^2}{2}\ e^{8A} \ , \label{ham_app}
\eeq
and making use of it turns the Einstein tensor into
\bea{}{}{}{}{}{}{}{}{}{}{}{}{}{}{}{}{}{}
\sqrt{-g}\ G^{(0)}{}_{\mu\nu} &=& g_{\mu\nu} \left[3\,A'' \ +\  5\,C''\ + \ H^2\, e^{8A}  \right]\ , \qquad
\sqrt{-g}\ G^{(0)}{}_{rr} \ = \ - \ g_{rr} \ H^2\, e^{8A} \ , \nonumber \\
\sqrt{-g}\ G^{(0)}{}_{ij} &=& g_{ij} \left[4\left(A''+C''\right) \ + \ H^2\, e^{8A}  \right] \ . \label{eqstot}
\eea
Note that the Einstein equations~\eqref{einst_conf} imply the useful relations
\beq
A'' \ = \ - \ C'' \ = \ \frac{h^2}{4\,\rho^2}\ e^{8 A} \ . \label{A2C2}
\eeq
Eqs.~\eqref{back_epos_fin2} solve the background equations, up to the contact terms localized on the boundaries discussed in~\cite{ms_23_3}.

There are some useful identities for the tensor background of eqs.~\eqref{back_epos_fin2}. The simplest ones are
\bea
\left({\cal H}_5^{(0)}\right)_{\mu\nu}^2 &=& -\  6\,\frac{h^2}{\rho^2} \ e^{2 A - 10 C} \ \eta_{\mu\nu} \ , \quad
\left({\cal H}_5^{(0)}\right)_{rr}^2 \,=\, -\ 6\,\frac{h^2}{\rho^2} \ e^{8 A} \ , \nonumber \\
\left({\cal H}_5^{(0)}\right)_{ij}^2 &=& 6\,\frac{h^2}{\rho^2} \ e^{ - 8 C} \ \delta_{ij} \ ,
\eea
but for studying perturbations one also needs to compute the components of
\beq
\left({\cal H}_5^{(0)}\right)_{MN,PQ}^2 \ = \  {g^{(0)}}^{R_1 S_1}\, {g^{(0)}}^{R_2 S_2}\, {g^{(0)}}^{R_3 S_3}\,{{\cal H}_5^{(0)}}_{MN R_1 R_2 R_3}\,  {{\cal H}_5^{(0)}}_{PQ S_1 S_2 S_3}\
\eeq
which are determined by symmetry and by the comparison with the preceding expressions, and read
\bea
\left({\cal H}_5^{(0)}\right)_{\mu\rho,\nu\sigma}^2 &=& - \ \frac{3}{2}\, \frac{h^2}{\rho^2} \ e^{4 A - 10 C} \left( \eta_{\mu\nu}\,\eta_{\rho\sigma} \ - \ \eta_{\mu\sigma}\,\eta_{\nu\rho} \right) \ , \nonumber \\
\left({\cal H}_5^{(0)}\right)_{\mu r,\nu r}^2 &=& - \ \frac{3}{2} \, \frac{h^2}{\rho^2} \ e^{10 A} \ \eta_{\mu\nu} \ , \qquad
\left({\cal H}_5^{(0)}\right)_{\mu i,\nu j}^2 \ = \  0  \ ,  \\
\left({\cal H}_5^{(0)}\right)_{ik,jl}^2 &=& \frac{3}{2}\, \frac{h^2}{\rho^2} \ e^{ - 6 C} \left( \delta_{ij}\,\delta_{kl} \ - \ \delta_{il}\,\delta_{jk} \right) \ , \qquad
\left({\cal H}_5^{(0)}\right)_{i r,j r}^2 \ = \  0 \ . \nonumber
\eea

In terms of the $z$ variable of eq.~\eqref{dzdr}, the equations for the background become
\bea
&& 3\left(A_z\right)^2 \ + \ 10\, A_z\, C_z \ + \ 5 \left(C_z\right)^2 \ = \ - \ 2\ {\cal W}_5^2 \ , \nonumber \\
&& A_{zz} \ = \ 4\,{\cal W}_5^2 \ -  \ \left(3 A_z + 5 C_z\right )A_z \ , \nonumber \\
&& C_{zz} \ = \ - \ 4\,{\cal W}_5^2 \ - \ \left(3 A_z + 5 C_z\right)C_z\ , \label{hamiltonian_F}
\eea
where we have introduced the convenient combinations
\beq{}{}
{\cal W}_5 \ = \ \frac{h}{4\,\rho}\ e^{A-5C} \ , \qquad {\cal K} \ = \ \left|\mathbf{k}\right| \,e^{A-C} \ , \label{W5K}
\eeq
which are used repeatedly in the main body of the paper.
Recalling also that
    \beq
h \ = \  2\,H\,\rho \ , \qquad z_0 \ = \  \left({2\,H\,\rho^3 }\right)^\frac{1}{2} \ = \ \rho\, h^\frac{1}{2} \ , \label{z0h_app}
\eeq
the limiting behavior of several useful quantities close to $z=0$ is
\bea
\frac{z}{z_0} &=& \frac{2}{3}\left(\frac{r}{\rho}\right)^\frac{3}{2} \ - \ \frac{1}{\sqrt{10}} \left(\frac{r}{\rho}\right)^\frac{5}{2} \ + \ \frac{19}{168} \left(\frac{r}{\rho}\right)^\frac{7}{2} + {\cal O}\left[\left(\frac{r}{\rho}\right)^\frac{9}{2}\right] \ , \nonumber \\
\frac{r}{\rho} &=& \left(\frac{3 \,z}{2\,z_0}\right)^\frac{2}{3} \left\{1\ + \ \frac{1}{\sqrt{10}}\left(\frac{3 \,z}{2\,z_0}\right)^\frac{2}{3} \ +\ \frac{47}{420} \left(\frac{3 \,z}{2\,z_0}\right)^\frac{4}{3}\ + \ {\cal O}\left[\left(\frac{z}{z_0}\right)^{2}\right]\right\} \ .
\eea
Consequently
\bea
e^A &=& \frac{1}{h^\frac{1}{4}}\left(\frac{2 \,z_0}{3 \,z}\right)^\frac{1}{6} \left\{1 \ - \ \frac{1}{4 \sqrt{10}}\left(\frac{3 \,z}{2\,z_0}\right)^\frac{2}{3} \ - \ \frac{121}{2240}\left(\frac{3 \,z}{2\,z_0}\right)^\frac{4}{3} \ + \ {\cal O}\left[\left(\frac{z}{z_0}\right)^\frac{7}{3}\right]\right\}\, \nonumber \\
e^C &=& h^\frac{1}{4}\left(\frac{3 \,z}{2 \,z_0}\right)^\frac{1}{6}
\left\{1 \ - \ \frac{1}{4 \sqrt{10}}\left(\frac{3 \,z}{2 \,z_0}\right)^\frac{2}{3}\ +\ \frac{23}{2240}\left(\frac{3 \,z}{2 \,z_0}\right)^\frac{4}{3} \ + \ {\cal O}\left[\left(\frac{z}{z_0}\right)^\frac{7}{3}\right]\right\} \,, \nonumber \\
A_z &=& - \ \frac{1}{6\,z} \ - \ \frac{1}{6\sqrt{10}}\left(\frac{3}{2\,z_0}\right)^\frac{2}{3}\ z^{-\frac{1}{3}} \ - \ \frac{8}{105}\left(\frac{3}{2\,z_0}\right)^\frac{4}{3}\ z^{\frac{1}{3}} \ + \ {\cal O}\left[\left(\frac{z}{z_0}\right)\right] \ , \nonumber \\
C_z &=&  \frac{1}{6\,z} \ - \ \frac{1}{6\sqrt{10}}\left(\frac{3}{2\,z_0}\right)^\frac{2}{3}\ z^{-\frac{1}{3}} \ + \ \frac{1}{105}\left(\frac{3}{2\,z_0}\right)^\frac{4}{3}\ z^{\frac{1}{3}} \ + \ {\cal O}\left[\left(\frac{z}{z_0}\right)\right] \ , \nonumber \\
{\cal W}_5^2 &=& \frac{1}{36\,z^2} \ + \ \frac{1}{18\,\sqrt{10}} \left(\frac{3}{2\,z_0}\right)^\frac{2}{3} z^{-\frac{4}{3}} \ + \ \frac{1}{2520}\left(\frac{3}{2\,z_0}\right)^\frac{4}{3} z^{-\frac{2}{3}} \ + \ {\cal O}\left[\left(\frac{z}{z_0}\right)^{0}\right]\, .
    \eea
The leading behavior of the metric, ${\cal K}^2$ and the five--form backgrounds is
\bea
&& ds^2 \ \sim \ \frac{dx^2 \ + \ dz^2}{\left(3\,\left|H\right|z\right)^\frac{1}{3}} \ + \ \left(3\,\left|H\right|z\right)^\frac{1}{3}\ d\,\vec{y}^{\,2} \ , \qquad  {\cal K}^2 \ \sim \  \frac{\left|\mathbf{k}\right|^2}{\left(3\,\left|H\right|\,z\right)^\frac{2}{3}} \ , \nonumber \\
&& {\cal H}_5 \ \sim \ H \left\{ \frac{dx^0 \wedge ...\wedge dx^3\wedge dz}{\left[3\left|H\right|z\right]^\frac{5}{3}} \ + \ dy^1 \wedge ... \wedge dy^5\right\} \ .
\eea

The limiting behavior for large $r$, and thus for $z$ close to the finite value $z_m$ of eq.~\eqref{zm_app} corresponding to the right end of the interval, for the quantities entering the non--supersymmetric backgrounds of eqs.~\eqref{back_epos_fin2}, are
{\small{
\bea{}{}
\!\!\!\!\!\!\!\!\!&& \frac{z_m - z}{z_0} \,  \sim \,
\frac{\sqrt{2}}{3}\left(\sqrt{10}\,+\,2\right)\  e^{\,-\,\frac{r}{4\rho}\left(\sqrt{10}\,-\,2\right)} \left[1 \ - \ \frac{4\sqrt{10}\,-\,11}{26} \ e^{\,-\,\frac{2\,r}{\rho}} \right] \ , \nonumber \\
\!\!\!\!\!\!\!\!\!&& e^{\,-\,\frac{r}{2\,\rho}} \,\sim\, \left[\frac{\sqrt{5}-\sqrt{2}}{2} \left(\frac{z_m-z}{z_0}\right)\right]^{\frac{\sqrt{10}\,+\,2}{3}}   \!\!\left\{1 + \frac{6-\sqrt{10}}{26}  \left[\frac{\sqrt{5}\,-\,\sqrt{2}}{2}  \left(\frac{z_m-z}{z_0}\right)\right]^{\frac{4\left(\sqrt{10}\,+\,2\right)}{3}} \!\! \right\}\,, \nonumber \\
\!\!\!\!\!\!\!\!\!&& e^{2A} \, \sim \, \sqrt{\frac{2}{h}} \left[\frac{\sqrt{5}-\sqrt{2}}{2} \left(\frac{z_m-z}{z_0}\right)\right]^{\frac{\sqrt{10}\,+\,2}{3}} \!\!\left\{1 + \frac{19-\sqrt{10}}{26}  \left[\frac{\sqrt{5}\,-\,\sqrt{2}}{2}  \left(\frac{z_m-z}{z_0}\right)\right]^{\frac{4\left(\sqrt{10}\,+\,2\right)}{3}} \!\! \right\}\,, \nonumber \\
\!\!\!\!\!\!\!\!\!&& e^{2C} \ \sim \ \sqrt{\frac{h}{2}}\left[\frac{\sqrt{5}-\sqrt{2}}{2} \left(\frac{z_m-z}{z_0}\right)\right]^{\,-\,\frac{\sqrt{10}}{5}} \!\!\left\{1 {+ \frac{-105+11 \sqrt{10}}{130} } \left[\frac{\sqrt{5}\,-\,\sqrt{2}}{2}  \left(\frac{z_m-z}{z_0}\right)\right]^{\frac{4\left(\sqrt{10}\,+\,2\right)}{3}} \!\! \right\} \ , \nonumber \\
\!\!\!\!\!\!\!\!\!&& A_z  \ \sim \ {-} \ \frac{1}{6} \ \frac{\sqrt{10}\,+\,2}{z_m - z} \ , \qquad\qquad
C_z \, \sim  \ \frac{1}{\sqrt{10}} \ \frac{1}{z_m - z} \ , \label{large_rn} \\
\!\!\!\!\!\!\!\!\!&& {\cal W}_5 \, \sim \,  \frac{\sqrt{2}}{2\,z_0}   \left[\frac{\sqrt{5}\,-\,\sqrt{2}}{2} \left(\frac{z_m-z}{z_0}\right)\right]^{\frac{2 \sqrt{10}\,+\,1}{3}} \!\!\!\!\!\!\!, \quad  {\cal K}^2 \, \sim \, \frac{2 \left(\left|\mathbf{k}\right|\, \rho\right)^2}{z_0^2}  \left[\frac{\sqrt{5}\,-\,\sqrt{2}}{2} \left(\frac{z_m-z}{z_0}\right)\right]^{\frac{2\sqrt{10}\,+\,16}{15}} \!\!\!\!\!\!\! .
\nonumber
\eea
}}

\section{\sc Intermediate Results for Tensor Perturbations} \label{app:fiveform}

In this Appendix we collect some intermediate results that are needed to obtain the tensor equations of Section~\ref{sec:perturbed_tensor}. As explained there, we parametrize the independent components of the tensor gauge field perturbations $\delta B_{M N P Q}$ according to eq.~\eqref{tensor_linear_H2}. The corresponding field strengths then take the form
\bea
\delta\,\,H_{\mu\nu\rho\sigma r} &=& \epsilon_{\mu\nu\rho\sigma}\,\left(\partial_r\,b \ - \ \partial_\tau\,b^\tau \right) \ , \qquad
{\delta\, H}_{\mu\nu\rho\sigma i}  \ = \  \epsilon_{\mu\nu\rho\sigma}\,\left(\partial_i\,b \ - \ \partial_\tau\,{b^\tau}_i \right) \ , \nonumber \\
{\delta\, H}_{\mu\nu\rho r i}  &=& \epsilon_{\mu\nu\rho\sigma}\left( \partial_i\,{b^\sigma} \ - \ \partial_r\,{b^{\sigma}}_i \ - \ \frac{1}{2}\, \epsilon^{\alpha\beta\gamma\sigma}\,\partial_\alpha\,b_{\beta\gamma i}\right) \ , \nonumber \\
{\delta\, H}_{\mu\nu\rho i j}  &=&  - \ \epsilon_{\mu\nu\rho\sigma}\left( \partial_{[i}\,{b^\sigma}_{j]} \ + \ \frac{1}{2}\, \epsilon^{\alpha\beta\gamma\sigma}\,\partial_\alpha\,b_{\beta\gamma i j} \right)\ , \nonumber \\
{\delta\, H}_{\mu\nu r i j}  &=& \partial_{[\mu}\,b^{(1)}_{\nu] i j} \ + \ \partial_r\, b_{\mu \nu i j }
\ - \ \partial_{[i}\,b_{\mu \nu | j]} \ , \label{tensorfieldstrenghts_app} \\
{\delta\, H}_{\mu \nu i j k}  &=& \frac{1}{2}\, \epsilon_{i j k l m}\left(\partial_{[\mu}\,b^{(2)}{{}_{\nu]}}^{l m} \ + \ \frac{1}{2}\,\epsilon^{p q r l m}\,\partial_p\,b_{\mu \nu q r} \right) \ , \nonumber \\
{\delta\, H}_{\mu r i j k}  &=& \frac{1}{2}\, \epsilon_{i j k l m}\left(\partial_\mu\,b^{l m} \ - \ \partial_r\,{b^{(2)}}{{}_\mu}^{l m}\ + \ \frac{1}{2}\,\epsilon^{p q s l m}\,\partial_p\, {b^{(1)}}_{\mu q s}\right) \ , \nonumber \\
{\delta\, H}_{\mu i j k l}  &=& \epsilon_{i j k l m}\left( \partial_\mu\,b^m \ - \ \partial_n \, b^{(2)}{{}_{\mu}}^{m n}\right) \ , \qquad \delta\,\,H_{r i j k l} \ = \  \ \epsilon_{i j k l m} \left( \partial_r\,b^m \ - \ \partial_n\,b^{m n} \right) \ , \nonumber \\
{\delta\, H}_{i j k l m}  &=&  \epsilon_{i j k l m}\, \partial_p\,b^p \ . \nonumber
\eea
Starting from the self--duality conditions
\beq
H_{M_1\cdots M_5} \ = \ \frac{\sqrt{-g}}{5!}\,\epsilon_{M_1\cdots M_{10}}\ g^{M_6 N_6}\cdots g^{M_{10} N_{10}}\, H_{N_6 \cdots N_{10}} \ ,
\eeq
and expanding them to first order in the tensor and gravity perturbations gives
\bea
\delta\,H_{M_1\cdots M_5}
&=& \frac{\sqrt{-g^{(0)}}}{5!}\,\epsilon_{M_1\cdots M_{10}}\ g^{(0) M_6 N_6}\cdots g^{(0) M_{10} N_{10}}\, \delta\,H_{N_6 \cdots N_{10}}  \,+\, \frac{1}{2}\,{h^M}_M\,H^{(0)}_{M_1\cdots M_5} \nonumber \\
&-& \frac{\sqrt{-g^{(0)}}}{4!}\,{\epsilon_{M_1\cdots M_6}}^{i_7 \cdots i_{10}}\ h^{M_6 i_6}\,  e^{-8C}\, H^{(0)}_{i_6 i_7 \cdots i_{10}}  \\
&-& \frac{\sqrt{-g^{(0)}}}{4!}\,{\epsilon_{M_1\cdots M_6}}^{ \mu_7 \cdots \mu_{10}}\ h^{M_6 r}\,  e^{-8A}\, H^{(0)}_{r \mu_7 \cdots \mu_{10}} \nonumber \\
&-& \frac{\sqrt{-g^{(0)}}}{3!}\,{\epsilon_{M_1\cdots M_6}}^{\mu_7 \cdots \mu_9 r}\ h^{M_6 \mu_6}\,  e^{-6A-2B}\, H^{(0)}_{\mu_6 \mu_7 \cdots \mu_9 r}  \ , \nonumber
\eea
where the $\epsilon$ tensors are flat, so that their indices are raised and lowered with the flat metric, $\epsilon_{0 \ldots 9}=1$, and the indices of the metric perturbations $h_{MN}$ are raised with the background metric.

One can now specialize the left--hand side to the independent cases, which leads to the five groups of independent tensor equations
\bea
\delta\,H_{\mu_1\cdots \mu_4 r} &=& \frac{e^{2B-10C}}{5!}\,\epsilon_{\mu_1\cdots \mu_4}\, \epsilon^{i_6 \cdots i_{10}}\ \delta\,H_{i_6 \cdots i_{10}} \nonumber \\ &+& \frac{h\,e^{8A}}{4\,\rho}\left( e^{-2A}\,{h^\mu}_\mu \,+\, e^{-2B}\, h_{rr}\,-\, e^{-2C}\, {h^k}_k \right)\epsilon_{\mu_1 \cdots \mu_4} \ , \nonumber \\
\delta\,H_{\mu_1\mu_2 \mu_3 r i} &=&  \frac{e^{2B-2A-8C}}{4!}\, {\epsilon_{\mu_1\cdots \mu_3}}^{\mu_4}\, {\epsilon_i}^{j_7 \cdots j_{10}}\, \delta\,H_{\mu_4 j_7 \cdots j_{10}} \nonumber \\
&-& \frac{h}{2\,\rho}\,e^{2B-2A-10C}\,{\epsilon_{\mu_1\cdots \mu_3}}^{\mu_4}\,h_{\mu_4 i} \ , \label{deltaH5} \\
\delta\,H_{\mu_1\mu_2 r i j} &=& \frac{e^{4(A+C)}}{2!\ 3!}\,{\epsilon_{\mu_1\mu_2}}^{\mu_3\mu_4}{\epsilon_{i j}}^{k l m}\ \delta\,H_{\mu_3\mu_4 k l m}  \ , \nonumber \\
\delta\,H_{\mu_1 r i j k}  &=&  \frac{e^{2A+6C}}{2!\ 3!}\,{\epsilon_{\mu_1}}^{\mu_2\mu_3 \mu_4}\,{\epsilon_{ijk}}^{lm}\, \delta\,H_{\mu_2\mu_3 \mu_4 l m} \ , \nonumber \\
\delta\,H_{r i j k l}  &=& \frac{e^{8C}}{4!}\,\epsilon^{\mu_1 \cdots \mu_4}\, {\epsilon_{i j k l}}^m\ \delta\,H_{\mu_1 \cdots \mu_4 m}  \ + \  \frac{h}{2\,\rho}\,{\epsilon_{i j k l}}^{m}\, h_{m r}\,  e^{-2C}\ ,  \nonumber
\eea
where now all indices are raised and lowered with the flat metric. Making use of eqs.~\eqref{tensorfieldstrenghts_app} leads finally to eqs.~\eqref{eqs_sd}.

With the gauge choice
\beq{}{}{}
B_{rMNP} \ = \ 0 \ ,
\eeq
which translates into the conditions
\beq{}{}{}
b_\mu \ = 0 \ , \qquad b_{\mu\nu i} = 0 \ , \qquad b_{\mu i j}^{(1)} = 0 \ , \qquad b_{ij} = 0 \ ,
\eeq
the field strengths finally reduce to
\bea
\delta\,{\cal H}_{\mu\nu\rho\sigma r} &=& \epsilon_{\mu\nu\rho\sigma}\,\partial_r\,b  \ , \qquad
{\delta\, {\cal H}}_{\mu\nu\rho\sigma i}  \ = \  \epsilon_{\mu\nu\rho\sigma}\,\left(\partial_i\,b \ - \ \partial_\tau\,{b^\tau}_i \right) \ , \nonumber \\
{\delta\, {\cal H}}_{\mu\nu\rho r i}  &=& - \ \epsilon_{\mu\nu\rho\sigma}\, \partial_r\,{b^{\sigma}}_i  \ , \qquad
{\delta\, {\cal H}}_{\mu\nu\rho i j}  \ = \  - \ \epsilon_{\mu\nu\rho\sigma}\left( \partial_{[i}\,{b^\sigma}_{j]} \ + \ \frac{1}{2}\, \epsilon^{\alpha\beta\gamma\sigma}\,\partial_\alpha\,b_{\beta\gamma i j} \right)\ , \nonumber \\
{\delta\, {\cal H}}_{\mu\nu r i j}  &=& \partial_r\, b_{\mu \nu i j }
 \ , \qquad
{\delta\, {\cal H}}_{\mu \nu i j k}  \ = \  \frac{1}{2}\, \epsilon_{i j k l m}\left(\partial_{[\mu}\,b^{(2)}{{}_{\nu]}}^{l m} \ + \ \frac{1}{2}\,\epsilon^{p q r l m}\,\partial_p\,b_{\mu \nu q r} \right) \ , \label{tensor_h_gaugefixed} \\
{\delta\, {\cal H}}_{\mu r i j k}  &=& - \ \frac{1}{2}\, \epsilon_{i j k l m}\, \partial_r\,{b^{(2)}}{{}_\mu}^{l m} \ , \qquad
{\delta\, {\cal H}}_{\mu i j k l}  \ = \  \epsilon_{i j k l m}\left( \partial_\mu\,b^m \ - \ \partial_n \, b^{(2)}{{}_{\mu}}^{m n}\right) \ , \nonumber \\
\delta\,{\cal H}_{r i j k l} &=& \ \epsilon_{i j k l m} \,\partial_r\,b^m \ , \qquad
{\delta\, {\cal H}}_{i j k l m}  \ = \  \epsilon_{i j k l m}\, \partial_p\,b^p \ . \nonumber
\eea

\section{\sc Intermediate Results for the Einstein Equations}\label{app:Einstein}
In this Appendix we collect some technical results needed to obtain the perturbed Einstein equations discussed in Section~\ref{sec:perturbed_tensor} starting from
\beq
R_{MN} \ = \ \frac{1}{24} \, \left({\cal H}_5{}^2\right)_{M N} \ .
\eeq

\subsection{\sc Intermediate Results for the Ricci Curvature}

The perturbed Christoffel symbols read
\beq
{\delta\,\Gamma^P}_{MN} \ = \ \frac{1}{2}\left(\nabla_M\,{h^P}_N \ + \  \nabla_N\,{h^P}_M \ - \ \nabla^P\,h_{MN} \right) \ , \label{deltagamma}
\eeq
where the gradient $\nabla$ refers to the background,
and from the Palatini identity
\beq
\delta\,{R_{MNP}}^Q \ = \ \nabla_N\, {\delta\,\Gamma^Q}_{MP} \ - \ \nabla_M\, {\delta\,\Gamma^Q}_{NP}
\eeq
one can deduce the perturbed Ricci curvature
\beq
- 2\,\delta\,R_{N R}\ = \  \Box_{10}\,h_{N R} \ +\ \nabla_N\,\nabla_R\,{h_S}^S \ - \ \nabla^P\left(\nabla_N\,h_{P R}\,+\,\nabla_R\,h_{P N} \right) \ . \label{pert_ricci}
\eeq

Using the results collected in Appendix~\ref{app:background}, one can compute the divergences of gradients that are needed in eq.~\eqref{pert_ricci}. Their general expression is
\beq
\nabla^P\,\nabla_N \,h_{P R} \ = \ g^{PQ} \left(  \partial_Q\,\nabla_N \,h_{P R} \,-\, {{\Gamma}^S}_{QN}\,\nabla_S\, h_{PR}\,-\,  {{\Gamma}^S}_{QP}\,\nabla_N\,h_{SR}\,-\, {{\Gamma}^S}_{QR}\,\nabla_N\,h_{PS}\right) \ ,
\eeq
where the metric $g$ and the connection $\Gamma$ refer to the background.
Some simplifications occur since the background satisfies the harmonic gauge conditions~\eqref{harmonicapp}, which are equivalent to
\beq
g^{PQ}\,{\Gamma^S}_{PQ} \ = \ 0 \ , \label{gauge_choice_h}
\eeq
and one is left with
\beq
\nabla^P\,\nabla_N \,h_{P R} \ = \ g^{PQ} \left(  \partial_Q\,\nabla_N \,h_{P R} \,-\, {{\Gamma}^S}_{QN}\,\nabla_S\, h_{PR}\,-\,  {{\Gamma}^S}_{QR}\,\nabla_N\,h_{PS}\right) \ .
\eeq

\subsection{\sc The Perturbed Energy--Momentum Tensor} \label{app:pert_emt}

The right--hand side of the perturbed Einstein equations involves the variation of $\left({\cal H}_5^2\right)_{MN}$ about the background values, which reads
\beq
\delta\left[\left({\cal H}_5^2\right)_{MN}\right] \ = \ \delta{\cal H}_{5 (M} \cdot {\cal H}_{5 N)}^{(0)} \ - \ 4 \, {\cal H}_{5 MK}^{(0)} \cdot {\cal H}_{5 NL}^{(0)} \ h^{KL} \ ,
\eeq
where $h_{KL}$ denotes, as before, the metric perturbation, and where indices are raised and lowered with the background metric. Taking into account the background in eqs.~\eqref{back_epos_fin2} and eqs.~\eqref{hH5}, one thus finds 
\bea
\delta{\cal H}_{5 (\mu} \cdot {\cal H}_{5 \nu)}^{(0)} &=& \frac{h}{2\,\rho}\, e^{2(A-B)}\,{{{\delta{\cal H}_5}^{\rho\sigma\tau}}_{r(\mu}}\,  \epsilon_{\nu) \rho\sigma\tau}\ , \nonumber \\
\delta{\cal H}_{5 (\mu} \cdot {\cal H}_{5 r)}^{(0)} &=& 0 \ , \nonumber \\
\delta{\cal H}_{5 (\mu} \cdot {\cal H}_{5 i)}^{(0)} &=& \frac{h}{\rho}\,e^{-8C}\,\epsilon_{i k_1\cdots k_4}\,{\delta{\cal H}_{5 \mu}}^{k_1 \cdots k_4} \ - \ 6 \, \frac{h^2}{\rho^2}\, e^{-10C}\,h_{\mu i} \ , \nonumber \\
\delta{\cal H}_{5 r} \cdot {\cal H}_{5 r}^{(0)} &=& \frac{h}{2\,\rho}\, {{{\delta{\cal H}_5}}^{\mu\rho\sigma\tau}}_r\,  \epsilon_{\mu \rho\sigma\tau} \ , \nonumber \\
\delta{\cal H}_{5 (r} \cdot {\cal H}_{5 i)}^{(0)} &=&
6\,\frac{h^2}{\rho^2}\,e^{-10C}\,h_{r i} \ + \ \frac{h}{\rho}\, \epsilon_{\beta_1\cdots \beta_4}\,{\delta{\cal H}_{5 i}}^{\beta_1 \cdots \beta_4} \ , \nonumber \\
\delta{\cal H}_{5 (i} \cdot {\cal H}_{5 j)}^{(0)} &=& \frac{h}{2\,\rho}\, e^{-8C}\,{{{\delta{\cal H}_5}^{klmn}}_{(i}}\,  \epsilon_{j) klmn} \ ,
\eea
where indices are now raised with the flat metrics $\eta^{\mu\nu}$ and $\delta^{ij}$. Finally, $\delta{\cal H}_5$ must be expressed in terms of the components of the four--form gauge field listed in eqs.~\eqref{tensor_linear_H2}, which lead to eqs.~\eqref{tensor_h_gaugefixed} after gauge fixing.

The final form of the perturbed energy--momentum tensor,
\beq
T^{(1)}{}_{MN} \ = \ \frac{1}{24}\, \left(\delta{\cal H}_{5 (M} \cdot {\cal H}_{5 N)}^{(0)} \ - \ 4 \, {\cal H}_{5 MK}^{(0)} \cdot {\cal H}_{5 NL}^{(0)} \ h^{KL} \right)
\eeq
is
\bea
T^{(1)}{}_{\mu\nu} &=&   -\,\frac{h}{4\,\rho}\,e^{2(A-B)}\,\partial_r\,b\ \eta_{\mu\nu} \, + \,\frac{h^2}{4\,\rho^2} \left[ e^{- 10 C} \left( \eta_{\mu\nu}\,{h_\rho}^\rho - h_{\mu\nu} \right)\,+\, e^{10A-4 B} \eta_{\mu\nu}\,h_{rr}\right] , \nonumber \\
T^{(1)}{}_{\mu r} &=&   - \ \frac{h^2}{4\,\rho^2}  \,e^{8A-2B}\, h_{\mu r} \ , \nonumber \\
T^{(1)}{}_{\mu i} &=&  e^{-8C} \left[ \frac{h}{\rho} \left( \partial_\mu\,b_i \ - \ \partial_n \, b^{(2)}{{}_{\mu i}}^{n}\right) \ - \ \frac{h^2}{4\,\rho^2 }\, e^{-2C}\,h_{\mu i} \right]\ , \nonumber \\
T^{(1)}{}_{rr} &=&   -\, \frac{h}{\rho}\,\partial_r\,b \,+\,  \frac{h^2}{4\,\rho^2}\, e^{6A}\, {h_\rho}^\rho  \nonumber\ , \\
T^{(1)}{}_{r i} &=&  \frac{h^2}{4\,\rho^2}\,e^{-10C}\,h_{r i} \ - \ \frac{h}{\rho}\left(\partial_i\,b\,-\,\partial_\mu\,{b^\mu}_i \right) \ , \nonumber \\
T^{(1)}{}_{ij} &=&   \frac{h}{\rho}\,\delta_{ij}\,e^{-8C}\,\partial_p\,b^p \,-\, \frac{h^2}{4\,\rho^2}\,e^{-10C}\left( \delta_{ij}\,{h_k}^k \ - \ h_{ij} \right)  \ . \label{g163m}
\eea
and the equations that we actually write are
\beq
- 2 \,\delta\,R_{MN} \ = \ - \ 2\, T^{(1)}{}_{MN} \ .
\eeq

\section{\sc The Non--Singlet Vector Modes as $\mathbf{k} \to 0$} \label{app:Mtildek0}

In this Appendix we discuss the possible self--adjoint boundary conditions for the non--singlet vector modes at the left end of the internal interval, focusing on the $\mathbf{k}$-independent part of the operator $\widetilde{\cal M}$ of eq.~\eqref{mtilde2}, or if you will on its small-$\mathbf{k}$ limit, which is diagonal. The potential in the $(1,1)$ entry,
\beq
V_1(r) \ = \ \frac{1}{320\,z_0^2 } \ \frac{e^{\sqrt{\frac{5}{2}}\ \frac{r}{\rho}}}{\sinh^3\left(\frac{r}{\rho}\right)}  \ \left[- \ 14 \sqrt{10}\ \sinh\left(\frac{2\,r}{\rho}\right) \ + \ 41 \cosh\left(\frac{2\,r}{\rho}\right)\ + \ 99\right] \label{potE1}
\eeq
only emerges for $\mathbf{k} \neq 0$, while the potential in the $(2,2)$ entry is the one already discussed in Section~\ref{sec:nonsingletk0vectors} for $\mathbf{k}=0$.
\begin{figure}[ht]
\centering
\includegraphics[width=65mm]{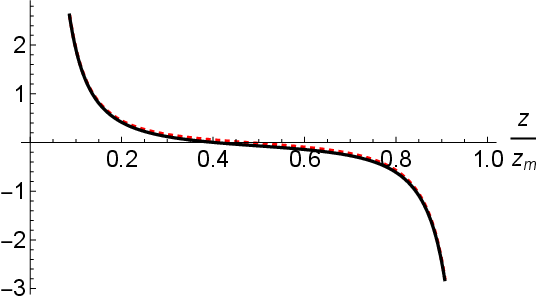}
\caption{\small The $\mathbf{k}=0$ portion of the Schr\"odinger potential for $W_1$ in eq.~\eqref{potE1} (black, solid) and its approximation~\eqref{pot_hyp} (red, dashed) with $(\mu,\tilde{\mu})=\left(\frac{2}{3},0.09\right)$, in units of $\frac{1}{{z_0}^2}$. $z_m$ and $z_0$ are defined in eqs.~\eqref{zm_app} and \eqref{z0}.}
\label{fig:pot_hmi_1_knot0}
\end{figure}

One can rely once more on the correspondence with the hypergeometric potentials od eq.~\eqref{pot_hyp} to identify stability regions in the small-$\mathbf{k}$ limit.
To this end, one needs to examine the limiting behavior of $\widetilde{M}$ as $z$ appraoches the two ends of the interval. From eq.~\eqref{alpha_beta_vec} and from Appendix~\ref{app:background}, one can see that, as $z\to 0$,
\beq
\widetilde{\cal M} \ \sim \ \left(-\,\partial_z^2 \ + \ \frac{7}{36\,z^2} \right) \ \mathbf{1} \, ,  \label{mtilde20}
\eeq
which is of the form~\eqref{HV0} with $\mu_1=\mu_2=\frac{2}{3}$. In a similar fashion, as $z \to z_m$, $\widetilde{\cal M}$ is of the form~\eqref{HVinf}, with $\tilde{\mu}_1=0.09$ and $\tilde{\mu}_2=1.1$. The resulting setting is the first case discussed in Section~\ref{sec:matrix_hyper}, and the self--adjoint boundary conditions given independently at the two ends thus depend on six real parameters. 

The dominant terms in eq.~\eqref{mtilde2} identify a two--dimensional vector space of normalizable zero modes spanned by
\beq
Z^{(1)} \ = \ \left(\begin{array}{c} e^\frac{A-C}{2} \\ 0 \end{array}\right) \ , \qquad Z^{(2)} \ = \ \left(\begin{array}{c} 0 \\ e^\frac{5A+3C}{2} \end{array} \right) \ . \label{E3}
\eeq
For the first zero mode
\beq
\frac{C_{11}}{C_{12}} \ = \ - \ 0.16 \ , \qquad \frac{C_{13}}{C_{14}} \ = \ 0.2 \ ,
\eeq
while for the second zero mode 
\beq
\frac{C_{21}}{C_{22}} \ = \ - \ 0.41 \ .
\eeq

\begin{figure}[ht]
\centering
\includegraphics[width=65mm]{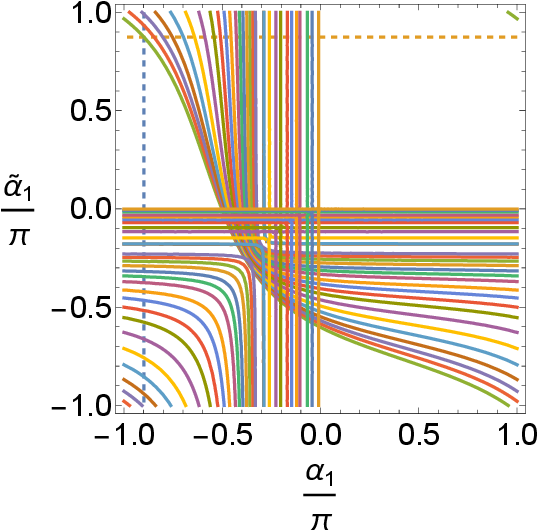}
\caption{\small The point $\left({\alpha_1},{\tilde{\alpha}_1}\right)=\pi\left(-0.9,0.87\right)$ identifies the special boundary conditions corresponding to the zero mode~$Z^{(1)}$. The shaded regions identify the boundary conditions leading to instabilities for the $\mathbf{k}$-independent portion of $\widetilde{\cal M}$. }
\label{fig:hmi_instabilities_knot0}
\end{figure}

As in the simpler one--component systems that we have analyzed, one can approach the $\mathbf{k}=0$ portion of $\widetilde{\cal M}$ by a pair of shifted hypergeometric potentials, with shifts
\beq
\Delta\,V_1 \ = \ \frac{\pi^2}{z_m^2}\, \left(0.1\right)^2 \ , \qquad \Delta\,V_2 \ = \ - \ \frac{\pi^2}{z_m^2}\, \left(0.71\right)^2\ ,
\eeq
relying once more on the one--component formalism of Section~\ref{sec:exactschrod}. The two potentials $V_1$ and $V_2$ are displayed in figs.~\ref{fig:pot_hmi_1_knot0} and~\ref{fig:hmi_pot}, together with their hypergeometric approximations. 
Letting, in the notation of eq.~\eqref{cot_definitions}
\beq
\frac{C_{11}}{C_{12}} \ = \ \cot\left( \frac{\alpha_1}{2}\right) \ , \qquad \frac{C_{13}}{C_{14}} \ = \ \cot\left( \frac{\tilde{\alpha}_1}{2}\right)  \ , \qquad  \frac{C_{22}}{C_{21}} \ = \  \tan\left( \frac{\alpha_2}{2}\right)\ .
\eeq
one can thus identify the stability regions in the $(\alpha_1,\tilde{\alpha}_1)$ plane displayed in fig.~\ref{fig:hmi_instabilities_knot0}, and the results of Section~\ref{sec:nonsingletk0vectors} translate, for $\alpha_2$, into the stability interval
\beq
- \ 2.42 \ < \ \tan\left( \frac{\alpha_2}{2}\right) \ < \ 0 \ .
\eeq

As we have seen, the matrix $\widetilde{\cal M}$ in eq.~\eqref{mtilde2} becomes diagonal as $\mathbf{k} \to 0$, and we could thus identify two zero modes. $Z^{(2)}$ is indeed the expected zero mode that we found in Section~\ref{sec:intvectexc}, while the presence of $Z^{(1)}$ is somewhat surprising, since it is also normalizable. The point is that the $\mathbf{k} \to 0$ limit is singular for the original theory, which can be seen from the need to perform the redefinition of eq.~\eqref{LAMBDA_vector_non_singlet}. Consequently, in this limit $Z^{(1)}$ would correspond to a field $W_1$ of vanishing norm with respect to $W_4$, while the original operator in eq.~\eqref{M_non_hermitian} is definitely not Hermitian for $\mathbf{k} =0$.

When the $\mathbf{k}$-dependent terms in $\widetilde{\cal M}$ are taken into account, their mean value computed with the vectors in eq.~\eqref{E3} diverges. As a result, the  dominant terms that we have analyzed are suitable to identify self--adjoint boundary conditions but are not a good starting point for perturbation theory. For this reason our variational tests relied on the $\mathbf{k}$-dependent corrections described in Section~\ref{sec:intvectexc}.

\section{\sc Self--Adjoint Variational Tests} \label{app:variationa2}

This Appendix provides some details on the variational estimates of the lowest $m^2$ eigenvalues described in Sections~\ref{sec:intvectexc} and \ref{sec:nonsingletscalrknot0}. 

 For our estimates of Section~\ref{sec:intvectexc}, we resorted on a pair of test functions $Z_1(z)$ and $Z_2(z)$ that comply to the limiting behavior at the two ends, taking into account the $\xi$--dependent corrections that we identified, which grant delicate compensations between singular contributions from kinetic and potential terms:
\bea{}{}{}{}{}{}{}
Z_1 \!\!\!&=&\!\!\!\!  \left\{C_{11} \left(\frac{z}{z_m}\right)^{\frac{7}{6}}  +  C_{12}\, \left(\frac{z}{z_m}\right)^{-\,\frac{1}{6}} +   \frac{\xi}{2}\left[C_{21} \left(\frac{z}{z_m}\right)^\frac{11}{6} - 3 \,C_{22} \left(\frac{z}{z_m}\right)^\frac{1}{2} \right]\right\} \exp\left[\,-\,\frac{a z^4}{z_m{}^4-z^4}\right]  \nonumber \\
&+&  \gamma_1 \left(\frac{z_m-z}{z_m}\right)^{\frac{1}{2}+\tilde{\mu}_1} \exp\left[\,-\,\frac{a \, \left(z_m\,-\,z\right)^4}{z_m{}^4\,-\,\left(z_m\,-\,z\right)^4}\right] \ , \nonumber \\
Z_2 \!\!\!&=&\!\!\!\!  \left\{C_{21}  \, \left(\frac{z}{z_m}\right)^{\frac{7}{6}} \, + \, C_{22} \left(\frac{z}{z_m}\right)^{-\,\frac{1}{6}} + \  \frac{\xi}{2}\left[C_{11} \, \left(\frac{z}{z_m}\right)^\frac{11}{6} - \ {3}\,C_{12} \left(\frac{z}{z_m}\right)^\frac{1}{2}\right]\right\} \exp\left[\,-\,\frac{a z^4}{z_m{}^4-z^4}\right] \nonumber \\
&+& \gamma_2 \left(\frac{z_m-z}{z_m}\right)^{\frac{1}{2}+\tilde{\mu}_2} \exp\left[\,-\,\frac{a \, \left(z_m\,-\,z\right)^4}{z_m{}^4\,-\,\left(z_m\,-\,z\right)^4}\right]
\label{Z_origin_vector}
\eea
The exponential factors separate the contributions from the two ends, which is instrumental to grant the needed cancellations.
The self--adjoint boundary conditions that we have explored are parametrized, in general, by an $SL(2,R)$ matrix and a phase $\beta$, so that
\bea
C_{22} &=& e^{i\beta}\left[\left(\cosh\rho\,\cos\theta_1\,+\,\sinh\rho\,\cos\theta_2\right)\,C_{11}\, + \, \left(- \cosh\rho\,\sin\theta_1\,+\,\sinh\rho\,\sin\theta_2\right) C_{12} \right] \ , \nonumber \\  C_{21} &=& e^{i\beta}\left[ \left( \cosh\rho\,\sin\theta_1\,+\,\sinh\rho\,\sin\theta_2\right)C_{11}\, + \, \left(\cosh\rho\,\cos\theta_1\,-\,\sinh\rho\,\cos\theta_2\right) C_{12}\right] \ .
\eea
The variational parameters for this case were thus $a$, $\gamma_1$, $\gamma_2$, $C_{11}$ and $C_{12}$. For simplicity, we have set $\beta=0$ in all our tests.

At the right end of the interval the Hamiltonian $\widetilde{\cal M}$ approaches a diagonal $Q^\dagger\,Q$ form, which is granted to be positive for the asymptotic behaviors compatible with the $L^2$ conditions, which are annihilated by $Q$, so that the contributions associated to $\gamma_{1,2}$ are not expected to lower significantly our estimates for the minimal $m^2$ eigenvalue. Therefore, we concentrated our efforts on test functions with $\gamma_{1}=0$ and $\gamma_{2}=0$. Furthermore, different choices for $a$ of order one gave similar results. In this fashion, $C_{11}$ and $C_{12}$ were our actual variational parameters, with resulting estimates of the form
\beq
m^2 \ = \ \frac{c^\dagger\, N \, c}{c^\dagger\, D \, c} \ ,
\eeq
where $c$ is a two--component vector collecting them and $N$ and $D$ two real symmetric matrices depending on $\xi$ and on the $SL(2,R)$ parameters, which we built numerically. In this setup, the best estimate is determined by the lowest eigenvalue of $D^{-1} N$.
\begin{figure}[ht]
\centering
\begin{tabular}{cc}
\includegraphics[width=65mm]{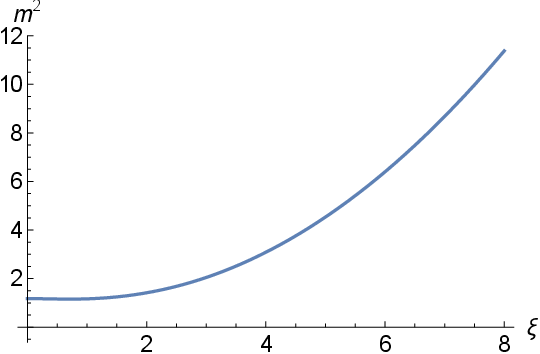} \qquad \qquad &
\includegraphics[width=65mm]{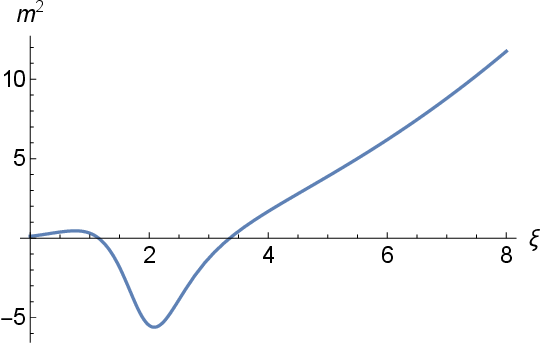} \\
\end{tabular}
\caption{\small Estimates for the lowest $m^2(\xi)$ (in units of $\frac{1}{z_0{}^2}$) for non--singlet vector modes obtained for $(\theta_1,\theta_2,\rho)=(\,-\,\frac{\pi}{2}, \frac{\pi}{2},2)$ (left panel), and for $(\theta_1,\theta_2,\rho)= (0,0,0)$ (right panel).}
\label{fig:vectornonsinglet}
\end{figure}

The main indications that we collected from our numerical tests are the following:
\begin{itemize}
    \item there are self--adjoint boundary conditions where no tachyons emerge for all values of $\mathbf{k}$. An example of this type has $(\rho,\theta_1,\theta_2,\beta)=(2,-\,\frac{\pi}{2}, \frac{\pi}{2},0)$, and a plot of the corresponding $m^2$ as a function of $\xi$ is shown in the left panel of fig.~\ref{fig:vectornonsinglet};

\item there are other boundary conditions where tachyonic modes are not present only for sufficiently small values of $R$, as explained in Section~\ref{sec:intvectexc}. An example of this type has $(\rho,\theta_1,\theta_2,\beta)= (0,0,0,0)$, and a plot of the corresponding $m^2$ as a function of $\xi$ is shown in the right panel of fig.~\ref{fig:vectornonsinglet}.
\end{itemize}

The variational tests for the scalar modes of Section~\ref{sec:nonsingletscalrknot0} proceeded along similar lines, but relied on the slightly more complicated functions
\bea
\widetilde{\Psi}{}_1 &=& \left\{\frac{C_{11}}{\sqrt{2}} \left(\frac{z}{z_m}\right)^\frac{7}{6} + \frac{C_{12}}{\sqrt{2}} \left(\frac{z}{z_m}\right)^{\,-\,\frac{1}{6}} \ + \ 2\,\xi\left[- \ \frac{3}{5}\ C_{21} \left(\frac{z}{z_m}\right)^\frac{3}{2} \ + \  C_{22} \left(\frac{z}{z_m}\right)^\frac{5}{6}\right]\right.\nonumber \\
&+& \left. \frac{3}{4 \sqrt{2}}\, \xi^2\left[ \frac{19}{40}\ C_{11} \left(\frac{z}{z_m}\right)^\frac{5}{2} \ - \ 3\,C_{12}\left(\frac{z}{z_m}\right)^\frac{7}{6}\, \log\left(\frac{z}{z_m}\right)\right] \right\} \exp\left[\,-\,\frac{a \, z^4}{z_m{}^4\,-z^4}\right] \nonumber \\
&+& \gamma_1 \left(1 \ - \ \frac{z}{z_m}\right)^{2.77}\exp\left[\,-\,\frac{a \, \left(z_m\,-\,z\right)^4}{z_m{}^4\,-\,\left(z_m\,-\,z\right)^4}\right]
\ , \nonumber \\
\widetilde{\Psi}{}_2 &=& \left\{C_{21} \left(\frac{z}{z_m}\right)^\frac{5}{6} + C_{22} \left(\frac{z}{z_m}\right)^\frac{1}{6} \ +\ \sqrt{2}\,\xi\left[-\, \frac{1}{5}\,C_{11}  \left(\frac{z}{z_m}\right)^\frac{11}{6} \ + 3\,C_{12}  \left(\frac{z}{z_m}\right)^\frac{1}{2}\right]\right.\nonumber \\
&+& \left.\frac{1}{8}\,\xi^2\left[\frac{27}{5}\,C_{21} \left(\frac{z}{z_m}\right)^\frac{13}{6} \ -\ C_{22} \left(\frac{z}{z_m}\right)^\frac{3}{2}\right]\right\} \exp\left[\,-\,\frac{a \, z^4}{z_m{}^4\,-z^4}\right]  \nonumber \\
&+& \gamma_2 \left(1 \ - \ \frac{z}{z_m}\right)^{\frac{1}{2}}\exp\left[\,-\,\frac{a \, \left(z_m\,-\,z\right)^4}{z_m{}^4\,-\,\left(z_m\,-\,z\right)^4}\right]
\ .
\eea
We were forced to proceed to second order in $\xi$ to eliminate all singular contributions to the mean value of the Hamiltonian. As before, and for similar reasons, we worked with $\gamma_1=\gamma_2=0$, exploring values of $a$ of order one.

\begin{figure}[ht]
\centering
\begin{tabular}{cc}
\includegraphics[width=65mm]{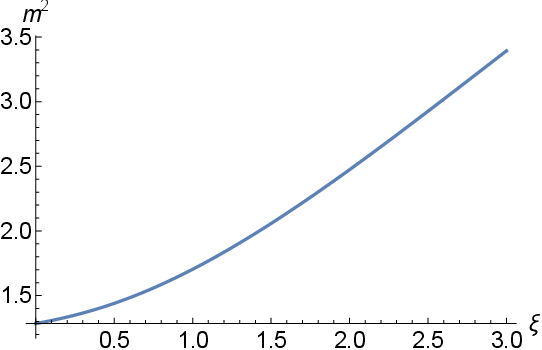} \qquad \qquad &
\includegraphics[width=65mm]{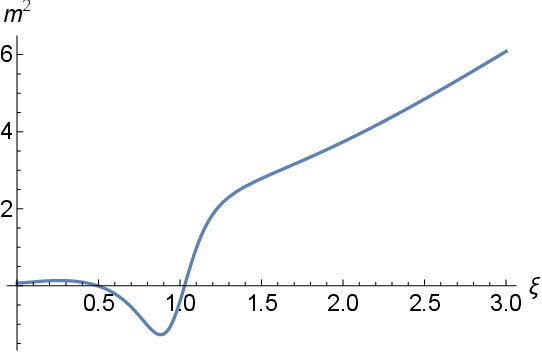} \\
\end{tabular}
\caption{\small Estimates for the lowest $m^2(\xi)$ (in units of $\frac{1}{z_0{}^2}$) for non--singlet scalar modes obtained for $(\theta_1,\theta_2,\rho)= \left(-\,\frac{\pi}{2},\frac{\pi}{2},0\right)$  (left panel), and for  $(\theta_1,\theta_2,\rho)= (1.2\,\pi,{0.3\,\pi},2)$ (right panel).}
\label{fig:scalarnonsinglet}
\end{figure}
In this sector we found again self--adjoint boundary conditions leading to no unstable modes for all values of $\xi_0$, as in the left panel of fig.~\ref{fig:scalarnonsinglet}, or for values of $\xi_0$ larger than a few units, as in the right panel of fig.~\ref{fig:scalarnonsinglet}. However, unstable boundary conditions were more difficult to find in this sector.

In all cases, our tests convey  useful information only for low--enough values of $\xi$, since the test functions contain higher--order corrections in $\xi$ that we left out. However, the formal arguments in Section~\ref{sec:intvectexc} indicate that the large--$\xi$ behavior leads to positive $m^2 \sim \left|\mathbf{k}\right|^2$, due to the diagonal terms in the Hamiltonians. 

\end{appendices}

 \end{document}